%% file: hats31-35.tex
\documentclass[apjl,iop]{emulateapj}
\usepackage{ifthen}
\usepackage{hyperref}
\usepackage{savesym}
\savesymbol{tablenum}
\usepackage{siunitx}
\restoresymbol{SIX}{tablenum}

\graphicspath{{img/}}

\DeclareSIUnit\msun{\ensuremath{\mathit{M_\odot}}}
\DeclareSIUnit\rsun{\ensuremath{\mathit{R_\odot}}}
\DeclareSIUnit\mjup{\ensuremath{\mathit{M_\mathrm{J}}}}
\DeclareSIUnit\rjup{\ensuremath{\mathit{R_\mathrm{J}}}}

\input{newcommand_gen.tex}

\newcommand{\loopcommanoperiod}{\ifnum\value{planetcounter}<2 ,\else \space\fi}

\input{newcommand_spec_multi.tex}

\newcounter{planetcounter}

\newboolean{emulateapj}
\setboolean{emulateapj}{true}

\newboolean{rvtablelong}
\setboolean{rvtablelong}{true}

\newlength{\plotwidthtwo}

\shortauthors{de Val-Borro et al.}
\shorttitle{\hatcur{31}\lowercase{b}--\hatcur{35}\lowercase{b}}

\begin{document}

\ifthenelse{\boolean{emulateapj}}{
    \newcommand{\titledag}{$\dagger$}
}{
    \newcommand{\titledag}{\dagger}
}

\title{
\hatcur{31}\lowercase{b} Through \hatcur{35}\lowercase{b}: Five
Transiting Hot Jupiters Discovered by the HATSouth
Survey\altaffilmark{\titledag}
}

\author{
    M.~de~Val-Borro\altaffilmark{1},
    G.~\'A.~Bakos\altaffilmark{1,$\star$,$\star\star$},
    R.~Brahm\altaffilmark{2,3},
    J.~D.~Hartman\altaffilmark{1},
    N.~Espinoza\altaffilmark{2,3},
    K.~Penev\altaffilmark{1},
    S.~Ciceri\altaffilmark{4,5},
    A.~Jord\'an\altaffilmark{2,3},
    W.~Bhatti\altaffilmark{1},
    Z.~Csubry\altaffilmark{1},
    D.~Bayliss\altaffilmark{6},
    J.~Bento\altaffilmark{7},
    G.~Zhou\altaffilmark{8},
    M.~Rabus\altaffilmark{2,4},
    L.~Mancini\altaffilmark{4},
    T.~Henning\altaffilmark{4},
    B.~Schmidt\altaffilmark{7},
    T.~G.~Tan\altaffilmark{9},
    C.~G.~Tinney\altaffilmark{10,11},
    D.~J.~Wright\altaffilmark{10,11},
    L.~Kedziora-Chudczer\altaffilmark{10,11},
    J.~Bailey\altaffilmark{10,11},
    V.~Suc\altaffilmark{2},
    S.~Durkan\altaffilmark{12},
    J.~L\'az\'ar\altaffilmark{13},
    I.~Papp\altaffilmark{13},
    P.~S\'ari\altaffilmark{13}
}

\altaffiltext{1}{Department of Astrophysical Sciences, Princeton
    University, NJ 08544, USA.}
\altaffiltext{2}{Instituto de Astrof\'isica, Facultad de F\'isica,
    Pontificia Universidad Cat\'olica de Chile, Av.\ Vicu\~na Mackenna
    4860, 7820436 Macul, Santiago, Chile}
\altaffiltext{3}{Millennium Institute of Astrophysics, Av.\ Vicu\~na Mackenna
    4860, 782-0436 Macul, Santiago, Chile.}
\altaffiltext{4}{Max Planck Institute for Astronomy, K\"{o}nigstuhl 17,
    69117 Heidelberg, Germany}
\altaffiltext{5}{Department of Astronomy, Stockholm University,
    10691 Stockholm, Sweden}
\altaffiltext{6}{Observatoire Astronomique de l'Universit\'e de Geneve, 51 ch.
  des Maillettes, 1290 Versoix, Switzerland.}
\altaffiltext{7}{Research School of Astronomy and Astrophysics, Australian
  National University, Canberra, ACT 2611, Australia.}
\altaffiltext{8}{Harvard-Smithsonian Center for Astrophysics, 60 Garden St.,
  Cambridge, MA 02138, USA.}
\altaffiltext{9}{Perth Exoplanet Survey Telescope, Perth, Australia.}
\altaffiltext{10}{Exoplanetary Science at UNSW, School of Physics,
  University of New South Wales, Sydney, NSW 2052, Australia.}
\altaffiltext{11}{Australian Centre for Astrobiology, 
  University of New South Wales, Sydney, NSW 2052, Australia.}
\altaffiltext{12}{Astrophysics Research Centre, School of Mathematics \&
Physics, Queen's University, Belfast BT7 1NN, UK}
\altaffiltext{13}{Hungarian Astronomical Association, 1451 Budapest, Hungary}
\altaffiltext{$\star$}{Alfred P.~Sloan Research Fellow}
\altaffiltext{$\star\star$}{Packard Fellow}
\altaffiltext{$\dagger$}{
The HATSouth network is operated by a collaboration consisting of
Princeton University (PU), the Max Planck Institute f\"ur Astronomie
(MPIA), the Australian National University (ANU), and the Pontificia
Universidad Cat\'olica de Chile (PUC).  The station at Las Campanas
Observatory (LCO) of the Carnegie Institute is operated by PU in
conjunction with PUC, the station at the High Energy Spectroscopic
Survey (H.E.S.S.) site is operated in conjunction with MPIA, and the
station at Siding Spring Observatory (SSO) is operated jointly with
ANU.
Based in part on data collected at Subaru Telescope, which is
operated by the National Astronomical Observatory of Japan. Based in
part on observations made with the MPG \SI{2.2}{\m}
and \textit{Euler} \SI{1.2}{\m} Telescopes at the ESO Observatory in La
Silla.
This paper uses observations obtained with
facilities of the Las Cumbres Observatory Global Telescope.
}

\begin{abstract}

\setcounter{footnote}{1}
We report the discovery of five new transiting hot Jupiter planets
discovered by the HATSouth survey, \hatcurb{31} through \hatcurb{35}.
These planets orbit moderately bright stars with $V$ magnitudes within
the range \hatcurCCtassmvshort{33}--\hatcurCCtassmvshort{32}\,mag while
the planets span a range of masses
\hatcurPPmshort{31}--\hatcurPPmshort{35}\,\mjup, and have somewhat
inflated radii between \hatcurPPrshort{33}--\hatcurPPrshort{31}\,\rjup.
These planets can be classified as typical hot Jupiters, with
\hatcurb{31} and \hatcurb{35} being moderately inflated gas giant
planets  with radii of \hatcurPPr{31}\,\rjup{} and
\hatcurPPr{35}\,\rjup, respectively, that can be used to constrain
inflation mechanisms.  All five systems present a higher Bayesian
evidence for a fixed circular orbit model than for an eccentric orbit.
The orbital periods range from \hatcurLCP{35}\,day for \hatcurb{35}) to
\hatcurLCP{31}\,day for \hatcurb{31}.
Additionally, \hatcurb{35} 
orbits a relatively young \hatcurISOspec{35}\
star with an age of \hatcurISOage{35}\,Gyr.  We discuss the analysis to
derive the properties of these systems and compare them in the context
of the sample of well characterized transiting hot Jupiters known to
date.

\setcounter{footnote}{0}
\end{abstract}

\keywords{
    planetary systems ---
    stars: individual (%
\hatcur{31},
\hatcurCCgsc{31}\loopcommanoperiod
\setcounter{planetcounter}{2}
\hatcur{32},
\hatcurCCgsc{32}\loopcommanoperiod
\setcounter{planetcounter}{3}
\hatcur{33},
\hatcurCCgsc{33}\loopcommanoperiod
\setcounter{planetcounter}{4}
\hatcur{34},
\hatcurCCgsc{34}\loopcommanoperiod
\setcounter{planetcounter}{5}
\hatcur{35},
\hatcurCCgsc{35}) ---
    techniques: spectroscopic, photometric
}

\section{Introduction}
\label{sec:introduction}





Planets that eclipse their host star during their orbit are key objects
for the study of exoplanetary systems.  The special geometry of
transiting extrasolar planets (TEPs) enables measurements of not only
the planet size but other important physical parameters, such as their
masses and densities, and the characterization of the alignment between
the orbital axis of a planet and the spin axis of its host star through
the Rossiter-McLaughlin effect.  The majority of well-characterized TEPs
have been discovered by wide-field photometric surveys, including Kepler
\citep{2010Sci...327..977B}, the Wide Angle Search for Planets
\citep[WASP;][]{2006PASP..118.1407P}, the Hungarian-made Automated
Telescope Network
\citep[HATNet;][]{2004PASP..116..266B,bakos:2013:hatsouth}, COnvection
ROtation and planetary Transits \citep[CoRoT;][]{2008A&A...482L..17B},
and the Kilodegree Extremely Little Telescope survey
\citep[KELT;][]{2012ApJ...761..123S}.

The known sample of exoplanets present a great diversity of orbital and
planetary parameters.  Extending the sample of close-orbiting TEPs
is a key goal of ground-based surveys as they allow for a large array of
additional observational measurements, such as information about the
chemical composition of the atmospheres of the planets using emission and
transmission spectroscopy for sufficiently bright targets.  The HATSouth
survey \citep{bakos:2013:hatsouth} has been designed to increase the
sample of well-characterized TEPs.  Some recent examples of planets
discovered by HATSouth are HATS-18b \citep{2016arXiv160600848P} and
HATS-25b through HATS-30b \citep{2016arXiv160600023E}.  A full list of
TEPs discovered by the HATSouth survey, along with all discovery and
follow-up light curves, can be found at
\url{http://hatsouth.org/}.

In this paper we present five new transiting planets
discovered by the HATSouth network around moderately bright stars:
\hatcurb{31} through \hatcurb{35}.
In Section~\ref{sec:obs} we describe the photometric transit detection
with HATSouth, as well as the data analysis methods and the procedures
used to confirm the planetary nature of the transit signal using
follow-up spectroscopic and photometric observations.  In
Section~\ref{sec:analysis} we describe the analysis carried out to rule
out false positive scenarios that could mimic a planetary signal, and to
ascertain the stellar and planetary parameters.  We discuss the
implication of our results and compare them with all known transiting hot Jupiters
to date in Section~\ref{sec:discussion}.

\section{Observations}
\label{sec:obs}

\subsection{Photometric detection}
\label{sec:detection}

The HATSouth survey is a global network of homogeneous, completely
automated wide-field telescopes located at three sites in the Southern
Hemisphere: Las Campanas Observatory (LCO) in Chile, the High Energy
Stereoscopic Survey (H.E.S.S.) site in Namibia, and Siding Spring
Observatory (SSO) in Australia.  Observations are performed using a
Sloan-$r$ filter with four-minute exposures.  The HATSouth network was
commissioned in 2009 and  since then has proved to be a robust system
for the monitoring of time-variable phenomena.  Each HATSouth unit
consists of four Takahashi E180 astrographs with an aperture of
\SI{18}{\cm} and f/2.8 focal ratio on a common mount, equipped with
Apogee \num{4096 x 4096} U16M ALTA cameras.  The observations and
aperture photometry reduction pipeline used by the HATSouth survey have
been described comprehensively in \citet{bakos:2013:hatsouth} and
\citet{penev:2013:hats1}.

Below, we describe specific details of the observations leading to the
discovery of \hatcurb{31}, \hatcurb{32}, \hatcurb{33}, \hatcurb{34} and
\hatcurb{35}.  The HATSouth raw data were reduced to trend-filtered light
curves using the External Parameter Decorrelation method
\citep[EPD;][]{bakos:2010:hat11} and the Trend Filtering Algorithm
\citep[TFA;][]{kovacs:2005:TFA} to correct for systematic variations in
the photometry before searching for transit signals.  We searched
the light curves for periodic box-shaped signals using the Box-fitting Least-Squares
\citep[BLS;][]{kovacs:2002:BLS} algorithm, and detected periodic transit
signals in the light curves as shown in Figure~\ref{fig:hatsouth}.  The reduced
data are available in Table~\ref{tab:phfu}.
We summarize below the transits detected in the light curves of the stars
\hatcur{31} through \hatcur{35}.

\begin{itemize}
\item \emph{\hatcur{31}} (\hatcurCCtwomass{31}; $\alpha =
  \hatcurCCra{31}$, $\delta = \hatcurCCdec{31}$; J2000; V=\hatcurCCtassmv{31}).
  A signal was detected with an apparent depth of
  $\sim$\hatcurLCdip{31}\,mmag at a period of $P=$\hatcurLCPshort{31}\,day.
\item \emph{\hatcur{32}} (\hatcurCCtwomass{32}; $\alpha =
  \hatcurCCra{32}$, $\delta = \hatcurCCdec{32}$; J2000; V=\hatcurCCtassmv{32}).
  A signal was detected with an apparent depth of  $\sim$\hatcurLCdip{32}\,mmag at a period of $P=$\hatcurLCPshort{32}\,day.

\item \emph{\hatcur{33}} (\hatcurCCtwomass{33}; $\alpha =
  \hatcurCCra{33}$, $\delta = \hatcurCCdec{33}$; J2000; V=\hatcurCCtassmv{33}).
  A signal was detected with an apparent depth of $\sim$\hatcurLCdip{33}\,mmag at a period of $P=$\hatcurLCPshort{33}\,day.

\item \emph{\hatcur{34}} (\hatcurCCtwomass{34}; $\alpha =
  \hatcurCCra{34}$, $\delta = \hatcurCCdec{34}$; J2000;
  V=\hatcurCCtassmv{34}).
  A signal was detected with an apparent depth of $\sim$\hatcurLCdip{34}\,mmag at a period of $P=$\hatcurLCPshort{34}\,day. 

\item \emph{\hatcur{35}} (\hatcurCCtwomass{35}; $\alpha =
  \hatcurCCra{35}$, $\delta = \hatcurCCdec{35}$; J2000; V=\hatcurCCtassmv{35}).
  A signal was detected with an apparent depth of $\sim$\hatcurLCdip{35}\,mmag at a period of $P=$\hatcurLCPshort{35}\,day.
  
\end{itemize}

Subsequent spectroscopic and photometric follow-up observations
for the five systems were carried out to confirm the transit signal
and the planetary nature of these objects as described in the
following sections.

\ifthenelse{\boolean{emulateapj}}{
    \begin{figure*}[!ht]
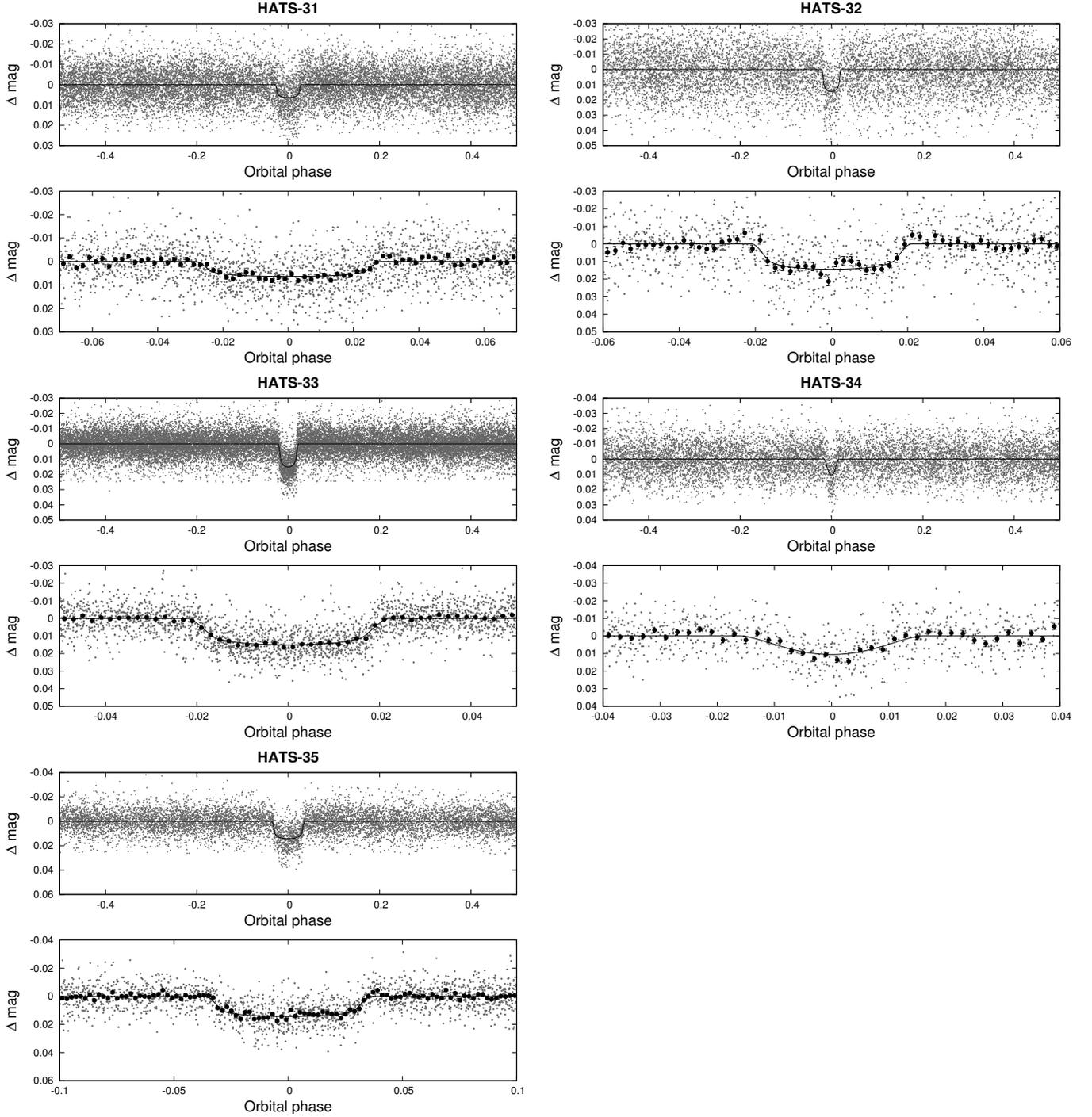

}{
    \begin{figure}[!ht]
}
\setlength{\plotwidthtwo}{0.495\linewidth}
\includegraphics[width={\plotwidthtwo}]{\hatcurhtr{31}-hs}
\hfil
\includegraphics[width={\plotwidthtwo}]{\hatcurhtr{32}-hs}
\includegraphics[width={\plotwidthtwo}]{\hatcurhtr{33}-hs}
\hfil
\includegraphics[width={\plotwidthtwo}]{\hatcurhtr{34}-hs}
\includegraphics[width={\plotwidthtwo}]{\hatcurhtr{35}-hs}
\caption[]{
    Phase-folded unbinned HATSouth light curves for the five new transiting
    planet systems. In each case we show two panels. The top panel shows the
    full light curve, while the bottom panel shows the light curve zoomed-in on
    the transit. The solid lines show the model fits to the light curves. The
    dark filled circles in the bottom panels show the light curves binned in
    phase with a bin size of 0.002.
\label{fig:hatsouth}}
\ifthenelse{\boolean{emulateapj}}{
    \end{figure*}
}{
    \end{figure}
}

\ifthenelse{\boolean{emulateapj}}{
    \begin{deluxetable*}{llrrrr}
}{
    \begin{deluxetable}{llrrrr}
}
\tablewidth{0pc}
\tabletypesize{\scriptsize}
\tablecaption{
    Summary of photometric observations
    \label{tab:photobs}
}
\tablehead{
    \multicolumn{1}{c}{Instrument/Field\tablenotemark{a}} &
    \multicolumn{1}{c}{Date(s)} &
    \multicolumn{1}{c}{\# Images} &
    \multicolumn{1}{c}{Cadence\tablenotemark{b}} &
    \multicolumn{1}{c}{Filter} &
    \multicolumn{1}{c}{Precision\tablenotemark{c}} \\
    \multicolumn{1}{c}{} &
    \multicolumn{1}{c}{} &
    \multicolumn{1}{c}{} &
    \multicolumn{1}{c}{(sec)} &
    \multicolumn{1}{c}{} &
    \multicolumn{1}{c}{(mmag)}
}
\startdata
\sidehead{\textbf{\hatcur{31}}}
~~~~HS-1.4/G565              & 2012 Dec--2013 Jun & 5750 & 282 & $r$ & 8.0 \\
~~~~HS-3.4/G565              & 2012 Dec--2013 Jul & 3850 & 280 & $r$ & 8.1 \\
~~~~HS-5.4/G565              & 2012 Dec--2013 Jul & 5187 & 287 & $r$ & 7.5 \\
~~~~LCOGT~1\,m+CTIO/sinistro & 2015 Feb 28        & 53   & 226 & $i$ & 1.1 \\
~~~~LCOGT~1\,m+SAAO/SBIG     & 2015 Mar 06        & 137  & 139 & $i$ & 3.6 \\
~~~~Swope~1\,m/e2v           & 2015 Apr 02        & 281  & 54  & $i$ & 4.0 \\
\sidehead{\textbf{\hatcur{32}}}
~~~~HS-2.3/G586      & 2010 Aug--2011 Nov & 4159 & 290 & $r$     & 14.9 \\
~~~~HS-4.3/G586      & 2010 Aug--2011 Nov & 4510 & 298 & $r$     & 14.1 \\
~~~~HS-6.3/G586      & 2010 Aug--2011 Nov & 498  & 293 & $r$     & 14.4 \\
~~~~PEST~0.3\,m      & 2014 Jul 09        & 145  & 133 & $R_{C}$ & 10.5 \\
~~~~DK~1.54\,m/DFOSC & 2014 Nov 04        & 120  & 145 & $R$     & 2.0  \\
~~~~Swope~1\,m/e2v   & 2015 May 28        & 53   & 189 & $i$     & 6.6  \\
\sidehead{\textbf{\hatcur{33}}}
~~~~HS-1.4/G747              & 2013 Mar--2013 Oct & 4271 & 287 & $r$ & 6.4 \\
~~~~HS-2.4/G747              & 2013 Sep--2013 Oct & 1280 & 287 & $r$ & 9.6 \\
~~~~HS-3.4/G747              & 2013 Apr--2013 Nov & 8813 & 297 & $r$ & 7.8 \\
~~~~HS-4.4/G747              & 2013 Sep--2013 Nov & 1531 & 297 & $r$ & 8.7 \\
~~~~HS-5.4/G747              & 2013 Mar--2013 Nov & 6049 & 297 & $r$ & 5.9 \\
~~~~HS-6.4/G747              & 2013 Sep--2013 Nov & 1557 & 290 & $r$ & 9.0 \\
~~~~LCOGT~1\,m+CTIO/sinistro & 2015 May 20        & 128  & 48  & $i$ & 4.7 \\
\sidehead{\textbf{\hatcur{34}}}
~~~~HS-2.4/G754 & 2012 Sep--2012 Dec & 3805 & 282 & $r$ & 9.1 \\
~~~~HS-4.4/G754 & 2012 Sep--2013 Jan & 2865 & 292 & $r$ & 10.0 \\
~~~~HS-6.4/G754 & 2012 Sep--2012 Dec & 2975 & 285 & $r$ & 9.9 \\
~~~~PEST~0.3\,m & 2014 Oct 26 & 54 & 211 & $R_{C}$ & 5.3 \\
~~~~DK~1.54\,m/DFOSC & 2014 Nov 03 & 97 & 125 & $R$ & 1.3 \\
~~~~AAT~3.9\,m/IRIS2 \tablenotemark{d} & 2015 Sep 25 & 715 & 10 & $K_{S}$ & 8.1 \\
\sidehead{\textbf{\hatcur{35}}}
~~~~HS-2.4/G778 & 2011 May--2012 Nov & 3013 & 287 & $r$ & 9.6 \\
~~~~HS-4.4/G778 & 2011 Jul--2012 Nov & 3699 & 298 & $r$ & 7.5 \\
~~~~HS-6.4/G778 & 2011 Apr--2012 Oct & 2294 & 298 & $r$ & 8.5 \\
~~~~LCOGT~1\,m+CTIO/sinistro & 2015 Jun 12 & 38 & 163 & $i$ & 1.2 \\
~~~~LCOGT~1\,m+SAAO/SBIG & 2015 Jul 14 & 19 & 144 & $i$ & 0.8 \\
~~~~LCOGT~1\,m+CTIO/sinistro & 2015 Jul 15 & 79 & 162 & $i$ & 1.1 \\
~~~~LCOGT~1\,m+SSO/SBIG & 2015 Jul 18 & 106 & 133 & $i$ & 1.6 \\
~~~~LCOGT~1\,m+CTIO/sinistro & 2015 Jul 24 & 105 & 162 & $i$ & 0.8 \\
\enddata
\tablenotetext{a}{
    For HATSouth data we list the HATSouth unit, CCD and field name
    from which the observations are taken. HS-1 and -2 are located at
    Las Campanas Observatory in Chile, HS-3 and -4 are located at the
    H.E.S.S. site in Namibia, and HS-5 and -6 are located at Siding
    Spring Observatory in Australia. Each unit has 4 ccds. Each field
    corresponds to one of 838 fixed pointings used to cover the full
    4$\pi$ celestial sphere. All data from a given HATSouth field and
    CCD number are reduced together, while detrending through External
    Parameter Decorrelation (EPD) is done independently for each
    unique unit+CCD+field combination.
}
\tablenotetext{b}{
    The median time between consecutive images rounded to the nearest
    second. Due to factors such as weather, the day--night cycle,
    guiding and focus corrections the cadence is only approximately
    uniform over short timescales.
}
\tablenotetext{c}{
    The RMS of the residuals from the best-fit model.
}
\tablenotetext{d}{
    This light curve covers a predicted secondary eclipse event, it is not included in the analysis carried out to determine the system parameters for \hatcur{34}, however it is included in the analysis carried out to exclude blend scenarios.
}
\ifthenelse{\boolean{emulateapj}}{
    \end{deluxetable*}
}{
    \end{deluxetable}
}

\subsection{Spectroscopic Observations}
\label{sec:obsspec}

In Table~\ref{tab:specobs} we summarize all  spectroscopic
observations taken for \hatcur{31} to \hatcur{35}.

\subsubsection{Reconnaissance Spectroscopy}
\label{sec:reconspec}

To exclude stellar binary false positives and confirm planetary
candidates detected by the HATSouth network, we carry out initial low-
and medium-resolution reconnaissance spectroscopy before
attempting higher precision observations to determine  orbital
parameters.  These reconnaissance observations consist of
spectral typing observations of all the objects using the Wide Field
Spectrograph (WiFeS) on the ANU 2.3m telescope at SSO.  The observing
strategy and data reduction procedure for WiFeS data are described in
detail in \citet{bayliss:2013:hats3}.  The number of medium- and 
low-resolution spectra obtained for each system are summarized in
Table~\ref{tab:specobs}.  \hatcur{31} through \hatcur{35} were
confirmed as single-lined stars by these WiFeS observations.  Using these
 low-resolution spectra we obtained approximate stellar atmospheric
parameters that indicate  \hatcur{31} is an F-type star, while
\hatcur{32} through \hatcur{35} are G-type stars. Medium-resolution
WiFeS observations with spectral resolution $R = \lambda/\Delta \lambda
= 7000$ are then used to rule out possible eclipsing stellar companions
in any of these systems by measuring no radial velocity variations in excess of
$\sim$ \SI{5}{\km\per\s}.

\subsubsection{High-Resolution Spectroscopy}
\label{sec:highrspec}

Following reconnaissance spectroscopy 
to reject possible false positives like blended binary systems and 
obtain first estimates of stellar parameters, stable and high-precision
spectroscopic measurements are obtained to collect high precision radial
velocity (RV) variations and line bisector (BS) time series for each of
the candidates.  Several high resolution spectra were acquired for these
objects with a combination of the FEROS \citep{1998SPIE.3355..844K},
HARPS \citep{2003Msngr.114...20M}, Coralie \citep{2001Msngr.105....1Q}
and CYCLOPS2+UCLES spectrographs \citep{2012SPIE.8446E..3AH}
between July 2014 to July 2015.

Altogether we obtained 11 spectra using CYCLOPS2+UCLES at the 3.9\,m Anglo-Australian Telescope (AAT), 11 spectra using HARPS at the ESO~3.6\,m telescope, 
10 spectra using CORALIE  at the Euler~1.2\,m telescope and 32 spectra with
FEROS at the MPG~2.2\,m telescope.  The data from the FEROS, HARPS and
Coralie instruments were reduced homogeneously with an automated
pipeline for echelle spectrographs described in detail in
\citet{jordan:2014:hats4}. The CYCLOPS2 observations were reduced and
analyzed following \citet{2013ApJ...774L...9A}.  Combined high-precision
RV and BS measurements are shown for each system folded
with the period of the transit signal in Figure~\ref{fig:rvbis}.  Note
that BS measurements from CYCLOPS2 for \hatcur{33} are missing due to
not having a BS pipeline for this instrument.  The high-resolution
spectroscopic data are provided in Table~\ref{tab:rvs} at the end of the
paper.

All the candidates show clear sinusoidal variation in RV that are
in phase with the observed transits.  From these observations we 
estimate  orbital parameters, as well as confirm the mass of the
companion for systems that host planets, and  measure precisely the
stellar atmospheric parameters.

\ifthenelse{\boolean{emulateapj}}{
    \begin{deluxetable*}{llrrrrr}
}{
    \begin{deluxetable}{llrrrrrrrr}
}
\tablewidth{0pc}
\tabletypesize{\scriptsize}
\tablecaption{
    Summary of spectroscopy observations
    \label{tab:specobs}
}
\tablehead{
    \multicolumn{1}{c}{Instrument}                         &
    \multicolumn{1}{c}{UT Date(s)}                         &
    \multicolumn{1}{c}{\# Spec.}                           &
    \multicolumn{1}{c}{Res.}                               &
    \multicolumn{1}{c}{S/N Range\tablenotemark{a}}         &
    \multicolumn{1}{c}{$\gamma_{\rm RV}$\tablenotemark{b}} &
    \multicolumn{1}{c}{RV Precision\tablenotemark{c}}      \\
                                                           &
                                                           &
                                                           &
    \multicolumn{1}{c}{$\Delta \lambda$/$\lambda$/1000}    &
                                                           &
    \multicolumn{1}{c}{(\kms)}                             &
    \multicolumn{1}{c}{(\ms)}
}
\startdata
\sidehead{\textbf{\hatcur{31}}}
ANU~2.3\,m/WiFeS & 2014 Dec 30--31 & 2 & 7   & 45--56 & -9.3     & 4000     \\
ANU~2.3\,m/WiFeS & 2015 Jan 1      & 1 & 3   & 77     & $\cdots$ & $\cdots$ \\
ESO~3.6\,m/HARPS & 2015 Feb 14--19 & 6 & 115 & 12--22 & -8.705   & 17       \\
\sidehead{\textbf{\hatcur{32}}}
ANU~2.3\,m/WiFeS & 2014 Jun 3--5      & 3 & 7  & 19--61 & 9.1      & 4000     \\
ANU~2.3\,m/WiFeS & 2014 Jun 4         & 1 & 3  & 69     & $\cdots$ & $\cdots$ \\
MPG~2.2\,m/FEROS & 2014 Jul--2015 Jun & 8 & 48 & 20--45 & 12.423   & 30       \\
\sidehead{\textbf{\hatcur{33}}}
ANU~2.3\,m/WiFeS     & 2014 Dec--2015 Mar & 4  & 7   & 51--71 & 11.420   & 4000     \\
ANU~2.3\,m/WiFeS     & 2015 Mar 4         & 1  & 3   & 44     & $\cdots$ & $\cdots$ \\
Euler~1.2\,m/Coralie & 2015 Mar--Jun      & 5  & 60  & 18--25 & 11.056   & 42       \\
ESO~3.6\,m/HARPS     & 2015 Apr 6--8      & 3  & 115 & 19--27 & 11.077   & 2        \\
AAT~3.9\,m/CYCLOPS2   & 2015 May 7--13     & 11 & 70  & 17--44 & 11.066   & 23       \\
MPG~2.2\,m/FEROS     & 2015 May--Jul      & 4  & 48  & 40--73 & 11.057   & 38       \\
\sidehead{\textbf{\hatcur{34}}}
ANU~2.3\,m/WiFeS & 2014 Oct 4 & 1 & 3 & 50 & $\cdots$ & $\cdots$ \\
ANU~2.3\,m/WiFeS & 2014 Oct 4--10 & 3 & 7 & 55--95 & 16.4 & 4000 \\
MPG~2.2\,m/FEROS\tablenotemark{d} & 2015 Jun--Jul & 10 & 48 & 16--43 & 17.734 & 23 \\
\sidehead{\textbf{\hatcur{35}}}
ANU~2.3\,m/WiFeS     & 2014 Oct 5         & 1  & 3   & 99     & $\cdots$ & $\cdots$ \\
ANU~2.3\,m/WiFeS     & 2014 Oct 11        & 1  & 7   & 102    & -14.3    & 4000     \\
Euler~1.2\,m/Coralie & 2014 Nov--2015 Jun & 5  & 60  & 16--21 & -14.173  & 110      \\
ESO~3.6\,m/HARPS     & 2015 Apr 7--8      & 2  & 115 & 15--23 & -14.245  & 42       \\
MPG~2.2\,m/FEROS     & 2015 Jun--Jul      & 10 & 48  & 41--63 & -14.185  & 20       \\
\enddata
\tablenotetext{a}{
    S/N per resolution element near 5180\,\AA.
}
\tablenotetext{b}{
    For high-precision radial velocity observations included in the orbit
    determination this is the zero-point velocity from the best-fit
    orbit. For other instruments it is the mean value. We do not
    provide this quantity for the lower resolution WiFeS observations
    which were only used to measure stellar atmospheric parameters.
}
\tablenotetext{c}{
    For high-precision radial velocity observations included in the orbit
    determination this is the scatter in the residuals from the
    best-fit orbit (which may include astrophysical jitter), for other
    instruments this is either an estimate of the precision (not
    including jitter), or the measured standard deviation. We do not
    provide this quantity for low-resolution observations from the
    ANU~2.3\,m/WiFeS.
}
\tablenotetext{d}{
    Three of the MPG~2.2\,m/FEROS observations of \hatcur{34} were
    excluded from the analysis due to having low S/N or high sky
    contamination.
}
\ifthenelse{\boolean{emulateapj}}{
    \end{deluxetable*}
}{
    \end{deluxetable}
}

\setcounter{planetcounter}{1}
\ifthenelse{\boolean{emulateapj}}{
    \begin{figure*} [ht]
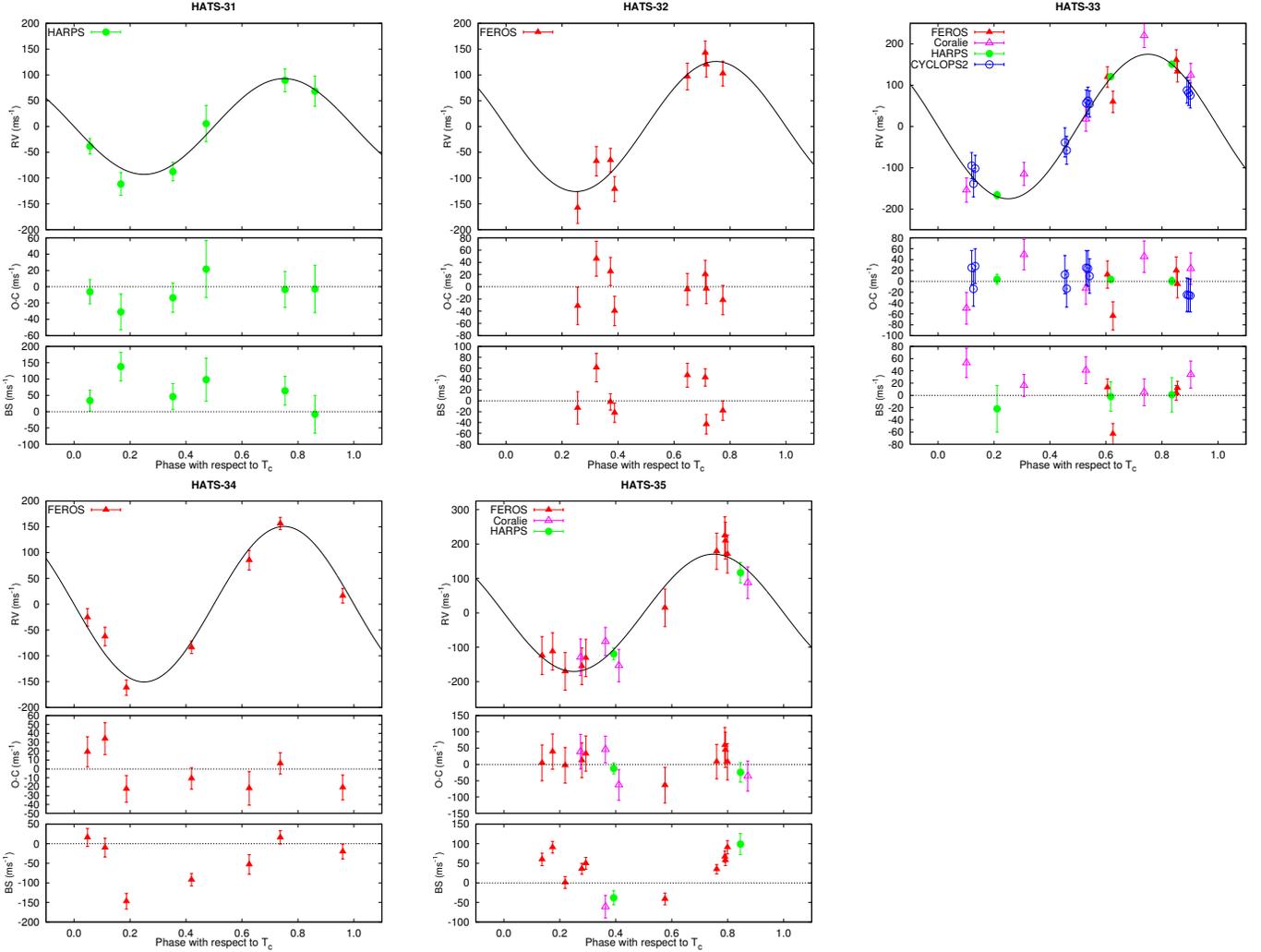

}{
    \begin{figure}[ht]
}
{
\centering
\setlength{\plotwidthtwo}{0.31\linewidth}
\includegraphics[width={\plotwidthtwo}]{\hatcurhtr{31}-rv}
\hfil
\includegraphics[width={\plotwidthtwo}]{\hatcurhtr{32}-rv}
\hfil
\includegraphics[width={\plotwidthtwo}]{\hatcurhtr{33}-rv}
}
{
\centering
\setlength{\plotwidthtwo}{0.31\linewidth}
\includegraphics[width={\plotwidthtwo}]{\hatcurhtr{34}-rv}
\hspace{4mm}
\includegraphics[width={\plotwidthtwo}]{\hatcurhtr{35}-rv}
}
\caption{
    Phased high-precision radial velocity measurements for the five new transiting
    planet systems. The instruments used are labelled in the plots. In
    each case we show three panels. The top panel shows the phased
    measurements together with our best-fit circular-orbit model for
    \hatcur{31} through \hatcur{35}
    (see \reftabl{planetparam}). Zero-phase corresponds to the time of
    mid-transit. The center-of-mass velocity has been subtracted. The
    second panel shows the velocity $O\!-\!C$ residuals from the best
    fit. The error bars include the jitter terms listed in
    Tables~\ref{tab:planetparam}~and~\ref{tab:planetparamtwo} added in
    quadrature to the formal errors for each instrument. The third panel
    shows the bisector spans. Note the different vertical scales of the
    panels.
}
\label{fig:rvbis}
\ifthenelse{\boolean{emulateapj}}{
    \end{figure*}
}{
    \end{figure}
}

\subsection{Photometric follow-up observations}
\label{sec:phot}

\begin{figure*}[!ht]
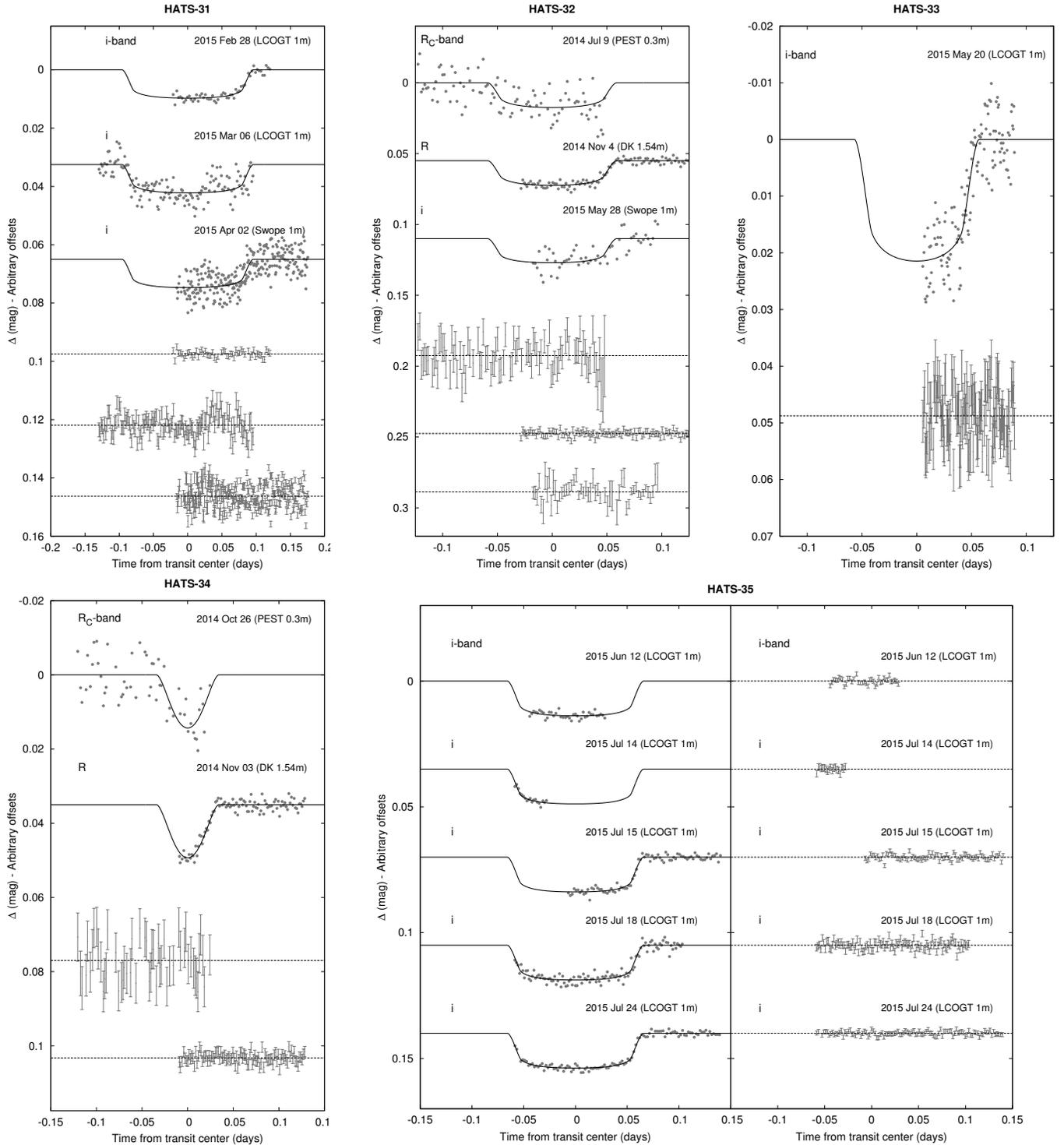

{
\centering
\setlength{\plotwidthtwo}{0.31\linewidth}
\includegraphics[width={\plotwidthtwo}]{\hatcurhtr{31}-lc}
\hfil
\includegraphics[width={\plotwidthtwo}]{\hatcurhtr{32}-lc}
\hfil
\includegraphics[width={\plotwidthtwo}]{\hatcurhtr{33}-lc}
}
{
\centering
\setlength{\plotwidthtwo}{0.31\linewidth}
\includegraphics[width={\plotwidthtwo}]{\hatcurhtr{34}-lc}
\hfil
\setlength{\plotwidthtwo}{0.62\linewidth}
\includegraphics[width={\plotwidthtwo}]{\hatcurhtr{35}-lc}
}
\caption{
    Unbinned transit light curves for \hatcur{31} through \hatcur{35}.  The
    light curves have been corrected for quadratic trends in time
    fitted simultaneously with the transit model, and for correlations
    with up to three parameters describing the shape of the PSF.
    The dates of the events, filters and instruments used are
    indicated.  Light curves following the first are displaced
    vertically for clarity.  Our best fit from the global modeling
    described in \refsecl{globmod} is shown by the solid lines. For
    \hatcur{31} through \hatcur{34} the residuals from the best-fit
    model are shown below in the same order as the original light
    curves, for \hatcur{35} the residuals are shown to the right of
    the light curves.  The error bars represent the photon and
    background shot noise, plus the readout noise. Note the differing
    vertical and horizontal scales used for each system.
}
\label{fig:lc}
\end{figure*}

To obtain higher precision light curves of the transit event,
we photometrically followed-up all the planets using
facilities with larger apertures than the HATSouth telescopes.
Photometric follow-up observations are summarized in
Table~\ref{tab:photobs}, including the cadence, filter and photometric
precision, and plotted in Figure~\ref{fig:lc}.  For all
objects the follow-up light curves were consistent with the discovery
observations.  These observations allow us to refine the transit
ephemeris of the systems and their physical parameters.

The egress of \hatcurb{31} was observed on 2015 February 28 and 2015
April 02 with the Las Cumbres Observatory Global Telescope (LCOGT)
1\,m telescope network \citep{brown:2013:lcogt} and the Swope 1\,m telescopes, respectively. Additionally, an
almost full transit of \hatcurb{31} was observed with LCOGT on 2015
March 6.  Another three partial transits of \hatcurb{32} were observed
with the PEST~0.3\,m, DK~1.54\,m and the Swope~1\,m telescopes.  The
egress of \hatcur{33} was measured with the 1\,m LCOGT at CTIO on 2015
May 20.  Both ingress and egress of \hatcurb{34} were observed by the
PEST~0.3\,m and DK~1.54\,m telescopes. Finally, five partial transit
events of \hatcurb{35} were obtained between 2015 June 12 and 2015 Jul
24 using the LCOGT network at CTIO, SAAO and SSO.  The data analysis
procedure of these photometric observations has been described
comprehensively in previous papers of HATSouth planet discoveries
\citep[see
e.g.,][]{2015AJ....149..166H,2015AJ....150...33B,2015A&A...580A..63M}.

We also monitored \hatcur{34} in the infrared $K_{S}$-band during the time of predicted secondary eclipse using the AAT+IRIS2.  Observations and data reduction were carried out in the manner described in \citet{2015MNRAS.454.3002Z}.  Details of this observation are set out in Table~\ref{tab:photobs}, and the observations are used to help rule out blend scenarios in Section~\ref{sec:blend}.

\subsection{Lucky imaging observations}
\label{sec:luckyimaging}

\begin{figure*}
  \centering
  \begin{tabular}{cc}
    \includegraphics[width=\columnwidth]{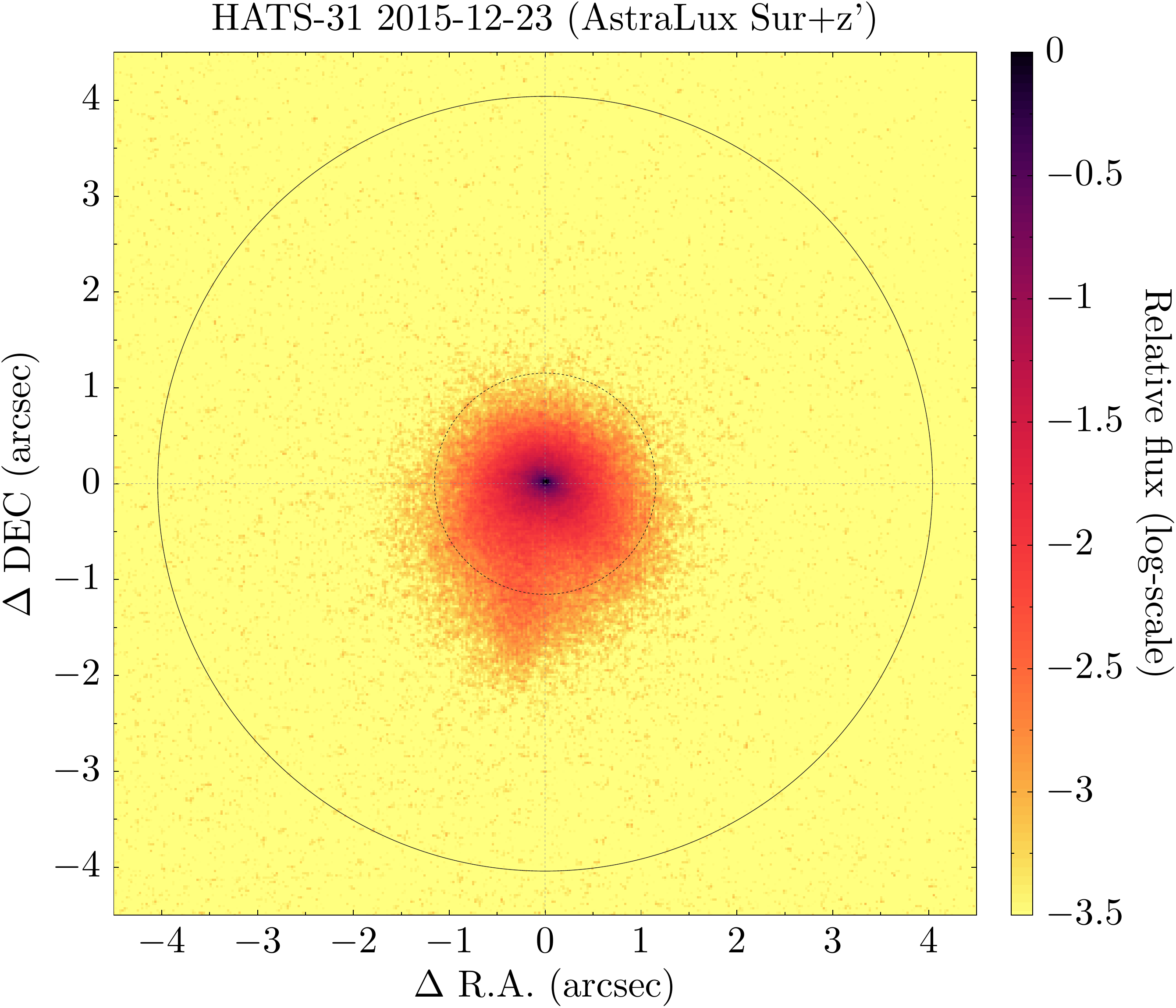} &
    \includegraphics[width=\columnwidth]{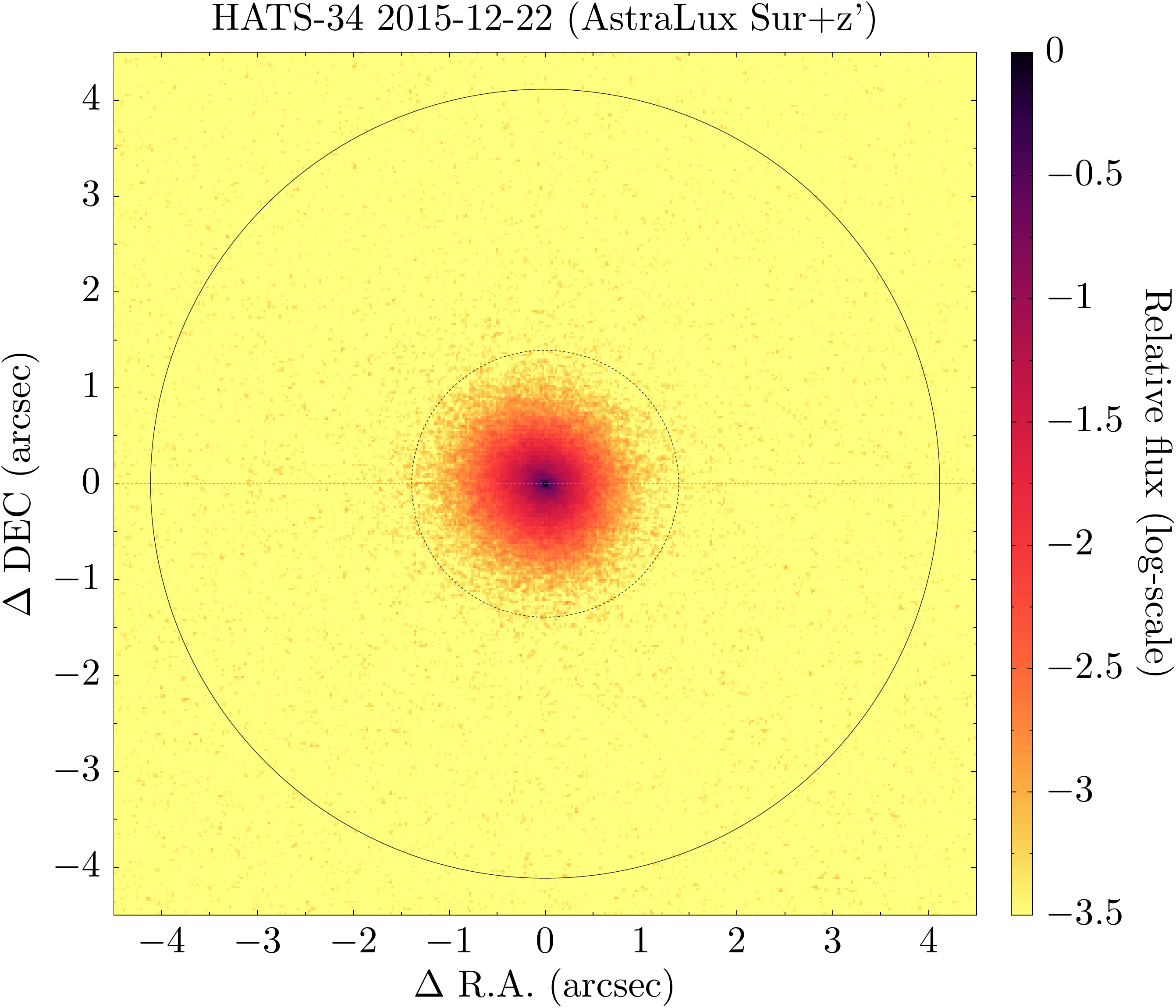}
  \end{tabular}
  \caption{Sloan $z'$-band images for \hatcur{31} and \hatcur{34}
    obtained with the AstraLux Sur camera. Circles of $1''$ radius
    and $5''$ radius are shown for reference on the images. Note the
    difference in the shape of the PSF.
}
\label{fig:astralux}
\end{figure*}

High-spatial-resolution (or ``lucky'') imaging observations of \hatcur{31}
and \hatcur{34} candidates were obtained using the Astralux Sur camera on the
New Technology Telescope (NTT) at La Silla Observatory \citep{2009Msngr.137...14H}. 
Data were reduced and contrast curves generated as described in \citet{2016arXiv160600023E}.
We show the resulting combination of the best $10\%$ of the images
acquired for each target for \hatcur{31} and \hatcur{34} in
Figure~\ref{fig:astralux}.  The resulting images show an asymmetric
extended profile for \hatcur{31} that is visible in all the Astralux
images.  This object was observed during twilight and the observations
were obtained out of focus.  The profile is more symmetric for
\hatcur{34}.

\begin{figure*}
  \centering
  \begin{tabular}{cc}
    \includegraphics[width=\columnwidth]{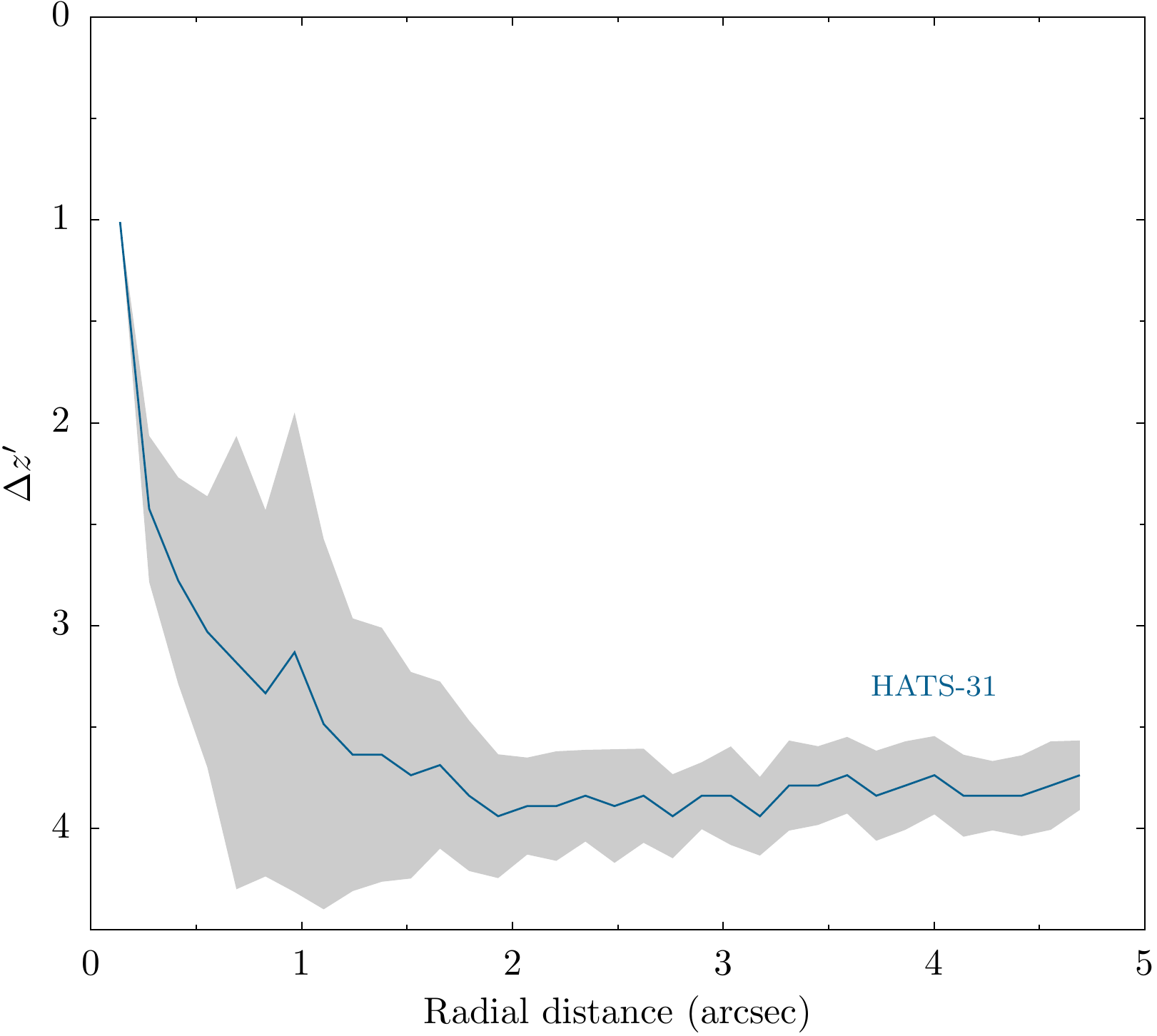} &
    \includegraphics[width=\columnwidth]{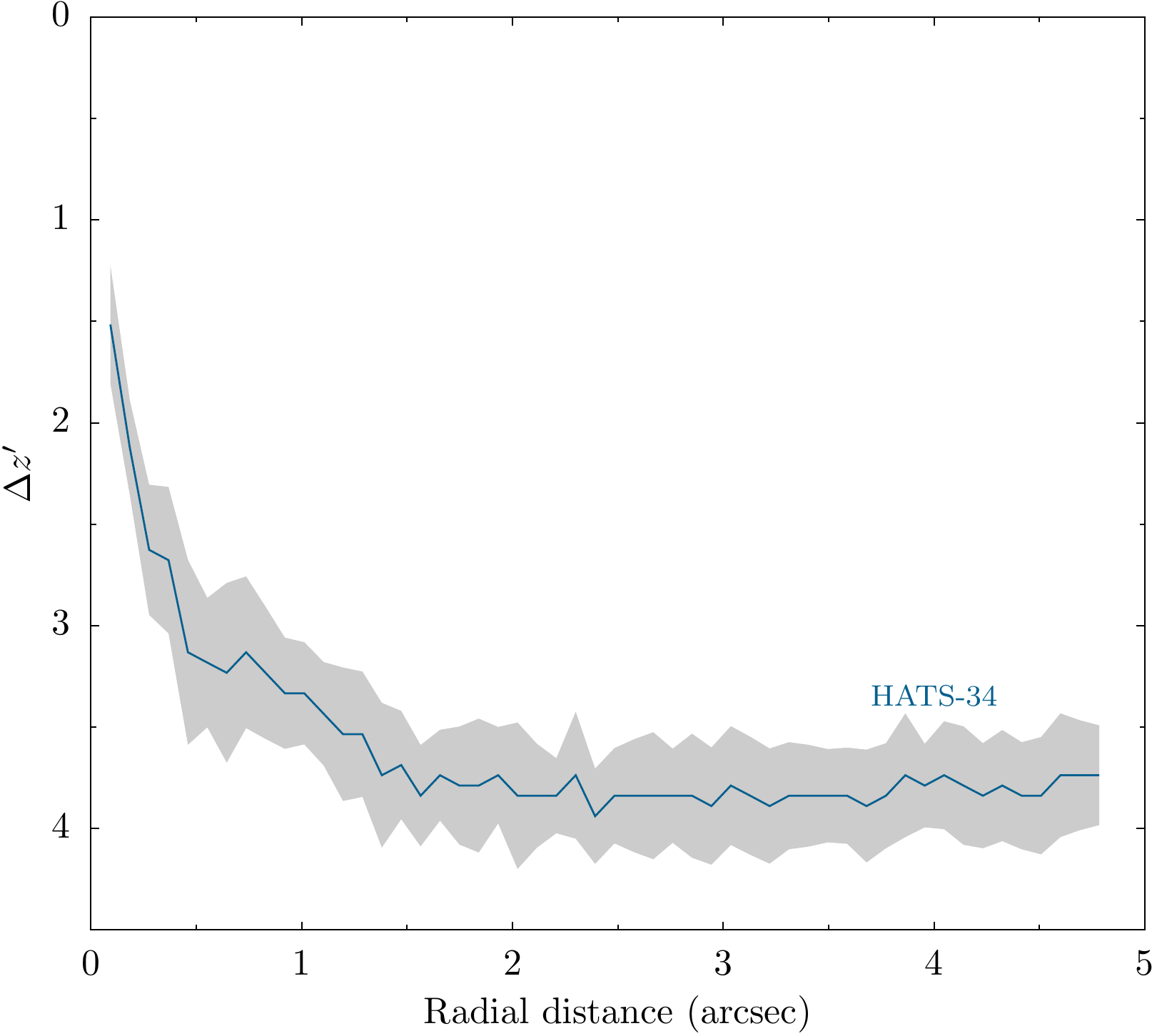}
  \end{tabular}
  \caption{Contrast curves for HATS-31 (left panel) and HATS-34 (right
    panel) based on observations with the Astralux Sur camera using the
    Sloan $z'$ filter shown in Figure~\ref{fig:astralux}.  The gray bar
    shows the 1-sigma uncertainty of the contrast at each radius.
}
\label{fig:contrast}
\end{figure*}

In Figure~\ref{fig:contrast} we show the generated $5-\sigma$ contrast
curves for \hatcur{31} and \hatcur{34}.  We simulate the Point Spread
Function (PSF) for our targets as a weighted sum of a Moffat profile
and an asymmetric Gaussian following the model description in
\citet{2016arXiv160600023E}.  The effective full width at half maximum
(FWHM) of this model was measured numerically at different angles by
finding the points at which the model has half of the peak flux. The
median of these measurements is taken as the resolution limit of our
observations.  For \hatcur{31}, the effective FWHM is $6.58 \pm 0.36$
pixels, which corresponds to a resolution limit of $151.4 \pm 8.3$
milli-arcseconds (mas).  In the case of \hatcur{34} the effective FWHM
is $4.17 \pm 0.33$ pixels, which gives a resolution limit of $96.0 \pm
7.5$ mas.

\clearpage

\ifthenelse{\boolean{emulateapj}}{
    \begin{deluxetable*}{llrrrrl}
}{
    \begin{deluxetable}{llrrrrl}
}
\tablewidth{0pc}
\tablecaption{
    Light curve data for \hatcur{31}--\hatcur{35}\label{tab:phfu}.
}
\tablehead{
    \colhead{Object\tablenotemark{a}} &
    \colhead{BJD\tablenotemark{b}} &
    \colhead{Mag\tablenotemark{c}} &
    \colhead{\ensuremath{\sigma_{\rm Mag}}} &
    \colhead{Mag(orig)\tablenotemark{d}} &
    \colhead{Filter} &
    \colhead{Instrument} \\
    \colhead{} &
    \colhead{\hbox{~~~~(2,400,000$+$)~~~~}} &
    \colhead{} &
    \colhead{} &
    \colhead{} &
    \colhead{} &
    \colhead{}
}
\startdata
\input{data/phfu_tab_combined_short.tex}
\enddata
\tablenotetext{a}{
    Either \hatcur{31}, \hatcur{32}, \hatcur{33}, \hatcur{34}, or \hatcur{35}.
}
\tablenotetext{b}{
    Barycentric Julian Date is computed directly from the UTC time
    without correction for leap seconds.
}
\tablenotetext{c}{
    The out-of-transit level has been subtracted. For observations
    made with the HATSouth instruments (identified by ``HS'' in the
    ``Instrument'' column) these magnitudes have been corrected for
    trends using the EPD and TFA procedures applied {\em prior} to
    fitting the transit model. This procedure may lead to an
    artificial dilution in the transit depths. The blend factors for
    the HATSouth light curves are listed in
    Tables~\ref{tab:planetparam}~and~\ref{tab:planetparamtwo}. For
    observations made with follow-up instruments (anything other than
    ``HS'' in the ``Instrument'' column), the magnitudes have been
    corrected for a quadratic trend in time, and for variations
    correlated with three PSF shape parameters, fit simultaneously
    with the transit.
}
\tablenotetext{d}{
    Raw magnitude values without correction for the quadratic trend in
    time, or for trends correlated with the shape of the PSF. These are only
    reported for the follow-up observations.
}
\tablecomments{
    This table is available in a machine-readable form in the online
    journal.  A portion is shown here for guidance regarding its form
    and content.
}
\ifthenelse{\boolean{emulateapj}}{
    \end{deluxetable*}
}{
    \end{deluxetable}
}

\section{Analysis}
\label{sec:analysis}

\subsection{Properties of the parent star}
\label{sec:stelparam}

To derive the physical properties of their planetary companions, we
first obtained the atmospheric parameters of the host stars.
We used high-resolution spectra of HATS-31 through HATS-35 obtained
with FEROS, together with the Zonal Atmospherical Stellar Parameter
Estimator (ZASPE) code (Brahm et al. 2016, in prep) to determine the
effective temperature (\teffstar), surface gravity (\loggstar), metallicity
(\feh), and projected equatorial rotation velocity (\vsini) for each star.

\teffstar\ and \feh\ values obtained using ZASPE were used with the
stellar density \rhostar{}, which was determined from the combined
light-curve and RV  analysis to determine a first estimate of the
stellar physical parameters following the method described in
\citet{sozzetti:2007}.  We used  the Yonsei-Yale isochrones
\citep[Y2;][]{yi:2001} to search for the parameters (stellar mass,
radius and age) that best match our estimated
\teffstar, \feh\ and \rhostar{} values.  Based on this comparison we
determine a revised value of \loggstar\ and then perform a second
iteration of ZASPE holding \loggstar\ fixed to this value while fitting
for \teffstar\, \feh\ and \vsini. These are then combined with \rhostar\
and once again compared to the Y2 isochrones to produce our final
adopted values for the physical stellar parameters.

The adopted parameters for \hatcur{31}, \hatcur{32} and \hatcur{33} are
given in \reftabl{stellar}, and for \hatcur{34}, and \hatcur{35} in
\reftabl{stellartwo}.  We show the locations of each of the stars on the
$\teffstar$--$\rhostar$ diagram (similar to a Hertzsprung-Russell
diagram) in Figure~\ref{fig:iso}.  This analysis
shows that \hatcur{31} has a
mass of \hatcurISOm{31}\,\msun, radius of \hatcurISOr{31}\,\rsun, and
age of \hatcurISOage{31}\,Gyr.  \hatcur{32} has a mass of
\hatcurISOm{32}\,\msun, radius of \hatcurISOr{32}\,\rsun, and age
of \hatcurISOage{32}\,Gyr.  \hatcur{33} has a mass of
\hatcurISOm{34}\,\msun, radius of \hatcurISOr{34}\,\rsun, and age
of \hatcurISOage{34}\,Gyr.  \hatcur{34} has a mass of
\hatcurISOm{34}\,\msun, radius of \hatcurISOr{34}\,\rsun, and age
of \hatcurISOage{34}\,Gyr.  Finally, \hatcur{35} has a mass of
\hatcurISOm{35}\,\msun, radius of \hatcurISOr{35}\,\rsun,
and age of \hatcurISOage{35}\,Gyr.  Distances for each star are
calculated by comparing the broad-band photometry of
\reftabl{stellar} to the predicted magnitudes in each filter from the
isochrones.  To determine the extinction we assumed a $R_{V} = 3.1$
extinction law \citep{cardelli:1989}. The distances for these systems
range between \hatcurXdistred{33}\,pc to \hatcurXdistred{31}\,pc for
\hatcur{33} and \hatcur{31}, respectively.

\ifthenelse{\boolean{emulateapj}}{
    \begin{figure*}[!ht]
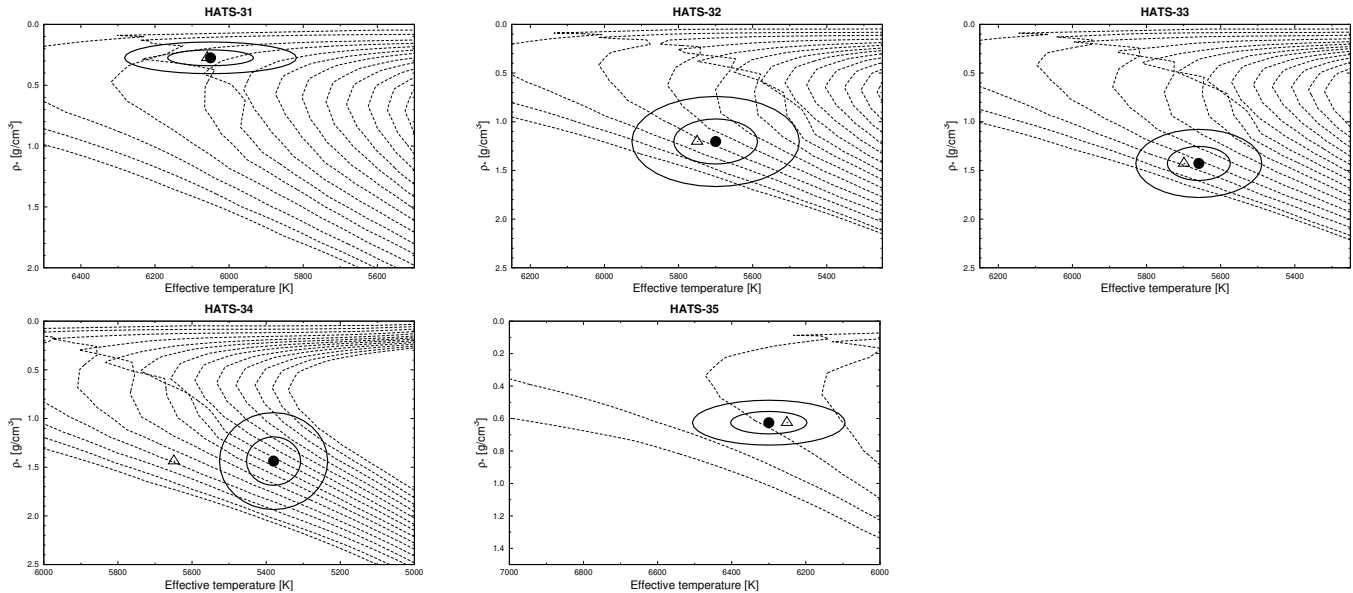

}{
    \begin{figure}[!ht]
}
{
\centering
\setlength{\plotwidthtwo}{0.31\linewidth}
\includegraphics[width={\plotwidthtwo}]{\hatcurhtr{31}-iso-rho}
\hfil
\includegraphics[width={\plotwidthtwo}]{\hatcurhtr{32}-iso-rho}
\hfil
\includegraphics[width={\plotwidthtwo}]{\hatcurhtr{33}-iso-rho}
}
{
\centering
\setlength{\plotwidthtwo}{0.31\linewidth}
\includegraphics[width={\plotwidthtwo}]{\hatcurhtr{34}-iso-rho}
\hspace{4mm}
\includegraphics[width={\plotwidthtwo}]{\hatcurhtr{35}-iso-rho}
}
\caption{
    Model isochrones from \cite{\hatcurisocite{31}} for the measured
    metallicities of each of the five new transiting planet host stars.
    We show models for ages of 0.2\,Gyr (leftmost dashed line), 1.0\,Gyr
    (second dashed line from left) and then models increasing in
    1.0\,Gyr increments (ages increasing from left to right). The
    adopted values of $\teffstar$ and \rhostar\ are shown by the black
    circles together with their 1$\sigma$ and 2$\sigma$ confidence
    ellipsoids.  The initial values of \teffstar\ and \rhostar\ from the
    first ZASPE and light curve analysis are represented with a
    triangle.  In the case of \hatcur{31} there was little change to the
    parameter values between the two iterations and the triangle lies
    partially under the black circle.
}
\label{fig:iso}
\ifthenelse{\boolean{emulateapj}}{
    \end{figure*}
}{
    \end{figure}
}

\ifthenelse{\boolean{emulateapj}}{
    \begin{deluxetable*}{lcccl}
}{
    \begin{deluxetable}{lcccl}
}
\tablewidth{0pc}
\tabletypesize{\footnotesize}
\tablecaption{
    Stellar parameters for \hatcur{31}, \hatcur{32} and \hatcur{33}
    \label{tab:stellar}
}
\tablehead{
    \multicolumn{1}{c}{} &
    \multicolumn{1}{c}{\bf HATS-31} &
    \multicolumn{1}{c}{\bf HATS-32} &
    \multicolumn{1}{c}{\bf HATS-33} &
    \multicolumn{1}{c}{} \\
    \multicolumn{1}{c}{~~~~~~~~Parameter~~~~~~~~} &
    \multicolumn{1}{c}{Value}                     &
    \multicolumn{1}{c}{Value}                     &
    \multicolumn{1}{c}{Value}                     &
    \multicolumn{1}{c}{Source}
}
\startdata
\noalign{\vskip -3pt}
\sidehead{Astrometric properties and cross-identifications}
~~~~2MASS-ID\dotfill               & \hatcurCCtwomass{31}  & \hatcurCCtwomass{32} & \hatcurCCtwomass{33} & \\
~~~~GSC-ID\dotfill                 & \hatcurCCgsc{31}      & \hatcurCCgsc{32}     & \hatcurCCgsc{33}     & \\
~~~~R.A. (J2000)\dotfill            & \hatcurCCra{31}       & \hatcurCCra{32}    & \hatcurCCra{33}    & 2MASS\\
~~~~Dec. (J2000)\dotfill            & \hatcurCCdec{31}      & \hatcurCCdec{32}   & \hatcurCCdec{33}   & 2MASS\\
~~~~$\mu_{\rm R.A.}$ (\masy)              & \hatcurCCpmra{31}     & \hatcurCCpmra{32} & \hatcurCCpmra{33} & UCAC4\\
~~~~$\mu_{\rm Dec.}$ (\masy)              & \hatcurCCpmdec{31}    & \hatcurCCpmdec{32} & \hatcurCCpmdec{33} & UCAC4\\
\sidehead{Spectroscopic properties}
~~~~$\teffstar$ (K)\dotfill         &  \hatcurSMEteff{31}   & \hatcurSMEteff{32} & \hatcurSMEteff{33} & ZASPE\tablenotemark{a}\\
~~~~$\feh$\dotfill                  &  \hatcurSMEzfeh{31}   & \hatcurSMEzfeh{32} & \hatcurSMEzfeh{33} & ZASPE               \\
~~~~$\vsini$ (\kms)\dotfill         &  \hatcurSMEvsin{31}   & \hatcurSMEvsin{32} & \hatcurSMEvsin{33} & ZASPE                \\
~~~~$\vmac$ (\kms)\dotfill          &  $3.90$   & $4.44$ & $5.01$ & Assumed              \\
~~~~$\vmic$ (\kms)\dotfill          &  $1.27$   & $1.04$ & $1.01$ & Assumed              \\
~~~~$\gamma_{\rm RV}$ (\ms)\dotfill&  \hatcurRVgammaabs{31}  & \hatcurRVgammaabs{32} & \hatcurRVgammaabs{33} & FEROS or HARPS\tablenotemark{b}  \\
\sidehead{Photometric properties}
~~~~$B$ (mag)\dotfill               &  \hatcurCCtassmB{31}  & \hatcurCCtassmB{32} & \hatcurCCtassmB{33} & APASS\tablenotemark{c} \\
~~~~$V$ (mag)\dotfill               &  \hatcurCCtassmv{31}  & \hatcurCCtassmv{32} & \hatcurCCtassmv{33} & APASS\tablenotemark{c} \\
~~~~$g$ (mag)\dotfill               &  \hatcurCCtassmg{31}  & \hatcurCCtassmg{32} & \hatcurCCtassmg{33} & APASS\tablenotemark{c} \\
~~~~$r$ (mag)\dotfill               &  \hatcurCCtassmr{31}  & \hatcurCCtassmr{32} & \hatcurCCtassmr{33} & APASS\tablenotemark{c} \\
~~~~$i$ (mag)\dotfill               &  \hatcurCCtassmi{31}  & \hatcurCCtassmi{32} & \hatcurCCtassmi{33} & APASS\tablenotemark{c} \\
~~~~$J$ (mag)\dotfill               &  \hatcurCCtwomassJmag{31} & \hatcurCCtwomassJmag{32} & \hatcurCCtwomassJmag{33} & 2MASS           \\
~~~~$H$ (mag)\dotfill               &  \hatcurCCtwomassHmag{31} & \hatcurCCtwomassHmag{32} & \hatcurCCtwomassHmag{33} & 2MASS           \\
~~~~$K_s$ (mag)\dotfill             &  \hatcurCCtwomassKmag{31} & \hatcurCCtwomassKmag{32} & \hatcurCCtwomassKmag{33} & 2MASS           \\
\sidehead{Derived properties}
~~~~$\mstar$ ($\msun$)\dotfill      &  \hatcurISOmlong{31}   & \hatcurISOmlong{32} & \hatcurISOmlong{33} & YY+$\rhostar$+ZASPE \tablenotemark{d}\\
~~~~$\rstar$ ($\rsun$)\dotfill      &  \hatcurISOrlong{31}   & \hatcurISOrlong{32} & \hatcurISOrlong{33} & YY+$\rhostar$+ZASPE         \\
~~~~$\loggstar$ (cgs)\dotfill       &  \hatcurISOlogg{31}    & \hatcurISOlogg{32} & \hatcurISOlogg{33} & YY+$\rhostar$+ZASPE         \\
~~~~$\rhostar$ (\gcmc)\dotfill       &  \hatcurLCrho{31}    & \hatcurLCrho{32} & \hatcurLCrho{33} & Light curves         \\
~~~~$\rhostar$ (\gcmc) \tablenotemark{e}\dotfill       &  \hatcurISOrho{31}    & \hatcurISOrho{32} & \hatcurISOrho{33} & YY+Light curves+ZASPE          \\
~~~~$\lstar$ ($\lsun$)\dotfill      &  \hatcurISOlum{31}     & \hatcurISOlum{32} & \hatcurISOlum{33} & YY+$\rhostar$+ZASPE         \\
~~~~$M_V$ (mag)\dotfill             &  \hatcurISOmv{31}      & \hatcurISOmv{32} & \hatcurISOmv{33} & YY+$\rhostar$+ZASPE         \\
~~~~$M_K$ (mag,\hatcurjhkfilset{31})\dotfill &  \hatcurISOMK{31} & \hatcurISOMK{32} & \hatcurISOMK{33} & YY+$\rhostar$+ZASPE         \\
~~~~Age (Gyr)\dotfill               &  \hatcurISOage{31}     & \hatcurISOage{32} & \hatcurISOage{33} & YY+$\rhostar$+ZASPE         \\
~~~~$A_{V}$ (mag)\dotfill               &  \hatcurXAv{31}     & \hatcurXAv{32} & \hatcurXAv{33} & YY+$\rhostar$+ZASPE         \\
~~~~Distance (pc)\dotfill           &  \hatcurXdistred{31}\phn  & \hatcurXdistred{32} & \hatcurXdistred{33} & YY+$\rhostar$+ZASPE\\ [-1.5ex]
\enddata
\tablecomments{
For all three systems the fixed-circular-orbit model has a higher Bayesian evidence than the eccentric-orbit model. We therefore assume a fixed circular orbit in generating the parameters listed for these systems.
}
\tablenotetext{a}{
    ZASPE = Zonal Atmospherical Stellar Parameter Estimator routine
    for the analysis of high-resolution spectra (Brahm et al.~2016, in
    preparation), applied to the HARPS spectra of \hatcur{31} and the
    FEROS spectra of the other systems. These parameters rely
    primarily on ZASPE, but have a small dependence also on the
    iterative analysis incorporating the isochrone search and global
    modeling of the data.
}
\tablenotetext{b}{
    From HARPS for \hatcur{31} and \hatcur{33} and from FEROS for
    \hatcur{32}. The error on $\gamma_{\rm RV}$ is
    determined from the orbital fit to the velocity measurements, and does
    not include the systematic uncertainty in transforming the
    velocities to the IAU standard system. The velocities have not
    been corrected for gravitational redshifts.
} \tablenotetext{c}{
    From APASS DR6 \citep{henden:2009} as
    listed in the UCAC 4 catalog \citep{zacharias:2012:ucac4}.
}
\tablenotetext{d}{
    \hatcurisoshort{31}+\rhostar+ZASPE = Based on the \hatcurisoshort{31}
    isochrones \citep{\hatcurisocite{31}}, \rhostar\ as a luminosity
    indicator, and the ZASPE results.
}
\tablenotetext{e}{
    In the case of $\rhostar$ we list two values. The first value is
    determined from the global fit to the light curves and RV data,
    without imposing a constraint that the parameters match the
    stellar evolution models. The second value results from
    restricting the posterior distribution to combinations of
    $\rhostar$+$\teffstar$+$\feh$ that match to a \hatcurisoshort{31}
    stellar model.
}
\ifthenelse{\boolean{emulateapj}}{
    \end{deluxetable*}
}{
    \end{deluxetable}
}

\ifthenelse{\boolean{emulateapj}}{
    \begin{deluxetable*}{lccl}
}{
    \begin{deluxetable}{lccl}
}
\tablewidth{0pc}
\tabletypesize{\footnotesize}
\tablecaption{
    Stellar parameters for \hatcur{34} and \hatcur{35}
    \label{tab:stellartwo}
}
\tablehead{
    \multicolumn{1}{c}{} &
    \multicolumn{1}{c}{\bf HATS-34} &
    \multicolumn{1}{c}{\bf HATS-35} &
    \multicolumn{1}{c}{} \\
    \multicolumn{1}{c}{~~~~~~~~Parameter~~~~~~~~} &
    \multicolumn{1}{c}{Value}                     &
    \multicolumn{1}{c}{Value}                     &
    \multicolumn{1}{c}{Source}
}
\startdata
\noalign{\vskip -3pt}
\sidehead{Astrometric properties and cross-identifications}
~~~~2MASS-ID\dotfill               & \hatcurCCtwomass{34}  & \hatcurCCtwomass{35} & \\
~~~~GSC-ID\dotfill                 & \hatcurCCgsc{34}      & \hatcurCCgsc{35}     & \\
~~~~R.A. (J2000)\dotfill            & \hatcurCCra{34}       & \hatcurCCra{35}    & 2MASS\\
~~~~Dec. (J2000)\dotfill            & \hatcurCCdec{34}      & \hatcurCCdec{35}   & 2MASS\\
~~~~$\mu_{\rm R.A.}$ (\masy)              & \hatcurCCpmra{34}     & \hatcurCCpmra{35} & UCAC4\\
~~~~$\mu_{\rm Dec.}$ (\masy)              & \hatcurCCpmdec{34}    & \hatcurCCpmdec{35} & UCAC4\\
\sidehead{Spectroscopic properties}
~~~~$\teffstar$ (K)\dotfill         &  \hatcurSMEteff{34}   & \hatcurSMEteff{35} & ZASPE\tablenotemark{a}\\
~~~~$\feh$\dotfill                  &  \hatcurSMEzfeh{34}   & \hatcurSMEzfeh{35} & ZASPE               \\
~~~~$\vsini$ (\kms)\dotfill         &  \hatcurSMEvsin{34}   & \hatcurSMEvsin{35} & ZASPE                \\
~~~~$\vmac$ (\kms)\dotfill          &  $3.56$   & $3.83$ & Assumed              \\
~~~~$\vmic$ (\kms)\dotfill          &  $0.88$   & $1.51$ & Assumed              \\
~~~~$\gamma_{\rm RV}$ (\ms)\dotfill&  \hatcurRVgammaabs{34}  & \hatcurRVgammaabs{35} & FEROS or HARPS\tablenotemark{b}  \\
\sidehead{Photometric properties}
~~~~$B$ (mag)\dotfill               &  \hatcurCCtassmB{34}  & \hatcurCCtassmB{35} & APASS\tablenotemark{c} \\
~~~~$V$ (mag)\dotfill               &  \hatcurCCtassmv{34}  & \hatcurCCtassmv{35} & APASS\tablenotemark{c} \\
~~~~$g$ (mag)\dotfill               &  \hatcurCCtassmg{34}  & $\cdots$ & APASS\tablenotemark{c} \\
~~~~$r$ (mag)\dotfill               &  \hatcurCCtassmr{34}  & $\cdots$ & APASS\tablenotemark{c} \\
~~~~$i$ (mag)\dotfill               &  \hatcurCCtassmi{34}  & $\cdots$ & APASS\tablenotemark{c} \\
~~~~$J$ (mag)\dotfill               &  \hatcurCCtwomassJmag{34} & \hatcurCCtwomassJmag{35} & 2MASS           \\
~~~~$H$ (mag)\dotfill               &  \hatcurCCtwomassHmag{34} & \hatcurCCtwomassHmag{35} & 2MASS           \\
~~~~$K_s$ (mag)\dotfill             &  \hatcurCCtwomassKmag{34} & \hatcurCCtwomassKmag{35} & 2MASS           \\
\sidehead{Derived properties}
~~~~$\mstar$ ($\msun$)\dotfill      &  \hatcurISOmlong{34}   & \hatcurISOmlong{35} & YY+$\rhostar$+ZASPE \tablenotemark{d}\\
~~~~$\rstar$ ($\rsun$)\dotfill      &  \hatcurISOrlong{34}   & \hatcurISOrlong{35} & YY+$\rhostar$+ZASPE         \\
~~~~$\loggstar$ (cgs)\dotfill       &  \hatcurISOlogg{34}    & \hatcurISOlogg{35} & YY+$\rhostar$+ZASPE         \\
~~~~$\rhostar$ (\gcmc)\dotfill       &  \hatcurLCrho{34}    & \hatcurLCrho{35} & Light curves         \\
~~~~$\rhostar$ (\gcmc) \tablenotemark{e}\dotfill       &  \hatcurISOrho{34}    & \hatcurISOrho{35} & YY+Light Curves+ZASPE         \\
~~~~$\lstar$ ($\lsun$)\dotfill      &  \hatcurISOlum{34}     & \hatcurISOlum{35} & YY+$\rhostar$+ZASPE         \\
~~~~$M_V$ (mag)\dotfill             &  \hatcurISOmv{34}      & \hatcurISOmv{35} & YY+$\rhostar$+ZASPE         \\
~~~~$M_K$ (mag,\hatcurjhkfilset{34})\dotfill &  \hatcurISOMK{34} & \hatcurISOMK{35} & YY+$\rhostar$+ZASPE         \\
~~~~Age (Gyr)\dotfill               &  \hatcurISOage{34}     & \hatcurISOage{35} & YY+$\rhostar$+ZASPE         \\
~~~~$A_{V}$ (mag)\dotfill               &  \hatcurXAv{34}     & \hatcurXAv{35} & YY+$\rhostar$+ZASPE         \\
~~~~Distance (pc)\dotfill           &  \hatcurXdistred{34}\phn  & \hatcurXdistred{35} & YY+$\rhostar$+ZASPE\\ [-1.5ex]
\enddata
\tablecomments{
For \hatcur{34} and \hatcur{35} the fixed-circular-orbit model has a
higher Bayesian evidence than the eccentric-orbit model,
We therefore assume a fixed circular orbit in generating the parameters
listed for both systems.
}
\tablenotetext{a}{
    ZASPE = Zonal Atmospherical Stellar Parameter Estimator routine
    for the analysis of high-resolution spectra (Brahm et al.~2016, in
    preparation), applied to the FEROS spectra of each star. These
    parameters rely primarily on ZASPE, but have a small dependence
    also on the iterative analysis incorporating the isochrone search
    and global modeling of the data.
}
\tablenotetext{b}{
    From FEROS for both objects. The error on $\gamma_{\rm RV}$ is
    determined from the orbital fit to the velocity measurements, and does
    not include the systematic uncertainty in transforming the
    velocities to the IAU standard system. The velocities have not
    been corrected for gravitational redshifts.
} \tablenotetext{c}{
    From APASS DR6 \citep{henden:2009} as
    listed in the UCAC 4 catalog \citep{zacharias:2012:ucac4}.
}
\tablenotetext{d}{
    \hatcurisoshort{31}+\rhostar+ZASPE = Based on the \hatcurisoshort{31}
    isochrones \citep{\hatcurisocite{31}}, \rhostar\ as a luminosity
    indicator, and the ZASPE results.
}
\tablenotetext{e}{
    In the case of $\rhostar$ we list two values. The first value is
    determined from the global fit to the light curves and RV data,
    without imposing a constraint that the parameters match the
    stellar evolution models. The second value results from
    restricting the posterior distribution to combinations of
    $\rhostar$+$\teffstar$+$\feh$ that match to a \hatcurisoshort{31}
    stellar model.
}
\ifthenelse{\boolean{emulateapj}}{
    \end{deluxetable*}
}{
    \end{deluxetable}
}

\subsection{Excluding blend scenarios}
\label{sec:blend}

In order to exclude blend scenarios we carried out an analysis following
\citet{hartman:2012:hat39hat41}. We attempt to model the available
photometric data (including light curves and catalog broad-band
photometric measurements) for each object as a blend between an
eclipsing binary star system and a third star along the line of
sight. The physical properties of the stars are constrained using the
Padova isochrones \citep{girardi:2000}, while we also require that the
brightest of the three stars in the blend have atmospheric parameters
consistent with those measured with ZASPE. We also simulate composite
cross-correlation functions (CCFs) and use them to predict velocities and BSs
for each blend scenario considered.

Based on this analysis we rule out blended stellar eclipsing binary
scenarios for all five systems. However, in general, we cannot rule out
the possibility that one or more of these objects may be an unresolved
binary star system with one component hosting a transiting planet. The
results for each object are as follows:
\begin{itemize}
\item {\em \hatcur{31}}: All blend models tested can be rejected with at
  least $3\sigma$ confidence based solely on the photometry. Those blend
  models which cannot be rejected based on the photometry with at least
  $5\sigma$ confidence predict large amplitude radial valocity and/or BS
  variations (i.e. greater than 1\,\kms, which is well above what is
  observed).
\item {\em \hatcur{32}}: All blend models tested yield higher $\chi^2$, based solely on the photometry, than the model of a single star with a transiting planet. Those blend models which cannot be rejected with $5\sigma$ confidence predict either velocity scatter above 200\,\ms\ (and a variation that does not look like the observed Keplerian variation), or a BS variation above 300\,\ms.
\item {\em \hatcur{33}}: All blend models tested can be rejected with at least $3\sigma$ confidence based solely on the photometry. Those models which are not rejected with at least $5\sigma$ confidence would have been easily identified as composite systems based on the CCFs computed from their spectra.
\item {\em \hatcur{34}}:  All blend models tested can be rejected with at lest $3\sigma$ confidence based solely on the photometry. In particular the best-fit blend model predicts a 5\,mmag secondary eclipse in $K_{S}$-band which was not seen in the AAT/IRIS2 observations. Our blend analysis allows for a quadratic trend in the follow-up light curves when fitting the data, and the best fit model includes a trend which cancels to some degree the predicted secondary eclipse. If we do not allow for such a trend in fitting the data, then the blend models are actually rejected with greater than $5\sigma$ confidence.  Some of the blend models that are rejected at 4--5$\sigma$ confidence (when the trend is included) do predict velocity and BS variations that have comparable amplitudes to the observed variations. In detail, however, the simulated blend velocities do not fit the data nearly as well as a single star with a planet. The BS variation is, however, captured somewhat better by the blend model. Nonetheless, given the constraints set by the photometry and radial velocities, we consider the blended stellar eclipsing binary model to be ruled out, and conclude that the observed BS variation must be due to some other cause (e.g., sky contamination or the presence of an unresolved star diluting the transiting planet system).
\item {\em \hatcur{35}}: All blend models tested can be rejected with greater than $8\sigma$ confidence based on the photometry alone. This is primarily driven by the large amplitude out-of-transit variation predicted for blend models capable of fitting the primary transit. The HATSouth light curve strongly excludes any such out-of-transit variation.
\end{itemize}

\subsection{Global modeling of the data}
\label{sec:globmod}

We modeled the HATSouth photometry, the follow-up photometry, and the
high-precision RV measurements following
\citet{pal:2008:hat7,bakos:2010:hat11,hartman:2012:hat39hat41}. We fit
\citet{mandel:2002} transit models to the light curves, allowing for a
dilution of the HATSouth transit depth as a result of blending from
neighboring stars and over-correction by the trend-filtering
method. For the follow-up light curves we include a quadratic trend in
time, and linear trends with up to three parameters describing the
shape of the PSF, in our model for each event to correct for
systematic errors in the photometry. We fit Keplerian orbits to the radial 
velocity data allowing the zero-point for each instrument to vary
independently in the fit, and allowing for RV noise which we also
vary as a free parameter for each instrument. We used a Differential
Evolution Markov Chain Monte Carlo procedure to explore the fitness
landscape and to determine the posterior distribution of the
parameters. Note that we tried fitting both fixed-circular-orbits and
free-eccentricity models to the data. We estimate the Bayesian
evidence for the fixed-circular and free-eccentricity models for each
system, and find that for \hatcurb{31} through \hatcurb{35} the
fixed-circular orbit models have higher evidence than the free
eccentricity models. For these systems we therefore adopt the
parameters that come from the fixed-circular orbit models.
The resulting parameters for \hatcurb{31}, \hatcurb{32} and \hatcurb{33}
are listed in \reftabl{planetparam}, while for \hatcurb{34} and
\hatcurb{35} they are listed in \reftabl{planetparamtwo}.

\hatcurb{31}, \hatcurb{32} and \hatcurb{34} have a mass that is
smaller than Jupiter, between \hatcurPPmlong{31}\,\mjup{} and
\hatcurPPmlong{34}\,\mjup, whereas the other two objects are slightly
more massive than Jupiter.  All planets have radii larger than Jupiter
within the range \hatcurPPrlong{33} to \hatcurPPrlong{31} \rjup.
These planets are moderately irradiated hot Jupiters with
\hatcurb{31} and \hatcurb{35} having relatively high  equilibrium
temperatures of \hatcurPPteff{31} K and \hatcurPPteff{35} K,
respectively.

\ifthenelse{\boolean{emulateapj}}{
    \begin{deluxetable*}{lccc}
}{
    \begin{deluxetable}{lccc}
}
\tabletypesize{\scriptsize}
\tablecaption{Orbital and planetary parameters for \hatcurb{31},
  \hatcurb{32} and \hatcurb{33}\label{tab:planetparam}}
\tablehead{
    \multicolumn{1}{c}{} &
    \multicolumn{1}{c}{\bf HATS-31b} &
    \multicolumn{1}{c}{\bf HATS-32b} &
    \multicolumn{1}{c}{\bf HATS-33b} \\
    \multicolumn{1}{c}{~~~~~~~~~~~~~~~Parameter~~~~~~~~~~~~~~~} &
    \multicolumn{1}{c}{Value} &
    \multicolumn{1}{c}{Value} &
    \multicolumn{1}{c}{Value}
}
\startdata
\noalign{\vskip -3pt}
\sidehead{\Lc{} parameters}
~~~$P$ (days)             \dotfill    & $\hatcurLCP{31}$ & $\hatcurLCP{32}$ & $\hatcurLCP{33}$ \\
~~~$T_c$ (${\rm BJD}$)
      \tablenotemark{a}   \dotfill    & $\hatcurLCT{31}$ & $\hatcurLCT{32}$ & $\hatcurLCT{33}$ \\
~~~$T_{14}$ (days)
      \tablenotemark{a}   \dotfill    & $\hatcurLCdur{31}$ & $\hatcurLCdur{32}$ & $\hatcurLCdur{33}$ \\
~~~$T_{12} = T_{34}$ (days)
      \tablenotemark{a}   \dotfill    & $\hatcurLCingdur{31}$ & $\hatcurLCingdur{32}$ & $\hatcurLCingdur{33}$ \\
~~~$\arstar$              \dotfill    & $\hatcurPPar{31}$ & $\hatcurPPar{32}$ & $\hatcurPPar{33}$ \\
~~~$\zrstar$ \tablenotemark{b}             \dotfill    & $\hatcurLCzeta{31}$\phn & $\hatcurLCzeta{32}$\phn & $\hatcurLCzeta{33}$\phn \\
~~~$\rpl/\rstar$          \dotfill    & $\hatcurLCrprstar{31}$ & $\hatcurLCrprstar{32}$ & $\hatcurLCrprstar{33}$ \\
~~~$b^2$                  \dotfill    & $\hatcurLCbsq{31}$ & $\hatcurLCbsq{32}$ & $\hatcurLCbsq{33}$ \\
~~~$b \equiv a \cos i/\rstar$
                          \dotfill    & $\hatcurLCimp{31}$ & $\hatcurLCimp{32}$ & $\hatcurLCimp{33}$ \\
~~~$i$ (deg)              \dotfill    & $\hatcurPPi{31}$\phn & $\hatcurPPi{32}$\phn & $\hatcurPPi{33}$\phn \\

\sidehead{HATSouth blend factors \tablenotemark{c}}
~~~Blend factor \dotfill & $\hatcurLCiblend{31}$ & $\hatcurLCiblend{32}$ & $\hatcurLCiblend{33}$ \\

\sidehead{Limb-darkening coefficients \tablenotemark{d}}
~~~$c_1,R$                  \dotfill    & $\cdots$ & $\hatcurLBiR{32}$ & $\cdots$ \\
~~~$c_2,R$                  \dotfill    & $\cdots$ & $\hatcurLBiiR{32}$ & $\cdots$ \\
~~~$c_1,r$                  \dotfill    & $\hatcurLBir{31}$ & $\hatcurLBir{32}$ & $\hatcurLBir{33}$ \\
~~~$c_2,r$                  \dotfill    & $\hatcurLBiir{31}$ & $\hatcurLBiir{32}$ & $\hatcurLBiir{33}$ \\
~~~$c_1,i$                  \dotfill    & $\hatcurLBii{31}$ & $\hatcurLBii{32}$ & $\hatcurLBii{33}$ \\
~~~$c_2,i$                  \dotfill    & $\hatcurLBiii{31}$ & $\hatcurLBiii{32}$ & $\hatcurLBiii{33}$ \\

\sidehead{RV parameters}
~~~$K$ (\ms)              \dotfill    & $\hatcurRVK{31}$\phn\phn & $\hatcurRVK{32}$\phn\phn & $\hatcurRVK{33}$\phn\phn \\
~~~$e$ \tablenotemark{e}               \dotfill    & $\hatcurRVeccentwosiglimeccen{31}$ & $\hatcurRVeccentwosiglimeccen{32}$ & $\hatcurRVeccentwosiglimeccen{33}$ \\
~~~RV jitter FEROS (\ms) \tablenotemark{f}       \dotfill    & $\cdots$ & \hatcurRVjitter{32} & \hatcurRVjittertwosiglimA{33} \\
~~~RV jitter HARPS (\ms)        \dotfill    & \hatcurRVjittertwosiglim{31} & $\cdots$ & \hatcurRVjittertwosiglimC{33} \\
~~~RV jitter Coralie (\ms)        \dotfill    & $\cdots$ & $\cdots$ & \hatcurRVjitterB{33} \\
~~~RV jitter CYCLOPS2 (\ms)        \dotfill    & $\cdots$ & $\cdots$ & \hatcurRVjitterD{33} \\

\sidehead{Planetary parameters}
~~~$\mpl$ ($\mjup$)       \dotfill    & $\hatcurPPmlong{31}$ & $\hatcurPPmlong{32}$ & $\hatcurPPmlong{33}$ \\
~~~$\rpl$ ($\rjup$)       \dotfill    & $\hatcurPPrlong{31}$ & $\hatcurPPrlong{32}$ & $\hatcurPPrlong{33}$ \\
~~~$C(\mpl,\rpl)$
    \tablenotemark{g}     \dotfill    & $\hatcurPPmrcorr{31}$ & $\hatcurPPmrcorr{32}$ & $\hatcurPPmrcorr{33}$ \\
~~~$\rhopl$ (\gcmc)       \dotfill    & $\hatcurPPrho{31}$ & $\hatcurPPrho{32}$ & $\hatcurPPrho{33}$ \\
~~~$\log g_p$ (cgs)       \dotfill    & $\hatcurPPlogg{31}$ & $\hatcurPPlogg{32}$ & $\hatcurPPlogg{33}$ \\
~~~$a$ (AU)               \dotfill    & $\hatcurPParel{31}$ & $\hatcurPParel{32}$ & $\hatcurPParel{33}$ \\
~~~$T_{\rm eq}$ (K)        \dotfill   & $\hatcurPPteff{31}$ & $\hatcurPPteff{32}$ & $\hatcurPPteff{33}$ \\
~~~$\Theta$ \tablenotemark{h} \dotfill & $\hatcurPPtheta{31}$ & $\hatcurPPtheta{32}$ & $\hatcurPPtheta{33}$ \\
~~~$\log_{10}\langle F \rangle$ (cgs) \tablenotemark{i}
                          \dotfill    & $\hatcurPPfluxavglog{31}$ & $\hatcurPPfluxavglog{32}$ & $\hatcurPPfluxavglog{33}$ \\ [-1.5ex]
\enddata
\tablenotetext{a}{
    Times are in Barycentric Julian Date calculated directly from UTC {\em without} correction for leap seconds.
    \ensuremath{T_c}: Reference epoch of
    mid transit that minimizes the correlation with the orbital
    period.
    \ensuremath{T_{14}}: total transit duration, time
    between first to last contact;
    \ensuremath{T_{12}=T_{34}}: ingress/egress time, time between first
    and second, or third and fourth contact.
}
\tablecomments{%
For all three systems the fixed-circular-orbit model has a higher Bayesian evidence than the eccentric-orbit model. We therefore assume a fixed circular orbit in generating the parameters listed for these systems.
}
\tablenotetext{b}{
   Reciprocal of the half duration of the transit used as a jump
   parameter in our Markov chain Monte Carlo (MCMC) analysis in place of
   $\arstar$. It is related to $\arstar$ by the expression $\zrstar =
   \arstar(2\pi(1+e\sin\omega))/(P\sqrt{1-b^2}\sqrt{1-e^2})$
   \citep{bakos:2010:hat11}.
}
\tablenotetext{c}{
    Scaling factor applied to the model transit that is fit to the HATSouth light curves. This factor accounts for dilution of the transit due to blending from neighboring stars and over-filtering of the light curve.  These factors are varied in the fit, and we allow independent factors for observations obtained with different HATSouth camera and field combinations.
}
\tablenotetext{d}{
    Values for a quadratic law, adopted from the tabulations by
    \cite{claret:2004} according to the spectroscopic (ZASPE) parameters
    listed in \reftabl{stellar}.
}
\tablenotetext{e}{
    For fixed circular orbit models we list
    the 95\% confidence upper limit on the eccentricity determined
    when $\sqrt{e}\cos\omega$ and $\sqrt{e}\sin\omega$ are allowed to
    vary in the fit.
}
\tablenotetext{f}{
    Term added in quadrature to the formal RV uncertainties for each
    instrument. This is treated as a free parameter in the fitting
    routine. In cases where the jitter is consistent with zero we list the 95\% confidence upper limit.
}
\tablenotetext{g}{
    Correlation coefficient between the planetary mass \mpl\ and radius
    \rpl\ estimated from the posterior parameter distribution.
  }
\tablenotetext{h}{
    The Safronov number is given by $\Theta = \frac{1}{2}(V_{\rm
    esc}/V_{\rm orb})^2 = (a/\rpl)(\mpl / \mstar )$
    \citep[see][]{hansen:2007}.
}
\tablenotetext{i}{
    Incoming flux per unit surface area, averaged over the orbit.
}
\ifthenelse{\boolean{emulateapj}}{
    \end{deluxetable*}
}{
    \end{deluxetable}
}

\begin{deluxetable*}{lcc}
\tabletypesize{\scriptsize}
\tablecaption{Orbital and planetary parameters for \hatcurb{34} and
  \hatcurb{35}\label{tab:planetparamtwo}}
\tablehead{
                      & \colhead{\textbf{HATS-34b}} & \colhead{\textbf{HATS-35b}} \\
  \colhead{Parameter} & \colhead{Value}             & \colhead{Value}
}
\startdata
\noalign{\vskip -3pt}
\sidehead{\Lc{} parameters}
~~~$P$ (days)             \dotfill    & $\hatcurLCP{34}$ & $\hatcurLCP{35}$ \\
~~~$T_c$ (${\rm BJD}$)
      \tablenotemark{a}   \dotfill    & $\hatcurLCT{34}$ & $\hatcurLCT{35}$ \\
~~~$T_{14}$ (days)
      \tablenotemark{a}   \dotfill    & $\hatcurLCdur{34}$ & $\hatcurLCdur{35}$ \\
~~~$T_{12} = T_{34}$ (days)
      \tablenotemark{a}   \dotfill    & $\hatcurLCingdur{34}$ & $\hatcurLCingdur{35}$ \\
~~~$\arstar$              \dotfill    & $\hatcurPPar{34}$ & $\hatcurPPar{35}$ \\
~~~$\zrstar$ \tablenotemark{b}             \dotfill    & $\hatcurLCzeta{34}$\phn & $\hatcurLCzeta{35}$\phn \\
~~~$\rpl/\rstar$          \dotfill    & $\hatcurLCrprstar{34}$ & $\hatcurLCrprstar{35}$ \\
~~~$b^2$                  \dotfill    & $\hatcurLCbsq{34}$ & $\hatcurLCbsq{35}$ \\
~~~$b \equiv a \cos i/\rstar$
                          \dotfill    & $\hatcurLCimp{34}$ & $\hatcurLCimp{35}$ \\
~~~$i$ (deg)              \dotfill    & $\hatcurPPi{34}$\phn & $\hatcurPPi{35}$\phn \\

\sidehead{HATSouth blend factors \tablenotemark{c}}
~~~Blend factor \dotfill & $\hatcurLCiblend{34}$ & $1 \pm 0$ \\

\sidehead{Limb-darkening coefficients \tablenotemark{d}}
~~~$c_1,R$                  \dotfill    & $\hatcurLBiR{34}$ & $\cdots$ \\
~~~$c_2,R$                  \dotfill    & $\hatcurLBiiR{34}$ & $\cdots$ \\
~~~$c_1,r$                  \dotfill    & $\hatcurLBir{34}$ & $\hatcurLBir{35}$ \\
~~~$c_2,r$                  \dotfill    & $\hatcurLBiir{34}$ & $\hatcurLBiir{35}$ \\
~~~$c_1,i$                  \dotfill    & $\cdots$ & $\hatcurLBii{35}$ \\
~~~$c_2,i$                  \dotfill    & $\cdots$ & $\hatcurLBiii{35}$ \\

\sidehead{RV parameters}
~~~$K$ (\ms)              \dotfill    & $\hatcurRVK{34}$\phn\phn & $\hatcurRVK{35}$\phn\phn \\
~~~$e$ \tablenotemark{e}               \dotfill    & $\hatcurRVeccentwosiglimeccen{34}$ & $\hatcurRVeccentwosiglimeccen{35}$ \\
~~~$\omega$ (deg)               \dotfill    & $\cdots$ & $\hatcurRVomega{35}$ \\
~~~$\sqrt{e}\cos\omega$ (deg)               \dotfill    & $\cdots$ & $\hatcurRVrk{35}$ \\
~~~$\sqrt{e}\sin\omega$ (deg)               \dotfill    & $\cdots$ & $\hatcurRVrh{35}$ \\
~~~$e\cos\omega$ (deg)               \dotfill    & $\cdots$ & $\hatcurRVk{35}$ \\
~~~$e\sin\omega$ (deg)               \dotfill    & $\cdots$ & $\hatcurRVh{35}$ \\
~~~RV jitter FEROS (\ms) \tablenotemark{f}       \dotfill    & \hatcurRVjittertwosiglim{34} & $\hatcurRVjittertwosiglimA{35}$ \\
~~~RV jitter HARPS (\ms)        \dotfill    & $\cdots$ & \hatcurRVjittertwosiglimC{35} \\
~~~RV jitter Coralie (\ms)        \dotfill    & $\cdots$ & \hatcurRVjitterB{35} \\

\sidehead{Planetary parameters}
~~~$\mpl$ ($\mjup$)       \dotfill    & $\hatcurPPmlong{34}$ & $\hatcurPPmlong{35}$ \\
~~~$\rpl$ ($\rjup$)       \dotfill    & $\hatcurPPrlong{34}$ & $\hatcurPPrlong{35}$ \\
~~~$C(\mpl,\rpl)$
    \tablenotemark{g}     \dotfill    & $\hatcurPPmrcorr{34}$ & $\hatcurPPmrcorr{35}$ \\
~~~$\rhopl$ (\gcmc)       \dotfill    & $\hatcurPPrho{34}$ & $\hatcurPPrho{35}$ \\
~~~$\log g_p$ (cgs)       \dotfill    & $\hatcurPPlogg{34}$ & $\hatcurPPlogg{35}$ \\
~~~$a$ (AU)               \dotfill    & $\hatcurPParel{34}$ & $\hatcurPParel{35}$ \\
~~~$T_{\rm eq}$ (K)        \dotfill   & $\hatcurPPteff{34}$ & $\hatcurPPteff{35}$ \\
~~~$\Theta$ \tablenotemark{h} \dotfill & $\hatcurPPtheta{34}$ & $\hatcurPPtheta{35}$ \\
~~~$\log_{10}\langle F \rangle$ (cgs) \tablenotemark{i}
                          \dotfill    & $\hatcurPPfluxavglog{34}$ & $\hatcurPPfluxavglog{35}$ \\ [-1.5ex]
\enddata
\tablenotetext{a}{
    Times are in Barycentric Julian Date calculated directly from UTC {\em without} correction for leap seconds.
    \ensuremath{T_c}: Reference epoch of
    mid transit that minimizes the correlation with the orbital
    period.
    \ensuremath{T_{14}}: total transit duration, time
    between first to last contact;
    \ensuremath{T_{12}=T_{34}}: ingress/egress time, time between first
    and second, or third and fourth contact.
}
\tablecomments{
For \hatcur{34} and \hatcur{35} the fixed-circular-orbit model has a higher Bayesian evidence than the eccentric-orbit model.
}
\tablenotetext{b}{
   Reciprocal of the half duration of the transit used as a jump parameter in our MCMC analysis in place of $\arstar$. It is related to $\arstar$ by the expression $\zrstar = \arstar(2\pi(1+e\sin\omega))/(P\sqrt{1-b^2}\sqrt{1-e^2})$ \citep{bakos:2010:hat11}.
}
\tablenotetext{c}{
    Scaling factor applied to the model transit that is fit to the HATSouth light curves. This factor accounts for dilution of the transit due to blending from neighboring stars and over-filtering of the light curve.  These factors are varied in the fit, and we allow independent factors for observations obtained with different HATSouth camera and field combinations. For \hatcur{35} we run TFA in signal-reconstruction mode, and have also confirmed that there are no diluting neighbors on the HATSouth images. We therefore fix the blend factor to unity for this system.
}
\tablenotetext{d}{
    Values for a quadratic law, adopted from the tabulations by
    \cite{claret:2004} according to the spectroscopic (ZASPE) parameters
    listed in Table~\ref{tab:stellartwo}.
}
\tablenotetext{e}{
    For fixed circular orbit models we list
    the 95\% confidence upper limit on the eccentricity determined
    when $\sqrt{e}\cos\omega$ and $\sqrt{e}\sin\omega$ are allowed to
    vary in the fit.
}
\tablenotetext{f}{
    Term added in quadrature to the formal radial velocity uncertainties for each
    instrument. This is treated as a free parameter in the fitting
    routine. In cases where thyis noise term is consistent with zero we list the 95\% confidence upper limit.
}
\tablenotetext{g}{
    Correlation coefficient between the planetary mass \mpl\ and radius
    \rpl\ estimated from the posterior parameter distribution.
}
\tablenotetext{h}{
    The Safronov number is given by $\Theta = \frac{1}{2}(V_{\rm
    esc}/V_{\rm orb})^2 = (a/\rpl)(\mpl / \mstar )$
    \citep[see][]{hansen:2007}.
}
\tablenotetext{i}{
    Incoming flux per unit surface area, averaged over the orbit.
}
\end{deluxetable*}

\section{Discussion}
\label{sec:discussion}

\begin{figure*}
  \centering
  \begin{tabular}{cc}
    \includegraphics[width=\columnwidth]{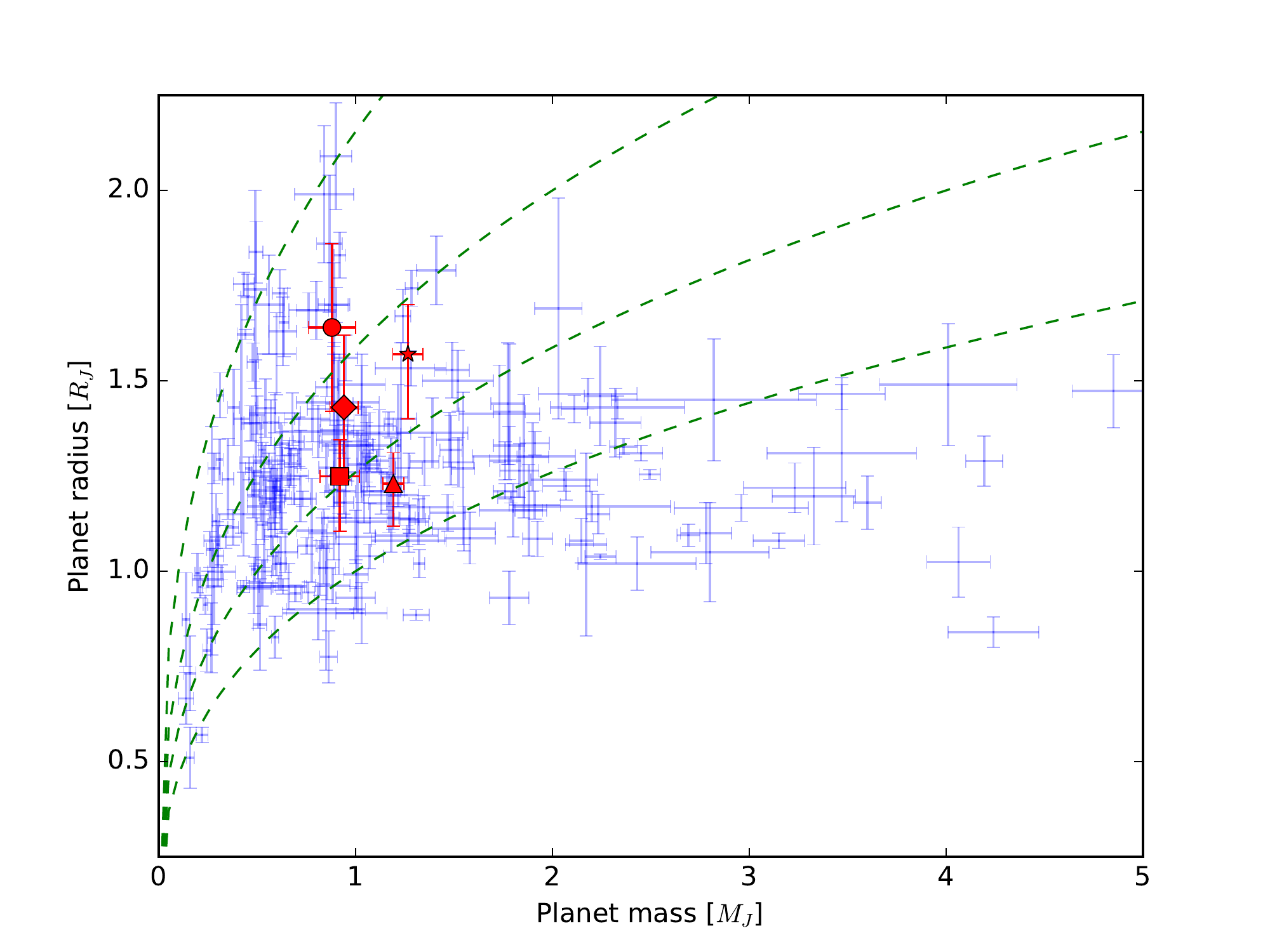} &
    \includegraphics[width=\columnwidth]{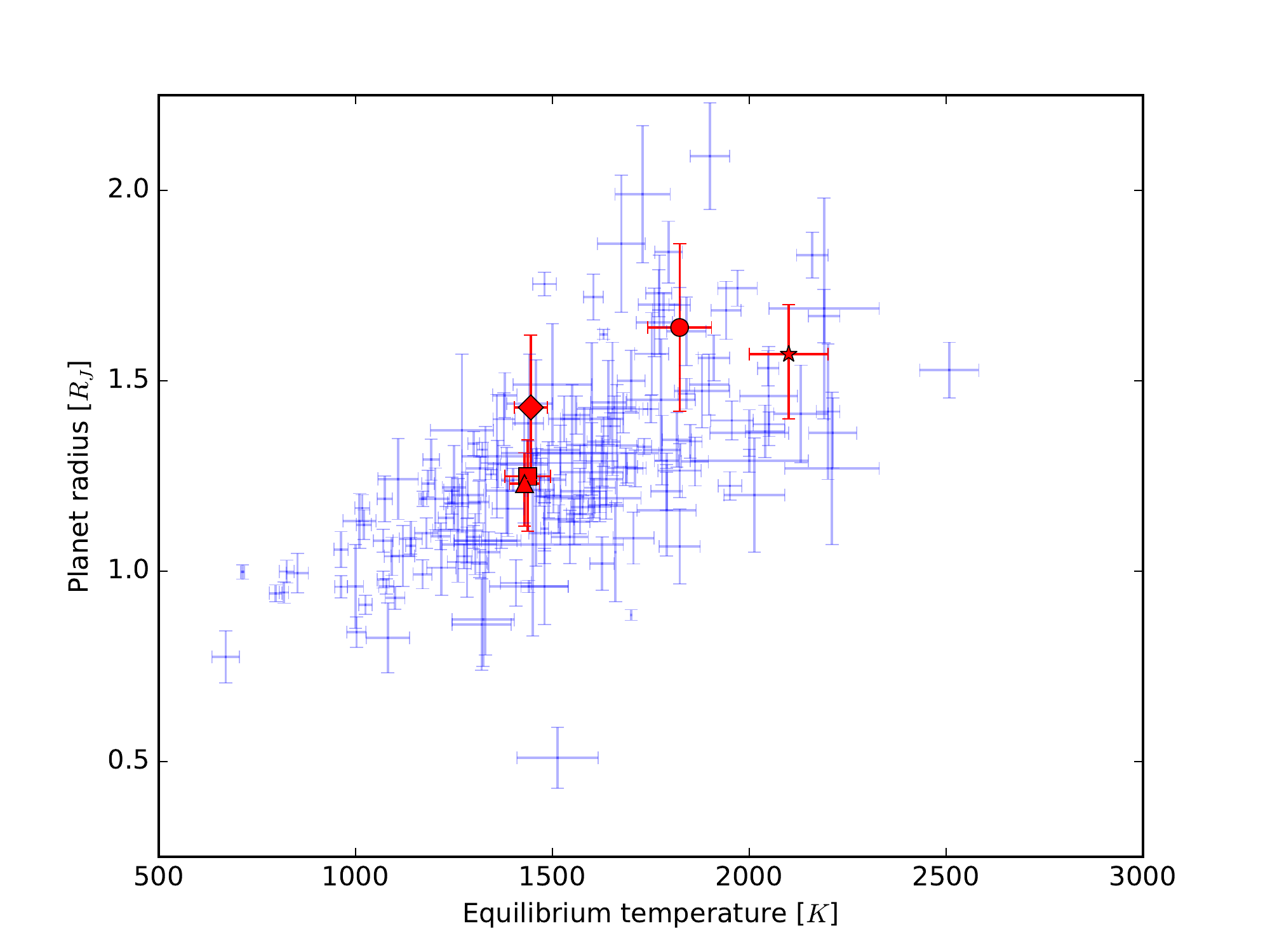}
  \end{tabular}
\caption{\emph{Left panel}: Mass-radius diagram of all
  known transiting hot Jupiters, i.e. planets with masses of
  $0.1M_J<M<5M_J$ and periods $P<10$ days, with precisely measured
  masses and radii.  \hatcurb{31} is
  shown with a red circle, \hatcurb{32} with a red square,
  \hatcurb{33} with a red triangle,
  \hatcurb{34} with a red diamond and
  \hatcurb{35} with a red star.
  Isodensity curves with density of 0.1, 0.25, 0.5 and 1
  $\rho_\mathrm{J}$ are shown by the dashed lines. \emph{Right panel}:
  the mass-density diagram \emph{Right panel}: Planet equilibrium
  temperature versus radius for the same sample of transiting hot
  Jupiters plotted in the mass-radius diagram. \hatcurb{31} through
  \hatcurb{35} are represented with the same symbols.
}
\label{fig:mrd}
\end{figure*}

\begin{figure}
  \centering
  \includegraphics[width=\columnwidth]{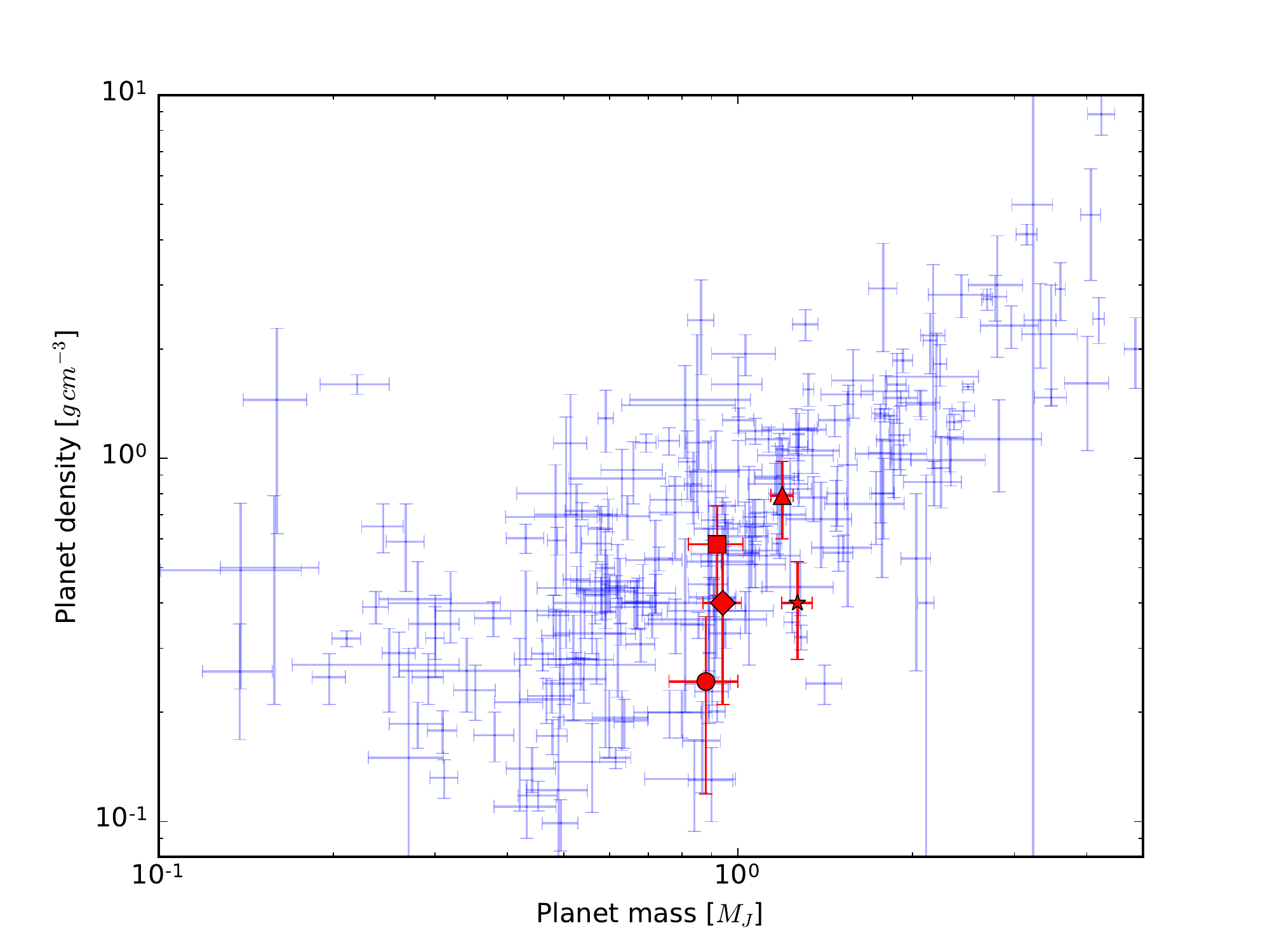}
  \caption{Mass-density diagram of all
  known transiting hot Jupiters, planets with masses of
  $0.1M_J<M<5M_J$ and periods $P<10$ days with well-characterized
  masses and radii, taken from the NASA Exoplanet Archive.
  Red data points as per Fig.~\ref{fig:mrd}.
}
\label{fig:dens-mass}
\end{figure}

We have presented five new transiting hot Jupiters,
\hatcurb{31}\ through to \hatcurb{35}, discovered by the HATSouth survey.  
Our analysis of the combined photometric and spectroscopic data  rules
out the possibility that these transit detections are blended stellar
eclipsing binary systems, and we conclude that these objects are transiting
planets.  In Figure~\ref{fig:mrd} we show the mass-radius and
equilibrium temperature versus radius diagrams of all known
transiting hot Jupiters with well determined masses and radii discovered
to date retrieved from the NASA Exoplanet
Archive\footnote{\url{http://exoplanetarchive.ipac.caltech.edu/}} on
2016 May 30, with \hatcurb{31} through
\hatcurb{35} superimposed in red. 

From the mass-radius diagram, the planets presented in this paper can be
classified as typical hot Jupiters.  \hatcurb{31}, \hatcurb{32} and
\hatcurb{34} are slightly less massive than Jupiter with
\hatcurPPmlong{31}\,\mjup, \hatcurPPmlong{32}\,\mjup{} and
\hatcurPPmlong{34}\,\mjup, respectively. However the radius values of the five
objects, all higher than that of Jupiter, vary between
\hatcurPPrlong{33}\,\rjup{} for \hatcurb{33} to
\hatcurPPrlong{31}\,\rjup{} for \hatcurb{31}.  The planet equilibrium
temperature versus radius diagram in the right panel of
Figure~\ref{fig:mrd} shows that the planets agree with  previously observed general trends. \hatcurb{31} and \hatcurb{35} have higher equilibrium
temperature, in the range \hatcurPPteff{31} K to
\hatcurPPteff{35} K, compared with the other three objects.

From the mass-radius and equilibrium temperature-radius diagrams it can
be seen that \hatcurb{31} and \hatcurb{35} reside in a different
region than the other three planets.
\hatcurb{31} and \hatcurb{35} have a radius of \hatcurPPrlong{31}\,\rjup{}
and \hatcurPPrlong{35}\,\rjup{}, respectively, and are therefore
moderately inflated planets, while \hatcurb{32}, \hatcurb{33} and
\hatcurb{34} have radii from \SIrange{1.2}{1.4}{\rjup}, which is close
to the mean radius of known hot Jupiters.
This indicates that the inflated radii are linked to the
increased irradiation from their parent star.  All of the discovered
planets have a period below the mean value of transiting hot Jupiters,
with the shortest period of the sample being \hatcurLCP{35}\, days for
\hatcurb{35}.

In Figure~\ref{fig:dens-mass} we show the planet density against mass
for \hatcurb{31}--\hatcurb{35} in the context of all known exoplanets
with well-characterized densities.  \hatcurb{31} is the lowest density
planet of the objects presented in this paper, with a mean density of
\hatcurPPrho{31}\,\gcmc,  while the other objects have typical densities
between \hatcurPPrho{35}\,\gcmc{} and \hatcurPPrho{33}\,\gcmc, for
objects of their mass and period.

\acknowledgements

\paragraph{Acknowledgements}
Development of the HATSouth project was funded by NSF MRI grant
NSF/AST-0723074, operations have been supported by NASA grants
NNX09AB29G and NNX12AH91H, and follow-up observations receive partial
support from grant NSF/AST-1108686.
A.J.\ acknowledges support from FONDECYT project 1130857, BASAL CATA
PFB-06, and project IC120009 ``Millennium Institute of Astrophysics
(MAS)'' of the Millenium Science Initiative, Chilean Ministry of
Economy. R.B.\ and N.E.\ are supported by CONICYT-PCHA/Doctorado
Nacional. R.B.\ and N.E.\ acknowledge additional support from project
IC120009 ``Millenium Institute of Astrophysics (MAS)'' of the Millennium
Science Initiative, Chilean Ministry of Economy.  V.S.\ acknowledges
support form BASAL CATA PFB-06.  M.R.\ acknowledges support from
FONDECYT postdoctoral fellowship 3120097.
This work is based on observations made with ESO Telescopes at the La
Silla Observatory.
This paper also uses observations obtained with facilities of the Las
Cumbres Observatory Global Telescope.
Work at the Australian National University is supported by ARC Laureate
Fellowship Grant FL0992131.
We acknowledge the use of the AAVSO Photometric All-Sky Survey (APASS),
funded by the Robert Martin Ayers Sciences Fund, and the SIMBAD
database, operated at CDS, Strasbourg, France.
Operations at the MPG~2.2\,m Telescope are jointly performed by the Max
Planck Gesellschaft and the European Southern Observatory.
We thank the MPG~2.2\,m Telescope support crew for their technical
assistance during observations.
We are grateful to P.Sackett for her help in the early phase of the
HATSouth project.
G.~B.~wishes to thank the warm hospitality of Adele and Joachim Cranz
at the farm Isabis, supporting the operations and service missions of
HATSouth.
Observing time were obtained through proposals CN2013A-171, CN2013B-55,
CN2014A-104, CN2014B-57, CN2015A-51 and ESO 096.C-0544.
This research has made use of the NASA Exoplanet Archive, which is
operated by the California Institute of Technology under contract with
the National Aeronautics and Space Administration under the Exoplanet
Exploration Program.

\clearpage
\bibliographystyle{aasjournal}
\bibliography{hatsbib}

\newpage
\tabletypesize{\scriptsize}
\ifthenelse{\boolean{emulateapj}}{
    \begin{deluxetable*}{llrrrrrl}
}{
    \begin{deluxetable}{llrrrrrl}
}
\tablewidth{0pc}
\tablecaption{
    Relative radial velocities and bisector spans for \hatcur{31}--\hatcur{35}.
    \label{tab:rvs}
}
\tablehead{
    \colhead{Star} &
    \colhead{BJD} &
    \colhead{RV\tablenotemark{a}} &
    \colhead{\ensuremath{\sigma_{\rm RV}}\tablenotemark{b}} &
    \colhead{BS} &
    \colhead{\ensuremath{\sigma_{\rm BS}}} &
    \colhead{Phase} &
    \colhead{Instrument}\\
    \colhead{} &
    \colhead{\hbox{(2,450,000$+$)}} &
    \colhead{(\ms)} &
    \colhead{(\ms)} &
    \colhead{(\ms)} &
    \colhead{(\ms)} &
    \colhead{} &
    \colhead{}
}
\startdata
\multicolumn{8}{c}{\bf HATS-31} \\
\hline\\
    \input{data/\hatcurhtr{31}_rvtable.tex}
\cutinhead{\bf HATS-32}
    \input{data/\hatcurhtr{32}_rvtable.tex}
\cutinhead{\bf HATS-33}
    \input{data/\hatcurhtr{33}_rvtable.tex}
\cutinhead{\bf HATS-34}
    \input{data/\hatcurhtr{34}_rvtable.tex}
\cutinhead{\bf HATS-35}
    \input{data/\hatcurhtr{35}_rvtable.tex}
\enddata
\tablenotetext{a}{
    The zero-point of these velocities is arbitrary. An overall offset
    $\gamma_{\rm rel}$ fitted independently to the velocities from
    each instrument has been subtracted.
}
\tablenotetext{b}{
    Internal errors excluding the component of astrophysical jitter
    considered in \refsecl{globmod}.
}
\ifthenelse{\boolean{rvtablelong}}{
}{
}
\ifthenelse{\boolean{emulateapj}}{
    \end{deluxetable*}
}{
    \end{deluxetable}
}

\end{document}

%% file: newcommand_gen.tex
\newcommand{\forloop}[5][1]%
{%
\setcounter{#2}{#3}%
\ifthenelse{#4}%
	{%
	#5%
	\addtocounter{#2}{#1}%
	\forloop[#1]{#2}{\value{#2}}{#4}{#5}%
	}%
	{%
	}%
}%

\newcommand{\ctbd}[1]{}

\newcommand{\Lc}{Light curve}

\newcommand{\masy}{\ensuremath{\rm mas\,yr^{-1}}}
\newcommand{\kms}{\ensuremath{\rm km\,s^{-1}}}
\newcommand{\ms}{\ensuremath{\rm m\,s^{-1}}}

\newcommand{\gcmc}{\ensuremath{\rm g\,cm^{-3}}}

\newcommand{\vsini}{\ensuremath{v \sin{i}}}
\newcommand{\feh}{\ensuremath{\rm [Fe/H]}}

\newcommand{\vmac}{\ensuremath{v_{\rm mac}}}
\newcommand{\vmic}{\ensuremath{v_{\rm mic}}}

\newcommand{\rsun}{\ensuremath{R_\sun}}
\newcommand{\msun}{\ensuremath{M_\sun}}
\newcommand{\lsun}{\ensuremath{L_\sun}}

\newcommand{\rstar}{\ensuremath{R_\star}}
\newcommand{\mstar}{\ensuremath{M_\star}}
\newcommand{\lstar}{\ensuremath{L_\star}}

\newcommand{\teffstar}{\ensuremath{T_{\rm eff\star}}}
\newcommand{\rhostar}{\ensuremath{\rho_\star}}
\newcommand{\loggstar}{\ensuremath{\log{g_{\star}}}}

\newcommand{\rpl}{\ensuremath{R_{p}}}
\newcommand{\mpl}{\ensuremath{M_{p}}}

\newcommand{\rhopl}{\ensuremath{\rho_{p}}}

\newcommand{\arstar}{\ensuremath{a/\rstar}}
\newcommand{\zrstar}{\ensuremath{\zeta/\rstar}}

\newcommand{\rjup}{\ensuremath{R_{\rm J}}}
\newcommand{\mjup}{\ensuremath{M_{\rm J}}}

\newcommand{\refsecl}[1]{\mbox{Section \ref{sec:#1}}}

\newcommand{\reftabl}[1]{Table~\ref{tab:#1}}



%% file: newcommand_spec_multi.tex
\input{planetinfo_multi.tex}
\input{planetinfo_multi_eccen.tex}
\input{newcommand_spec_HATS566-004.tex}
\input{newcommand_spec_HATS586-004.tex}
\input{newcommand_spec_HATS747-032.tex}
\input{newcommand_spec_HATS754-010.tex}
\input{newcommand_spec_HATS779-003.tex}
\newcommand{\hatcur}[1]{\ifnum#1=31 %
\hatcurxxxxxA
\else
\ifnum#1=32 %
\hatcurxxxxxB
\else
\ifnum#1=33 %
\hatcurxxxxxC
\else
\ifnum#1=34 %
\hatcurxxxxxD
\else
\ifnum#1=35 %
\hatcurxxxxxE
\else
??????\fi
\fi
\fi
\fi
\fi
}
\newcommand{\hatcurb}[1]{\ifnum#1=31 %
\hatcurbxxxxxA
\else
\ifnum#1=32 %
\hatcurbxxxxxB
\else
\ifnum#1=33 %
\hatcurbxxxxxC
\else
\ifnum#1=34 %
\hatcurbxxxxxD
\else
\ifnum#1=35 %
\hatcurbxxxxxE
\else
??????\fi
\fi
\fi
\fi
\fi
}
\newcommand{\hatcurc}[1]{\ifnum#1=31 %
\hatcurcxxxxxA
\else
\ifnum#1=32 %
\hatcurcxxxxxB
\else
\ifnum#1=33 %
\hatcurcxxxxxC
\else
\ifnum#1=34 %
\hatcurcxxxxxD
\else
\ifnum#1=35 %
\hatcurcxxxxxE
\else
??????\fi
\fi
\fi
\fi
\fi
}
\newcommand{\hatcurCCtassvi}[1]{\ifnum#1=31 %
\hatcurCCtassvixxxxxA
\else
\ifnum#1=32 %
\hatcurCCtassvixxxxxB
\else
\ifnum#1=33 %
\hatcurCCtassvixxxxxC
\else
\ifnum#1=34 %
\hatcurCCtassvixxxxxD
\else
\ifnum#1=35 %
\hatcurCCtassvixxxxxE
\else
??????\fi
\fi
\fi
\fi
\fi
}
\newcommand{\hatcurisocite}[1]{\ifnum#1=31 %
\hatcurisocitexxxxxA
\else
\ifnum#1=32 %
\hatcurisocitexxxxxB
\else
\ifnum#1=33 %
\hatcurisocitexxxxxC
\else
\ifnum#1=34 %
\hatcurisocitexxxxxD
\else
\ifnum#1=35 %
\hatcurisocitexxxxxE
\else
??????\fi
\fi
\fi
\fi
\fi
}
\newcommand{\hatcurisofull}[1]{\ifnum#1=31 %
\hatcurisofullxxxxxA
\else
\ifnum#1=32 %
\hatcurisofullxxxxxB
\else
\ifnum#1=33 %
\hatcurisofullxxxxxC
\else
\ifnum#1=34 %
\hatcurisofullxxxxxD
\else
\ifnum#1=35 %
\hatcurisofullxxxxxE
\else
??????\fi
\fi
\fi
\fi
\fi
}
\newcommand{\hatcurisoshort}[1]{\ifnum#1=31 %
\hatcurisoshortxxxxxA
\else
\ifnum#1=32 %
\hatcurisoshortxxxxxB
\else
\ifnum#1=33 %
\hatcurisoshortxxxxxC
\else
\ifnum#1=34 %
\hatcurisoshortxxxxxD
\else
\ifnum#1=35 %
\hatcurisoshortxxxxxE
\else
??????\fi
\fi
\fi
\fi
\fi
}
\newcommand{\hatcurjhkfilset}[1]{\ifnum#1=31 %
\hatcurjhkfilsetxxxxxA
\else
\ifnum#1=32 %
\hatcurjhkfilsetxxxxxB
\else
\ifnum#1=33 %
\hatcurjhkfilsetxxxxxC
\else
\ifnum#1=34 %
\hatcurjhkfilsetxxxxxD
\else
\ifnum#1=35 %
\hatcurjhkfilsetxxxxxE
\else
??????\fi
\fi
\fi
\fi
\fi
}
\newcommand{\hatcurlumind}[1]{\ifnum#1=31 %
\hatcurlumindxxxxxA
\else
\ifnum#1=32 %
\hatcurlumindxxxxxB
\else
\ifnum#1=33 %
\hatcurlumindxxxxxC
\else
\ifnum#1=34 %
\hatcurlumindxxxxxD
\else
\ifnum#1=35 %
\hatcurlumindxxxxxE
\else
??????\fi
\fi
\fi
\fi
\fi
}
\newcommand{\hatcurplanetnum}[1]{\ifnum#1=31 %
\hatcurplanetnumxxxxxA
\else
\ifnum#1=32 %
\hatcurplanetnumxxxxxB
\else
\ifnum#1=33 %
\hatcurplanetnumxxxxxC
\else
\ifnum#1=34 %
\hatcurplanetnumxxxxxD
\else
\ifnum#1=35 %
\hatcurplanetnumxxxxxE
\else
??????\fi
\fi
\fi
\fi
\fi
}
\newcommand{\hatcurRVgammaabs}[1]{\ifnum#1=31 %
\hatcurRVgammaabsxxxxxA
\else
\ifnum#1=32 %
\hatcurRVgammaabsxxxxxB
\else
\ifnum#1=33 %
\hatcurRVgammaabsxxxxxC
\else
\ifnum#1=34 %
\hatcurRVgammaabsxxxxxD
\else
\ifnum#1=35 %
\hatcurRVgammaabsxxxxxE
\else
??????\fi
\fi
\fi
\fi
\fi
}
\newcommand{\hatcurRVgammaabsinst}[1]{\ifnum#1=31 %
\hatcurRVgammaabsinstxxxxxA
\else
\ifnum#1=32 %
\hatcurRVgammaabsinstxxxxxB
\else
\ifnum#1=33 %
\hatcurRVgammaabsinstxxxxxC
\else
\ifnum#1=34 %
\hatcurRVgammaabsinstxxxxxD
\else
\ifnum#1=35 %
\hatcurRVgammaabsinstxxxxxE
\else
??????\fi
\fi
\fi
\fi
\fi
}
\newcommand{\hatcurRVgammarel}[1]{\ifnum#1=31 %
\hatcurRVgammarelxxxxxA
\else
\ifnum#1=32 %
\hatcurRVgammarelxxxxxB
\else
\ifnum#1=33 %
\hatcurRVgammarelxxxxxC
\else
\ifnum#1=34 %
\hatcurRVgammarelxxxxxD
\else
\ifnum#1=35 %
\hatcurRVgammarelxxxxxE
\else
??????\fi
\fi
\fi
\fi
\fi
}
\newcommand{\hatcurSMElogg}[1]{\ifnum#1=31 %
\hatcurSMEloggxxxxxA
\else
\ifnum#1=32 %
\hatcurSMEloggxxxxxB
\else
\ifnum#1=33 %
\hatcurSMEloggxxxxxC
\else
\ifnum#1=34 %
\hatcurSMEloggxxxxxD
\else
\ifnum#1=35 %
\hatcurSMEloggxxxxxE
\else
??????\fi
\fi
\fi
\fi
\fi
}
\newcommand{\hatcurSMEteff}[1]{\ifnum#1=31 %
\hatcurSMEteffxxxxxA
\else
\ifnum#1=32 %
\hatcurSMEteffxxxxxB
\else
\ifnum#1=33 %
\hatcurSMEteffxxxxxC
\else
\ifnum#1=34 %
\hatcurSMEteffxxxxxD
\else
\ifnum#1=35 %
\hatcurSMEteffxxxxxE
\else
??????\fi
\fi
\fi
\fi
\fi
}
\newcommand{\hatcurSMEversion}[1]{\ifnum#1=31 %
\hatcurSMEversionxxxxxA
\else
\ifnum#1=32 %
\hatcurSMEversionxxxxxB
\else
\ifnum#1=33 %
\hatcurSMEversionxxxxxC
\else
\ifnum#1=34 %
\hatcurSMEversionxxxxxD
\else
\ifnum#1=35 %
\hatcurSMEversionxxxxxE
\else
??????\fi
\fi
\fi
\fi
\fi
}
\newcommand{\hatcurSMEvmac}[1]{\ifnum#1=31 %
\hatcurSMEvmacxxxxxA
\else
\ifnum#1=32 %
\hatcurSMEvmacxxxxxB
\else
\ifnum#1=33 %
\hatcurSMEvmacxxxxxC
\else
\ifnum#1=34 %
\hatcurSMEvmacxxxxxD
\else
\ifnum#1=35 %
\hatcurSMEvmacxxxxxE
\else
??????\fi
\fi
\fi
\fi
\fi
}
\newcommand{\hatcurSMEvmic}[1]{\ifnum#1=31 %
\hatcurSMEvmicxxxxxA
\else
\ifnum#1=32 %
\hatcurSMEvmicxxxxxB
\else
\ifnum#1=33 %
\hatcurSMEvmicxxxxxC
\else
\ifnum#1=34 %
\hatcurSMEvmicxxxxxD
\else
\ifnum#1=35 %
\hatcurSMEvmicxxxxxE
\else
??????\fi
\fi
\fi
\fi
\fi
}
\newcommand{\hatcurSMEvsin}[1]{\ifnum#1=31 %
\hatcurSMEvsinxxxxxA
\else
\ifnum#1=32 %
\hatcurSMEvsinxxxxxB
\else
\ifnum#1=33 %
\hatcurSMEvsinxxxxxC
\else
\ifnum#1=34 %
\hatcurSMEvsinxxxxxD
\else
\ifnum#1=35 %
\hatcurSMEvsinxxxxxE
\else
??????\fi
\fi
\fi
\fi
\fi
}
\newcommand{\hatcurSMEzfeh}[1]{\ifnum#1=31 %
\hatcurSMEzfehxxxxxA
\else
\ifnum#1=32 %
\hatcurSMEzfehxxxxxB
\else
\ifnum#1=33 %
\hatcurSMEzfehxxxxxC
\else
\ifnum#1=34 %
\hatcurSMEzfehxxxxxD
\else
\ifnum#1=35 %
\hatcurSMEzfehxxxxxE
\else
??????\fi
\fi
\fi
\fi
\fi
}
\newcommand{\hatcurSMEzfehshort}[1]{\ifnum#1=31 %
\hatcurSMEzfehshortxxxxxA
\else
\ifnum#1=32 %
\hatcurSMEzfehshortxxxxxB
\else
\ifnum#1=33 %
\hatcurSMEzfehshortxxxxxC
\else
\ifnum#1=34 %
\hatcurSMEzfehshortxxxxxD
\else
\ifnum#1=35 %
\hatcurSMEzfehshortxxxxxE
\else
??????\fi
\fi
\fi
\fi
\fi
}

%% file: planetinfo_multi.tex
\input{HATS566-004.tex}
\input{HATS586-004.tex}
\input{HATS747-032.tex}
\input{HATS754-010.tex}
\input{HATS779-003.tex}
\newcommand{\hatcurCCbbHmag}[1]{\ifnum#1=31 %
\hatcurCCbbHmagxxxxxA
\else
\ifnum#1=32 %
\hatcurCCbbHmagxxxxxB
\else
\ifnum#1=33 %
\hatcurCCbbHmagxxxxxC
\else
\ifnum#1=34 %
\hatcurCCbbHmagxxxxxD
\else
\ifnum#1=35 %
\hatcurCCbbHmagxxxxxE
\else
??????\fi
\fi
\fi
\fi
\fi
}
\newcommand{\hatcurCCbbJmag}[1]{\ifnum#1=31 %
\hatcurCCbbJmagxxxxxA
\else
\ifnum#1=32 %
\hatcurCCbbJmagxxxxxB
\else
\ifnum#1=33 %
\hatcurCCbbJmagxxxxxC
\else
\ifnum#1=34 %
\hatcurCCbbJmagxxxxxD
\else
\ifnum#1=35 %
\hatcurCCbbJmagxxxxxE
\else
??????\fi
\fi
\fi
\fi
\fi
}
\newcommand{\hatcurCCbbKmag}[1]{\ifnum#1=31 %
\hatcurCCbbKmagxxxxxA
\else
\ifnum#1=32 %
\hatcurCCbbKmagxxxxxB
\else
\ifnum#1=33 %
\hatcurCCbbKmagxxxxxC
\else
\ifnum#1=34 %
\hatcurCCbbKmagxxxxxD
\else
\ifnum#1=35 %
\hatcurCCbbKmagxxxxxE
\else
??????\fi
\fi
\fi
\fi
\fi
}
\newcommand{\hatcurCCcitHmag}[1]{\ifnum#1=31 %
\hatcurCCcitHmagxxxxxA
\else
\ifnum#1=32 %
\hatcurCCcitHmagxxxxxB
\else
\ifnum#1=33 %
\hatcurCCcitHmagxxxxxC
\else
\ifnum#1=34 %
\hatcurCCcitHmagxxxxxD
\else
\ifnum#1=35 %
\hatcurCCcitHmagxxxxxE
\else
??????\fi
\fi
\fi
\fi
\fi
}
\newcommand{\hatcurCCcitJmag}[1]{\ifnum#1=31 %
\hatcurCCcitJmagxxxxxA
\else
\ifnum#1=32 %
\hatcurCCcitJmagxxxxxB
\else
\ifnum#1=33 %
\hatcurCCcitJmagxxxxxC
\else
\ifnum#1=34 %
\hatcurCCcitJmagxxxxxD
\else
\ifnum#1=35 %
\hatcurCCcitJmagxxxxxE
\else
??????\fi
\fi
\fi
\fi
\fi
}
\newcommand{\hatcurCCcitKmag}[1]{\ifnum#1=31 %
\hatcurCCcitKmagxxxxxA
\else
\ifnum#1=32 %
\hatcurCCcitKmagxxxxxB
\else
\ifnum#1=33 %
\hatcurCCcitKmagxxxxxC
\else
\ifnum#1=34 %
\hatcurCCcitKmagxxxxxD
\else
\ifnum#1=35 %
\hatcurCCcitKmagxxxxxE
\else
??????\fi
\fi
\fi
\fi
\fi
}
\newcommand{\hatcurCCdec}[1]{\ifnum#1=31 %
\hatcurCCdecxxxxxA
\else
\ifnum#1=32 %
\hatcurCCdecxxxxxB
\else
\ifnum#1=33 %
\hatcurCCdecxxxxxC
\else
\ifnum#1=34 %
\hatcurCCdecxxxxxD
\else
\ifnum#1=35 %
\hatcurCCdecxxxxxE
\else
??????\fi
\fi
\fi
\fi
\fi
}
\newcommand{\hatcurCCesoHKmag}[1]{\ifnum#1=31 %
\hatcurCCesoHKmagxxxxxA
\else
\ifnum#1=32 %
\hatcurCCesoHKmagxxxxxB
\else
\ifnum#1=33 %
\hatcurCCesoHKmagxxxxxC
\else
\ifnum#1=34 %
\hatcurCCesoHKmagxxxxxD
\else
\ifnum#1=35 %
\hatcurCCesoHKmagxxxxxE
\else
??????\fi
\fi
\fi
\fi
\fi
}
\newcommand{\hatcurCCesoHmag}[1]{\ifnum#1=31 %
\hatcurCCesoHmagxxxxxA
\else
\ifnum#1=32 %
\hatcurCCesoHmagxxxxxB
\else
\ifnum#1=33 %
\hatcurCCesoHmagxxxxxC
\else
\ifnum#1=34 %
\hatcurCCesoHmagxxxxxD
\else
\ifnum#1=35 %
\hatcurCCesoHmagxxxxxE
\else
??????\fi
\fi
\fi
\fi
\fi
}
\newcommand{\hatcurCCesoJHmag}[1]{\ifnum#1=31 %
\hatcurCCesoJHmagxxxxxA
\else
\ifnum#1=32 %
\hatcurCCesoJHmagxxxxxB
\else
\ifnum#1=33 %
\hatcurCCesoJHmagxxxxxC
\else
\ifnum#1=34 %
\hatcurCCesoJHmagxxxxxD
\else
\ifnum#1=35 %
\hatcurCCesoJHmagxxxxxE
\else
??????\fi
\fi
\fi
\fi
\fi
}
\newcommand{\hatcurCCesoJKmag}[1]{\ifnum#1=31 %
\hatcurCCesoJKmagxxxxxA
\else
\ifnum#1=32 %
\hatcurCCesoJKmagxxxxxB
\else
\ifnum#1=33 %
\hatcurCCesoJKmagxxxxxC
\else
\ifnum#1=34 %
\hatcurCCesoJKmagxxxxxD
\else
\ifnum#1=35 %
\hatcurCCesoJKmagxxxxxE
\else
??????\fi
\fi
\fi
\fi
\fi
}
\newcommand{\hatcurCCesoJmag}[1]{\ifnum#1=31 %
\hatcurCCesoJmagxxxxxA
\else
\ifnum#1=32 %
\hatcurCCesoJmagxxxxxB
\else
\ifnum#1=33 %
\hatcurCCesoJmagxxxxxC
\else
\ifnum#1=34 %
\hatcurCCesoJmagxxxxxD
\else
\ifnum#1=35 %
\hatcurCCesoJmagxxxxxE
\else
??????\fi
\fi
\fi
\fi
\fi
}
\newcommand{\hatcurCCesoKmag}[1]{\ifnum#1=31 %
\hatcurCCesoKmagxxxxxA
\else
\ifnum#1=32 %
\hatcurCCesoKmagxxxxxB
\else
\ifnum#1=33 %
\hatcurCCesoKmagxxxxxC
\else
\ifnum#1=34 %
\hatcurCCesoKmagxxxxxD
\else
\ifnum#1=35 %
\hatcurCCesoKmagxxxxxE
\else
??????\fi
\fi
\fi
\fi
\fi
}
\newcommand{\hatcurCCgsc}[1]{\ifnum#1=31 %
\hatcurCCgscxxxxxA
\else
\ifnum#1=32 %
\hatcurCCgscxxxxxB
\else
\ifnum#1=33 %
\hatcurCCgscxxxxxC
\else
\ifnum#1=34 %
\hatcurCCgscxxxxxD
\else
\ifnum#1=35 %
\hatcurCCgscxxxxxE
\else
??????\fi
\fi
\fi
\fi
\fi
}
\newcommand{\hatcurCCmag}[1]{\ifnum#1=31 %
\hatcurCCmagxxxxxA
\else
\ifnum#1=32 %
\hatcurCCmagxxxxxB
\else
\ifnum#1=33 %
\hatcurCCmagxxxxxC
\else
\ifnum#1=34 %
\hatcurCCmagxxxxxD
\else
\ifnum#1=35 %
\hatcurCCmagxxxxxE
\else
??????\fi
\fi
\fi
\fi
\fi
}
\newcommand{\hatcurCCpm}[1]{\ifnum#1=31 %
\hatcurCCpmxxxxxA
\else
\ifnum#1=32 %
\hatcurCCpmxxxxxB
\else
\ifnum#1=33 %
\hatcurCCpmxxxxxC
\else
\ifnum#1=34 %
\hatcurCCpmxxxxxD
\else
\ifnum#1=35 %
\hatcurCCpmxxxxxE
\else
??????\fi
\fi
\fi
\fi
\fi
}
\newcommand{\hatcurCCpmdec}[1]{\ifnum#1=31 %
\hatcurCCpmdecxxxxxA
\else
\ifnum#1=32 %
\hatcurCCpmdecxxxxxB
\else
\ifnum#1=33 %
\hatcurCCpmdecxxxxxC
\else
\ifnum#1=34 %
\hatcurCCpmdecxxxxxD
\else
\ifnum#1=35 %
\hatcurCCpmdecxxxxxE
\else
??????\fi
\fi
\fi
\fi
\fi
}
\newcommand{\hatcurCCpmra}[1]{\ifnum#1=31 %
\hatcurCCpmraxxxxxA
\else
\ifnum#1=32 %
\hatcurCCpmraxxxxxB
\else
\ifnum#1=33 %
\hatcurCCpmraxxxxxC
\else
\ifnum#1=34 %
\hatcurCCpmraxxxxxD
\else
\ifnum#1=35 %
\hatcurCCpmraxxxxxE
\else
??????\fi
\fi
\fi
\fi
\fi
}
\newcommand{\hatcurCCra}[1]{\ifnum#1=31 %
\hatcurCCraxxxxxA
\else
\ifnum#1=32 %
\hatcurCCraxxxxxB
\else
\ifnum#1=33 %
\hatcurCCraxxxxxC
\else
\ifnum#1=34 %
\hatcurCCraxxxxxD
\else
\ifnum#1=35 %
\hatcurCCraxxxxxE
\else
??????\fi
\fi
\fi
\fi
\fi
}
\newcommand{\hatcurCCtassmB}[1]{\ifnum#1=31 %
\hatcurCCtassmBxxxxxA
\else
\ifnum#1=32 %
\hatcurCCtassmBxxxxxB
\else
\ifnum#1=33 %
\hatcurCCtassmBxxxxxC
\else
\ifnum#1=34 %
\hatcurCCtassmBxxxxxD
\else
\ifnum#1=35 %
\hatcurCCtassmBxxxxxE
\else
??????\fi
\fi
\fi
\fi
\fi
}
\newcommand{\hatcurCCtassmBshort}[1]{\ifnum#1=31 %
\hatcurCCtassmBshortxxxxxA
\else
\ifnum#1=32 %
\hatcurCCtassmBshortxxxxxB
\else
\ifnum#1=33 %
\hatcurCCtassmBshortxxxxxC
\else
\ifnum#1=34 %
\hatcurCCtassmBshortxxxxxD
\else
\ifnum#1=35 %
\hatcurCCtassmBshortxxxxxE
\else
??????\fi
\fi
\fi
\fi
\fi
}
\newcommand{\hatcurCCtassmg}[1]{\ifnum#1=31 %
\hatcurCCtassmgxxxxxA
\else
\ifnum#1=32 %
\hatcurCCtassmgxxxxxB
\else
\ifnum#1=33 %
\hatcurCCtassmgxxxxxC
\else
\ifnum#1=34 %
\hatcurCCtassmgxxxxxD
\else
\ifnum#1=35 %
\hatcurCCtassmgxxxxxE
\else
??????\fi
\fi
\fi
\fi
\fi
}
\newcommand{\hatcurCCtassmgshort}[1]{\ifnum#1=31 %
\hatcurCCtassmgshortxxxxxA
\else
\ifnum#1=32 %
\hatcurCCtassmgshortxxxxxB
\else
\ifnum#1=33 %
\hatcurCCtassmgshortxxxxxC
\else
\ifnum#1=34 %
\hatcurCCtassmgshortxxxxxD
\else
\ifnum#1=35 %
\hatcurCCtassmgshortxxxxxE
\else
??????\fi
\fi
\fi
\fi
\fi
}
\newcommand{\hatcurCCtassmi}[1]{\ifnum#1=31 %
\hatcurCCtassmixxxxxA
\else
\ifnum#1=32 %
\hatcurCCtassmixxxxxB
\else
\ifnum#1=33 %
\hatcurCCtassmixxxxxC
\else
\ifnum#1=34 %
\hatcurCCtassmixxxxxD
\else
\ifnum#1=35 %
\hatcurCCtassmixxxxxE
\else
??????\fi
\fi
\fi
\fi
\fi
}
\newcommand{\hatcurCCtassmI}[1]{\ifnum#1=31 %
\hatcurCCtassmIxxxxxA
\else
\ifnum#1=32 %
\hatcurCCtassmIxxxxxB
\else
\ifnum#1=33 %
\hatcurCCtassmIxxxxxC
\else
\ifnum#1=34 %
\hatcurCCtassmIxxxxxD
\else
\ifnum#1=35 %
\hatcurCCtassmIxxxxxE
\else
??????\fi
\fi
\fi
\fi
\fi
}
\newcommand{\hatcurCCtassmishort}[1]{\ifnum#1=31 %
\hatcurCCtassmishortxxxxxA
\else
\ifnum#1=32 %
\hatcurCCtassmishortxxxxxB
\else
\ifnum#1=33 %
\hatcurCCtassmishortxxxxxC
\else
\ifnum#1=34 %
\hatcurCCtassmishortxxxxxD
\else
\ifnum#1=35 %
\hatcurCCtassmishortxxxxxE
\else
??????\fi
\fi
\fi
\fi
\fi
}
\newcommand{\hatcurCCtassmIshort}[1]{\ifnum#1=31 %
\hatcurCCtassmIshortxxxxxA
\else
\ifnum#1=32 %
\hatcurCCtassmIshortxxxxxB
\else
\ifnum#1=33 %
\hatcurCCtassmIshortxxxxxC
\else
\ifnum#1=34 %
\hatcurCCtassmIshortxxxxxD
\else
\ifnum#1=35 %
\hatcurCCtassmIshortxxxxxE
\else
??????\fi
\fi
\fi
\fi
\fi
}
\newcommand{\hatcurCCtassmr}[1]{\ifnum#1=31 %
\hatcurCCtassmrxxxxxA
\else
\ifnum#1=32 %
\hatcurCCtassmrxxxxxB
\else
\ifnum#1=33 %
\hatcurCCtassmrxxxxxC
\else
\ifnum#1=34 %
\hatcurCCtassmrxxxxxD
\else
\ifnum#1=35 %
\hatcurCCtassmrxxxxxE
\else
??????\fi
\fi
\fi
\fi
\fi
}
\newcommand{\hatcurCCtassmrshort}[1]{\ifnum#1=31 %
\hatcurCCtassmrshortxxxxxA
\else
\ifnum#1=32 %
\hatcurCCtassmrshortxxxxxB
\else
\ifnum#1=33 %
\hatcurCCtassmrshortxxxxxC
\else
\ifnum#1=34 %
\hatcurCCtassmrshortxxxxxD
\else
\ifnum#1=35 %
\hatcurCCtassmrshortxxxxxE
\else
??????\fi
\fi
\fi
\fi
\fi
}
\newcommand{\hatcurCCtassmv}[1]{\ifnum#1=31 %
\hatcurCCtassmvxxxxxA
\else
\ifnum#1=32 %
\hatcurCCtassmvxxxxxB
\else
\ifnum#1=33 %
\hatcurCCtassmvxxxxxC
\else
\ifnum#1=34 %
\hatcurCCtassmvxxxxxD
\else
\ifnum#1=35 %
\hatcurCCtassmvxxxxxE
\else
??????\fi
\fi
\fi
\fi
\fi
}
\newcommand{\hatcurCCtassmvshort}[1]{\ifnum#1=31 %
\hatcurCCtassmvshortxxxxxA
\else
\ifnum#1=32 %
\hatcurCCtassmvshortxxxxxB
\else
\ifnum#1=33 %
\hatcurCCtassmvshortxxxxxC
\else
\ifnum#1=34 %
\hatcurCCtassmvshortxxxxxD
\else
\ifnum#1=35 %
\hatcurCCtassmvshortxxxxxE
\else
??????\fi
\fi
\fi
\fi
\fi
}
\newcommand{\hatcurCCtwomass}[1]{\ifnum#1=31 %
\hatcurCCtwomassxxxxxA
\else
\ifnum#1=32 %
\hatcurCCtwomassxxxxxB
\else
\ifnum#1=33 %
\hatcurCCtwomassxxxxxC
\else
\ifnum#1=34 %
\hatcurCCtwomassxxxxxD
\else
\ifnum#1=35 %
\hatcurCCtwomassxxxxxE
\else
??????\fi
\fi
\fi
\fi
\fi
}
\newcommand{\hatcurCCtwomassHmag}[1]{\ifnum#1=31 %
\hatcurCCtwomassHmagxxxxxA
\else
\ifnum#1=32 %
\hatcurCCtwomassHmagxxxxxB
\else
\ifnum#1=33 %
\hatcurCCtwomassHmagxxxxxC
\else
\ifnum#1=34 %
\hatcurCCtwomassHmagxxxxxD
\else
\ifnum#1=35 %
\hatcurCCtwomassHmagxxxxxE
\else
??????\fi
\fi
\fi
\fi
\fi
}
\newcommand{\hatcurCCtwomassJmag}[1]{\ifnum#1=31 %
\hatcurCCtwomassJmagxxxxxA
\else
\ifnum#1=32 %
\hatcurCCtwomassJmagxxxxxB
\else
\ifnum#1=33 %
\hatcurCCtwomassJmagxxxxxC
\else
\ifnum#1=34 %
\hatcurCCtwomassJmagxxxxxD
\else
\ifnum#1=35 %
\hatcurCCtwomassJmagxxxxxE
\else
??????\fi
\fi
\fi
\fi
\fi
}
\newcommand{\hatcurCCtwomassKmag}[1]{\ifnum#1=31 %
\hatcurCCtwomassKmagxxxxxA
\else
\ifnum#1=32 %
\hatcurCCtwomassKmagxxxxxB
\else
\ifnum#1=33 %
\hatcurCCtwomassKmagxxxxxC
\else
\ifnum#1=34 %
\hatcurCCtwomassKmagxxxxxD
\else
\ifnum#1=35 %
\hatcurCCtwomassKmagxxxxxE
\else
??????\fi
\fi
\fi
\fi
\fi
}
\newcommand{\hatcurfield}[1]{\ifnum#1=31 %
\hatcurfieldxxxxxA
\else
\ifnum#1=32 %
\hatcurfieldxxxxxB
\else
\ifnum#1=33 %
\hatcurfieldxxxxxC
\else
\ifnum#1=34 %
\hatcurfieldxxxxxD
\else
\ifnum#1=35 %
\hatcurfieldxxxxxE
\else
??????\fi
\fi
\fi
\fi
\fi
}
\newcommand{\hatcurhtr}[1]{\ifnum#1=31 %
\hatcurhtrxxxxxA
\else
\ifnum#1=32 %
\hatcurhtrxxxxxB
\else
\ifnum#1=33 %
\hatcurhtrxxxxxC
\else
\ifnum#1=34 %
\hatcurhtrxxxxxD
\else
\ifnum#1=35 %
\hatcurhtrxxxxxE
\else
??????\fi
\fi
\fi
\fi
\fi
}
\newcommand{\hatcurISOage}[1]{\ifnum#1=31 %
\hatcurISOagexxxxxA
\else
\ifnum#1=32 %
\hatcurISOagexxxxxB
\else
\ifnum#1=33 %
\hatcurISOagexxxxxC
\else
\ifnum#1=34 %
\hatcurISOagexxxxxD
\else
\ifnum#1=35 %
\hatcurISOagexxxxxE
\else
??????\fi
\fi
\fi
\fi
\fi
}
\newcommand{\hatcurISOJK}[1]{\ifnum#1=31 %
\hatcurISOJKxxxxxA
\else
\ifnum#1=32 %
\hatcurISOJKxxxxxB
\else
\ifnum#1=33 %
\hatcurISOJKxxxxxC
\else
\ifnum#1=34 %
\hatcurISOJKxxxxxD
\else
\ifnum#1=35 %
\hatcurISOJKxxxxxE
\else
??????\fi
\fi
\fi
\fi
\fi
}
\newcommand{\hatcurISOlogg}[1]{\ifnum#1=31 %
\hatcurISOloggxxxxxA
\else
\ifnum#1=32 %
\hatcurISOloggxxxxxB
\else
\ifnum#1=33 %
\hatcurISOloggxxxxxC
\else
\ifnum#1=34 %
\hatcurISOloggxxxxxD
\else
\ifnum#1=35 %
\hatcurISOloggxxxxxE
\else
??????\fi
\fi
\fi
\fi
\fi
}
\newcommand{\hatcurISOlum}[1]{\ifnum#1=31 %
\hatcurISOlumxxxxxA
\else
\ifnum#1=32 %
\hatcurISOlumxxxxxB
\else
\ifnum#1=33 %
\hatcurISOlumxxxxxC
\else
\ifnum#1=34 %
\hatcurISOlumxxxxxD
\else
\ifnum#1=35 %
\hatcurISOlumxxxxxE
\else
??????\fi
\fi
\fi
\fi
\fi
}
\newcommand{\hatcurISOlumshort}[1]{\ifnum#1=31 %
\hatcurISOlumshortxxxxxA
\else
\ifnum#1=32 %
\hatcurISOlumshortxxxxxB
\else
\ifnum#1=33 %
\hatcurISOlumshortxxxxxC
\else
\ifnum#1=34 %
\hatcurISOlumshortxxxxxD
\else
\ifnum#1=35 %
\hatcurISOlumshortxxxxxE
\else
??????\fi
\fi
\fi
\fi
\fi
}
\newcommand{\hatcurISOm}[1]{\ifnum#1=31 %
\hatcurISOmxxxxxA
\else
\ifnum#1=32 %
\hatcurISOmxxxxxB
\else
\ifnum#1=33 %
\hatcurISOmxxxxxC
\else
\ifnum#1=34 %
\hatcurISOmxxxxxD
\else
\ifnum#1=35 %
\hatcurISOmxxxxxE
\else
??????\fi
\fi
\fi
\fi
\fi
}
\newcommand{\hatcurISOMH}[1]{\ifnum#1=31 %
\hatcurISOMHxxxxxA
\else
\ifnum#1=32 %
\hatcurISOMHxxxxxB
\else
\ifnum#1=33 %
\hatcurISOMHxxxxxC
\else
\ifnum#1=34 %
\hatcurISOMHxxxxxD
\else
\ifnum#1=35 %
\hatcurISOMHxxxxxE
\else
??????\fi
\fi
\fi
\fi
\fi
}
\newcommand{\hatcurISOMJ}[1]{\ifnum#1=31 %
\hatcurISOMJxxxxxA
\else
\ifnum#1=32 %
\hatcurISOMJxxxxxB
\else
\ifnum#1=33 %
\hatcurISOMJxxxxxC
\else
\ifnum#1=34 %
\hatcurISOMJxxxxxD
\else
\ifnum#1=35 %
\hatcurISOMJxxxxxE
\else
??????\fi
\fi
\fi
\fi
\fi
}
\newcommand{\hatcurISOMK}[1]{\ifnum#1=31 %
\hatcurISOMKxxxxxA
\else
\ifnum#1=32 %
\hatcurISOMKxxxxxB
\else
\ifnum#1=33 %
\hatcurISOMKxxxxxC
\else
\ifnum#1=34 %
\hatcurISOMKxxxxxD
\else
\ifnum#1=35 %
\hatcurISOMKxxxxxE
\else
??????\fi
\fi
\fi
\fi
\fi
}
\newcommand{\hatcurISOmlong}[1]{\ifnum#1=31 %
\hatcurISOmlongxxxxxA
\else
\ifnum#1=32 %
\hatcurISOmlongxxxxxB
\else
\ifnum#1=33 %
\hatcurISOmlongxxxxxC
\else
\ifnum#1=34 %
\hatcurISOmlongxxxxxD
\else
\ifnum#1=35 %
\hatcurISOmlongxxxxxE
\else
??????\fi
\fi
\fi
\fi
\fi
}
\newcommand{\hatcurISOmshort}[1]{\ifnum#1=31 %
\hatcurISOmshortxxxxxA
\else
\ifnum#1=32 %
\hatcurISOmshortxxxxxB
\else
\ifnum#1=33 %
\hatcurISOmshortxxxxxC
\else
\ifnum#1=34 %
\hatcurISOmshortxxxxxD
\else
\ifnum#1=35 %
\hatcurISOmshortxxxxxE
\else
??????\fi
\fi
\fi
\fi
\fi
}
\newcommand{\hatcurISOmv}[1]{\ifnum#1=31 %
\hatcurISOmvxxxxxA
\else
\ifnum#1=32 %
\hatcurISOmvxxxxxB
\else
\ifnum#1=33 %
\hatcurISOmvxxxxxC
\else
\ifnum#1=34 %
\hatcurISOmvxxxxxD
\else
\ifnum#1=35 %
\hatcurISOmvxxxxxE
\else
??????\fi
\fi
\fi
\fi
\fi
}
\newcommand{\hatcurISOr}[1]{\ifnum#1=31 %
\hatcurISOrxxxxxA
\else
\ifnum#1=32 %
\hatcurISOrxxxxxB
\else
\ifnum#1=33 %
\hatcurISOrxxxxxC
\else
\ifnum#1=34 %
\hatcurISOrxxxxxD
\else
\ifnum#1=35 %
\hatcurISOrxxxxxE
\else
??????\fi
\fi
\fi
\fi
\fi
}
\newcommand{\hatcurISOrho}[1]{\ifnum#1=31 %
\hatcurISOrhoxxxxxA
\else
\ifnum#1=32 %
\hatcurISOrhoxxxxxB
\else
\ifnum#1=33 %
\hatcurISOrhoxxxxxC
\else
\ifnum#1=34 %
\hatcurISOrhoxxxxxD
\else
\ifnum#1=35 %
\hatcurISOrhoxxxxxE
\else
??????\fi
\fi
\fi
\fi
\fi
}
\newcommand{\hatcurISOrholong}[1]{\ifnum#1=31 %
\hatcurISOrholongxxxxxA
\else
\ifnum#1=32 %
\hatcurISOrholongxxxxxB
\else
\ifnum#1=33 %
\hatcurISOrholongxxxxxC
\else
\ifnum#1=34 %
\hatcurISOrholongxxxxxD
\else
\ifnum#1=35 %
\hatcurISOrholongxxxxxE
\else
??????\fi
\fi
\fi
\fi
\fi
}
\newcommand{\hatcurISOrlong}[1]{\ifnum#1=31 %
\hatcurISOrlongxxxxxA
\else
\ifnum#1=32 %
\hatcurISOrlongxxxxxB
\else
\ifnum#1=33 %
\hatcurISOrlongxxxxxC
\else
\ifnum#1=34 %
\hatcurISOrlongxxxxxD
\else
\ifnum#1=35 %
\hatcurISOrlongxxxxxE
\else
??????\fi
\fi
\fi
\fi
\fi
}
\newcommand{\hatcurISOrshort}[1]{\ifnum#1=31 %
\hatcurISOrshortxxxxxA
\else
\ifnum#1=32 %
\hatcurISOrshortxxxxxB
\else
\ifnum#1=33 %
\hatcurISOrshortxxxxxC
\else
\ifnum#1=34 %
\hatcurISOrshortxxxxxD
\else
\ifnum#1=35 %
\hatcurISOrshortxxxxxE
\else
??????\fi
\fi
\fi
\fi
\fi
}
\newcommand{\hatcurISOsigma}[1]{\ifnum#1=31 %
\hatcurISOsigmaxxxxxA
\else
\ifnum#1=32 %
\hatcurISOsigmaxxxxxB
\else
\ifnum#1=33 %
\hatcurISOsigmaxxxxxC
\else
\ifnum#1=34 %
\hatcurISOsigmaxxxxxD
\else
\ifnum#1=35 %
\hatcurISOsigmaxxxxxE
\else
??????\fi
\fi
\fi
\fi
\fi
}
\newcommand{\hatcurISOspec}[1]{\ifnum#1=31 %
\hatcurISOspecxxxxxA
\else
\ifnum#1=32 %
\hatcurISOspecxxxxxB
\else
\ifnum#1=33 %
\hatcurISOspecxxxxxC
\else
\ifnum#1=34 %
\hatcurISOspecxxxxxD
\else
\ifnum#1=35 %
\hatcurISOspecxxxxxE
\else
??????\fi
\fi
\fi
\fi
\fi
}
\newcommand{\hatcurISOvi}[1]{\ifnum#1=31 %
\hatcurISOvixxxxxA
\else
\ifnum#1=32 %
\hatcurISOvixxxxxB
\else
\ifnum#1=33 %
\hatcurISOvixxxxxC
\else
\ifnum#1=34 %
\hatcurISOvixxxxxD
\else
\ifnum#1=35 %
\hatcurISOvixxxxxE
\else
??????\fi
\fi
\fi
\fi
\fi
}
\newcommand{\hatcurLBig}[1]{\ifnum#1=31 %
\hatcurLBigxxxxxA
\else
\ifnum#1=32 %
\hatcurLBigxxxxxB
\else
\ifnum#1=33 %
\hatcurLBigxxxxxC
\else
\ifnum#1=34 %
\hatcurLBigxxxxxD
\else
\ifnum#1=35 %
\hatcurLBigxxxxxE
\else
??????\fi
\fi
\fi
\fi
\fi
}
\newcommand{\hatcurLBii}[1]{\ifnum#1=31 %
\hatcurLBiixxxxxA
\else
\ifnum#1=32 %
\hatcurLBiixxxxxB
\else
\ifnum#1=33 %
\hatcurLBiixxxxxC
\else
\ifnum#1=34 %
\hatcurLBiixxxxxD
\else
\ifnum#1=35 %
\hatcurLBiixxxxxE
\else
??????\fi
\fi
\fi
\fi
\fi
}
\newcommand{\hatcurLBiI}[1]{\ifnum#1=31 %
\hatcurLBiIxxxxxA
\else
\ifnum#1=32 %
\hatcurLBiIxxxxxB
\else
\ifnum#1=33 %
\hatcurLBiIxxxxxC
\else
\ifnum#1=34 %
\hatcurLBiIxxxxxD
\else
\ifnum#1=35 %
\hatcurLBiIxxxxxE
\else
??????\fi
\fi
\fi
\fi
\fi
}
\newcommand{\hatcurLBiig}[1]{\ifnum#1=31 %
\hatcurLBiigxxxxxA
\else
\ifnum#1=32 %
\hatcurLBiigxxxxxB
\else
\ifnum#1=33 %
\hatcurLBiigxxxxxC
\else
\ifnum#1=34 %
\hatcurLBiigxxxxxD
\else
\ifnum#1=35 %
\hatcurLBiigxxxxxE
\else
??????\fi
\fi
\fi
\fi
\fi
}
\newcommand{\hatcurLBiii}[1]{\ifnum#1=31 %
\hatcurLBiiixxxxxA
\else
\ifnum#1=32 %
\hatcurLBiiixxxxxB
\else
\ifnum#1=33 %
\hatcurLBiiixxxxxC
\else
\ifnum#1=34 %
\hatcurLBiiixxxxxD
\else
\ifnum#1=35 %
\hatcurLBiiixxxxxE
\else
??????\fi
\fi
\fi
\fi
\fi
}
\newcommand{\hatcurLBiiI}[1]{\ifnum#1=31 %
\hatcurLBiiIxxxxxA
\else
\ifnum#1=32 %
\hatcurLBiiIxxxxxB
\else
\ifnum#1=33 %
\hatcurLBiiIxxxxxC
\else
\ifnum#1=34 %
\hatcurLBiiIxxxxxD
\else
\ifnum#1=35 %
\hatcurLBiiIxxxxxE
\else
??????\fi
\fi
\fi
\fi
\fi
}
\newcommand{\hatcurLBiikep}[1]{\ifnum#1=31 %
\hatcurLBiikepxxxxxA
\else
\ifnum#1=33 %
\hatcurLBiikepxxxxxC
\else
\ifnum#1=34 %
\hatcurLBiikepxxxxxD
\else
\ifnum#1=35 %
\hatcurLBiikepxxxxxE
\else
??????\fi
\fi
\fi
\fi
}
\newcommand{\hatcurLBiir}[1]{\ifnum#1=31 %
\hatcurLBiirxxxxxA
\else
\ifnum#1=32 %
\hatcurLBiirxxxxxB
\else
\ifnum#1=33 %
\hatcurLBiirxxxxxC
\else
\ifnum#1=34 %
\hatcurLBiirxxxxxD
\else
\ifnum#1=35 %
\hatcurLBiirxxxxxE
\else
??????\fi
\fi
\fi
\fi
\fi
}
\newcommand{\hatcurLBiiR}[1]{\ifnum#1=31 %
\hatcurLBiiRxxxxxA
\else
\ifnum#1=32 %
\hatcurLBiiRxxxxxB
\else
\ifnum#1=33 %
\hatcurLBiiRxxxxxC
\else
\ifnum#1=34 %
\hatcurLBiiRxxxxxD
\else
\ifnum#1=35 %
\hatcurLBiiRxxxxxE
\else
??????\fi
\fi
\fi
\fi
\fi
}
\newcommand{\hatcurLBiiz}[1]{\ifnum#1=31 %
\hatcurLBiizxxxxxA
\else
\ifnum#1=32 %
\hatcurLBiizxxxxxB
\else
\ifnum#1=33 %
\hatcurLBiizxxxxxC
\else
\ifnum#1=34 %
\hatcurLBiizxxxxxD
\else
\ifnum#1=35 %
\hatcurLBiizxxxxxE
\else
??????\fi
\fi
\fi
\fi
\fi
}
\newcommand{\hatcurLBikep}[1]{\ifnum#1=31 %
\hatcurLBikepxxxxxA
\else
\ifnum#1=33 %
\hatcurLBikepxxxxxC
\else
\ifnum#1=34 %
\hatcurLBikepxxxxxD
\else
\ifnum#1=35 %
\hatcurLBikepxxxxxE
\else
??????\fi
\fi
\fi
\fi
}
\newcommand{\hatcurLBir}[1]{\ifnum#1=31 %
\hatcurLBirxxxxxA
\else
\ifnum#1=32 %
\hatcurLBirxxxxxB
\else
\ifnum#1=33 %
\hatcurLBirxxxxxC
\else
\ifnum#1=34 %
\hatcurLBirxxxxxD
\else
\ifnum#1=35 %
\hatcurLBirxxxxxE
\else
??????\fi
\fi
\fi
\fi
\fi
}
\newcommand{\hatcurLBiR}[1]{\ifnum#1=31 %
\hatcurLBiRxxxxxA
\else
\ifnum#1=32 %
\hatcurLBiRxxxxxB
\else
\ifnum#1=33 %
\hatcurLBiRxxxxxC
\else
\ifnum#1=34 %
\hatcurLBiRxxxxxD
\else
\ifnum#1=35 %
\hatcurLBiRxxxxxE
\else
??????\fi
\fi
\fi
\fi
\fi
}
\newcommand{\hatcurLBiz}[1]{\ifnum#1=31 %
\hatcurLBizxxxxxA
\else
\ifnum#1=32 %
\hatcurLBizxxxxxB
\else
\ifnum#1=33 %
\hatcurLBizxxxxxC
\else
\ifnum#1=34 %
\hatcurLBizxxxxxD
\else
\ifnum#1=35 %
\hatcurLBizxxxxxE
\else
??????\fi
\fi
\fi
\fi
\fi
}
\newcommand{\hatcurLCbsq}[1]{\ifnum#1=31 %
\hatcurLCbsqxxxxxA
\else
\ifnum#1=32 %
\hatcurLCbsqxxxxxB
\else
\ifnum#1=33 %
\hatcurLCbsqxxxxxC
\else
\ifnum#1=34 %
\hatcurLCbsqxxxxxD
\else
\ifnum#1=35 %
\hatcurLCbsqxxxxxE
\else
??????\fi
\fi
\fi
\fi
\fi
}
\newcommand{\hatcurLCdip}[1]{\ifnum#1=31 %
\hatcurLCdipxxxxxA
\else
\ifnum#1=32 %
\hatcurLCdipxxxxxB
\else
\ifnum#1=33 %
\hatcurLCdipxxxxxC
\else
\ifnum#1=34 %
\hatcurLCdipxxxxxD
\else
\ifnum#1=35 %
\hatcurLCdipxxxxxE
\else
??????\fi
\fi
\fi
\fi
\fi
}
\newcommand{\hatcurLCdur}[1]{\ifnum#1=31 %
\hatcurLCdurxxxxxA
\else
\ifnum#1=32 %
\hatcurLCdurxxxxxB
\else
\ifnum#1=33 %
\hatcurLCdurxxxxxC
\else
\ifnum#1=34 %
\hatcurLCdurxxxxxD
\else
\ifnum#1=35 %
\hatcurLCdurxxxxxE
\else
??????\fi
\fi
\fi
\fi
\fi
}
\newcommand{\hatcurLCdurhr}[1]{\ifnum#1=31 %
\hatcurLCdurhrxxxxxA
\else
\ifnum#1=32 %
\hatcurLCdurhrxxxxxB
\else
\ifnum#1=33 %
\hatcurLCdurhrxxxxxC
\else
\ifnum#1=34 %
\hatcurLCdurhrxxxxxD
\else
\ifnum#1=35 %
\hatcurLCdurhrxxxxxE
\else
??????\fi
\fi
\fi
\fi
\fi
}
\newcommand{\hatcurLCdurhrshort}[1]{\ifnum#1=31 %
\hatcurLCdurhrshortxxxxxA
\else
\ifnum#1=32 %
\hatcurLCdurhrshortxxxxxB
\else
\ifnum#1=33 %
\hatcurLCdurhrshortxxxxxC
\else
\ifnum#1=34 %
\hatcurLCdurhrshortxxxxxD
\else
\ifnum#1=35 %
\hatcurLCdurhrshortxxxxxE
\else
??????\fi
\fi
\fi
\fi
\fi
}
\newcommand{\hatcurLCdurshort}[1]{\ifnum#1=31 %
\hatcurLCdurshortxxxxxA
\else
\ifnum#1=32 %
\hatcurLCdurshortxxxxxB
\else
\ifnum#1=33 %
\hatcurLCdurshortxxxxxC
\else
\ifnum#1=34 %
\hatcurLCdurshortxxxxxD
\else
\ifnum#1=35 %
\hatcurLCdurshortxxxxxE
\else
??????\fi
\fi
\fi
\fi
\fi
}
\newcommand{\hatcurLChatnetm}[1]{\ifnum#1=31 %
\hatcurLChatnetmxxxxxA
\else
\ifnum#1=32 %
\hatcurLChatnetmxxxxxB
\else
\ifnum#1=33 %
\hatcurLChatnetmxxxxxC
\else
\ifnum#1=34 %
\hatcurLChatnetmxxxxxD
\else
\ifnum#1=35 %
\hatcurLChatnetmxxxxxE
\else
??????\fi
\fi
\fi
\fi
\fi
}
\newcommand{\hatcurLCiblend}[1]{\ifnum#1=31 %
\hatcurLCiblendxxxxxA
\else
\ifnum#1=32 %
\hatcurLCiblendxxxxxB
\else
\ifnum#1=33 %
\hatcurLCiblendxxxxxC
\else
\ifnum#1=34 %
\hatcurLCiblendxxxxxD
\else
\ifnum#1=35 %
\hatcurLCiblendxxxxxE
\else
??????\fi
\fi
\fi
\fi
\fi
}
\newcommand{\hatcurLCimp}[1]{\ifnum#1=31 %
\hatcurLCimpxxxxxA
\else
\ifnum#1=32 %
\hatcurLCimpxxxxxB
\else
\ifnum#1=33 %
\hatcurLCimpxxxxxC
\else
\ifnum#1=34 %
\hatcurLCimpxxxxxD
\else
\ifnum#1=35 %
\hatcurLCimpxxxxxE
\else
??????\fi
\fi
\fi
\fi
\fi
}
\newcommand{\hatcurLCingdur}[1]{\ifnum#1=31 %
\hatcurLCingdurxxxxxA
\else
\ifnum#1=32 %
\hatcurLCingdurxxxxxB
\else
\ifnum#1=33 %
\hatcurLCingdurxxxxxC
\else
\ifnum#1=34 %
\hatcurLCingdurxxxxxD
\else
\ifnum#1=35 %
\hatcurLCingdurxxxxxE
\else
??????\fi
\fi
\fi
\fi
\fi
}
\newcommand{\hatcurLCP}[1]{\ifnum#1=31 %
\hatcurLCPxxxxxA
\else
\ifnum#1=32 %
\hatcurLCPxxxxxB
\else
\ifnum#1=33 %
\hatcurLCPxxxxxC
\else
\ifnum#1=34 %
\hatcurLCPxxxxxD
\else
\ifnum#1=35 %
\hatcurLCPxxxxxE
\else
??????\fi
\fi
\fi
\fi
\fi
}
\newcommand{\hatcurLCPprec}[1]{\ifnum#1=31 %
\hatcurLCPprecxxxxxA
\else
\ifnum#1=32 %
\hatcurLCPprecxxxxxB
\else
\ifnum#1=33 %
\hatcurLCPprecxxxxxC
\else
\ifnum#1=34 %
\hatcurLCPprecxxxxxD
\else
\ifnum#1=35 %
\hatcurLCPprecxxxxxE
\else
??????\fi
\fi
\fi
\fi
\fi
}
\newcommand{\hatcurLCPshort}[1]{\ifnum#1=31 %
\hatcurLCPshortxxxxxA
\else
\ifnum#1=32 %
\hatcurLCPshortxxxxxB
\else
\ifnum#1=33 %
\hatcurLCPshortxxxxxC
\else
\ifnum#1=34 %
\hatcurLCPshortxxxxxD
\else
\ifnum#1=35 %
\hatcurLCPshortxxxxxE
\else
??????\fi
\fi
\fi
\fi
\fi
}
\newcommand{\hatcurLCq}[1]{\ifnum#1=31 %
\hatcurLCqxxxxxA
\else
\ifnum#1=32 %
\hatcurLCqxxxxxB
\else
\ifnum#1=33 %
\hatcurLCqxxxxxC
\else
\ifnum#1=34 %
\hatcurLCqxxxxxD
\else
\ifnum#1=35 %
\hatcurLCqxxxxxE
\else
??????\fi
\fi
\fi
\fi
\fi
}
\newcommand{\hatcurLCqshort}[1]{\ifnum#1=31 %
\hatcurLCqshortxxxxxA
\else
\ifnum#1=32 %
\hatcurLCqshortxxxxxB
\else
\ifnum#1=33 %
\hatcurLCqshortxxxxxC
\else
\ifnum#1=34 %
\hatcurLCqshortxxxxxD
\else
\ifnum#1=35 %
\hatcurLCqshortxxxxxE
\else
??????\fi
\fi
\fi
\fi
\fi
}
\newcommand{\hatcurLCrho}[1]{\ifnum#1=31 %
\hatcurLCrhoxxxxxA
\else
\ifnum#1=32 %
\hatcurLCrhoxxxxxB
\else
\ifnum#1=33 %
\hatcurLCrhoxxxxxC
\else
\ifnum#1=34 %
\hatcurLCrhoxxxxxD
\else
\ifnum#1=35 %
\hatcurLCrhoxxxxxE
\else
??????\fi
\fi
\fi
\fi
\fi
}
\newcommand{\hatcurLCrprstar}[1]{\ifnum#1=31 %
\hatcurLCrprstarxxxxxA
\else
\ifnum#1=32 %
\hatcurLCrprstarxxxxxB
\else
\ifnum#1=33 %
\hatcurLCrprstarxxxxxC
\else
\ifnum#1=34 %
\hatcurLCrprstarxxxxxD
\else
\ifnum#1=35 %
\hatcurLCrprstarxxxxxE
\else
??????\fi
\fi
\fi
\fi
\fi
}
\newcommand{\hatcurLCT}[1]{\ifnum#1=31 %
\hatcurLCTxxxxxA
\else
\ifnum#1=32 %
\hatcurLCTxxxxxB
\else
\ifnum#1=33 %
\hatcurLCTxxxxxC
\else
\ifnum#1=34 %
\hatcurLCTxxxxxD
\else
\ifnum#1=35 %
\hatcurLCTxxxxxE
\else
??????\fi
\fi
\fi
\fi
\fi
}
\newcommand{\hatcurLCTA}[1]{\ifnum#1=31 %
\hatcurLCTAxxxxxA
\else
\ifnum#1=32 %
\hatcurLCTAxxxxxB
\else
\ifnum#1=33 %
\hatcurLCTAxxxxxC
\else
\ifnum#1=34 %
\hatcurLCTAxxxxxD
\else
\ifnum#1=35 %
\hatcurLCTAxxxxxE
\else
??????\fi
\fi
\fi
\fi
\fi
}
\newcommand{\hatcurLCTB}[1]{\ifnum#1=31 %
\hatcurLCTBxxxxxA
\else
\ifnum#1=32 %
\hatcurLCTBxxxxxB
\else
\ifnum#1=33 %
\hatcurLCTBxxxxxC
\else
\ifnum#1=34 %
\hatcurLCTBxxxxxD
\else
\ifnum#1=35 %
\hatcurLCTBxxxxxE
\else
??????\fi
\fi
\fi
\fi
\fi
}
\newcommand{\hatcurLCzeta}[1]{\ifnum#1=31 %
\hatcurLCzetaxxxxxA
\else
\ifnum#1=32 %
\hatcurLCzetaxxxxxB
\else
\ifnum#1=33 %
\hatcurLCzetaxxxxxC
\else
\ifnum#1=34 %
\hatcurLCzetaxxxxxD
\else
\ifnum#1=35 %
\hatcurLCzetaxxxxxE
\else
??????\fi
\fi
\fi
\fi
\fi
}
\newcommand{\hatcurPPaequiv}[1]{\ifnum#1=31 %
\hatcurPPaequivxxxxxA
\else
\ifnum#1=32 %
\hatcurPPaequivxxxxxB
\else
\ifnum#1=33 %
\hatcurPPaequivxxxxxC
\else
\ifnum#1=34 %
\hatcurPPaequivxxxxxD
\else
\ifnum#1=35 %
\hatcurPPaequivxxxxxE
\else
??????\fi
\fi
\fi
\fi
\fi
}
\newcommand{\hatcurPPar}[1]{\ifnum#1=31 %
\hatcurPParxxxxxA
\else
\ifnum#1=32 %
\hatcurPParxxxxxB
\else
\ifnum#1=33 %
\hatcurPParxxxxxC
\else
\ifnum#1=34 %
\hatcurPParxxxxxD
\else
\ifnum#1=35 %
\hatcurPParxxxxxE
\else
??????\fi
\fi
\fi
\fi
\fi
}
\newcommand{\hatcurPParel}[1]{\ifnum#1=31 %
\hatcurPParelxxxxxA
\else
\ifnum#1=32 %
\hatcurPParelxxxxxB
\else
\ifnum#1=33 %
\hatcurPParelxxxxxC
\else
\ifnum#1=34 %
\hatcurPParelxxxxxD
\else
\ifnum#1=35 %
\hatcurPParelxxxxxE
\else
??????\fi
\fi
\fi
\fi
\fi
}
\newcommand{\hatcurPPfluxap}[1]{\ifnum#1=31 %
\hatcurPPfluxapxxxxxA
\else
\ifnum#1=32 %
\hatcurPPfluxapxxxxxB
\else
\ifnum#1=33 %
\hatcurPPfluxapxxxxxC
\else
\ifnum#1=34 %
\hatcurPPfluxapxxxxxD
\else
\ifnum#1=35 %
\hatcurPPfluxapxxxxxE
\else
??????\fi
\fi
\fi
\fi
\fi
}
\newcommand{\hatcurPPfluxapdim}[1]{\ifnum#1=31 %
\hatcurPPfluxapdimxxxxxA
\else
\ifnum#1=32 %
\hatcurPPfluxapdimxxxxxB
\else
\ifnum#1=33 %
\hatcurPPfluxapdimxxxxxC
\else
\ifnum#1=34 %
\hatcurPPfluxapdimxxxxxD
\else
\ifnum#1=35 %
\hatcurPPfluxapdimxxxxxE
\else
??????\fi
\fi
\fi
\fi
\fi
}
\newcommand{\hatcurPPfluxavg}[1]{\ifnum#1=31 %
\hatcurPPfluxavgxxxxxA
\else
\ifnum#1=32 %
\hatcurPPfluxavgxxxxxB
\else
\ifnum#1=33 %
\hatcurPPfluxavgxxxxxC
\else
\ifnum#1=34 %
\hatcurPPfluxavgxxxxxD
\else
\ifnum#1=35 %
\hatcurPPfluxavgxxxxxE
\else
??????\fi
\fi
\fi
\fi
\fi
}
\newcommand{\hatcurPPfluxavgdim}[1]{\ifnum#1=31 %
\hatcurPPfluxavgdimxxxxxA
\else
\ifnum#1=32 %
\hatcurPPfluxavgdimxxxxxB
\else
\ifnum#1=33 %
\hatcurPPfluxavgdimxxxxxC
\else
\ifnum#1=34 %
\hatcurPPfluxavgdimxxxxxD
\else
\ifnum#1=35 %
\hatcurPPfluxavgdimxxxxxE
\else
??????\fi
\fi
\fi
\fi
\fi
}
\newcommand{\hatcurPPfluxavglog}[1]{\ifnum#1=31 %
\hatcurPPfluxavglogxxxxxA
\else
\ifnum#1=32 %
\hatcurPPfluxavglogxxxxxB
\else
\ifnum#1=33 %
\hatcurPPfluxavglogxxxxxC
\else
\ifnum#1=34 %
\hatcurPPfluxavglogxxxxxD
\else
\ifnum#1=35 %
\hatcurPPfluxavglogxxxxxE
\else
??????\fi
\fi
\fi
\fi
\fi
}
\newcommand{\hatcurPPfluxperi}[1]{\ifnum#1=31 %
\hatcurPPfluxperixxxxxA
\else
\ifnum#1=32 %
\hatcurPPfluxperixxxxxB
\else
\ifnum#1=33 %
\hatcurPPfluxperixxxxxC
\else
\ifnum#1=34 %
\hatcurPPfluxperixxxxxD
\else
\ifnum#1=35 %
\hatcurPPfluxperixxxxxE
\else
??????\fi
\fi
\fi
\fi
\fi
}
\newcommand{\hatcurPPfluxperidim}[1]{\ifnum#1=31 %
\hatcurPPfluxperidimxxxxxA
\else
\ifnum#1=32 %
\hatcurPPfluxperidimxxxxxB
\else
\ifnum#1=33 %
\hatcurPPfluxperidimxxxxxC
\else
\ifnum#1=34 %
\hatcurPPfluxperidimxxxxxD
\else
\ifnum#1=35 %
\hatcurPPfluxperidimxxxxxE
\else
??????\fi
\fi
\fi
\fi
\fi
}
\newcommand{\hatcurPPg}[1]{\ifnum#1=31 %
\hatcurPPgxxxxxA
\else
\ifnum#1=32 %
\hatcurPPgxxxxxB
\else
\ifnum#1=33 %
\hatcurPPgxxxxxC
\else
\ifnum#1=34 %
\hatcurPPgxxxxxD
\else
\ifnum#1=35 %
\hatcurPPgxxxxxE
\else
??????\fi
\fi
\fi
\fi
\fi
}
\newcommand{\hatcurPPi}[1]{\ifnum#1=31 %
\hatcurPPixxxxxA
\else
\ifnum#1=32 %
\hatcurPPixxxxxB
\else
\ifnum#1=33 %
\hatcurPPixxxxxC
\else
\ifnum#1=34 %
\hatcurPPixxxxxD
\else
\ifnum#1=35 %
\hatcurPPixxxxxE
\else
??????\fi
\fi
\fi
\fi
\fi
}
\newcommand{\hatcurPPlogg}[1]{\ifnum#1=31 %
\hatcurPPloggxxxxxA
\else
\ifnum#1=32 %
\hatcurPPloggxxxxxB
\else
\ifnum#1=33 %
\hatcurPPloggxxxxxC
\else
\ifnum#1=34 %
\hatcurPPloggxxxxxD
\else
\ifnum#1=35 %
\hatcurPPloggxxxxxE
\else
??????\fi
\fi
\fi
\fi
\fi
}
\newcommand{\hatcurPPm}[1]{\ifnum#1=31 %
\hatcurPPmxxxxxA
\else
\ifnum#1=32 %
\hatcurPPmxxxxxB
\else
\ifnum#1=33 %
\hatcurPPmxxxxxC
\else
\ifnum#1=34 %
\hatcurPPmxxxxxD
\else
\ifnum#1=35 %
\hatcurPPmxxxxxE
\else
??????\fi
\fi
\fi
\fi
\fi
}
\newcommand{\hatcurPPme}[1]{\ifnum#1=31 %
\hatcurPPmexxxxxA
\else
\ifnum#1=32 %
\hatcurPPmexxxxxB
\else
\ifnum#1=33 %
\hatcurPPmexxxxxC
\else
\ifnum#1=34 %
\hatcurPPmexxxxxD
\else
\ifnum#1=35 %
\hatcurPPmexxxxxE
\else
??????\fi
\fi
\fi
\fi
\fi
}
\newcommand{\hatcurPPmelong}[1]{\ifnum#1=31 %
\hatcurPPmelongxxxxxA
\else
\ifnum#1=32 %
\hatcurPPmelongxxxxxB
\else
\ifnum#1=33 %
\hatcurPPmelongxxxxxC
\else
\ifnum#1=34 %
\hatcurPPmelongxxxxxD
\else
\ifnum#1=35 %
\hatcurPPmelongxxxxxE
\else
??????\fi
\fi
\fi
\fi
\fi
}
\newcommand{\hatcurPPmeshort}[1]{\ifnum#1=31 %
\hatcurPPmeshortxxxxxA
\else
\ifnum#1=32 %
\hatcurPPmeshortxxxxxB
\else
\ifnum#1=33 %
\hatcurPPmeshortxxxxxC
\else
\ifnum#1=34 %
\hatcurPPmeshortxxxxxD
\else
\ifnum#1=35 %
\hatcurPPmeshortxxxxxE
\else
??????\fi
\fi
\fi
\fi
\fi
}
\newcommand{\hatcurPPmlong}[1]{\ifnum#1=31 %
\hatcurPPmlongxxxxxA
\else
\ifnum#1=32 %
\hatcurPPmlongxxxxxB
\else
\ifnum#1=33 %
\hatcurPPmlongxxxxxC
\else
\ifnum#1=34 %
\hatcurPPmlongxxxxxD
\else
\ifnum#1=35 %
\hatcurPPmlongxxxxxE
\else
??????\fi
\fi
\fi
\fi
\fi
}
\newcommand{\hatcurPPmrcorr}[1]{\ifnum#1=31 %
\hatcurPPmrcorrxxxxxA
\else
\ifnum#1=32 %
\hatcurPPmrcorrxxxxxB
\else
\ifnum#1=33 %
\hatcurPPmrcorrxxxxxC
\else
\ifnum#1=34 %
\hatcurPPmrcorrxxxxxD
\else
\ifnum#1=35 %
\hatcurPPmrcorrxxxxxE
\else
??????\fi
\fi
\fi
\fi
\fi
}
\newcommand{\hatcurPPmshort}[1]{\ifnum#1=31 %
\hatcurPPmshortxxxxxA
\else
\ifnum#1=32 %
\hatcurPPmshortxxxxxB
\else
\ifnum#1=33 %
\hatcurPPmshortxxxxxC
\else
\ifnum#1=34 %
\hatcurPPmshortxxxxxD
\else
\ifnum#1=35 %
\hatcurPPmshortxxxxxE
\else
??????\fi
\fi
\fi
\fi
\fi
}
\newcommand{\hatcurPPperi}[1]{\ifnum#1=31 %
\hatcurPPperixxxxxA
\else
\ifnum#1=32 %
\hatcurPPperixxxxxB
\else
\ifnum#1=33 %
\hatcurPPperixxxxxC
\else
\ifnum#1=34 %
\hatcurPPperixxxxxD
\else
\ifnum#1=35 %
\hatcurPPperixxxxxE
\else
??????\fi
\fi
\fi
\fi
\fi
}
\newcommand{\hatcurPPphiconj}[1]{\ifnum#1=31 %
\hatcurPPphiconjxxxxxA
\else
\ifnum#1=32 %
\hatcurPPphiconjxxxxxB
\else
\ifnum#1=33 %
\hatcurPPphiconjxxxxxC
\else
\ifnum#1=34 %
\hatcurPPphiconjxxxxxD
\else
\ifnum#1=35 %
\hatcurPPphiconjxxxxxE
\else
??????\fi
\fi
\fi
\fi
\fi
}
\newcommand{\hatcurPPr}[1]{\ifnum#1=31 %
\hatcurPPrxxxxxA
\else
\ifnum#1=32 %
\hatcurPPrxxxxxB
\else
\ifnum#1=33 %
\hatcurPPrxxxxxC
\else
\ifnum#1=34 %
\hatcurPPrxxxxxD
\else
\ifnum#1=35 %
\hatcurPPrxxxxxE
\else
??????\fi
\fi
\fi
\fi
\fi
}
\newcommand{\hatcurPPre}[1]{\ifnum#1=31 %
\hatcurPPrexxxxxA
\else
\ifnum#1=32 %
\hatcurPPrexxxxxB
\else
\ifnum#1=33 %
\hatcurPPrexxxxxC
\else
\ifnum#1=34 %
\hatcurPPrexxxxxD
\else
\ifnum#1=35 %
\hatcurPPrexxxxxE
\else
??????\fi
\fi
\fi
\fi
\fi
}
\newcommand{\hatcurPPrelong}[1]{\ifnum#1=31 %
\hatcurPPrelongxxxxxA
\else
\ifnum#1=32 %
\hatcurPPrelongxxxxxB
\else
\ifnum#1=33 %
\hatcurPPrelongxxxxxC
\else
\ifnum#1=34 %
\hatcurPPrelongxxxxxD
\else
\ifnum#1=35 %
\hatcurPPrelongxxxxxE
\else
??????\fi
\fi
\fi
\fi
\fi
}
\newcommand{\hatcurPPreshort}[1]{\ifnum#1=31 %
\hatcurPPreshortxxxxxA
\else
\ifnum#1=32 %
\hatcurPPreshortxxxxxB
\else
\ifnum#1=33 %
\hatcurPPreshortxxxxxC
\else
\ifnum#1=34 %
\hatcurPPreshortxxxxxD
\else
\ifnum#1=35 %
\hatcurPPreshortxxxxxE
\else
??????\fi
\fi
\fi
\fi
\fi
}
\newcommand{\hatcurPPrho}[1]{\ifnum#1=31 %
\hatcurPPrhoxxxxxA
\else
\ifnum#1=32 %
\hatcurPPrhoxxxxxB
\else
\ifnum#1=33 %
\hatcurPPrhoxxxxxC
\else
\ifnum#1=34 %
\hatcurPPrhoxxxxxD
\else
\ifnum#1=35 %
\hatcurPPrhoxxxxxE
\else
??????\fi
\fi
\fi
\fi
\fi
}
\newcommand{\hatcurPPrlong}[1]{\ifnum#1=31 %
\hatcurPPrlongxxxxxA
\else
\ifnum#1=32 %
\hatcurPPrlongxxxxxB
\else
\ifnum#1=33 %
\hatcurPPrlongxxxxxC
\else
\ifnum#1=34 %
\hatcurPPrlongxxxxxD
\else
\ifnum#1=35 %
\hatcurPPrlongxxxxxE
\else
??????\fi
\fi
\fi
\fi
\fi
}
\newcommand{\hatcurPPrshort}[1]{\ifnum#1=31 %
\hatcurPPrshortxxxxxA
\else
\ifnum#1=32 %
\hatcurPPrshortxxxxxB
\else
\ifnum#1=33 %
\hatcurPPrshortxxxxxC
\else
\ifnum#1=34 %
\hatcurPPrshortxxxxxD
\else
\ifnum#1=35 %
\hatcurPPrshortxxxxxE
\else
??????\fi
\fi
\fi
\fi
\fi
}
\newcommand{\hatcurPPtcirc}[1]{\ifnum#1=31 %
\hatcurPPtcircxxxxxA
\else
\ifnum#1=32 %
\hatcurPPtcircxxxxxB
\else
\ifnum#1=33 %
\hatcurPPtcircxxxxxC
\else
\ifnum#1=34 %
\hatcurPPtcircxxxxxD
\else
\ifnum#1=35 %
\hatcurPPtcircxxxxxE
\else
??????\fi
\fi
\fi
\fi
\fi
}
\newcommand{\hatcurPPteff}[1]{\ifnum#1=31 %
\hatcurPPteffxxxxxA
\else
\ifnum#1=32 %
\hatcurPPteffxxxxxB
\else
\ifnum#1=33 %
\hatcurPPteffxxxxxC
\else
\ifnum#1=34 %
\hatcurPPteffxxxxxD
\else
\ifnum#1=35 %
\hatcurPPteffxxxxxE
\else
??????\fi
\fi
\fi
\fi
\fi
}
\newcommand{\hatcurPPtheta}[1]{\ifnum#1=31 %
\hatcurPPthetaxxxxxA
\else
\ifnum#1=32 %
\hatcurPPthetaxxxxxB
\else
\ifnum#1=33 %
\hatcurPPthetaxxxxxC
\else
\ifnum#1=34 %
\hatcurPPthetaxxxxxD
\else
\ifnum#1=35 %
\hatcurPPthetaxxxxxE
\else
??????\fi
\fi
\fi
\fi
\fi
}
\newcommand{\hatcurPPtinfall}[1]{\ifnum#1=31 %
\hatcurPPtinfallxxxxxA
\else
\ifnum#1=32 %
\hatcurPPtinfallxxxxxB
\else
\ifnum#1=33 %
\hatcurPPtinfallxxxxxC
\else
\ifnum#1=34 %
\hatcurPPtinfallxxxxxD
\else
\ifnum#1=35 %
\hatcurPPtinfallxxxxxE
\else
??????\fi
\fi
\fi
\fi
\fi
}
\newcommand{\hatcurRVeccen}[1]{\ifnum#1=31 %
\hatcurRVeccenxxxxxA
\else
\ifnum#1=32 %
\hatcurRVeccenxxxxxB
\else
\ifnum#1=33 %
\hatcurRVeccenxxxxxC
\else
\ifnum#1=34 %
\hatcurRVeccenxxxxxD
\else
\ifnum#1=35 %
\hatcurRVeccenxxxxxE
\else
??????\fi
\fi
\fi
\fi
\fi
}
\newcommand{\hatcurRVeccentwosiglim}[1]{\ifnum#1=31 %
\hatcurRVeccentwosiglimxxxxxA
\else
\ifnum#1=32 %
\hatcurRVeccentwosiglimxxxxxB
\else
\ifnum#1=33 %
\hatcurRVeccentwosiglimxxxxxC
\else
\ifnum#1=34 %
\hatcurRVeccentwosiglimxxxxxD
\else
\ifnum#1=35 %
\hatcurRVeccentwosiglimxxxxxE
\else
??????\fi
\fi
\fi
\fi
\fi
}
\newcommand{\hatcurRVfitrms}[1]{\ifnum#1=31 %
\hatcurRVfitrmsxxxxxA
\else
\ifnum#1=32 %
\hatcurRVfitrmsxxxxxB
\else
\ifnum#1=34 %
\hatcurRVfitrmsxxxxxD
\else
??????\fi
\fi
\fi
}
\newcommand{\hatcurRVfitrmsA}[1]{\ifnum#1=33 %
\hatcurRVfitrmsAxxxxxC
\else
\ifnum#1=35 %
\hatcurRVfitrmsAxxxxxE
\else
??????\fi
\fi
}
\newcommand{\hatcurRVfitrmsB}[1]{\ifnum#1=33 %
\hatcurRVfitrmsBxxxxxC
\else
\ifnum#1=35 %
\hatcurRVfitrmsBxxxxxE
\else
??????\fi
\fi
}
\newcommand{\hatcurRVfitrmsC}[1]{\ifnum#1=33 %
\hatcurRVfitrmsCxxxxxC
\else
\ifnum#1=35 %
\hatcurRVfitrmsCxxxxxE
\else
??????\fi
\fi
}
\newcommand{\hatcurRVfitrmsD}[1]{\ifnum#1=33 %
\hatcurRVfitrmsDxxxxxC
\else
??????\fi
}
\newcommand{\hatcurRVgamma}[1]{\ifnum#1=31 %
\hatcurRVgammaxxxxxA
\else
\ifnum#1=32 %
\hatcurRVgammaxxxxxB
\else
\ifnum#1=34 %
\hatcurRVgammaxxxxxD
\else
??????\fi
\fi
\fi
}
\newcommand{\hatcurRVgammaA}[1]{\ifnum#1=33 %
\hatcurRVgammaAxxxxxC
\else
\ifnum#1=35 %
\hatcurRVgammaAxxxxxE
\else
??????\fi
\fi
}
\newcommand{\hatcurRVgammaB}[1]{\ifnum#1=33 %
\hatcurRVgammaBxxxxxC
\else
\ifnum#1=35 %
\hatcurRVgammaBxxxxxE
\else
??????\fi
\fi
}
\newcommand{\hatcurRVgammaC}[1]{\ifnum#1=33 %
\hatcurRVgammaCxxxxxC
\else
\ifnum#1=35 %
\hatcurRVgammaCxxxxxE
\else
??????\fi
\fi
}
\newcommand{\hatcurRVgammaD}[1]{\ifnum#1=33 %
\hatcurRVgammaDxxxxxC
\else
??????\fi
}
\newcommand{\hatcurRVh}[1]{\ifnum#1=31 %
\hatcurRVhxxxxxA
\else
\ifnum#1=32 %
\hatcurRVhxxxxxB
\else
\ifnum#1=33 %
\hatcurRVhxxxxxC
\else
\ifnum#1=34 %
\hatcurRVhxxxxxD
\else
\ifnum#1=35 %
\hatcurRVhxxxxxE
\else
??????\fi
\fi
\fi
\fi
\fi
}
\newcommand{\hatcurRVjitter}[1]{\ifnum#1=31 %
\hatcurRVjitterxxxxxA
\else
\ifnum#1=32 %
\hatcurRVjitterxxxxxB
\else
\ifnum#1=34 %
\hatcurRVjitterxxxxxD
\else
??????\fi
\fi
\fi
}
\newcommand{\hatcurRVjitterA}[1]{\ifnum#1=33 %
\hatcurRVjitterAxxxxxC
\else
\ifnum#1=35 %
\hatcurRVjitterAxxxxxE
\else
??????\fi
\fi
}
\newcommand{\hatcurRVjitterB}[1]{\ifnum#1=33 %
\hatcurRVjitterBxxxxxC
\else
\ifnum#1=35 %
\hatcurRVjitterBxxxxxE
\else
??????\fi
\fi
}
\newcommand{\hatcurRVjitterC}[1]{\ifnum#1=33 %
\hatcurRVjitterCxxxxxC
\else
\ifnum#1=35 %
\hatcurRVjitterCxxxxxE
\else
??????\fi
\fi
}
\newcommand{\hatcurRVjitterD}[1]{\ifnum#1=33 %
\hatcurRVjitterDxxxxxC
\else
??????\fi
}
\newcommand{\hatcurRVjittertwosiglim}[1]{\ifnum#1=31 %
\hatcurRVjittertwosiglimxxxxxA
\else
\ifnum#1=32 %
\hatcurRVjittertwosiglimxxxxxB
\else
\ifnum#1=34 %
\hatcurRVjittertwosiglimxxxxxD
\else
??????\fi
\fi
\fi
}
\newcommand{\hatcurRVjittertwosiglimA}[1]{\ifnum#1=33 %
\hatcurRVjittertwosiglimAxxxxxC
\else
\ifnum#1=35 %
\hatcurRVjittertwosiglimAxxxxxE
\else
??????\fi
\fi
}
\newcommand{\hatcurRVjittertwosiglimB}[1]{\ifnum#1=33 %
\hatcurRVjittertwosiglimBxxxxxC
\else
\ifnum#1=35 %
\hatcurRVjittertwosiglimBxxxxxE
\else
??????\fi
\fi
}
\newcommand{\hatcurRVjittertwosiglimC}[1]{\ifnum#1=33 %
\hatcurRVjittertwosiglimCxxxxxC
\else
\ifnum#1=35 %
\hatcurRVjittertwosiglimCxxxxxE
\else
??????\fi
\fi
}
\newcommand{\hatcurRVjittertwosiglimD}[1]{\ifnum#1=33 %
\hatcurRVjittertwosiglimDxxxxxC
\else
??????\fi
}
\newcommand{\hatcurRVk}[1]{\ifnum#1=31 %
\hatcurRVkxxxxxA
\else
\ifnum#1=32 %
\hatcurRVkxxxxxB
\else
\ifnum#1=33 %
\hatcurRVkxxxxxC
\else
\ifnum#1=34 %
\hatcurRVkxxxxxD
\else
\ifnum#1=35 %
\hatcurRVkxxxxxE
\else
??????\fi
\fi
\fi
\fi
\fi
}
\newcommand{\hatcurRVK}[1]{\ifnum#1=31 %
\hatcurRVKxxxxxA
\else
\ifnum#1=32 %
\hatcurRVKxxxxxB
\else
\ifnum#1=33 %
\hatcurRVKxxxxxC
\else
\ifnum#1=34 %
\hatcurRVKxxxxxD
\else
\ifnum#1=35 %
\hatcurRVKxxxxxE
\else
??????\fi
\fi
\fi
\fi
\fi
}
\newcommand{\hatcurRVomega}[1]{\ifnum#1=31 %
\hatcurRVomegaxxxxxA
\else
\ifnum#1=32 %
\hatcurRVomegaxxxxxB
\else
\ifnum#1=33 %
\hatcurRVomegaxxxxxC
\else
\ifnum#1=34 %
\hatcurRVomegaxxxxxD
\else
\ifnum#1=35 %
\hatcurRVomegaxxxxxE
\else
??????\fi
\fi
\fi
\fi
\fi
}
\newcommand{\hatcurRVrh}[1]{\ifnum#1=31 %
\hatcurRVrhxxxxxA
\else
\ifnum#1=32 %
\hatcurRVrhxxxxxB
\else
\ifnum#1=33 %
\hatcurRVrhxxxxxC
\else
\ifnum#1=34 %
\hatcurRVrhxxxxxD
\else
\ifnum#1=35 %
\hatcurRVrhxxxxxE
\else
??????\fi
\fi
\fi
\fi
\fi
}
\newcommand{\hatcurRVrk}[1]{\ifnum#1=31 %
\hatcurRVrkxxxxxA
\else
\ifnum#1=32 %
\hatcurRVrkxxxxxB
\else
\ifnum#1=33 %
\hatcurRVrkxxxxxC
\else
\ifnum#1=34 %
\hatcurRVrkxxxxxD
\else
\ifnum#1=35 %
\hatcurRVrkxxxxxE
\else
??????\fi
\fi
\fi
\fi
\fi
}
\newcommand{\hatcurRVtrone}[1]{\ifnum#1=31 %
\hatcurRVtronexxxxxA
\else
\ifnum#1=32 %
\hatcurRVtronexxxxxB
\else
\ifnum#1=33 %
\hatcurRVtronexxxxxC
\else
\ifnum#1=34 %
\hatcurRVtronexxxxxD
\else
\ifnum#1=35 %
\hatcurRVtronexxxxxE
\else
??????\fi
\fi
\fi
\fi
\fi
}
\newcommand{\hatcurRVtrtwo}[1]{\ifnum#1=31 %
\hatcurRVtrtwoxxxxxA
\else
\ifnum#1=32 %
\hatcurRVtrtwoxxxxxB
\else
\ifnum#1=33 %
\hatcurRVtrtwoxxxxxC
\else
\ifnum#1=34 %
\hatcurRVtrtwoxxxxxD
\else
\ifnum#1=35 %
\hatcurRVtrtwoxxxxxE
\else
??????\fi
\fi
\fi
\fi
\fi
}
\newcommand{\hatcurSMEiilogg}[1]{\ifnum#1=31 %
\hatcurSMEiiloggxxxxxA
\else
\ifnum#1=32 %
\hatcurSMEiiloggxxxxxB
\else
\ifnum#1=33 %
\hatcurSMEiiloggxxxxxC
\else
\ifnum#1=34 %
\hatcurSMEiiloggxxxxxD
\else
\ifnum#1=35 %
\hatcurSMEiiloggxxxxxE
\else
??????\fi
\fi
\fi
\fi
\fi
}
\newcommand{\hatcurSMEiiteff}[1]{\ifnum#1=31 %
\hatcurSMEiiteffxxxxxA
\else
\ifnum#1=32 %
\hatcurSMEiiteffxxxxxB
\else
\ifnum#1=33 %
\hatcurSMEiiteffxxxxxC
\else
\ifnum#1=34 %
\hatcurSMEiiteffxxxxxD
\else
\ifnum#1=35 %
\hatcurSMEiiteffxxxxxE
\else
??????\fi
\fi
\fi
\fi
\fi
}
\newcommand{\hatcurSMEiivsin}[1]{\ifnum#1=31 %
\hatcurSMEiivsinxxxxxA
\else
\ifnum#1=32 %
\hatcurSMEiivsinxxxxxB
\else
\ifnum#1=33 %
\hatcurSMEiivsinxxxxxC
\else
\ifnum#1=34 %
\hatcurSMEiivsinxxxxxD
\else
\ifnum#1=35 %
\hatcurSMEiivsinxxxxxE
\else
??????\fi
\fi
\fi
\fi
\fi
}
\newcommand{\hatcurSMEiizfeh}[1]{\ifnum#1=31 %
\hatcurSMEiizfehxxxxxA
\else
\ifnum#1=32 %
\hatcurSMEiizfehxxxxxB
\else
\ifnum#1=33 %
\hatcurSMEiizfehxxxxxC
\else
\ifnum#1=34 %
\hatcurSMEiizfehxxxxxD
\else
\ifnum#1=35 %
\hatcurSMEiizfehxxxxxE
\else
??????\fi
\fi
\fi
\fi
\fi
}
\newcommand{\hatcurSMEiizfehshort}[1]{\ifnum#1=31 %
\hatcurSMEiizfehshortxxxxxA
\else
\ifnum#1=32 %
\hatcurSMEiizfehshortxxxxxB
\else
\ifnum#1=33 %
\hatcurSMEiizfehshortxxxxxC
\else
\ifnum#1=34 %
\hatcurSMEiizfehshortxxxxxD
\else
\ifnum#1=35 %
\hatcurSMEiizfehshortxxxxxE
\else
??????\fi
\fi
\fi
\fi
\fi
}
\newcommand{\hatcurSMEilogg}[1]{\ifnum#1=31 %
\hatcurSMEiloggxxxxxA
\else
\ifnum#1=32 %
\hatcurSMEiloggxxxxxB
\else
\ifnum#1=33 %
\hatcurSMEiloggxxxxxC
\else
\ifnum#1=34 %
\hatcurSMEiloggxxxxxD
\else
\ifnum#1=35 %
\hatcurSMEiloggxxxxxE
\else
??????\fi
\fi
\fi
\fi
\fi
}
\newcommand{\hatcurSMEiteff}[1]{\ifnum#1=31 %
\hatcurSMEiteffxxxxxA
\else
\ifnum#1=32 %
\hatcurSMEiteffxxxxxB
\else
\ifnum#1=33 %
\hatcurSMEiteffxxxxxC
\else
\ifnum#1=34 %
\hatcurSMEiteffxxxxxD
\else
\ifnum#1=35 %
\hatcurSMEiteffxxxxxE
\else
??????\fi
\fi
\fi
\fi
\fi
}
\newcommand{\hatcurSMEivmac}[1]{\ifnum#1=31 %
\hatcurSMEivmacxxxxxA
\else
\ifnum#1=32 %
\hatcurSMEivmacxxxxxB
\else
\ifnum#1=33 %
\hatcurSMEivmacxxxxxC
\else
\ifnum#1=34 %
\hatcurSMEivmacxxxxxD
\else
\ifnum#1=35 %
\hatcurSMEivmacxxxxxE
\else
??????\fi
\fi
\fi
\fi
\fi
}
\newcommand{\hatcurSMEivmic}[1]{\ifnum#1=31 %
\hatcurSMEivmicxxxxxA
\else
\ifnum#1=32 %
\hatcurSMEivmicxxxxxB
\else
\ifnum#1=33 %
\hatcurSMEivmicxxxxxC
\else
\ifnum#1=34 %
\hatcurSMEivmicxxxxxD
\else
\ifnum#1=35 %
\hatcurSMEivmicxxxxxE
\else
??????\fi
\fi
\fi
\fi
\fi
}
\newcommand{\hatcurSMEivsin}[1]{\ifnum#1=31 %
\hatcurSMEivsinxxxxxA
\else
\ifnum#1=32 %
\hatcurSMEivsinxxxxxB
\else
\ifnum#1=33 %
\hatcurSMEivsinxxxxxC
\else
\ifnum#1=34 %
\hatcurSMEivsinxxxxxD
\else
\ifnum#1=35 %
\hatcurSMEivsinxxxxxE
\else
??????\fi
\fi
\fi
\fi
\fi
}
\newcommand{\hatcurSMEizfeh}[1]{\ifnum#1=31 %
\hatcurSMEizfehxxxxxA
\else
\ifnum#1=32 %
\hatcurSMEizfehxxxxxB
\else
\ifnum#1=33 %
\hatcurSMEizfehxxxxxC
\else
\ifnum#1=34 %
\hatcurSMEizfehxxxxxD
\else
\ifnum#1=35 %
\hatcurSMEizfehxxxxxE
\else
??????\fi
\fi
\fi
\fi
\fi
}
\newcommand{\hatcurSMEizfehshort}[1]{\ifnum#1=31 %
\hatcurSMEizfehshortxxxxxA
\else
\ifnum#1=32 %
\hatcurSMEizfehshortxxxxxB
\else
\ifnum#1=33 %
\hatcurSMEizfehshortxxxxxC
\else
\ifnum#1=34 %
\hatcurSMEizfehshortxxxxxD
\else
\ifnum#1=35 %
\hatcurSMEizfehshortxxxxxE
\else
??????\fi
\fi
\fi
\fi
\fi
}
\newcommand{\hatcurXAv}[1]{\ifnum#1=31 %
\hatcurXAvxxxxxA
\else
\ifnum#1=32 %
\hatcurXAvxxxxxB
\else
\ifnum#1=33 %
\hatcurXAvxxxxxC
\else
\ifnum#1=34 %
\hatcurXAvxxxxxD
\else
\ifnum#1=35 %
\hatcurXAvxxxxxE
\else
??????\fi
\fi
\fi
\fi
\fi
}
\newcommand{\hatcurXdist}[1]{\ifnum#1=31 %
\hatcurXdistxxxxxA
\else
\ifnum#1=32 %
\hatcurXdistxxxxxB
\else
\ifnum#1=33 %
\hatcurXdistxxxxxC
\else
\ifnum#1=34 %
\hatcurXdistxxxxxD
\else
\ifnum#1=35 %
\hatcurXdistxxxxxE
\else
??????\fi
\fi
\fi
\fi
\fi
}
\newcommand{\hatcurXdistred}[1]{\ifnum#1=31 %
\hatcurXdistredxxxxxA
\else
\ifnum#1=32 %
\hatcurXdistredxxxxxB
\else
\ifnum#1=33 %
\hatcurXdistredxxxxxC
\else
\ifnum#1=34 %
\hatcurXdistredxxxxxD
\else
\ifnum#1=35 %
\hatcurXdistredxxxxxE
\else
??????\fi
\fi
\fi
\fi
\fi
}
\newcommand{\hatcurXEBV}[1]{\ifnum#1=31 %
\hatcurXEBVxxxxxA
\else
\ifnum#1=32 %
\hatcurXEBVxxxxxB
\else
\ifnum#1=33 %
\hatcurXEBVxxxxxC
\else
\ifnum#1=34 %
\hatcurXEBVxxxxxD
\else
\ifnum#1=35 %
\hatcurXEBVxxxxxE
\else
??????\fi
\fi
\fi
\fi
\fi
}
\newcommand{\hatcurXjhisored}[1]{\ifnum#1=31 %
\hatcurXjhisoredxxxxxA
\else
\ifnum#1=32 %
\hatcurXjhisoredxxxxxB
\else
\ifnum#1=33 %
\hatcurXjhisoredxxxxxC
\else
\ifnum#1=34 %
\hatcurXjhisoredxxxxxD
\else
\ifnum#1=35 %
\hatcurXjhisoredxxxxxE
\else
??????\fi
\fi
\fi
\fi
\fi
}
\newcommand{\hatcurXjkisored}[1]{\ifnum#1=31 %
\hatcurXjkisoredxxxxxA
\else
\ifnum#1=32 %
\hatcurXjkisoredxxxxxB
\else
\ifnum#1=33 %
\hatcurXjkisoredxxxxxC
\else
\ifnum#1=34 %
\hatcurXjkisoredxxxxxD
\else
\ifnum#1=35 %
\hatcurXjkisoredxxxxxE
\else
??????\fi
\fi
\fi
\fi
\fi
}
\newcommand{\hatcurXmhisored}[1]{\ifnum#1=31 %
\hatcurXmhisoredxxxxxA
\else
\ifnum#1=32 %
\hatcurXmhisoredxxxxxB
\else
\ifnum#1=33 %
\hatcurXmhisoredxxxxxC
\else
\ifnum#1=34 %
\hatcurXmhisoredxxxxxD
\else
\ifnum#1=35 %
\hatcurXmhisoredxxxxxE
\else
??????\fi
\fi
\fi
\fi
\fi
}
\newcommand{\hatcurXmiisored}[1]{\ifnum#1=31 %
\hatcurXmiisoredxxxxxA
\else
\ifnum#1=32 %
\hatcurXmiisoredxxxxxB
\else
\ifnum#1=33 %
\hatcurXmiisoredxxxxxC
\else
\ifnum#1=34 %
\hatcurXmiisoredxxxxxD
\else
\ifnum#1=35 %
\hatcurXmiisoredxxxxxE
\else
??????\fi
\fi
\fi
\fi
\fi
}
\newcommand{\hatcurXmjisored}[1]{\ifnum#1=31 %
\hatcurXmjisoredxxxxxA
\else
\ifnum#1=32 %
\hatcurXmjisoredxxxxxB
\else
\ifnum#1=33 %
\hatcurXmjisoredxxxxxC
\else
\ifnum#1=34 %
\hatcurXmjisoredxxxxxD
\else
\ifnum#1=35 %
\hatcurXmjisoredxxxxxE
\else
??????\fi
\fi
\fi
\fi
\fi
}
\newcommand{\hatcurXmkisored}[1]{\ifnum#1=31 %
\hatcurXmkisoredxxxxxA
\else
\ifnum#1=32 %
\hatcurXmkisoredxxxxxB
\else
\ifnum#1=33 %
\hatcurXmkisoredxxxxxC
\else
\ifnum#1=34 %
\hatcurXmkisoredxxxxxD
\else
\ifnum#1=35 %
\hatcurXmkisoredxxxxxE
\else
??????\fi
\fi
\fi
\fi
\fi
}
\newcommand{\hatcurXmvisored}[1]{\ifnum#1=31 %
\hatcurXmvisoredxxxxxA
\else
\ifnum#1=32 %
\hatcurXmvisoredxxxxxB
\else
\ifnum#1=33 %
\hatcurXmvisoredxxxxxC
\else
\ifnum#1=34 %
\hatcurXmvisoredxxxxxD
\else
\ifnum#1=35 %
\hatcurXmvisoredxxxxxE
\else
??????\fi
\fi
\fi
\fi
\fi
}
\newcommand{\hatcurXsecdur}[1]{\ifnum#1=31 %
\hatcurXsecdurxxxxxA
\else
\ifnum#1=32 %
\hatcurXsecdurxxxxxB
\else
\ifnum#1=33 %
\hatcurXsecdurxxxxxC
\else
\ifnum#1=34 %
\hatcurXsecdurxxxxxD
\else
\ifnum#1=35 %
\hatcurXsecdurxxxxxE
\else
??????\fi
\fi
\fi
\fi
\fi
}
\newcommand{\hatcurXsecingdur}[1]{\ifnum#1=31 %
\hatcurXsecingdurxxxxxA
\else
\ifnum#1=32 %
\hatcurXsecingdurxxxxxB
\else
\ifnum#1=33 %
\hatcurXsecingdurxxxxxC
\else
\ifnum#1=34 %
\hatcurXsecingdurxxxxxD
\else
\ifnum#1=35 %
\hatcurXsecingdurxxxxxE
\else
??????\fi
\fi
\fi
\fi
\fi
}
\newcommand{\hatcurXsecondary}[1]{\ifnum#1=31 %
\hatcurXsecondaryxxxxxA
\else
\ifnum#1=32 %
\hatcurXsecondaryxxxxxB
\else
\ifnum#1=33 %
\hatcurXsecondaryxxxxxC
\else
\ifnum#1=34 %
\hatcurXsecondaryxxxxxD
\else
\ifnum#1=35 %
\hatcurXsecondaryxxxxxE
\else
??????\fi
\fi
\fi
\fi
\fi
}
\newcommand{\hatcurXsecphase}[1]{\ifnum#1=31 %
\hatcurXsecphasexxxxxA
\else
\ifnum#1=32 %
\hatcurXsecphasexxxxxB
\else
\ifnum#1=33 %
\hatcurXsecphasexxxxxC
\else
\ifnum#1=34 %
\hatcurXsecphasexxxxxD
\else
\ifnum#1=35 %
\hatcurXsecphasexxxxxE
\else
??????\fi
\fi
\fi
\fi
\fi
}
\newcommand{\hatcurXviisored}[1]{\ifnum#1=31 %
\hatcurXviisoredxxxxxA
\else
\ifnum#1=32 %
\hatcurXviisoredxxxxxB
\else
\ifnum#1=33 %
\hatcurXviisoredxxxxxC
\else
\ifnum#1=34 %
\hatcurXviisoredxxxxxD
\else
\ifnum#1=35 %
\hatcurXviisoredxxxxxE
\else
??????\fi
\fi
\fi
\fi
\fi
}
\newcommand{\hatcurXvkisored}[1]{\ifnum#1=31 %
\hatcurXvkisoredxxxxxA
\else
\ifnum#1=32 %
\hatcurXvkisoredxxxxxB
\else
\ifnum#1=33 %
\hatcurXvkisoredxxxxxC
\else
\ifnum#1=34 %
\hatcurXvkisoredxxxxxD
\else
\ifnum#1=35 %
\hatcurXvkisoredxxxxxE
\else
??????\fi
\fi
\fi
\fi
\fi
}

%% file: HATS566-004.tex
\newcommand{\hatcurhtrxxxxxA}{HATS566-004}                             
\newcommand{\hatcurfieldxxxxxA}{\ensuremath{string}}                   
\newcommand{\hatcurCCraxxxxxA}{\ensuremath{12^{\mathrm h}46^{\mathrm m}48.72{\mathrm s}}}                            
\newcommand{\hatcurCCdecxxxxxA}{\ensuremath{-24{\arcdeg}25{\arcmin}38.5{\arcsec}}}                           
\newcommand{\hatcurCCmagxxxxxA}{13.105}                                
\newcommand{\hatcurCCtwomassxxxxxA}{2MASS~12464866-2425385}            
\newcommand{\hatcurCCgscxxxxxA}{GSC~6688-00298}                        
\newcommand{\hatcurCCtassmvxxxxxA}{\ensuremath{13.105\pm0.030}}        
\newcommand{\hatcurCCtassmvshortxxxxxA}{\ensuremath{13.1}}             
\newcommand{\hatcurCCtassmBxxxxxA}{\ensuremath{13.687\pm0.030}}        
\newcommand{\hatcurCCtassmBshortxxxxxA}{\ensuremath{13.7}}             
\newcommand{\hatcurCCtassmIxxxxxA}{\ensuremath{100\pm1000}}            
\newcommand{\hatcurCCtassmIshortxxxxxA}{\ensuremath{100.0}}            
\newcommand{\hatcurCCtassmgxxxxxA}{\ensuremath{13.356\pm0.030}}        
\newcommand{\hatcurCCtassmgshortxxxxxA}{\ensuremath{13.4}}             
\newcommand{\hatcurCCtassmrxxxxxA}{\ensuremath{12.950\pm0.030}}        
\newcommand{\hatcurCCtassmrshortxxxxxA}{\ensuremath{12.9}}             
\newcommand{\hatcurCCtassmixxxxxA}{\ensuremath{12.775\pm0.080}}        
\newcommand{\hatcurCCtassmishortxxxxxA}{\ensuremath{12.8}}             
\newcommand{\hatcurCCtwomassJmagxxxxxA}{\ensuremath{11.912\pm0.021}}   
\newcommand{\hatcurCCtwomassHmagxxxxxA}{\ensuremath{11.618\pm0.025}}   
\newcommand{\hatcurCCtwomassKmagxxxxxA}{\ensuremath{11.572\pm0.023}}   
\newcommand{\hatcurCCcitJmagxxxxxA}{\ensuremath{11.931\pm0.021}}       
\newcommand{\hatcurCCcitHmagxxxxxA}{\ensuremath{11.613\pm0.025}}       
\newcommand{\hatcurCCcitKmagxxxxxA}{\ensuremath{11.596\pm0.023}}       
\newcommand{\hatcurCCbbJmagxxxxxA}{\ensuremath{11.977\pm0.023}}        
\newcommand{\hatcurCCbbHmagxxxxxA}{\ensuremath{11.634\pm0.026}}        
\newcommand{\hatcurCCbbKmagxxxxxA}{\ensuremath{11.616\pm0.023}}        
\newcommand{\hatcurCCesoJmagxxxxxA}{\ensuremath{11.979\pm0.024}}       
\newcommand{\hatcurCCesoHmagxxxxxA}{\ensuremath{11.628\pm0.029}}       
\newcommand{\hatcurCCesoKmagxxxxxA}{\ensuremath{11.615\pm0.024}}       
\newcommand{\hatcurCCesoJHmagxxxxxA}{\ensuremath{0.351\pm0.035}}       
\newcommand{\hatcurCCesoJKmagxxxxxA}{\ensuremath{0.365\pm0.034}}       
\newcommand{\hatcurCCesoHKmagxxxxxA}{\ensuremath{0.014\pm0.038}}       
\newcommand{\hatcurLCdipxxxxxA}{\ensuremath{6.1}}                      
\newcommand{\hatcurLCrprstarxxxxxA}{\ensuremath{0.0908\pm0.0049}}      
\newcommand{\hatcurLCbsqxxxxxA}{\ensuremath{0.23_{-0.16}^{+0.11}}}     
\newcommand{\hatcurLCimpxxxxxA}{\ensuremath{0.48_{-0.21}^{+0.11}}}     
\newcommand{\hatcurLCzetaxxxxxA}{\ensuremath{11.67\pm0.16}}            
\newcommand{\hatcurLCdurxxxxxA}{\ensuremath{0.1914\pm0.0043}}          
\newcommand{\hatcurLCdurshortxxxxxA}{\ensuremath{0.1914}}              
\newcommand{\hatcurLCdurhrxxxxxA}{\ensuremath{4.59\pm0.10}}            
\newcommand{\hatcurLCdurhrshortxxxxxA}{\ensuremath{4.594}}             
\newcommand{\hatcurLCqxxxxxA}{\ensuremath{0.0567\pm0.0013}}            
\newcommand{\hatcurLCqshortxxxxxA}{\ensuremath{0.057}}                 
\newcommand{\hatcurLCingdurxxxxxA}{\ensuremath{0.0202\pm0.0042}}       
\newcommand{\hatcurLCPxxxxxA}{\ensuremath{3.377960\pm0.000012}}        
\newcommand{\hatcurLCPprecxxxxxA}{\ensuremath{3.3779595}}              
\newcommand{\hatcurLCPshortxxxxxA}{\ensuremath{3.3780}}                
\newcommand{\hatcurLCTxxxxxA}{\ensuremath{2456960.1476\pm0.0011}}      
\newcommand{\hatcurLCTAxxxxxA}{\ensuremath{2456281.1779\pm0.0025}}     
\newcommand{\hatcurLCTBxxxxxA}{\ensuremath{2457115.5338\pm0.0012}}     
\newcommand{\hatcurLChatnetmxxxxxA}{\ensuremath{12.943640\pm0.000069}} 
\newcommand{\hatcurLCiblendxxxxxA}{\ensuremath{0.648\pm0.071}}         
\newcommand{\hatcurLCrhoxxxxxA}{\ensuremath{0.275_{-0.061}^{+0.082}}}  
\newcommand{\hatcurSMEiteffxxxxxA}{\ensuremath{6060\pm120}}            
\newcommand{\hatcurSMEizfehxxxxxA}{\ensuremath{0.000\pm0.070}}         
\newcommand{\hatcurSMEizfehshortxxxxxA}{\ensuremath{0.00}}             
\newcommand{\hatcurSMEiloggxxxxxA}{\ensuremath{4.05\pm0.26}}           
\newcommand{\hatcurSMEivsinxxxxxA}{\ensuremath{7.00\pm0.50}}           
\newcommand{\hatcurSMEivmacxxxxxA}{\ensuremath{0.0}}                   
\newcommand{\hatcurSMEivmicxxxxxA}{\ensuremath{0.0}}                   
\newcommand{\hatcurSMEiiteffxxxxxA}{\ensuremath{6050\pm120}}           
\newcommand{\hatcurSMEiizfehxxxxxA}{\ensuremath{0.000\pm0.070}}        
\newcommand{\hatcurSMEiizfehshortxxxxxA}{\ensuremath{0.0}}             
\newcommand{\hatcurSMEiiloggxxxxxA}{\ensuremath{4.012\pm0.060}}        
\newcommand{\hatcurSMEiivsinxxxxxA}{\ensuremath{7.01\pm0.50}}          
\newcommand{\hatcurLBizxxxxxA}{\ensuremath{0.1658}}                    
\newcommand{\hatcurLBiizxxxxxA}{\ensuremath{0.3498}}                   
\newcommand{\hatcurLBiixxxxxA}{\ensuremath{0.2195}}                    
\newcommand{\hatcurLBiiixxxxxA}{\ensuremath{0.3558}}                   
\newcommand{\hatcurLBiIxxxxxA}{\ensuremath{0.2003}}                    
\newcommand{\hatcurLBiiIxxxxxA}{\ensuremath{0.3553}}                   
\newcommand{\hatcurLBigxxxxxA}{\ensuremath{0.4810}}                    
\newcommand{\hatcurLBiigxxxxxA}{\ensuremath{0.2874}}                   
\newcommand{\hatcurLBirxxxxxA}{\ensuremath{0.3000}}                    
\newcommand{\hatcurLBiirxxxxxA}{\ensuremath{0.3587}}                   
\newcommand{\hatcurLBiRxxxxxA}{\ensuremath{0.2774}}                    
\newcommand{\hatcurLBiiRxxxxxA}{\ensuremath{0.3592}}                   
\newcommand{\hatcurLBikepxxxxxA}{\ensuremath{0.1000}}                  
\newcommand{\hatcurLBiikepxxxxxA}{\ensuremath{0.1000}}                 
\newcommand{\hatcurISOmxxxxxA}{\ensuremath{1.275\pm0.096}}             
\newcommand{\hatcurISOmshortxxxxxA}{\ensuremath{1.27}}                 
\newcommand{\hatcurISOmlongxxxxxA}{\ensuremath{1.275\pm0.096}}         
\newcommand{\hatcurISOrxxxxxA}{\ensuremath{1.87\pm0.18}}               
\newcommand{\hatcurISOrshortxxxxxA}{\ensuremath{1.87}}                 
\newcommand{\hatcurISOrlongxxxxxA}{\ensuremath{1.87\pm0.18}}           
\newcommand{\hatcurISOrhoxxxxxA}{\ensuremath{0.275_{-0.059}^{+0.082}}} 
\newcommand{\hatcurISOrholongxxxxxA}{\ensuremath{0.275_{-0.059}^{+0.082}}} 
\newcommand{\hatcurISOloggxxxxxA}{\ensuremath{4.000\pm0.065}}          
\newcommand{\hatcurISOlumxxxxxA}{\ensuremath{4.16\pm0.95}}             
\newcommand{\hatcurISOlumshortxxxxxA}{\ensuremath{4.16}}               
\newcommand{\hatcurISOmvxxxxxA}{\ensuremath{3.25\pm0.24}}              
\newcommand{\hatcurISOvixxxxxA}{\ensuremath{0.615\pm0.033}}            
\newcommand{\hatcurISOagexxxxxA}{\ensuremath{4.3\pm1.1}}               
\newcommand{\hatcurISOsigmaxxxxxA}{\ensuremath{0.000300\pm0.000092}}   
\newcommand{\hatcurISOMJxxxxxA}{\ensuremath{2.23\pm0.21}}              
\newcommand{\hatcurISOMHxxxxxA}{\ensuremath{1.93\pm0.21}}              
\newcommand{\hatcurISOMKxxxxxA}{\ensuremath{1.88\pm0.21}}              
\newcommand{\hatcurISOJKxxxxxA}{\ensuremath{0.350\pm0.020}}            
\newcommand{\hatcurISOspecxxxxxA}{F}                                   
\newcommand{\hatcurRVKxxxxxA}{\ensuremath{102\pm13}}                   
\newcommand{\hatcurRVrkxxxxxA}{\ensuremath{0\pm0}}                     
\newcommand{\hatcurRVrhxxxxxA}{\ensuremath{0\pm0}}                     
\newcommand{\hatcurRVkxxxxxA}{\ensuremath{0\pm0}}                      
\newcommand{\hatcurRVhxxxxxA}{\ensuremath{0\pm0}}                      
\newcommand{\hatcurRVtronexxxxxA}{\ensuremath{0\pm0}}                  
\newcommand{\hatcurRVtrtwoxxxxxA}{\ensuremath{0\pm0}}                  
\newcommand{\hatcurRVgammaxxxxxA}{\ensuremath{-8704.5\pm8.9}}          
\newcommand{\hatcurRVjitterxxxxxA}{\ensuremath{0.00\pm0.11}}           
\newcommand{\hatcurRVjittertwosiglimxxxxxA}{\ensuremath{<0.1}}         
\newcommand{\hatcurRVfitrmsxxxxxA}{\ensuremath{.1fym}}                 %
\newcommand{\hatcurRVeccenxxxxxA}{\ensuremath{0\pm0}}                  
\newcommand{\hatcurRVeccentwosiglimxxxxxA}{\ensuremath{<0.000}}        
\newcommand{\hatcurRVomegaxxxxxA}{\ensuremath{0\pm0}}                  
\newcommand{\hatcurPPixxxxxA}{\ensuremath{85.0_{-1.6}^{+2.5}}}         
\newcommand{\hatcurPPgxxxxxA}{\ensuremath{8.0\pm2.3}}                  
\newcommand{\hatcurPPloggxxxxxA}{\ensuremath{2.90\pm0.12}}             
\newcommand{\hatcurPParxxxxxA}{\ensuremath{5.49\pm0.43}}               
\newcommand{\hatcurPParelxxxxxA}{\ensuremath{0.0478\pm0.0012}}         
\newcommand{\hatcurPPrhoxxxxxA}{\ensuremath{0.243_{-0.088}^{+0.124}}}  
\newcommand{\hatcurPPmxxxxxA}{\ensuremath{0.88\pm0.12}}                
\newcommand{\hatcurPPmshortxxxxxA}{\ensuremath{0.88}}                  
\newcommand{\hatcurPPmlongxxxxxA}{\ensuremath{0.88\pm0.12}}            
\newcommand{\hatcurPPmexxxxxA}{\ensuremath{280\pm38}}                  
\newcommand{\hatcurPPmeshortxxxxxA}{\ensuremath{280.1}}                
\newcommand{\hatcurPPmelongxxxxxA}{\ensuremath{280\pm38}}              
\newcommand{\hatcurPPrxxxxxA}{\ensuremath{1.64\pm0.22}}                
\newcommand{\hatcurPPrshortxxxxxA}{\ensuremath{1.64}}                  
\newcommand{\hatcurPPrlongxxxxxA}{\ensuremath{1.64\pm0.22}}            
\newcommand{\hatcurPPrexxxxxA}{\ensuremath{18.4\pm2.4}}                
\newcommand{\hatcurPPreshortxxxxxA}{\ensuremath{18.4}}                 
\newcommand{\hatcurPPrelongxxxxxA}{\ensuremath{18.4\pm2.4}}            
\newcommand{\hatcurPPmrcorrxxxxxA}{\ensuremath{0.10}}                  
\newcommand{\hatcurPPteffxxxxxA}{\ensuremath{1823\pm81}}               
\newcommand{\hatcurPPthetaxxxxxA}{\ensuremath{0.0396\pm0.0079}}        
\newcommand{\hatcurPPfluxperixxxxxA}{\ensuremath{2.49\pm0.46}}         
\newcommand{\hatcurPPfluxperidimxxxxxA}{\ensuremath{9}}                
\newcommand{\hatcurPPfluxapxxxxxA}{\ensuremath{2.49\pm0.46}}           
\newcommand{\hatcurPPfluxapdimxxxxxA}{\ensuremath{9}}                  
\newcommand{\hatcurPPfluxavgxxxxxA}{\ensuremath{2.49\pm0.46}}          
\newcommand{\hatcurPPfluxavgdimxxxxxA}{\ensuremath{9}}                 
\newcommand{\hatcurPPfluxavglogxxxxxA}{\ensuremath{9.397\pm0.076}}     
\newcommand{\hatcurXsecphasexxxxxA}{\ensuremath{0\pm0}}                
\newcommand{\hatcurXsecondaryxxxxxA}{\ensuremath{2456961.8366\pm0.0011}} 
\newcommand{\hatcurXsecdurxxxxxA}{\ensuremath{0.1914\pm0.0043}}        
\newcommand{\hatcurXsecingdurxxxxxA}{\ensuremath{0.0202\pm0.0042}}     
\newcommand{\hatcurPPphiconjxxxxxA}{\ensuremath{0\pm0}}                
\newcommand{\hatcurPPperixxxxxA}{\ensuremath{2456959.3031\pm0.0011}}   
\newcommand{\hatcurPPaequivxxxxxA}{\ensuremath{0.0234\pm0.0021}}       
\newcommand{\hatcurPPtcircxxxxxA}{\ensuremath{51_{-27}^{+44}}}         
\newcommand{\hatcurPPtinfallxxxxxA}{\ensuremath{236_{-77}^{+139}}}     
\newcommand{\hatcurXdistxxxxxA}{\ensuremath{886\pm86}}                 
\newcommand{\hatcurXAvxxxxxA}{\ensuremath{0.154\pm0.092}}              
\newcommand{\hatcurXdistredxxxxxA}{\ensuremath{872\pm84}}              
\newcommand{\hatcurXEBVxxxxxA}{\ensuremath{0.050\pm0.030}}             
\newcommand{\hatcurXmvisoredxxxxxA}{\ensuremath{13.105\pm0.029}}       
\newcommand{\hatcurXmiisoredxxxxxA}{\ensuremath{12.407\pm0.018}}       
\newcommand{\hatcurXmjisoredxxxxxA}{\ensuremath{11.972\pm0.015}}       
\newcommand{\hatcurXmhisoredxxxxxA}{\ensuremath{11.662\pm0.016}}       
\newcommand{\hatcurXmkisoredxxxxxA}{\ensuremath{11.598\pm0.016}}       
\newcommand{\hatcurXviisoredxxxxxA}{\ensuremath{0.697\pm0.024}}        
\newcommand{\hatcurXvkisoredxxxxxA}{\ensuremath{1.506\pm0.036}}        
\newcommand{\hatcurXjhisoredxxxxxA}{\ensuremath{0.310\pm0.012}}        
\newcommand{\hatcurXjkisoredxxxxxA}{\ensuremath{0.374\pm0.010}}        
\newcommand{\hatcurCCpmraxxxxxA}{\ensuremath{1.2\pm1.0}}               
\newcommand{\hatcurCCpmdecxxxxxA}{\ensuremath{-0.4\pm1.1}}             
\newcommand{\hatcurCCpmxxxxxA}{\ensuremath{1.3\pm1.5}}                 

%% file: HATS586-004.tex
\newcommand{\hatcurhtrxxxxxB}{HATS586-004}                             
\newcommand{\hatcurfieldxxxxxB}{\ensuremath{string}}                   
\newcommand{\hatcurCCraxxxxxB}{\ensuremath{23^{\mathrm h}04^{\mathrm m}18.12{\mathrm s}}}                            
\newcommand{\hatcurCCdecxxxxxB}{\ensuremath{-21{\arcdeg}16{\arcmin}19.0{\arcsec}}}                           
\newcommand{\hatcurCCmagxxxxxB}{14.384}                                
\newcommand{\hatcurCCtwomassxxxxxB}{2MASS~23041801-2116189}            
\newcommand{\hatcurCCgscxxxxxB}{GSC~6400-00924}                        
\newcommand{\hatcurCCtassmvxxxxxB}{\ensuremath{14.384\pm0.010}}        
\newcommand{\hatcurCCtassmvshortxxxxxB}{\ensuremath{14.4}}             
\newcommand{\hatcurCCtassmBxxxxxB}{\ensuremath{15.106\pm0.030}}        
\newcommand{\hatcurCCtassmBshortxxxxxB}{\ensuremath{15.1}}             
\newcommand{\hatcurCCtassmIxxxxxB}{\ensuremath{100\pm1000}}            
\newcommand{\hatcurCCtassmIshortxxxxxB}{\ensuremath{100.0}}            
\newcommand{\hatcurCCtassmgxxxxxB}{\ensuremath{14.694\pm0.010}}        
\newcommand{\hatcurCCtassmgshortxxxxxB}{\ensuremath{14.7}}             
\newcommand{\hatcurCCtassmrxxxxxB}{\ensuremath{14.130\pm0.010}}        
\newcommand{\hatcurCCtassmrshortxxxxxB}{\ensuremath{14.1}}             
\newcommand{\hatcurCCtassmixxxxxB}{\ensuremath{14.035\pm0.010}}        
\newcommand{\hatcurCCtassmishortxxxxxB}{\ensuremath{14.0}}             
\newcommand{\hatcurCCtwomassJmagxxxxxB}{\ensuremath{13.090\pm0.022}}   
\newcommand{\hatcurCCtwomassHmagxxxxxB}{\ensuremath{12.768\pm0.027}}   
\newcommand{\hatcurCCtwomassKmagxxxxxB}{\ensuremath{12.699\pm0.033}}   
\newcommand{\hatcurCCcitJmagxxxxxB}{\ensuremath{13.106\pm0.022}}       
\newcommand{\hatcurCCcitHmagxxxxxB}{\ensuremath{12.763\pm0.027}}       
\newcommand{\hatcurCCcitKmagxxxxxB}{\ensuremath{12.723\pm0.033}}       
\newcommand{\hatcurCCbbJmagxxxxxB}{\ensuremath{13.157\pm0.024}}        
\newcommand{\hatcurCCbbHmagxxxxxB}{\ensuremath{12.784\pm0.028}}        
\newcommand{\hatcurCCbbKmagxxxxxB}{\ensuremath{12.743\pm0.033}}        
\newcommand{\hatcurCCesoJmagxxxxxB}{\ensuremath{13.159\pm0.026}}       
\newcommand{\hatcurCCesoHmagxxxxxB}{\ensuremath{12.779\pm0.031}}       
\newcommand{\hatcurCCesoKmagxxxxxB}{\ensuremath{12.742\pm0.034}}       
\newcommand{\hatcurCCesoJHmagxxxxxB}{\ensuremath{0.380\pm0.038}}       
\newcommand{\hatcurCCesoJKmagxxxxxB}{\ensuremath{0.418\pm0.042}}       
\newcommand{\hatcurCCesoHKmagxxxxxB}{\ensuremath{0.038\pm0.047}}       
\newcommand{\hatcurLCdipxxxxxB}{\ensuremath{15.3}}                     
\newcommand{\hatcurLCrprstarxxxxxB}{\ensuremath{0.1171\pm0.0039}}      
\newcommand{\hatcurLCbsqxxxxxB}{\ensuremath{0.16_{-0.10}^{+0.12}}}     
\newcommand{\hatcurLCimpxxxxxB}{\ensuremath{0.40_{-0.16}^{+0.13}}}     
\newcommand{\hatcurLCzetaxxxxxB}{\ensuremath{19.18\pm0.52}}            
\newcommand{\hatcurLCdurxxxxxB}{\ensuremath{0.1189\pm0.0036}}          
\newcommand{\hatcurLCdurshortxxxxxB}{\ensuremath{0.1189}}              
\newcommand{\hatcurLCdurhrxxxxxB}{\ensuremath{2.853\pm0.086}}          
\newcommand{\hatcurLCdurhrshortxxxxxB}{\ensuremath{2.853}}             
\newcommand{\hatcurLCqxxxxxB}{\ensuremath{0.0423\pm0.0013}}            
\newcommand{\hatcurLCqshortxxxxxB}{\ensuremath{0.042}}                 
\newcommand{\hatcurLCingdurxxxxxB}{\ensuremath{0.0145\pm0.0025}}       
\newcommand{\hatcurLCPxxxxxB}{\ensuremath{2.8126548\pm0.0000055}}      
\newcommand{\hatcurLCPprecxxxxxB}{\ensuremath{2.8126548}}              
\newcommand{\hatcurLCPshortxxxxxB}{\ensuremath{2.8127}}                
\newcommand{\hatcurLCTxxxxxB}{\ensuremath{2456454.6252\pm0.0011}}      
\newcommand{\hatcurLCTAxxxxxB}{\ensuremath{2455416.7556\pm0.0023}}     
\newcommand{\hatcurLCTBxxxxxB}{\ensuremath{2457171.8521\pm0.0018}}     
\newcommand{\hatcurLChatnetmxxxxxB}{\ensuremath{14.17329\pm0.00014}}   
\newcommand{\hatcurLCiblendxxxxxB}{\ensuremath{0.822\pm0.068}}         
\newcommand{\hatcurLCrhoxxxxxB}{\ensuremath{1.19\pm0.23}}              
\newcommand{\hatcurSMEiteffxxxxxB}{\ensuremath{5750\pm110}}            
\newcommand{\hatcurSMEizfehxxxxxB}{\ensuremath{0.380\pm0.070}}         
\newcommand{\hatcurSMEizfehshortxxxxxB}{\ensuremath{0.38}}             
\newcommand{\hatcurSMEiloggxxxxxB}{\ensuremath{4.49\pm0.21}}           
\newcommand{\hatcurSMEivsinxxxxxB}{\ensuremath{3.3\pm1.1}}             
\newcommand{\hatcurSMEivmacxxxxxB}{\ensuremath{0.0}}                   
\newcommand{\hatcurSMEivmicxxxxxB}{\ensuremath{0.0}}                   
\newcommand{\hatcurSMEiiteffxxxxxB}{\ensuremath{5700\pm110}}           
\newcommand{\hatcurSMEiizfehxxxxxB}{\ensuremath{0.390\pm0.050}}        
\newcommand{\hatcurSMEiizfehshortxxxxxB}{\ensuremath{0.39}}            
\newcommand{\hatcurSMEiiloggxxxxxB}{\ensuremath{4.395\pm0.055}}        
\newcommand{\hatcurSMEiivsinxxxxxB}{\ensuremath{3.56\pm0.69}}          
\newcommand{\hatcurLBizxxxxxB}{\ensuremath{0.2272}}                    
\newcommand{\hatcurLBiizxxxxxB}{\ensuremath{0.3299}}                   
\newcommand{\hatcurLBiixxxxxB}{\ensuremath{0.2989}}                    
\newcommand{\hatcurLBiiixxxxxB}{\ensuremath{0.3243}}                   
\newcommand{\hatcurLBiIxxxxxB}{\ensuremath{0.2749}}                    
\newcommand{\hatcurLBiiIxxxxxB}{\ensuremath{0.3270}}                   
\newcommand{\hatcurLBigxxxxxB}{\ensuremath{0.6151}}                    
\newcommand{\hatcurLBiigxxxxxB}{\ensuremath{0.1921}}                   
\newcommand{\hatcurLBirxxxxxB}{\ensuremath{0.4005}}                    
\newcommand{\hatcurLBiirxxxxxB}{\ensuremath{0.3061}}                   
\newcommand{\hatcurLBiRxxxxxB}{\ensuremath{0.3723}}                    
\newcommand{\hatcurLBiiRxxxxxB}{\ensuremath{0.3122}}                   
\newcommand{\hatcurISOmxxxxxB}{\ensuremath{1.099\pm0.044}}             
\newcommand{\hatcurISOmshortxxxxxB}{\ensuremath{1.10}}                 
\newcommand{\hatcurISOmlongxxxxxB}{\ensuremath{1.099\pm0.044}}         
\newcommand{\hatcurISOrxxxxxB}{\ensuremath{1.097_{-0.063}^{+0.098}}}   
\newcommand{\hatcurISOrshortxxxxxB}{\ensuremath{1.10}}                 
\newcommand{\hatcurISOrlongxxxxxB}{\ensuremath{1.097_{-0.063}^{+0.098}}} 
\newcommand{\hatcurISOrhoxxxxxB}{\ensuremath{1.17\pm0.22}}             
\newcommand{\hatcurISOrholongxxxxxB}{\ensuremath{1.17\pm0.22}}         
\newcommand{\hatcurISOloggxxxxxB}{\ensuremath{4.396\pm0.057}}          
\newcommand{\hatcurISOlumxxxxxB}{\ensuremath{1.14_{-0.18}^{+0.24}}}    
\newcommand{\hatcurISOlumshortxxxxxB}{\ensuremath{1.14}}               
\newcommand{\hatcurISOmvxxxxxB}{\ensuremath{4.69\pm0.21}}              
\newcommand{\hatcurISOvixxxxxB}{\ensuremath{0.734\pm0.035}}            
\newcommand{\hatcurISOagexxxxxB}{\ensuremath{3.5\pm1.8}}               
\newcommand{\hatcurISOsigmaxxxxxB}{\ensuremath{0.00100\pm0.00018}}     
\newcommand{\hatcurISOMJxxxxxB}{\ensuremath{3.50\pm0.18}}              
\newcommand{\hatcurISOMHxxxxxB}{\ensuremath{3.16\pm0.17}}              
\newcommand{\hatcurISOMKxxxxxB}{\ensuremath{3.10\pm0.17}}              
\newcommand{\hatcurISOJKxxxxxB}{\ensuremath{0.400\pm0.020}}            
\newcommand{\hatcurISOspecxxxxxB}{G}                                   
\newcommand{\hatcurRVKxxxxxB}{\ensuremath{124\pm14}}                   
\newcommand{\hatcurRVrkxxxxxB}{\ensuremath{0\pm0}}                     
\newcommand{\hatcurRVrhxxxxxB}{\ensuremath{0\pm0}}                     
\newcommand{\hatcurRVkxxxxxB}{\ensuremath{0\pm0}}                      
\newcommand{\hatcurRVhxxxxxB}{\ensuremath{0\pm0}}                      
\newcommand{\hatcurRVtronexxxxxB}{\ensuremath{0\pm0}}                  
\newcommand{\hatcurRVtrtwoxxxxxB}{\ensuremath{0\pm0}}                  
\newcommand{\hatcurRVgammaxxxxxB}{\ensuremath{12423\pm13}}             
\newcommand{\hatcurRVjitterxxxxxB}{\ensuremath{27\pm14}}               
\newcommand{\hatcurRVjittertwosiglimxxxxxB}{\ensuremath{<54.7}}        
\newcommand{\hatcurRVfitrmsxxxxxB}{\ensuremath{.1fym}}                 %
\newcommand{\hatcurRVeccenxxxxxB}{\ensuremath{0\pm0}}                  
\newcommand{\hatcurRVeccentwosiglimxxxxxB}{\ensuremath{<0.000}}        
\newcommand{\hatcurRVomegaxxxxxB}{\ensuremath{0\pm0}}                  
\newcommand{\hatcurPPixxxxxB}{\ensuremath{87.1\pm1.2}}                 
\newcommand{\hatcurPPgxxxxxB}{\ensuremath{14.4\pm2.9}}                 
\newcommand{\hatcurPPloggxxxxxB}{\ensuremath{3.158\pm0.092}}           
\newcommand{\hatcurPParxxxxxB}{\ensuremath{7.88_{-0.61}^{+0.46}}}      
\newcommand{\hatcurPParelxxxxxB}{\ensuremath{0.04024\pm0.00053}}       
\newcommand{\hatcurPPrhoxxxxxB}{\ensuremath{0.58\pm0.16}}              
\newcommand{\hatcurPPmxxxxxB}{\ensuremath{0.92\pm0.10}}                
\newcommand{\hatcurPPmshortxxxxxB}{\ensuremath{0.92}}                  
\newcommand{\hatcurPPmlongxxxxxB}{\ensuremath{0.92\pm0.10}}            
\newcommand{\hatcurPPmexxxxxB}{\ensuremath{291\pm33}}                  
\newcommand{\hatcurPPmeshortxxxxxB}{\ensuremath{291.1}}                
\newcommand{\hatcurPPmelongxxxxxB}{\ensuremath{291\pm33}}              
\newcommand{\hatcurPPrxxxxxB}{\ensuremath{1.249_{-0.096}^{+0.144}}}    
\newcommand{\hatcurPPrshortxxxxxB}{\ensuremath{1.25}}                  
\newcommand{\hatcurPPrlongxxxxxB}{\ensuremath{1.249_{-0.096}^{+0.144}}} 
\newcommand{\hatcurPPrexxxxxB}{\ensuremath{14.0_{-1.1}^{+1.6}}}        
\newcommand{\hatcurPPreshortxxxxxB}{\ensuremath{14.0}}                 
\newcommand{\hatcurPPrelongxxxxxB}{\ensuremath{14.0_{-1.1}^{+1.6}}}    
\newcommand{\hatcurPPmrcorrxxxxxB}{\ensuremath{0.12}}                  
\newcommand{\hatcurPPteffxxxxxB}{\ensuremath{1437\pm58}}               
\newcommand{\hatcurPPthetaxxxxxB}{\ensuremath{0.0531\pm0.0076}}        
\newcommand{\hatcurPPfluxperixxxxxB}{\ensuremath{9.6_{-1.3}^{+1.8}}}   
\newcommand{\hatcurPPfluxperidimxxxxxB}{\ensuremath{8}}                
\newcommand{\hatcurPPfluxapxxxxxB}{\ensuremath{9.6_{-1.3}^{+1.8}}}     
\newcommand{\hatcurPPfluxapdimxxxxxB}{\ensuremath{8}}                  
\newcommand{\hatcurPPfluxavgxxxxxB}{\ensuremath{9.6_{-1.3}^{+1.8}}}    
\newcommand{\hatcurPPfluxavgdimxxxxxB}{\ensuremath{8}}                 
\newcommand{\hatcurPPfluxavglogxxxxxB}{\ensuremath{8.984\pm0.070}}     
\newcommand{\hatcurXsecphasexxxxxB}{\ensuremath{0\pm0}}                
\newcommand{\hatcurXsecondaryxxxxxB}{\ensuremath{2456456.0315\pm0.0011}} 
\newcommand{\hatcurXsecdurxxxxxB}{\ensuremath{0.1189\pm0.0036}}        
\newcommand{\hatcurXsecingdurxxxxxB}{\ensuremath{0.0145\pm0.0025}}     
\newcommand{\hatcurPPphiconjxxxxxB}{\ensuremath{0\pm0}}                
\newcommand{\hatcurPPperixxxxxB}{\ensuremath{2456453.9220\pm0.0011}}   
\newcommand{\hatcurPPaequivxxxxxB}{\ensuremath{0.0376\pm0.0030}}       
\newcommand{\hatcurPPtcircxxxxxB}{\ensuremath{85\pm38}}                
\newcommand{\hatcurPPtinfallxxxxxB}{\ensuremath{970\pm320}}            
\newcommand{\hatcurXdistxxxxxB}{\ensuremath{848_{-55}^{+78}}}          
\newcommand{\hatcurXAvxxxxxB}{\ensuremath{0.064_{-0.064}^{+0.085}}}    
\newcommand{\hatcurXdistredxxxxxB}{\ensuremath{839_{-55}^{+77}}}       
\newcommand{\hatcurXEBVxxxxxB}{\ensuremath{0.021\pm0.022}}             
\newcommand{\hatcurXmvisoredxxxxxB}{\ensuremath{14.387\pm0.012}}       
\newcommand{\hatcurXmiisoredxxxxxB}{\ensuremath{13.614\pm0.012}}       
\newcommand{\hatcurXmjisoredxxxxxB}{\ensuremath{13.141\pm0.020}}       
\newcommand{\hatcurXmhisoredxxxxxB}{\ensuremath{12.793\pm0.031}}       
\newcommand{\hatcurXmkisoredxxxxxB}{\ensuremath{12.731\pm0.032}}       
\newcommand{\hatcurXviisoredxxxxxB}{\ensuremath{0.772\pm0.015}}        
\newcommand{\hatcurXvkisoredxxxxxB}{\ensuremath{1.654_{-0.024}^{+0.034}}} 
\newcommand{\hatcurXjhisoredxxxxxB}{\ensuremath{0.349_{-0.011}^{+0.016}}} 
\newcommand{\hatcurXjkisoredxxxxxB}{\ensuremath{0.4100_{-0.0090}^{+0.0160}}} 
\newcommand{\hatcurCCpmraxxxxxB}{\ensuremath{2.9\pm1.9}}               
\newcommand{\hatcurCCpmdecxxxxxB}{\ensuremath{-20.3\pm1.9}}            
\newcommand{\hatcurCCpmxxxxxB}{\ensuremath{20.5\pm2.7}}                

%% file: HATS747-032.tex
\newcommand{\hatcurhtrxxxxxC}{HATS747-032}                      
\newcommand{\hatcurfieldxxxxxC}{\ensuremath{string}}            
\newcommand{\hatcurCCraxxxxxC}{\ensuremath{19^{\mathrm h}38^{\mathrm m}31.92{\mathrm s}}}                     
\newcommand{\hatcurCCdecxxxxxC}{\ensuremath{-55{\arcdeg}19{\arcmin}48.4{\arcsec}}}                    
\newcommand{\hatcurCCmagxxxxxC}{11.911}                         
\newcommand{\hatcurCCtwomassxxxxxC}{2MASS~19383207-5519483}     
\newcommand{\hatcurCCgscxxxxxC}{GSC~8778-01635}                 
\newcommand{\hatcurCCtassmvxxxxxC}{\ensuremath{11.911\pm0.070}} 
\newcommand{\hatcurCCtassmvshortxxxxxC}{\ensuremath{11.9}}      
\newcommand{\hatcurCCtassmBxxxxxC}{\ensuremath{12.633\pm0.070}} 
\newcommand{\hatcurCCtassmBshortxxxxxC}{\ensuremath{12.6}}      
\newcommand{\hatcurCCtassmIxxxxxC}{\ensuremath{100\pm1000}}     
\newcommand{\hatcurCCtassmIshortxxxxxC}{\ensuremath{100.0}}     
\newcommand{\hatcurCCtassmgxxxxxC}{\ensuremath{12.174\pm0.040}} 
\newcommand{\hatcurCCtassmgshortxxxxxC}{\ensuremath{12.2}}      
\newcommand{\hatcurCCtassmrxxxxxC}{\ensuremath{11.704\pm0.040}} 
\newcommand{\hatcurCCtassmrshortxxxxxC}{\ensuremath{11.7}}      
\newcommand{\hatcurCCtassmixxxxxC}{\ensuremath{11.52\pm0.11}}   
\newcommand{\hatcurCCtassmishortxxxxxC}{\ensuremath{11.5}}      
\newcommand{\hatcurCCtwomassJmagxxxxxC}{\ensuremath{10.659\pm0.024}} 
\newcommand{\hatcurCCtwomassHmagxxxxxC}{\ensuremath{10.337\pm0.026}} 
\newcommand{\hatcurCCtwomassKmagxxxxxC}{\ensuremath{10.287\pm0.027}} 
\newcommand{\hatcurCCcitJmagxxxxxC}{\ensuremath{10.676\pm0.024}} 
\newcommand{\hatcurCCcitHmagxxxxxC}{\ensuremath{10.332\pm0.026}} 
\newcommand{\hatcurCCcitKmagxxxxxC}{\ensuremath{10.311\pm0.027}} 
\newcommand{\hatcurCCbbJmagxxxxxC}{\ensuremath{10.725\pm0.026}} 
\newcommand{\hatcurCCbbHmagxxxxxC}{\ensuremath{10.353\pm0.027}} 
\newcommand{\hatcurCCbbKmagxxxxxC}{\ensuremath{10.331\pm0.027}} 
\newcommand{\hatcurCCesoJmagxxxxxC}{\ensuremath{10.727\pm0.027}} 
\newcommand{\hatcurCCesoHmagxxxxxC}{\ensuremath{10.347\pm0.030}} 
\newcommand{\hatcurCCesoKmagxxxxxC}{\ensuremath{10.330\pm0.028}} 
\newcommand{\hatcurCCesoJHmagxxxxxC}{\ensuremath{0.380\pm0.038}} 
\newcommand{\hatcurCCesoJKmagxxxxxC}{\ensuremath{0.398\pm0.039}} 
\newcommand{\hatcurCCesoHKmagxxxxxC}{\ensuremath{0.018\pm0.041}} 
\newcommand{\hatcurLCdipxxxxxC}{\ensuremath{14.0}}              
\newcommand{\hatcurLCrprstarxxxxxC}{\ensuremath{0.1237\pm0.0080}} 
\newcommand{\hatcurLCbsqxxxxxC}{\ensuremath{0.107_{-0.065}^{+0.081}}} 
\newcommand{\hatcurLCimpxxxxxC}{\ensuremath{0.33_{-0.12}^{+0.11}}} 
\newcommand{\hatcurLCzetaxxxxxC}{\ensuremath{20.45\pm0.23}}     
\newcommand{\hatcurLCdurxxxxxC}{\ensuremath{0.1115\pm0.0018}}   
\newcommand{\hatcurLCdurshortxxxxxC}{\ensuremath{0.1115}}       
\newcommand{\hatcurLCdurhrxxxxxC}{\ensuremath{2.676\pm0.044}}   
\newcommand{\hatcurLCdurhrshortxxxxxC}{\ensuremath{2.676}}      
\newcommand{\hatcurLCqxxxxxC}{\ensuremath{0.04370\pm0.00072}}   
\newcommand{\hatcurLCqshortxxxxxC}{\ensuremath{0.044}}          
\newcommand{\hatcurLCingdurxxxxxC}{\ensuremath{0.0135\pm0.0016}} 
\newcommand{\hatcurLCPxxxxxC}{\ensuremath{2.5495551\pm0.0000061}} 
\newcommand{\hatcurLCPprecxxxxxC}{\ensuremath{2.5495551}}       
\newcommand{\hatcurLCPshortxxxxxC}{\ensuremath{2.5496}}         
\newcommand{\hatcurLCTxxxxxC}{\ensuremath{2456497.23181\pm0.00050}} 
\newcommand{\hatcurLCTAxxxxxC}{\ensuremath{2456367.20452\pm0.00067}} 
\newcommand{\hatcurLCTBxxxxxC}{\ensuremath{2457162.6658\pm0.0015}} 
\newcommand{\hatcurLChatnetmxxxxxC}{\ensuremath{11.783440\pm0.000048}} 
\newcommand{\hatcurLCiblendxxxxxC}{\ensuremath{0.769\pm0.099}}  
\newcommand{\hatcurLCrhoxxxxxC}{\ensuremath{1.42\pm0.17}}       
\newcommand{\hatcurSMEiteffxxxxxC}{\ensuremath{5700\pm100}}     
\newcommand{\hatcurSMEizfehxxxxxC}{\ensuremath{0.280\pm0.050}}  
\newcommand{\hatcurSMEizfehshortxxxxxC}{\ensuremath{0.28}}      
\newcommand{\hatcurSMEiloggxxxxxC}{\ensuremath{4.54\pm0.15}}    
\newcommand{\hatcurSMEivsinxxxxxC}{\ensuremath{3.42\pm0.66}}    
\newcommand{\hatcurSMEivmacxxxxxC}{\ensuremath{0.0}}            
\newcommand{\hatcurSMEivmicxxxxxC}{\ensuremath{0.0}}            
\newcommand{\hatcurSMEiiteffxxxxxC}{\ensuremath{5659\pm85}}     
\newcommand{\hatcurSMEiizfehxxxxxC}{\ensuremath{0.290\pm0.050}} 
\newcommand{\hatcurSMEiizfehshortxxxxxC}{\ensuremath{0.29}}     
\newcommand{\hatcurSMEiiloggxxxxxC}{\ensuremath{4.443\pm0.041}} 
\newcommand{\hatcurSMEiivsinxxxxxC}{\ensuremath{3.87\pm0.42}}   
\newcommand{\hatcurLBizxxxxxC}{\ensuremath{0.2308}}             
\newcommand{\hatcurLBiizxxxxxC}{\ensuremath{0.3242}}            
\newcommand{\hatcurLBiixxxxxC}{\ensuremath{0.3013}}             
\newcommand{\hatcurLBiiixxxxxC}{\ensuremath{0.3193}}            
\newcommand{\hatcurLBiIxxxxxC}{\ensuremath{0.2778}}             
\newcommand{\hatcurLBiiIxxxxxC}{\ensuremath{0.3216}}            
\newcommand{\hatcurLBigxxxxxC}{\ensuremath{0.6140}}             
\newcommand{\hatcurLBiigxxxxxC}{\ensuremath{0.1919}}            
\newcommand{\hatcurLBirxxxxxC}{\ensuremath{0.4016}}             
\newcommand{\hatcurLBiirxxxxxC}{\ensuremath{0.3030}}            
\newcommand{\hatcurLBiRxxxxxC}{\ensuremath{0.3737}}             
\newcommand{\hatcurLBiiRxxxxxC}{\ensuremath{0.3085}}            
\newcommand{\hatcurLBikepxxxxxC}{\ensuremath{0.1000}}           
\newcommand{\hatcurLBiikepxxxxxC}{\ensuremath{0.1000}}          
\newcommand{\hatcurISOmxxxxxC}{\ensuremath{1.062\pm0.032}}      
\newcommand{\hatcurISOmshortxxxxxC}{\ensuremath{1.06}}          
\newcommand{\hatcurISOmlongxxxxxC}{\ensuremath{1.062\pm0.032}}  
\newcommand{\hatcurISOrxxxxxC}{\ensuremath{1.022_{-0.037}^{+0.050}}} 
\newcommand{\hatcurISOrshortxxxxxC}{\ensuremath{1.02}}          
\newcommand{\hatcurISOrlongxxxxxC}{\ensuremath{1.022_{-0.037}^{+0.050}}} 
\newcommand{\hatcurISOrhoxxxxxC}{\ensuremath{1.40\pm0.16}}      
\newcommand{\hatcurISOrholongxxxxxC}{\ensuremath{1.40\pm0.16}}  
\newcommand{\hatcurISOloggxxxxxC}{\ensuremath{4.445\pm0.036}}   
\newcommand{\hatcurISOlumxxxxxC}{\ensuremath{0.96\pm0.12}}      
\newcommand{\hatcurISOlumshortxxxxxC}{\ensuremath{0.96}}        
\newcommand{\hatcurISOmvxxxxxC}{\ensuremath{4.89\pm0.14}}       
\newcommand{\hatcurISOvixxxxxC}{\ensuremath{0.744\pm0.026}}     
\newcommand{\hatcurISOagexxxxxC}{\ensuremath{3.0\pm1.7}}        
\newcommand{\hatcurISOsigmaxxxxxC}{\ensuremath{0.00150\pm0.00015}} 
\newcommand{\hatcurISOMJxxxxxC}{\ensuremath{3.68\pm0.11}}       
\newcommand{\hatcurISOMHxxxxxC}{\ensuremath{3.33\pm0.10}}       
\newcommand{\hatcurISOMKxxxxxC}{\ensuremath{3.270\pm0.099}}     
\newcommand{\hatcurISOJKxxxxxC}{\ensuremath{0.410\pm0.020}}     
\newcommand{\hatcurISOspecxxxxxC}{G}                            
\newcommand{\hatcurRVKxxxxxC}{\ensuremath{170.1\pm6.8}}         
\newcommand{\hatcurRVrkxxxxxC}{\ensuremath{0\pm0}}              
\newcommand{\hatcurRVrhxxxxxC}{\ensuremath{0\pm0}}              
\newcommand{\hatcurRVkxxxxxC}{\ensuremath{0\pm0}}               
\newcommand{\hatcurRVhxxxxxC}{\ensuremath{0\pm0}}               
\newcommand{\hatcurRVtronexxxxxC}{\ensuremath{0\pm0}}           
\newcommand{\hatcurRVtrtwoxxxxxC}{\ensuremath{0\pm0}}           
\newcommand{\hatcurRVgammaAxxxxxC}{\ensuremath{11057\pm24}}     
\newcommand{\hatcurRVjitterAxxxxxC}{\ensuremath{36\pm22}}       
\newcommand{\hatcurRVjittertwosiglimAxxxxxC}{\ensuremath{<86.5}} 
\newcommand{\hatcurRVfitrmsAxxxxxC}{\ensuremath{0.0}}           
\newcommand{\hatcurRVgammaBxxxxxC}{\ensuremath{11056\pm24}}     
\newcommand{\hatcurRVjitterBxxxxxC}{\ensuremath{46\pm31}}       
\newcommand{\hatcurRVjittertwosiglimBxxxxxC}{\ensuremath{<106.2}} 
\newcommand{\hatcurRVfitrmsBxxxxxC}{\ensuremath{0.0}}           
\newcommand{\hatcurRVgammaCxxxxxC}{\ensuremath{11077\pm12}}     
\newcommand{\hatcurRVjitterCxxxxxC}{\ensuremath{0\pm12}}        
\newcommand{\hatcurRVjittertwosiglimCxxxxxC}{\ensuremath{<19.2}} 
\newcommand{\hatcurRVfitrmsCxxxxxC}{\ensuremath{0.0}}           
\newcommand{\hatcurRVgammaDxxxxxC}{\ensuremath{11066.1\pm7.2}}  
\newcommand{\hatcurRVjitterDxxxxxC}{\ensuremath{19.4\pm7.2}}    
\newcommand{\hatcurRVjittertwosiglimDxxxxxC}{\ensuremath{<32.6}} 
\newcommand{\hatcurRVfitrmsDxxxxxC}{\ensuremath{.1fym}}         %
\newcommand{\hatcurRVeccenxxxxxC}{\ensuremath{0\pm0}}           
\newcommand{\hatcurRVeccentwosiglimxxxxxC}{\ensuremath{<0.000}} 
\newcommand{\hatcurRVomegaxxxxxC}{\ensuremath{0\pm0}}           
\newcommand{\hatcurPPixxxxxC}{\ensuremath{87.62\pm0.92}}        
\newcommand{\hatcurPPgxxxxxC}{\ensuremath{19.4\pm3.2}}          
\newcommand{\hatcurPPloggxxxxxC}{\ensuremath{3.289\pm0.072}}    
\newcommand{\hatcurPParxxxxxC}{\ensuremath{7.83\pm0.32}}        
\newcommand{\hatcurPParelxxxxxC}{\ensuremath{0.03727\pm0.00037}} 
\newcommand{\hatcurPPrhoxxxxxC}{\ensuremath{0.79\pm0.19}}       
\newcommand{\hatcurPPmxxxxxC}{\ensuremath{1.192\pm0.053}}       
\newcommand{\hatcurPPmshortxxxxxC}{\ensuremath{1.19}}           
\newcommand{\hatcurPPmlongxxxxxC}{\ensuremath{1.192\pm0.053}}   
\newcommand{\hatcurPPmexxxxxC}{\ensuremath{379\pm17}}           
\newcommand{\hatcurPPmeshortxxxxxC}{\ensuremath{379.0}}         
\newcommand{\hatcurPPmelongxxxxxC}{\ensuremath{379\pm17}}       
\newcommand{\hatcurPPrxxxxxC}{\ensuremath{1.230_{-0.081}^{+0.112}}} 
\newcommand{\hatcurPPrshortxxxxxC}{\ensuremath{1.23}}           
\newcommand{\hatcurPPrlongxxxxxC}{\ensuremath{1.230_{-0.081}^{+0.112}}} 
\newcommand{\hatcurPPrexxxxxC}{\ensuremath{13.79_{-0.90}^{+1.25}}} 
\newcommand{\hatcurPPreshortxxxxxC}{\ensuremath{13.8}}          
\newcommand{\hatcurPPrelongxxxxxC}{\ensuremath{13.79_{-0.90}^{+1.25}}} 
\newcommand{\hatcurPPmrcorrxxxxxC}{\ensuremath{0.03}}           
\newcommand{\hatcurPPteffxxxxxC}{\ensuremath{1429\pm38}}        
\newcommand{\hatcurPPthetaxxxxxC}{\ensuremath{0.0675\pm0.0061}} 
\newcommand{\hatcurPPfluxperixxxxxC}{\ensuremath{9.4\pm1.0}}    
\newcommand{\hatcurPPfluxperidimxxxxxC}{\ensuremath{8}}         
\newcommand{\hatcurPPfluxapxxxxxC}{\ensuremath{9.4\pm1.0}}      
\newcommand{\hatcurPPfluxapdimxxxxxC}{\ensuremath{8}}           
\newcommand{\hatcurPPfluxavgxxxxxC}{\ensuremath{9.4\pm1.0}}     
\newcommand{\hatcurPPfluxavgdimxxxxxC}{\ensuremath{8}}          
\newcommand{\hatcurPPfluxavglogxxxxxC}{\ensuremath{8.973\pm0.046}} 
\newcommand{\hatcurXsecphasexxxxxC}{\ensuremath{0\pm0}}         
\newcommand{\hatcurXsecondaryxxxxxC}{\ensuremath{2456498.50659\pm0.00050}} 
\newcommand{\hatcurXsecdurxxxxxC}{\ensuremath{0.1115\pm0.0018}} 
\newcommand{\hatcurXsecingdurxxxxxC}{\ensuremath{0.0135\pm0.0016}} 
\newcommand{\hatcurPPphiconjxxxxxC}{\ensuremath{0\pm0}}         
\newcommand{\hatcurPPperixxxxxC}{\ensuremath{2456496.59442\pm0.00050}} 
\newcommand{\hatcurPPaequivxxxxxC}{\ensuremath{0.0381\pm0.0020}} 
\newcommand{\hatcurPPtcircxxxxxC}{\ensuremath{77\pm32}}         
\newcommand{\hatcurPPtinfallxxxxxC}{\ensuremath{640\pm120}}     
\newcommand{\hatcurXdistxxxxxC}{\ensuremath{258\pm12}}          
\newcommand{\hatcurXAvxxxxxC}{\ensuremath{0.000\pm0.058}}       
\newcommand{\hatcurXdistredxxxxxC}{\ensuremath{255\pm12}}       
\newcommand{\hatcurXEBVxxxxxC}{\ensuremath{0.000\pm0.019}}      
\newcommand{\hatcurXmvisoredxxxxxC}{\ensuremath{11.954\pm0.050}} 
\newcommand{\hatcurXmiisoredxxxxxC}{\ensuremath{11.196\pm0.028}} 
\newcommand{\hatcurXmjisoredxxxxxC}{\ensuremath{10.726\pm0.018}} 
\newcommand{\hatcurXmhisoredxxxxxC}{\ensuremath{10.367\pm0.017}} 
\newcommand{\hatcurXmkisoredxxxxxC}{\ensuremath{10.307\pm0.019}} 
\newcommand{\hatcurXviisoredxxxxxC}{\ensuremath{0.757\pm0.026}} 
\newcommand{\hatcurXvkisoredxxxxxC}{\ensuremath{1.646\pm0.057}} 
\newcommand{\hatcurXjhisoredxxxxxC}{\ensuremath{0.358\pm0.015}} 
\newcommand{\hatcurXjkisoredxxxxxC}{\ensuremath{0.418\pm0.017}} 
\newcommand{\hatcurCCpmraxxxxxC}{\ensuremath{7.1\pm1.1}}        
\newcommand{\hatcurCCpmdecxxxxxC}{\ensuremath{-40.6\pm1.3}}     
\newcommand{\hatcurCCpmxxxxxC}{\ensuremath{41.2\pm1.7}}         

%% file: HATS754-010.tex
\newcommand{\hatcurhtrxxxxxD}{HATS754-010}                             
\newcommand{\hatcurfieldxxxxxD}{\ensuremath{string}}                   
\newcommand{\hatcurCCraxxxxxD}{\ensuremath{00^{\mathrm h}03^{\mathrm m}05.88{\mathrm s}}}                            
\newcommand{\hatcurCCdecxxxxxD}{\ensuremath{-62{\arcdeg}28{\arcmin}09.6{\arcsec}}}                           
\newcommand{\hatcurCCmagxxxxxD}{13.849}                                
\newcommand{\hatcurCCtwomassxxxxxD}{2MASS~00030587-6228096}            
\newcommand{\hatcurCCgscxxxxxD}{GSC~8840-01777}                        
\newcommand{\hatcurCCtassmvxxxxxD}{\ensuremath{13.849\pm0.010}}        
\newcommand{\hatcurCCtassmvshortxxxxxD}{\ensuremath{13.8}}             
\newcommand{\hatcurCCtassmBxxxxxD}{\ensuremath{14.595\pm0.010}}        
\newcommand{\hatcurCCtassmBshortxxxxxD}{\ensuremath{14.6}}             
\newcommand{\hatcurCCtassmIxxxxxD}{\ensuremath{nff\pmnff}}             
\newcommand{\hatcurCCtassmIshortxxxxxD}{\ensuremath{0.0}}              
\newcommand{\hatcurCCtassmgxxxxxD}{\ensuremath{14.182\pm0.010}}        
\newcommand{\hatcurCCtassmgshortxxxxxD}{\ensuremath{14.2}}             
\newcommand{\hatcurCCtassmrxxxxxD}{\ensuremath{13.650\pm0.010}}        
\newcommand{\hatcurCCtassmrshortxxxxxD}{\ensuremath{13.7}}             
\newcommand{\hatcurCCtassmixxxxxD}{\ensuremath{13.459\pm0.050}}        
\newcommand{\hatcurCCtassmishortxxxxxD}{\ensuremath{13.5}}             
\newcommand{\hatcurCCtwomassJmagxxxxxD}{\ensuremath{12.513\pm0.024}}   
\newcommand{\hatcurCCtwomassHmagxxxxxD}{\ensuremath{12.139\pm0.022}}   
\newcommand{\hatcurCCtwomassKmagxxxxxD}{\ensuremath{12.095\pm0.019}}   
\newcommand{\hatcurCCcitJmagxxxxxD}{\ensuremath{12.527\pm0.024}}       
\newcommand{\hatcurCCcitHmagxxxxxD}{\ensuremath{12.134\pm0.023}}       
\newcommand{\hatcurCCcitKmagxxxxxD}{\ensuremath{12.119\pm0.019}}       
\newcommand{\hatcurCCbbJmagxxxxxD}{\ensuremath{12.580\pm0.026}}        
\newcommand{\hatcurCCbbHmagxxxxxD}{\ensuremath{12.155\pm0.023}}        
\newcommand{\hatcurCCbbKmagxxxxxD}{\ensuremath{12.139\pm0.019}}        
\newcommand{\hatcurCCesoJmagxxxxxD}{\ensuremath{12.583\pm0.028}}       
\newcommand{\hatcurCCesoHmagxxxxxD}{\ensuremath{12.149\pm0.026}}       
\newcommand{\hatcurCCesoKmagxxxxxD}{\ensuremath{12.138\pm0.020}}       
\newcommand{\hatcurCCesoJHmagxxxxxD}{\ensuremath{0.434\pm0.035}}       
\newcommand{\hatcurCCesoJKmagxxxxxD}{\ensuremath{0.446\pm0.033}}       
\newcommand{\hatcurCCesoHKmagxxxxxD}{\ensuremath{0.011\pm0.032}}       
\newcommand{\hatcurLCdipxxxxxD}{\ensuremath{13.4}}                     
\newcommand{\hatcurLCrprstarxxxxxD}{\ensuremath{0.150\pm0.014}}        
\newcommand{\hatcurLCbsqxxxxxD}{\ensuremath{0.879_{-0.063}^{+0.041}}}  
\newcommand{\hatcurLCimpxxxxxD}{\ensuremath{0.937_{-0.034}^{+0.022}}}  
\newcommand{\hatcurLCzetaxxxxxD}{\ensuremath{61\pm11}}                 
\newcommand{\hatcurLCdurxxxxxD}{\ensuremath{0.0648\pm0.0029}}          
\newcommand{\hatcurLCdurshortxxxxxD}{\ensuremath{0.0648}}              
\newcommand{\hatcurLCdurhrxxxxxD}{\ensuremath{1.555\pm0.069}}          
\newcommand{\hatcurLCdurhrshortxxxxxD}{\ensuremath{1.555}}             
\newcommand{\hatcurLCqxxxxxD}{\ensuremath{0.0308\pm0.0014}}            
\newcommand{\hatcurLCqshortxxxxxD}{\ensuremath{0.031}}                 
\newcommand{\hatcurLCingdurxxxxxD}{\ensuremath{0.081\pm0.011}}         
\newcommand{\hatcurLCPxxxxxD}{\ensuremath{2.1061607\pm0.0000047}}      
\newcommand{\hatcurLCPprecxxxxxD}{\ensuremath{2.1061607}}              
\newcommand{\hatcurLCPshortxxxxxD}{\ensuremath{2.1062}}                
\newcommand{\hatcurLCTxxxxxD}{\ensuremath{2456634.85732\pm0.00075}}    
\newcommand{\hatcurLCTAxxxxxD}{\ensuremath{2456179.9267\pm0.0013}}     
\newcommand{\hatcurLCTBxxxxxD}{\ensuremath{2457289.8733\pm0.0016}}     
\newcommand{\hatcurLChatnetmxxxxxD}{\ensuremath{13.687650\pm0.000096}} 
\newcommand{\hatcurLCiblendxxxxxD}{\ensuremath{0.822\pm0.086}}         
\newcommand{\hatcurLCrhoxxxxxD}{\ensuremath{1.44\pm0.25}}              
\newcommand{\hatcurSMEiteffxxxxxD}{\ensuremath{5650\pm140}}            
\newcommand{\hatcurSMEizfehxxxxxD}{\ensuremath{0.380\pm0.070}}         
\newcommand{\hatcurSMEizfehshortxxxxxD}{\ensuremath{0.38}}             
\newcommand{\hatcurSMEiloggxxxxxD}{\ensuremath{4.69\pm0.17}}           
\newcommand{\hatcurSMEivsinxxxxxD}{\ensuremath{3.78\pm0.76}}           
\newcommand{\hatcurSMEivmacxxxxxD}{\ensuremath{0.0}}                   
\newcommand{\hatcurSMEivmicxxxxxD}{\ensuremath{0.0}}                   
\newcommand{\hatcurSMEiiteffxxxxxD}{\ensuremath{5380\pm73}}            
\newcommand{\hatcurSMEiizfehxxxxxD}{\ensuremath{0.250\pm0.070}}        
\newcommand{\hatcurSMEiizfehshortxxxxxD}{\ensuremath{0.25}}            
\newcommand{\hatcurSMEiiloggxxxxxD}{\ensuremath{4.425\pm0.046}}        
\newcommand{\hatcurSMEiivsinxxxxxD}{\ensuremath{4.07\pm0.58}}          
\newcommand{\hatcurLBizxxxxxD}{\ensuremath{0.2673}}                    
\newcommand{\hatcurLBiizxxxxxD}{\ensuremath{0.3043}}                   
\newcommand{\hatcurLBiixxxxxD}{\ensuremath{0.3462}}                    
\newcommand{\hatcurLBiiixxxxxD}{\ensuremath{0.2910}}                   
\newcommand{\hatcurLBiIxxxxxD}{\ensuremath{0.3203}}                    
\newcommand{\hatcurLBiiIxxxxxD}{\ensuremath{0.2955}}                   
\newcommand{\hatcurLBigxxxxxD}{\ensuremath{0.6869}}                    
\newcommand{\hatcurLBiigxxxxxD}{\ensuremath{0.1335}}                   
\newcommand{\hatcurLBirxxxxxD}{\ensuremath{0.4588}}                    
\newcommand{\hatcurLBiirxxxxxD}{\ensuremath{0.2646}}                   
\newcommand{\hatcurLBiRxxxxxD}{\ensuremath{0.4277}}                    
\newcommand{\hatcurLBiiRxxxxxD}{\ensuremath{0.2728}}                   
\newcommand{\hatcurLBikepxxxxxD}{\ensuremath{0.1000}}                  
\newcommand{\hatcurLBiikepxxxxxD}{\ensuremath{0.1000}}                 
\newcommand{\hatcurISOmxxxxxD}{\ensuremath{0.955\pm0.031}}             
\newcommand{\hatcurISOmshortxxxxxD}{\ensuremath{0.95}}                 
\newcommand{\hatcurISOmlongxxxxxD}{\ensuremath{0.955\pm0.031}}         
\newcommand{\hatcurISOrxxxxxD}{\ensuremath{0.980\pm0.047}}             
\newcommand{\hatcurISOrshortxxxxxD}{\ensuremath{0.98}}                 
\newcommand{\hatcurISOrlongxxxxxD}{\ensuremath{0.980\pm0.047}}         
\newcommand{\hatcurISOrhoxxxxxD}{\ensuremath{1.44\pm0.22}}             
\newcommand{\hatcurISOrholongxxxxxD}{\ensuremath{1.44\pm0.22}}         
\newcommand{\hatcurISOloggxxxxxD}{\ensuremath{4.435\pm0.043}}          
\newcommand{\hatcurISOlumxxxxxD}{\ensuremath{0.724\pm0.089}}           
\newcommand{\hatcurISOlumshortxxxxxD}{\ensuremath{0.72}}               
\newcommand{\hatcurISOmvxxxxxD}{\ensuremath{5.26\pm0.15}}              
\newcommand{\hatcurISOvixxxxxD}{\ensuremath{0.821\pm0.019}}            
\newcommand{\hatcurISOagexxxxxD}{\ensuremath{7.7\pm2.7}}               
\newcommand{\hatcurISOsigmaxxxxxD}{\ensuremath{0.00130\pm0.00019}}     
\newcommand{\hatcurISOMJxxxxxD}{\ensuremath{3.90\pm0.12}}              
\newcommand{\hatcurISOMHxxxxxD}{\ensuremath{3.48\pm0.11}}              
\newcommand{\hatcurISOMKxxxxxD}{\ensuremath{3.41\pm0.11}}              
\newcommand{\hatcurISOJKxxxxxD}{\ensuremath{0.25\pm0.24}}              
\newcommand{\hatcurISOspecxxxxxD}{G}                                   
\newcommand{\hatcurRVKxxxxxD}{\ensuremath{152\pm11}}                   
\newcommand{\hatcurRVrkxxxxxD}{\ensuremath{0\pm0}}                     
\newcommand{\hatcurRVrhxxxxxD}{\ensuremath{0\pm0}}                     
\newcommand{\hatcurRVkxxxxxD}{\ensuremath{0\pm0}}                      
\newcommand{\hatcurRVhxxxxxD}{\ensuremath{0\pm0}}                      
\newcommand{\hatcurRVtronexxxxxD}{\ensuremath{0\pm0}}                  
\newcommand{\hatcurRVtrtwoxxxxxD}{\ensuremath{0\pm0}}                  
\newcommand{\hatcurRVgammaxxxxxD}{\ensuremath{17735.1\pm9.3}}          
\newcommand{\hatcurRVjitterxxxxxD}{\ensuremath{0\pm15}}                
\newcommand{\hatcurRVjittertwosiglimxxxxxD}{\ensuremath{<31.3}}        
\newcommand{\hatcurRVfitrmsxxxxxD}{\ensuremath{.1fym}}                 %
\newcommand{\hatcurRVeccenxxxxxD}{\ensuremath{0\pm0}}                  
\newcommand{\hatcurRVeccentwosiglimxxxxxD}{\ensuremath{<0.000}}        
\newcommand{\hatcurRVomegaxxxxxD}{\ensuremath{0\pm0}}                  
\newcommand{\hatcurPPixxxxxD}{\ensuremath{82.28_{-0.59}^{+0.43}}}      
\newcommand{\hatcurPPgxxxxxD}{\ensuremath{11.3_{-2.6}^{+3.6}}}         
\newcommand{\hatcurPPloggxxxxxD}{\ensuremath{3.05\pm0.12}}             
\newcommand{\hatcurPParxxxxxD}{\ensuremath{6.96\pm0.34}}               
\newcommand{\hatcurPParelxxxxxD}{\ensuremath{0.03166\pm0.00034}}       
\newcommand{\hatcurPPrhoxxxxxD}{\ensuremath{0.40_{-0.13}^{+0.19}}}     
\newcommand{\hatcurPPmxxxxxD}{\ensuremath{0.941\pm0.072}}              
\newcommand{\hatcurPPmshortxxxxxD}{\ensuremath{0.94}}                  
\newcommand{\hatcurPPmlongxxxxxD}{\ensuremath{0.941\pm0.072}}          
\newcommand{\hatcurPPmexxxxxD}{\ensuremath{299\pm23}}                  
\newcommand{\hatcurPPmeshortxxxxxD}{\ensuremath{298.9}}                
\newcommand{\hatcurPPmelongxxxxxD}{\ensuremath{299\pm23}}              
\newcommand{\hatcurPPrxxxxxD}{\ensuremath{1.43\pm0.19}}                
\newcommand{\hatcurPPrshortxxxxxD}{\ensuremath{1.43}}                  
\newcommand{\hatcurPPrlongxxxxxD}{\ensuremath{1.43\pm0.19}}            
\newcommand{\hatcurPPrexxxxxD}{\ensuremath{16.0\pm2.1}}                
\newcommand{\hatcurPPreshortxxxxxD}{\ensuremath{16.0}}                 
\newcommand{\hatcurPPrelongxxxxxD}{\ensuremath{16.0\pm2.1}}            
\newcommand{\hatcurPPmrcorrxxxxxD}{\ensuremath{-0.00}}                 
\newcommand{\hatcurPPteffxxxxxD}{\ensuremath{1445\pm42}}               
\newcommand{\hatcurPPthetaxxxxxD}{\ensuremath{0.0430\pm0.0068}}        
\newcommand{\hatcurPPfluxperixxxxxD}{\ensuremath{9.9\pm1.1}}           
\newcommand{\hatcurPPfluxperidimxxxxxD}{\ensuremath{8}}                
\newcommand{\hatcurPPfluxapxxxxxD}{\ensuremath{9.9\pm1.1}}             
\newcommand{\hatcurPPfluxapdimxxxxxD}{\ensuremath{8}}                  
\newcommand{\hatcurPPfluxavgxxxxxD}{\ensuremath{9.9\pm1.1}}            
\newcommand{\hatcurPPfluxavgdimxxxxxD}{\ensuremath{8}}                 
\newcommand{\hatcurPPfluxavglogxxxxxD}{\ensuremath{8.994\pm0.050}}     
\newcommand{\hatcurXsecphasexxxxxD}{\ensuremath{0\pm0}}                
\newcommand{\hatcurXsecondaryxxxxxD}{\ensuremath{2456635.91040\pm0.00075}} 
\newcommand{\hatcurXsecdurxxxxxD}{\ensuremath{0.0743\pm0.0063}}        
\newcommand{\hatcurXsecingdurxxxxxD}{\ensuremath{0.0372\pm0.0038}}     
\newcommand{\hatcurPPphiconjxxxxxD}{\ensuremath{0\pm0}}                
\newcommand{\hatcurPPperixxxxxD}{\ensuremath{2456634.33078\pm0.00075}} 
\newcommand{\hatcurPPaequivxxxxxD}{\ensuremath{0.0372\pm0.0022}}       
\newcommand{\hatcurPPtcircxxxxxD}{\ensuremath{11.8_{-5.6}^{+10.1}}}    
\newcommand{\hatcurPPtinfallxxxxxD}{\ensuremath{325_{-74}^{+98}}}      
\newcommand{\hatcurXdistxxxxxD}{\ensuremath{556\pm29}}                 
\newcommand{\hatcurXAvxxxxxD}{\ensuremath{0.0000\pm0.0063}}            
\newcommand{\hatcurXdistredxxxxxD}{\ensuremath{532\pm32}}              
\newcommand{\hatcurXEBVxxxxxD}{\ensuremath{0.0000\pm0.0020}}           
\newcommand{\hatcurXmvisoredxxxxxD}{\ensuremath{13.887\pm0.020}}       
\newcommand{\hatcurXmiisoredxxxxxD}{\ensuremath{13.0660\pm0.0087}}     
\newcommand{\hatcurXmjisoredxxxxxD}{\ensuremath{12.525\pm0.024}}       
\newcommand{\hatcurXmhisoredxxxxxD}{\ensuremath{12.108\pm0.038}}       
\newcommand{\hatcurXmkisoredxxxxxD}{\ensuremath{12.038\pm0.040}}       
\newcommand{\hatcurXviisoredxxxxxD}{\ensuremath{0.821\pm0.018}}        
\newcommand{\hatcurXvkisoredxxxxxD}{\ensuremath{1.849\pm0.058}}        
\newcommand{\hatcurXjhisoredxxxxxD}{\ensuremath{0.417\pm0.016}}        
\newcommand{\hatcurXjkisoredxxxxxD}{\ensuremath{0.486\pm0.018}}        
\newcommand{\hatcurCCpmraxxxxxD}{\ensuremath{2.1\pm1.4}}               
\newcommand{\hatcurCCpmdecxxxxxD}{\ensuremath{4.7\pm1.5}}              
\newcommand{\hatcurCCpmxxxxxD}{\ensuremath{5.1\pm2.1}}                 

%% file: HATS779-003.tex
\newcommand{\hatcurhtrxxxxxE}{HATS779-003}                      
\newcommand{\hatcurfieldxxxxxE}{\ensuremath{string}}            
\newcommand{\hatcurCCraxxxxxE}{\ensuremath{19^{\mathrm h}46^{\mathrm m}45.12{\mathrm s}}}                     
\newcommand{\hatcurCCdecxxxxxE}{\ensuremath{-63{\arcdeg}33{\arcmin}56.2{\arcsec}}}                    
\newcommand{\hatcurCCmagxxxxxE}{12.56}                          
\newcommand{\hatcurCCtwomassxxxxxE}{2MASS~19464518-6333561}     
\newcommand{\hatcurCCgscxxxxxE}{GSC~9089-00775}                 
\newcommand{\hatcurCCtassmvxxxxxE}{\ensuremath{12.56\pm0.10}}   
\newcommand{\hatcurCCtassmvshortxxxxxE}{\ensuremath{12.6}}      
\newcommand{\hatcurCCtassmBxxxxxE}{\ensuremath{13.26\pm0.10}}   
\newcommand{\hatcurCCtassmBshortxxxxxE}{\ensuremath{13.3}}      
\newcommand{\hatcurCCtassmIxxxxxE}{\ensuremath{nff\pmnff}}      
\newcommand{\hatcurCCtassmIshortxxxxxE}{\ensuremath{0.0}}       
\newcommand{\hatcurCCtassmgxxxxxE}{\ensuremath{nff\pmnff}}      
\newcommand{\hatcurCCtassmgshortxxxxxE}{\ensuremath{0.0}}       
\newcommand{\hatcurCCtassmrxxxxxE}{\ensuremath{nff\pmnff}}      
\newcommand{\hatcurCCtassmrshortxxxxxE}{\ensuremath{0.0}}       
\newcommand{\hatcurCCtassmixxxxxE}{\ensuremath{nff\pmnff}}      
\newcommand{\hatcurCCtassmishortxxxxxE}{\ensuremath{0.0}}       
\newcommand{\hatcurCCtwomassJmagxxxxxE}{\ensuremath{11.439\pm0.025}} 
\newcommand{\hatcurCCtwomassHmagxxxxxE}{\ensuremath{11.192\pm0.023}} 
\newcommand{\hatcurCCtwomassKmagxxxxxE}{\ensuremath{11.118\pm0.025}} 
\newcommand{\hatcurCCcitJmagxxxxxE}{\ensuremath{11.458\pm0.026}} 
\newcommand{\hatcurCCcitHmagxxxxxE}{\ensuremath{11.187\pm0.023}} 
\newcommand{\hatcurCCcitKmagxxxxxE}{\ensuremath{11.142\pm0.025}} 
\newcommand{\hatcurCCbbJmagxxxxxE}{\ensuremath{11.504\pm0.026}} 
\newcommand{\hatcurCCbbHmagxxxxxE}{\ensuremath{11.208\pm0.024}} 
\newcommand{\hatcurCCbbKmagxxxxxE}{\ensuremath{11.162\pm0.025}} 
\newcommand{\hatcurCCesoJmagxxxxxE}{\ensuremath{11.505\pm0.027}} 
\newcommand{\hatcurCCesoHmagxxxxxE}{\ensuremath{11.203\pm0.027}} 
\newcommand{\hatcurCCesoKmagxxxxxE}{\ensuremath{11.161\pm0.026}} 
\newcommand{\hatcurCCesoJHmagxxxxxE}{\ensuremath{0.302\pm0.037}} 
\newcommand{\hatcurCCesoJKmagxxxxxE}{\ensuremath{0.3440\pm0.0090}} 
\newcommand{\hatcurCCesoHKmagxxxxxE}{\ensuremath{0.042\pm0.038}} 
\newcommand{\hatcurLCdipxxxxxE}{\ensuremath{13.1}}              
\newcommand{\hatcurLCrprstarxxxxxE}{\ensuremath{0.1051\pm0.0012}} 
\newcommand{\hatcurLCbsqxxxxxE}{\ensuremath{0.068_{-0.054}^{+0.066}}} 
\newcommand{\hatcurLCimpxxxxxE}{\ensuremath{0.26_{-0.14}^{+0.11}}} 
\newcommand{\hatcurLCzetaxxxxxE}{\ensuremath{17.13\pm0.11}}     
\newcommand{\hatcurLCdurxxxxxE}{\ensuremath{0.1299\pm0.0010}}   
\newcommand{\hatcurLCdurshortxxxxxE}{\ensuremath{0.1299}}       
\newcommand{\hatcurLCdurhrxxxxxE}{\ensuremath{3.118\pm0.024}}   
\newcommand{\hatcurLCdurhrshortxxxxxE}{\ensuremath{3.118}}      
\newcommand{\hatcurLCqxxxxxE}{\ensuremath{0.07130\pm0.00056}}   
\newcommand{\hatcurLCqshortxxxxxE}{\ensuremath{0.071}}          
\newcommand{\hatcurLCingdurxxxxxE}{\ensuremath{0.01314\pm0.00083}} 
\newcommand{\hatcurLCPxxxxxE}{\ensuremath{1.8209993\pm0.0000016}} 
\newcommand{\hatcurLCPprecxxxxxE}{\ensuremath{1.8209993}}       
\newcommand{\hatcurLCPshortxxxxxE}{\ensuremath{1.8210}}         
\newcommand{\hatcurLCTxxxxxE}{\ensuremath{2456981.80199\pm0.00041}} 
\newcommand{\hatcurLCTAxxxxxE}{\ensuremath{2455679.7875\pm0.0012}} 
\newcommand{\hatcurLCTBxxxxxE}{\ensuremath{2457227.63687\pm0.00049}} 
\newcommand{\hatcurLCrhoxxxxxE}{\ensuremath{0.627_{-0.060}^{+0.046}}} 
\newcommand{\hatcurSMEiteffxxxxxE}{\ensuremath{6250\pm160}}     
\newcommand{\hatcurSMEizfehxxxxxE}{\ensuremath{0.180\pm0.090}}  
\newcommand{\hatcurSMEizfehshortxxxxxE}{\ensuremath{0.18}}      
\newcommand{\hatcurSMEiloggxxxxxE}{\ensuremath{4.20\pm0.23}}    
\newcommand{\hatcurSMEivsinxxxxxE}{\ensuremath{8.66\pm0.32}}    
\newcommand{\hatcurSMEivmacxxxxxE}{\ensuremath{0.0}}            
\newcommand{\hatcurSMEivmicxxxxxE}{\ensuremath{0.0}}            
\newcommand{\hatcurSMEiiteffxxxxxE}{\ensuremath{6300\pm100}}    
\newcommand{\hatcurSMEiizfehxxxxxE}{\ensuremath{0.210\pm0.060}} 
\newcommand{\hatcurSMEiizfehshortxxxxxE}{\ensuremath{0.21}}     
\newcommand{\hatcurSMEiiloggxxxxxE}{\ensuremath{4.241\pm0.020}} 
\newcommand{\hatcurSMEiivsinxxxxxE}{\ensuremath{8.66\pm0.34}}   
\newcommand{\hatcurLBizxxxxxE}{\ensuremath{0.1430}}             
\newcommand{\hatcurLBiizxxxxxE}{\ensuremath{0.3653}}            
\newcommand{\hatcurLBiixxxxxE}{\ensuremath{0.1962}}             
\newcommand{\hatcurLBiiixxxxxE}{\ensuremath{0.3739}}            
\newcommand{\hatcurLBiIxxxxxE}{\ensuremath{0.1770}}             
\newcommand{\hatcurLBiiIxxxxxE}{\ensuremath{0.3724}}            
\newcommand{\hatcurLBigxxxxxE}{\ensuremath{0.4494}}             
\newcommand{\hatcurLBiigxxxxxE}{\ensuremath{0.3131}}            
\newcommand{\hatcurLBirxxxxxE}{\ensuremath{0.2746}}             
\newcommand{\hatcurLBiirxxxxxE}{\ensuremath{0.3780}}            
\newcommand{\hatcurLBiRxxxxxE}{\ensuremath{0.2524}}             
\newcommand{\hatcurLBiiRxxxxxE}{\ensuremath{0.3784}}            
\newcommand{\hatcurLBikepxxxxxE}{\ensuremath{0.1000}}           
\newcommand{\hatcurLBiikepxxxxxE}{\ensuremath{0.1000}}          
\newcommand{\hatcurISOmxxxxxE}{\ensuremath{1.317\pm0.040}}      
\newcommand{\hatcurISOmshortxxxxxE}{\ensuremath{1.32}}          
\newcommand{\hatcurISOmlongxxxxxE}{\ensuremath{1.317\pm0.040}}  
\newcommand{\hatcurISOrxxxxxE}{\ensuremath{1.433_{-0.038}^{+0.056}}} 
\newcommand{\hatcurISOrshortxxxxxE}{\ensuremath{1.43}}          
\newcommand{\hatcurISOrlongxxxxxE}{\ensuremath{1.433_{-0.038}^{+0.056}}} 
\newcommand{\hatcurISOrhoxxxxxE}{\ensuremath{0.628_{-0.061}^{+0.045}}} 
\newcommand{\hatcurISOrholongxxxxxE}{\ensuremath{0.628_{-0.061}^{+0.045}}} 
\newcommand{\hatcurISOloggxxxxxE}{\ensuremath{4.244\pm0.023}}   
\newcommand{\hatcurISOlumxxxxxE}{\ensuremath{2.92\pm0.31}}      
\newcommand{\hatcurISOlumshortxxxxxE}{\ensuremath{2.92}}        
\newcommand{\hatcurISOmvxxxxxE}{\ensuremath{3.58\pm0.12}}       
\newcommand{\hatcurISOvixxxxxE}{\ensuremath{0.550\pm0.027}}     
\newcommand{\hatcurISOagexxxxxE}{\ensuremath{2.13\pm0.51}}      
\newcommand{\hatcurISOsigmaxxxxxE}{\ensuremath{0.000800\pm0.000068}} 
\newcommand{\hatcurISOMJxxxxxE}{\ensuremath{2.690\pm0.084}}     
\newcommand{\hatcurISOMHxxxxxE}{\ensuremath{2.443\pm0.074}}     
\newcommand{\hatcurISOMKxxxxxE}{\ensuremath{2.399\pm0.073}}     
\newcommand{\hatcurISOJKxxxxxE}{\ensuremath{0.290\pm0.020}}     
\newcommand{\hatcurISOspecxxxxxE}{F}                            
\newcommand{\hatcurRVKxxxxxE}{\ensuremath{170\pm10}}            
\newcommand{\hatcurRVrkxxxxxE}{\ensuremath{0\pm0}}              
\newcommand{\hatcurRVrhxxxxxE}{\ensuremath{0\pm0}}              
\newcommand{\hatcurRVkxxxxxE}{\ensuremath{0\pm0}}               
\newcommand{\hatcurRVhxxxxxE}{\ensuremath{0\pm0}}               
\newcommand{\hatcurRVtronexxxxxE}{\ensuremath{0\pm0}}           
\newcommand{\hatcurRVtrtwoxxxxxE}{\ensuremath{0\pm0}}           
\newcommand{\hatcurRVgammaAxxxxxE}{\ensuremath{-14175\pm13}}    
\newcommand{\hatcurRVjitterAxxxxxE}{\ensuremath{20\pm25}}       
\newcommand{\hatcurRVjittertwosiglimAxxxxxE}{\ensuremath{<50.5}} 
\newcommand{\hatcurRVfitrmsAxxxxxE}{\ensuremath{0.0}}           
\newcommand{\hatcurRVgammaBxxxxxE}{\ensuremath{-14224\pm31}}    
\newcommand{\hatcurRVjitterBxxxxxE}{\ensuremath{2\pm52}}        
\newcommand{\hatcurRVjittertwosiglimBxxxxxE}{\ensuremath{<180.3}} 
\newcommand{\hatcurRVfitrmsBxxxxxE}{\ensuremath{0.0}}           
\newcommand{\hatcurRVgammaCxxxxxE}{\ensuremath{-14242\pm32}}    
\newcommand{\hatcurRVjitterCxxxxxE}{\ensuremath{0\pm49}}        
\newcommand{\hatcurRVjittertwosiglimCxxxxxE}{\ensuremath{<96.3}} 
\newcommand{\hatcurRVfitrmsCxxxxxE}{\ensuremath{0.0}}           
\newcommand{\hatcurRVeccenxxxxxE}{\ensuremath{0\pm0}}           
\newcommand{\hatcurRVeccentwosiglimxxxxxE}{\ensuremath{<0.000}} 
\newcommand{\hatcurRVomegaxxxxxE}{\ensuremath{0\pm0}}           
\newcommand{\hatcurPPixxxxxE}{\ensuremath{86.9\pm1.3}}          
\newcommand{\hatcurPPgxxxxxE}{\ensuremath{14.16_{-1.62}^{+0.82}}} 
\newcommand{\hatcurPPloggxxxxxE}{\ensuremath{3.151_{-0.053}^{+0.025}}} 
\newcommand{\hatcurPParxxxxxE}{\ensuremath{4.79_{-0.16}^{+0.11}}} 
\newcommand{\hatcurPParelxxxxxE}{\ensuremath{0.03199\pm0.00033}} 
\newcommand{\hatcurPPrhoxxxxxE}{\ensuremath{0.484_{-0.070}^{+0.042}}} 
\newcommand{\hatcurPPmxxxxxE}{\ensuremath{1.222\pm0.078}}       
\newcommand{\hatcurPPmshortxxxxxE}{\ensuremath{1.22}}           
\newcommand{\hatcurPPmlongxxxxxE}{\ensuremath{1.222\pm0.078}}   
\newcommand{\hatcurPPmexxxxxE}{\ensuremath{388\pm25}}           
\newcommand{\hatcurPPmeshortxxxxxE}{\ensuremath{388.2}}         
\newcommand{\hatcurPPmelongxxxxxE}{\ensuremath{388\pm25}}       
\newcommand{\hatcurPPrxxxxxE}{\ensuremath{1.464_{-0.044}^{+0.069}}} 
\newcommand{\hatcurPPrshortxxxxxE}{\ensuremath{1.46}}           
\newcommand{\hatcurPPrlongxxxxxE}{\ensuremath{1.464_{-0.044}^{+0.069}}} 
\newcommand{\hatcurPPrexxxxxE}{\ensuremath{16.41_{-0.49}^{+0.77}}} 
\newcommand{\hatcurPPreshortxxxxxE}{\ensuremath{16.4}}          
\newcommand{\hatcurPPrelongxxxxxE}{\ensuremath{16.41_{-0.49}^{+0.77}}} 
\newcommand{\hatcurPPmrcorrxxxxxE}{\ensuremath{0.18}}           
\newcommand{\hatcurPPteffxxxxxE}{\ensuremath{2037\pm43}}        
\newcommand{\hatcurPPthetaxxxxxE}{\ensuremath{0.0405_{-0.0036}^{+0.0020}}} 
\newcommand{\hatcurPPfluxperixxxxxE}{\ensuremath{3.89\pm0.34}}  
\newcommand{\hatcurPPfluxperidimxxxxxE}{\ensuremath{9}}         
\newcommand{\hatcurPPfluxapxxxxxE}{\ensuremath{3.89\pm0.34}}    
\newcommand{\hatcurPPfluxapdimxxxxxE}{\ensuremath{9}}           
\newcommand{\hatcurPPfluxavgxxxxxE}{\ensuremath{3.89\pm0.34}}   
\newcommand{\hatcurPPfluxavgdimxxxxxE}{\ensuremath{9}}          
\newcommand{\hatcurPPfluxavglogxxxxxE}{\ensuremath{9.590\pm0.037}} 
\newcommand{\hatcurXsecphasexxxxxE}{\ensuremath{0\pm0}}         
\newcommand{\hatcurXsecondaryxxxxxE}{\ensuremath{2456982.71249\pm0.00041}} 
\newcommand{\hatcurXsecdurxxxxxE}{\ensuremath{0.1299\pm0.0010}} 
\newcommand{\hatcurXsecingdurxxxxxE}{\ensuremath{0.01314\pm0.00083}} 
\newcommand{\hatcurPPphiconjxxxxxE}{\ensuremath{0\pm0}}         
\newcommand{\hatcurPPperixxxxxE}{\ensuremath{2456981.34674\pm0.00042}} 
\newcommand{\hatcurPPaequivxxxxxE}{\ensuremath{0.01870\pm0.00080}} 
\newcommand{\hatcurPPtcircxxxxxE}{\ensuremath{8.9_{-1.8}^{+1.2}}} 
\newcommand{\hatcurPPtinfallxxxxxE}{\ensuremath{47.4\pm7.0}}    
\newcommand{\hatcurXdistxxxxxE}{\ensuremath{566\pm20}}          
\newcommand{\hatcurXAvxxxxxE}{\ensuremath{0.25\pm0.14}}         
\newcommand{\hatcurXdistredxxxxxE}{\ensuremath{557_{-17}^{+22}}} 
\newcommand{\hatcurXEBVxxxxxE}{\ensuremath{0.082\pm0.045}}      
\newcommand{\hatcurXmvisoredxxxxxE}{\ensuremath{12.571\pm0.085}} 
\newcommand{\hatcurXmiisoredxxxxxE}{\ensuremath{11.888\pm0.035}} 
\newcommand{\hatcurXmjisoredxxxxxE}{\ensuremath{11.493\pm0.018}} 
\newcommand{\hatcurXmhisoredxxxxxE}{\ensuremath{11.220\pm0.016}} 
\newcommand{\hatcurXmkisoredxxxxxE}{\ensuremath{11.155\pm0.018}} 
\newcommand{\hatcurXviisoredxxxxxE}{\ensuremath{0.683\pm0.059}} 
\newcommand{\hatcurXvkisoredxxxxxE}{\ensuremath{1.416\pm0.094}} 
\newcommand{\hatcurXjhisoredxxxxxE}{\ensuremath{0.273\pm0.015}} 
\newcommand{\hatcurXjkisoredxxxxxE}{\ensuremath{0.337\pm0.019}} 
\newcommand{\hatcurCCpmraxxxxxE}{\ensuremath{16.7\pm1.3}}       
\newcommand{\hatcurCCpmdecxxxxxE}{\ensuremath{-12.1\pm1.3}}     
\newcommand{\hatcurCCpmxxxxxE}{\ensuremath{20.6\pm1.8}}         

%% file: planetinfo_multi_eccen.tex
\input{HATS566-004.eccen.tex}
\input{HATS586-004.eccen.tex}
\input{HATS747-032.eccen.tex}
\input{HATS754-010.eccen.tex}
\input{HATS779-003.eccen.tex}
\newcommand{\hatcurCCbbHmageccen}[1]{\ifnum#1=31 %
\hatcurCCbbHmageccenxxxxxA
\else
\ifnum#1=32 %
\hatcurCCbbHmageccenxxxxxB
\else
\ifnum#1=33 %
\hatcurCCbbHmageccenxxxxxC
\else
\ifnum#1=34 %
\hatcurCCbbHmageccenxxxxxD
\else
\ifnum#1=35 %
\hatcurCCbbHmageccenxxxxxE
\else
??????\fi
\fi
\fi
\fi
\fi
}
\newcommand{\hatcurCCbbJmageccen}[1]{\ifnum#1=31 %
\hatcurCCbbJmageccenxxxxxA
\else
\ifnum#1=32 %
\hatcurCCbbJmageccenxxxxxB
\else
\ifnum#1=33 %
\hatcurCCbbJmageccenxxxxxC
\else
\ifnum#1=34 %
\hatcurCCbbJmageccenxxxxxD
\else
\ifnum#1=35 %
\hatcurCCbbJmageccenxxxxxE
\else
??????\fi
\fi
\fi
\fi
\fi
}
\newcommand{\hatcurCCbbKmageccen}[1]{\ifnum#1=31 %
\hatcurCCbbKmageccenxxxxxA
\else
\ifnum#1=32 %
\hatcurCCbbKmageccenxxxxxB
\else
\ifnum#1=33 %
\hatcurCCbbKmageccenxxxxxC
\else
\ifnum#1=34 %
\hatcurCCbbKmageccenxxxxxD
\else
\ifnum#1=35 %
\hatcurCCbbKmageccenxxxxxE
\else
??????\fi
\fi
\fi
\fi
\fi
}
\newcommand{\hatcurCCcitHmageccen}[1]{\ifnum#1=31 %
\hatcurCCcitHmageccenxxxxxA
\else
\ifnum#1=32 %
\hatcurCCcitHmageccenxxxxxB
\else
\ifnum#1=33 %
\hatcurCCcitHmageccenxxxxxC
\else
\ifnum#1=34 %
\hatcurCCcitHmageccenxxxxxD
\else
\ifnum#1=35 %
\hatcurCCcitHmageccenxxxxxE
\else
??????\fi
\fi
\fi
\fi
\fi
}
\newcommand{\hatcurCCcitJmageccen}[1]{\ifnum#1=31 %
\hatcurCCcitJmageccenxxxxxA
\else
\ifnum#1=32 %
\hatcurCCcitJmageccenxxxxxB
\else
\ifnum#1=33 %
\hatcurCCcitJmageccenxxxxxC
\else
\ifnum#1=34 %
\hatcurCCcitJmageccenxxxxxD
\else
\ifnum#1=35 %
\hatcurCCcitJmageccenxxxxxE
\else
??????\fi
\fi
\fi
\fi
\fi
}
\newcommand{\hatcurCCcitKmageccen}[1]{\ifnum#1=31 %
\hatcurCCcitKmageccenxxxxxA
\else
\ifnum#1=32 %
\hatcurCCcitKmageccenxxxxxB
\else
\ifnum#1=33 %
\hatcurCCcitKmageccenxxxxxC
\else
\ifnum#1=34 %
\hatcurCCcitKmageccenxxxxxD
\else
\ifnum#1=35 %
\hatcurCCcitKmageccenxxxxxE
\else
??????\fi
\fi
\fi
\fi
\fi
}
\newcommand{\hatcurCCdececcen}[1]{\ifnum#1=31 %
\hatcurCCdececcenxxxxxA
\else
\ifnum#1=32 %
\hatcurCCdececcenxxxxxB
\else
\ifnum#1=33 %
\hatcurCCdececcenxxxxxC
\else
\ifnum#1=34 %
\hatcurCCdececcenxxxxxD
\else
\ifnum#1=35 %
\hatcurCCdececcenxxxxxE
\else
??????\fi
\fi
\fi
\fi
\fi
}
\newcommand{\hatcurCCesoHKmageccen}[1]{\ifnum#1=31 %
\hatcurCCesoHKmageccenxxxxxA
\else
\ifnum#1=32 %
\hatcurCCesoHKmageccenxxxxxB
\else
\ifnum#1=33 %
\hatcurCCesoHKmageccenxxxxxC
\else
\ifnum#1=34 %
\hatcurCCesoHKmageccenxxxxxD
\else
\ifnum#1=35 %
\hatcurCCesoHKmageccenxxxxxE
\else
??????\fi
\fi
\fi
\fi
\fi
}
\newcommand{\hatcurCCesoHmageccen}[1]{\ifnum#1=31 %
\hatcurCCesoHmageccenxxxxxA
\else
\ifnum#1=32 %
\hatcurCCesoHmageccenxxxxxB
\else
\ifnum#1=33 %
\hatcurCCesoHmageccenxxxxxC
\else
\ifnum#1=34 %
\hatcurCCesoHmageccenxxxxxD
\else
\ifnum#1=35 %
\hatcurCCesoHmageccenxxxxxE
\else
??????\fi
\fi
\fi
\fi
\fi
}
\newcommand{\hatcurCCesoJHmageccen}[1]{\ifnum#1=31 %
\hatcurCCesoJHmageccenxxxxxA
\else
\ifnum#1=32 %
\hatcurCCesoJHmageccenxxxxxB
\else
\ifnum#1=33 %
\hatcurCCesoJHmageccenxxxxxC
\else
\ifnum#1=34 %
\hatcurCCesoJHmageccenxxxxxD
\else
\ifnum#1=35 %
\hatcurCCesoJHmageccenxxxxxE
\else
??????\fi
\fi
\fi
\fi
\fi
}
\newcommand{\hatcurCCesoJKmageccen}[1]{\ifnum#1=31 %
\hatcurCCesoJKmageccenxxxxxA
\else
\ifnum#1=32 %
\hatcurCCesoJKmageccenxxxxxB
\else
\ifnum#1=33 %
\hatcurCCesoJKmageccenxxxxxC
\else
\ifnum#1=34 %
\hatcurCCesoJKmageccenxxxxxD
\else
\ifnum#1=35 %
\hatcurCCesoJKmageccenxxxxxE
\else
??????\fi
\fi
\fi
\fi
\fi
}
\newcommand{\hatcurCCesoJmageccen}[1]{\ifnum#1=31 %
\hatcurCCesoJmageccenxxxxxA
\else
\ifnum#1=32 %
\hatcurCCesoJmageccenxxxxxB
\else
\ifnum#1=33 %
\hatcurCCesoJmageccenxxxxxC
\else
\ifnum#1=34 %
\hatcurCCesoJmageccenxxxxxD
\else
\ifnum#1=35 %
\hatcurCCesoJmageccenxxxxxE
\else
??????\fi
\fi
\fi
\fi
\fi
}
\newcommand{\hatcurCCesoKmageccen}[1]{\ifnum#1=31 %
\hatcurCCesoKmageccenxxxxxA
\else
\ifnum#1=32 %
\hatcurCCesoKmageccenxxxxxB
\else
\ifnum#1=33 %
\hatcurCCesoKmageccenxxxxxC
\else
\ifnum#1=34 %
\hatcurCCesoKmageccenxxxxxD
\else
\ifnum#1=35 %
\hatcurCCesoKmageccenxxxxxE
\else
??????\fi
\fi
\fi
\fi
\fi
}
\newcommand{\hatcurCCgsceccen}[1]{\ifnum#1=31 %
\hatcurCCgsceccenxxxxxA
\else
\ifnum#1=32 %
\hatcurCCgsceccenxxxxxB
\else
\ifnum#1=33 %
\hatcurCCgsceccenxxxxxC
\else
\ifnum#1=34 %
\hatcurCCgsceccenxxxxxD
\else
\ifnum#1=35 %
\hatcurCCgsceccenxxxxxE
\else
??????\fi
\fi
\fi
\fi
\fi
}
\newcommand{\hatcurCCmageccen}[1]{\ifnum#1=31 %
\hatcurCCmageccenxxxxxA
\else
\ifnum#1=32 %
\hatcurCCmageccenxxxxxB
\else
\ifnum#1=33 %
\hatcurCCmageccenxxxxxC
\else
\ifnum#1=34 %
\hatcurCCmageccenxxxxxD
\else
\ifnum#1=35 %
\hatcurCCmageccenxxxxxE
\else
??????\fi
\fi
\fi
\fi
\fi
}
\newcommand{\hatcurCCpmdececcen}[1]{\ifnum#1=31 %
\hatcurCCpmdececcenxxxxxA
\else
\ifnum#1=32 %
\hatcurCCpmdececcenxxxxxB
\else
\ifnum#1=33 %
\hatcurCCpmdececcenxxxxxC
\else
\ifnum#1=34 %
\hatcurCCpmdececcenxxxxxD
\else
\ifnum#1=35 %
\hatcurCCpmdececcenxxxxxE
\else
??????\fi
\fi
\fi
\fi
\fi
}
\newcommand{\hatcurCCpmeccen}[1]{\ifnum#1=31 %
\hatcurCCpmeccenxxxxxA
\else
\ifnum#1=32 %
\hatcurCCpmeccenxxxxxB
\else
\ifnum#1=33 %
\hatcurCCpmeccenxxxxxC
\else
\ifnum#1=34 %
\hatcurCCpmeccenxxxxxD
\else
\ifnum#1=35 %
\hatcurCCpmeccenxxxxxE
\else
??????\fi
\fi
\fi
\fi
\fi
}
\newcommand{\hatcurCCpmraeccen}[1]{\ifnum#1=31 %
\hatcurCCpmraeccenxxxxxA
\else
\ifnum#1=32 %
\hatcurCCpmraeccenxxxxxB
\else
\ifnum#1=33 %
\hatcurCCpmraeccenxxxxxC
\else
\ifnum#1=34 %
\hatcurCCpmraeccenxxxxxD
\else
\ifnum#1=35 %
\hatcurCCpmraeccenxxxxxE
\else
??????\fi
\fi
\fi
\fi
\fi
}
\newcommand{\hatcurCCraeccen}[1]{\ifnum#1=31 %
\hatcurCCraeccenxxxxxA
\else
\ifnum#1=32 %
\hatcurCCraeccenxxxxxB
\else
\ifnum#1=33 %
\hatcurCCraeccenxxxxxC
\else
\ifnum#1=34 %
\hatcurCCraeccenxxxxxD
\else
\ifnum#1=35 %
\hatcurCCraeccenxxxxxE
\else
??????\fi
\fi
\fi
\fi
\fi
}
\newcommand{\hatcurCCtassmBeccen}[1]{\ifnum#1=31 %
\hatcurCCtassmBeccenxxxxxA
\else
\ifnum#1=32 %
\hatcurCCtassmBeccenxxxxxB
\else
\ifnum#1=33 %
\hatcurCCtassmBeccenxxxxxC
\else
\ifnum#1=34 %
\hatcurCCtassmBeccenxxxxxD
\else
\ifnum#1=35 %
\hatcurCCtassmBeccenxxxxxE
\else
??????\fi
\fi
\fi
\fi
\fi
}
\newcommand{\hatcurCCtassmBshorteccen}[1]{\ifnum#1=31 %
\hatcurCCtassmBshorteccenxxxxxA
\else
\ifnum#1=32 %
\hatcurCCtassmBshorteccenxxxxxB
\else
\ifnum#1=33 %
\hatcurCCtassmBshorteccenxxxxxC
\else
\ifnum#1=34 %
\hatcurCCtassmBshorteccenxxxxxD
\else
\ifnum#1=35 %
\hatcurCCtassmBshorteccenxxxxxE
\else
??????\fi
\fi
\fi
\fi
\fi
}
\newcommand{\hatcurCCtassmgeccen}[1]{\ifnum#1=31 %
\hatcurCCtassmgeccenxxxxxA
\else
\ifnum#1=32 %
\hatcurCCtassmgeccenxxxxxB
\else
\ifnum#1=33 %
\hatcurCCtassmgeccenxxxxxC
\else
\ifnum#1=34 %
\hatcurCCtassmgeccenxxxxxD
\else
\ifnum#1=35 %
\hatcurCCtassmgeccenxxxxxE
\else
??????\fi
\fi
\fi
\fi
\fi
}
\newcommand{\hatcurCCtassmgshorteccen}[1]{\ifnum#1=31 %
\hatcurCCtassmgshorteccenxxxxxA
\else
\ifnum#1=32 %
\hatcurCCtassmgshorteccenxxxxxB
\else
\ifnum#1=33 %
\hatcurCCtassmgshorteccenxxxxxC
\else
\ifnum#1=34 %
\hatcurCCtassmgshorteccenxxxxxD
\else
\ifnum#1=35 %
\hatcurCCtassmgshorteccenxxxxxE
\else
??????\fi
\fi
\fi
\fi
\fi
}
\newcommand{\hatcurCCtassmieccen}[1]{\ifnum#1=31 %
\hatcurCCtassmieccenxxxxxA
\else
\ifnum#1=32 %
\hatcurCCtassmieccenxxxxxB
\else
\ifnum#1=33 %
\hatcurCCtassmieccenxxxxxC
\else
\ifnum#1=34 %
\hatcurCCtassmieccenxxxxxD
\else
\ifnum#1=35 %
\hatcurCCtassmieccenxxxxxE
\else
??????\fi
\fi
\fi
\fi
\fi
}
\newcommand{\hatcurCCtassmIeccen}[1]{\ifnum#1=31 %
\hatcurCCtassmIeccenxxxxxA
\else
\ifnum#1=32 %
\hatcurCCtassmIeccenxxxxxB
\else
\ifnum#1=33 %
\hatcurCCtassmIeccenxxxxxC
\else
\ifnum#1=34 %
\hatcurCCtassmIeccenxxxxxD
\else
\ifnum#1=35 %
\hatcurCCtassmIeccenxxxxxE
\else
??????\fi
\fi
\fi
\fi
\fi
}
\newcommand{\hatcurCCtassmishorteccen}[1]{\ifnum#1=31 %
\hatcurCCtassmishorteccenxxxxxA
\else
\ifnum#1=32 %
\hatcurCCtassmishorteccenxxxxxB
\else
\ifnum#1=33 %
\hatcurCCtassmishorteccenxxxxxC
\else
\ifnum#1=34 %
\hatcurCCtassmishorteccenxxxxxD
\else
\ifnum#1=35 %
\hatcurCCtassmishorteccenxxxxxE
\else
??????\fi
\fi
\fi
\fi
\fi
}
\newcommand{\hatcurCCtassmIshorteccen}[1]{\ifnum#1=31 %
\hatcurCCtassmIshorteccenxxxxxA
\else
\ifnum#1=32 %
\hatcurCCtassmIshorteccenxxxxxB
\else
\ifnum#1=33 %
\hatcurCCtassmIshorteccenxxxxxC
\else
\ifnum#1=34 %
\hatcurCCtassmIshorteccenxxxxxD
\else
\ifnum#1=35 %
\hatcurCCtassmIshorteccenxxxxxE
\else
??????\fi
\fi
\fi
\fi
\fi
}
\newcommand{\hatcurCCtassmreccen}[1]{\ifnum#1=31 %
\hatcurCCtassmreccenxxxxxA
\else
\ifnum#1=32 %
\hatcurCCtassmreccenxxxxxB
\else
\ifnum#1=33 %
\hatcurCCtassmreccenxxxxxC
\else
\ifnum#1=34 %
\hatcurCCtassmreccenxxxxxD
\else
\ifnum#1=35 %
\hatcurCCtassmreccenxxxxxE
\else
??????\fi
\fi
\fi
\fi
\fi
}
\newcommand{\hatcurCCtassmrshorteccen}[1]{\ifnum#1=31 %
\hatcurCCtassmrshorteccenxxxxxA
\else
\ifnum#1=32 %
\hatcurCCtassmrshorteccenxxxxxB
\else
\ifnum#1=33 %
\hatcurCCtassmrshorteccenxxxxxC
\else
\ifnum#1=34 %
\hatcurCCtassmrshorteccenxxxxxD
\else
\ifnum#1=35 %
\hatcurCCtassmrshorteccenxxxxxE
\else
??????\fi
\fi
\fi
\fi
\fi
}
\newcommand{\hatcurCCtassmveccen}[1]{\ifnum#1=31 %
\hatcurCCtassmveccenxxxxxA
\else
\ifnum#1=32 %
\hatcurCCtassmveccenxxxxxB
\else
\ifnum#1=33 %
\hatcurCCtassmveccenxxxxxC
\else
\ifnum#1=34 %
\hatcurCCtassmveccenxxxxxD
\else
\ifnum#1=35 %
\hatcurCCtassmveccenxxxxxE
\else
??????\fi
\fi
\fi
\fi
\fi
}
\newcommand{\hatcurCCtassmvshorteccen}[1]{\ifnum#1=31 %
\hatcurCCtassmvshorteccenxxxxxA
\else
\ifnum#1=32 %
\hatcurCCtassmvshorteccenxxxxxB
\else
\ifnum#1=33 %
\hatcurCCtassmvshorteccenxxxxxC
\else
\ifnum#1=34 %
\hatcurCCtassmvshorteccenxxxxxD
\else
\ifnum#1=35 %
\hatcurCCtassmvshorteccenxxxxxE
\else
??????\fi
\fi
\fi
\fi
\fi
}
\newcommand{\hatcurCCtwomasseccen}[1]{\ifnum#1=31 %
\hatcurCCtwomasseccenxxxxxA
\else
\ifnum#1=32 %
\hatcurCCtwomasseccenxxxxxB
\else
\ifnum#1=33 %
\hatcurCCtwomasseccenxxxxxC
\else
\ifnum#1=34 %
\hatcurCCtwomasseccenxxxxxD
\else
\ifnum#1=35 %
\hatcurCCtwomasseccenxxxxxE
\else
??????\fi
\fi
\fi
\fi
\fi
}
\newcommand{\hatcurCCtwomassHmageccen}[1]{\ifnum#1=31 %
\hatcurCCtwomassHmageccenxxxxxA
\else
\ifnum#1=32 %
\hatcurCCtwomassHmageccenxxxxxB
\else
\ifnum#1=33 %
\hatcurCCtwomassHmageccenxxxxxC
\else
\ifnum#1=34 %
\hatcurCCtwomassHmageccenxxxxxD
\else
\ifnum#1=35 %
\hatcurCCtwomassHmageccenxxxxxE
\else
??????\fi
\fi
\fi
\fi
\fi
}
\newcommand{\hatcurCCtwomassJmageccen}[1]{\ifnum#1=31 %
\hatcurCCtwomassJmageccenxxxxxA
\else
\ifnum#1=32 %
\hatcurCCtwomassJmageccenxxxxxB
\else
\ifnum#1=33 %
\hatcurCCtwomassJmageccenxxxxxC
\else
\ifnum#1=34 %
\hatcurCCtwomassJmageccenxxxxxD
\else
\ifnum#1=35 %
\hatcurCCtwomassJmageccenxxxxxE
\else
??????\fi
\fi
\fi
\fi
\fi
}
\newcommand{\hatcurCCtwomassKmageccen}[1]{\ifnum#1=31 %
\hatcurCCtwomassKmageccenxxxxxA
\else
\ifnum#1=32 %
\hatcurCCtwomassKmageccenxxxxxB
\else
\ifnum#1=33 %
\hatcurCCtwomassKmageccenxxxxxC
\else
\ifnum#1=34 %
\hatcurCCtwomassKmageccenxxxxxD
\else
\ifnum#1=35 %
\hatcurCCtwomassKmageccenxxxxxE
\else
??????\fi
\fi
\fi
\fi
\fi
}
\newcommand{\hatcurfieldeccen}[1]{\ifnum#1=31 %
\hatcurfieldeccenxxxxxA
\else
\ifnum#1=32 %
\hatcurfieldeccenxxxxxB
\else
\ifnum#1=33 %
\hatcurfieldeccenxxxxxC
\else
\ifnum#1=34 %
\hatcurfieldeccenxxxxxD
\else
\ifnum#1=35 %
\hatcurfieldeccenxxxxxE
\else
??????\fi
\fi
\fi
\fi
\fi
}
\newcommand{\hatcurhtreccen}[1]{\ifnum#1=31 %
\hatcurhtreccenxxxxxA
\else
\ifnum#1=32 %
\hatcurhtreccenxxxxxB
\else
\ifnum#1=33 %
\hatcurhtreccenxxxxxC
\else
\ifnum#1=34 %
\hatcurhtreccenxxxxxD
\else
\ifnum#1=35 %
\hatcurhtreccenxxxxxE
\else
??????\fi
\fi
\fi
\fi
\fi
}
\newcommand{\hatcurISOageeccen}[1]{\ifnum#1=31 %
\hatcurISOageeccenxxxxxA
\else
\ifnum#1=32 %
\hatcurISOageeccenxxxxxB
\else
\ifnum#1=33 %
\hatcurISOageeccenxxxxxC
\else
\ifnum#1=34 %
\hatcurISOageeccenxxxxxD
\else
\ifnum#1=35 %
\hatcurISOageeccenxxxxxE
\else
??????\fi
\fi
\fi
\fi
\fi
}
\newcommand{\hatcurISOJKeccen}[1]{\ifnum#1=31 %
\hatcurISOJKeccenxxxxxA
\else
\ifnum#1=32 %
\hatcurISOJKeccenxxxxxB
\else
\ifnum#1=33 %
\hatcurISOJKeccenxxxxxC
\else
\ifnum#1=34 %
\hatcurISOJKeccenxxxxxD
\else
\ifnum#1=35 %
\hatcurISOJKeccenxxxxxE
\else
??????\fi
\fi
\fi
\fi
\fi
}
\newcommand{\hatcurISOloggeccen}[1]{\ifnum#1=31 %
\hatcurISOloggeccenxxxxxA
\else
\ifnum#1=32 %
\hatcurISOloggeccenxxxxxB
\else
\ifnum#1=33 %
\hatcurISOloggeccenxxxxxC
\else
\ifnum#1=34 %
\hatcurISOloggeccenxxxxxD
\else
\ifnum#1=35 %
\hatcurISOloggeccenxxxxxE
\else
??????\fi
\fi
\fi
\fi
\fi
}
\newcommand{\hatcurISOlumeccen}[1]{\ifnum#1=31 %
\hatcurISOlumeccenxxxxxA
\else
\ifnum#1=32 %
\hatcurISOlumeccenxxxxxB
\else
\ifnum#1=33 %
\hatcurISOlumeccenxxxxxC
\else
\ifnum#1=34 %
\hatcurISOlumeccenxxxxxD
\else
\ifnum#1=35 %
\hatcurISOlumeccenxxxxxE
\else
??????\fi
\fi
\fi
\fi
\fi
}
\newcommand{\hatcurISOlumshorteccen}[1]{\ifnum#1=31 %
\hatcurISOlumshorteccenxxxxxA
\else
\ifnum#1=32 %
\hatcurISOlumshorteccenxxxxxB
\else
\ifnum#1=33 %
\hatcurISOlumshorteccenxxxxxC
\else
\ifnum#1=34 %
\hatcurISOlumshorteccenxxxxxD
\else
\ifnum#1=35 %
\hatcurISOlumshorteccenxxxxxE
\else
??????\fi
\fi
\fi
\fi
\fi
}
\newcommand{\hatcurISOmeccen}[1]{\ifnum#1=31 %
\hatcurISOmeccenxxxxxA
\else
\ifnum#1=32 %
\hatcurISOmeccenxxxxxB
\else
\ifnum#1=33 %
\hatcurISOmeccenxxxxxC
\else
\ifnum#1=34 %
\hatcurISOmeccenxxxxxD
\else
\ifnum#1=35 %
\hatcurISOmeccenxxxxxE
\else
??????\fi
\fi
\fi
\fi
\fi
}
\newcommand{\hatcurISOMHeccen}[1]{\ifnum#1=31 %
\hatcurISOMHeccenxxxxxA
\else
\ifnum#1=32 %
\hatcurISOMHeccenxxxxxB
\else
\ifnum#1=33 %
\hatcurISOMHeccenxxxxxC
\else
\ifnum#1=34 %
\hatcurISOMHeccenxxxxxD
\else
\ifnum#1=35 %
\hatcurISOMHeccenxxxxxE
\else
??????\fi
\fi
\fi
\fi
\fi
}
\newcommand{\hatcurISOMJeccen}[1]{\ifnum#1=31 %
\hatcurISOMJeccenxxxxxA
\else
\ifnum#1=32 %
\hatcurISOMJeccenxxxxxB
\else
\ifnum#1=33 %
\hatcurISOMJeccenxxxxxC
\else
\ifnum#1=34 %
\hatcurISOMJeccenxxxxxD
\else
\ifnum#1=35 %
\hatcurISOMJeccenxxxxxE
\else
??????\fi
\fi
\fi
\fi
\fi
}
\newcommand{\hatcurISOMKeccen}[1]{\ifnum#1=31 %
\hatcurISOMKeccenxxxxxA
\else
\ifnum#1=32 %
\hatcurISOMKeccenxxxxxB
\else
\ifnum#1=33 %
\hatcurISOMKeccenxxxxxC
\else
\ifnum#1=34 %
\hatcurISOMKeccenxxxxxD
\else
\ifnum#1=35 %
\hatcurISOMKeccenxxxxxE
\else
??????\fi
\fi
\fi
\fi
\fi
}
\newcommand{\hatcurISOmlongeccen}[1]{\ifnum#1=31 %
\hatcurISOmlongeccenxxxxxA
\else
\ifnum#1=32 %
\hatcurISOmlongeccenxxxxxB
\else
\ifnum#1=33 %
\hatcurISOmlongeccenxxxxxC
\else
\ifnum#1=34 %
\hatcurISOmlongeccenxxxxxD
\else
\ifnum#1=35 %
\hatcurISOmlongeccenxxxxxE
\else
??????\fi
\fi
\fi
\fi
\fi
}
\newcommand{\hatcurISOmshorteccen}[1]{\ifnum#1=31 %
\hatcurISOmshorteccenxxxxxA
\else
\ifnum#1=32 %
\hatcurISOmshorteccenxxxxxB
\else
\ifnum#1=33 %
\hatcurISOmshorteccenxxxxxC
\else
\ifnum#1=34 %
\hatcurISOmshorteccenxxxxxD
\else
\ifnum#1=35 %
\hatcurISOmshorteccenxxxxxE
\else
??????\fi
\fi
\fi
\fi
\fi
}
\newcommand{\hatcurISOmveccen}[1]{\ifnum#1=31 %
\hatcurISOmveccenxxxxxA
\else
\ifnum#1=32 %
\hatcurISOmveccenxxxxxB
\else
\ifnum#1=33 %
\hatcurISOmveccenxxxxxC
\else
\ifnum#1=34 %
\hatcurISOmveccenxxxxxD
\else
\ifnum#1=35 %
\hatcurISOmveccenxxxxxE
\else
??????\fi
\fi
\fi
\fi
\fi
}
\newcommand{\hatcurISOreccen}[1]{\ifnum#1=31 %
\hatcurISOreccenxxxxxA
\else
\ifnum#1=32 %
\hatcurISOreccenxxxxxB
\else
\ifnum#1=33 %
\hatcurISOreccenxxxxxC
\else
\ifnum#1=34 %
\hatcurISOreccenxxxxxD
\else
\ifnum#1=35 %
\hatcurISOreccenxxxxxE
\else
??????\fi
\fi
\fi
\fi
\fi
}
\newcommand{\hatcurISOrhoeccen}[1]{\ifnum#1=31 %
\hatcurISOrhoeccenxxxxxA
\else
\ifnum#1=32 %
\hatcurISOrhoeccenxxxxxB
\else
\ifnum#1=33 %
\hatcurISOrhoeccenxxxxxC
\else
\ifnum#1=34 %
\hatcurISOrhoeccenxxxxxD
\else
\ifnum#1=35 %
\hatcurISOrhoeccenxxxxxE
\else
??????\fi
\fi
\fi
\fi
\fi
}
\newcommand{\hatcurISOrholongeccen}[1]{\ifnum#1=31 %
\hatcurISOrholongeccenxxxxxA
\else
\ifnum#1=32 %
\hatcurISOrholongeccenxxxxxB
\else
\ifnum#1=33 %
\hatcurISOrholongeccenxxxxxC
\else
\ifnum#1=34 %
\hatcurISOrholongeccenxxxxxD
\else
\ifnum#1=35 %
\hatcurISOrholongeccenxxxxxE
\else
??????\fi
\fi
\fi
\fi
\fi
}
\newcommand{\hatcurISOrlongeccen}[1]{\ifnum#1=31 %
\hatcurISOrlongeccenxxxxxA
\else
\ifnum#1=32 %
\hatcurISOrlongeccenxxxxxB
\else
\ifnum#1=33 %
\hatcurISOrlongeccenxxxxxC
\else
\ifnum#1=34 %
\hatcurISOrlongeccenxxxxxD
\else
\ifnum#1=35 %
\hatcurISOrlongeccenxxxxxE
\else
??????\fi
\fi
\fi
\fi
\fi
}
\newcommand{\hatcurISOrshorteccen}[1]{\ifnum#1=31 %
\hatcurISOrshorteccenxxxxxA
\else
\ifnum#1=32 %
\hatcurISOrshorteccenxxxxxB
\else
\ifnum#1=33 %
\hatcurISOrshorteccenxxxxxC
\else
\ifnum#1=34 %
\hatcurISOrshorteccenxxxxxD
\else
\ifnum#1=35 %
\hatcurISOrshorteccenxxxxxE
\else
??????\fi
\fi
\fi
\fi
\fi
}
\newcommand{\hatcurISOsigmaeccen}[1]{\ifnum#1=31 %
\hatcurISOsigmaeccenxxxxxA
\else
\ifnum#1=32 %
\hatcurISOsigmaeccenxxxxxB
\else
\ifnum#1=33 %
\hatcurISOsigmaeccenxxxxxC
\else
\ifnum#1=34 %
\hatcurISOsigmaeccenxxxxxD
\else
\ifnum#1=35 %
\hatcurISOsigmaeccenxxxxxE
\else
??????\fi
\fi
\fi
\fi
\fi
}
\newcommand{\hatcurISOspececcen}[1]{\ifnum#1=31 %
\hatcurISOspececcenxxxxxA
\else
\ifnum#1=32 %
\hatcurISOspececcenxxxxxB
\else
\ifnum#1=33 %
\hatcurISOspececcenxxxxxC
\else
\ifnum#1=34 %
\hatcurISOspececcenxxxxxD
\else
\ifnum#1=35 %
\hatcurISOspececcenxxxxxE
\else
??????\fi
\fi
\fi
\fi
\fi
}
\newcommand{\hatcurISOvieccen}[1]{\ifnum#1=31 %
\hatcurISOvieccenxxxxxA
\else
\ifnum#1=32 %
\hatcurISOvieccenxxxxxB
\else
\ifnum#1=33 %
\hatcurISOvieccenxxxxxC
\else
\ifnum#1=34 %
\hatcurISOvieccenxxxxxD
\else
\ifnum#1=35 %
\hatcurISOvieccenxxxxxE
\else
??????\fi
\fi
\fi
\fi
\fi
}
\newcommand{\hatcurLBigeccen}[1]{\ifnum#1=31 %
\hatcurLBigeccenxxxxxA
\else
\ifnum#1=32 %
\hatcurLBigeccenxxxxxB
\else
\ifnum#1=33 %
\hatcurLBigeccenxxxxxC
\else
\ifnum#1=34 %
\hatcurLBigeccenxxxxxD
\else
\ifnum#1=35 %
\hatcurLBigeccenxxxxxE
\else
??????\fi
\fi
\fi
\fi
\fi
}
\newcommand{\hatcurLBiieccen}[1]{\ifnum#1=31 %
\hatcurLBiieccenxxxxxA
\else
\ifnum#1=32 %
\hatcurLBiieccenxxxxxB
\else
\ifnum#1=33 %
\hatcurLBiieccenxxxxxC
\else
\ifnum#1=34 %
\hatcurLBiieccenxxxxxD
\else
\ifnum#1=35 %
\hatcurLBiieccenxxxxxE
\else
??????\fi
\fi
\fi
\fi
\fi
}
\newcommand{\hatcurLBiIeccen}[1]{\ifnum#1=31 %
\hatcurLBiIeccenxxxxxA
\else
\ifnum#1=32 %
\hatcurLBiIeccenxxxxxB
\else
\ifnum#1=33 %
\hatcurLBiIeccenxxxxxC
\else
\ifnum#1=34 %
\hatcurLBiIeccenxxxxxD
\else
\ifnum#1=35 %
\hatcurLBiIeccenxxxxxE
\else
??????\fi
\fi
\fi
\fi
\fi
}
\newcommand{\hatcurLBiigeccen}[1]{\ifnum#1=31 %
\hatcurLBiigeccenxxxxxA
\else
\ifnum#1=32 %
\hatcurLBiigeccenxxxxxB
\else
\ifnum#1=33 %
\hatcurLBiigeccenxxxxxC
\else
\ifnum#1=34 %
\hatcurLBiigeccenxxxxxD
\else
\ifnum#1=35 %
\hatcurLBiigeccenxxxxxE
\else
??????\fi
\fi
\fi
\fi
\fi
}
\newcommand{\hatcurLBiiieccen}[1]{\ifnum#1=31 %
\hatcurLBiiieccenxxxxxA
\else
\ifnum#1=32 %
\hatcurLBiiieccenxxxxxB
\else
\ifnum#1=33 %
\hatcurLBiiieccenxxxxxC
\else
\ifnum#1=34 %
\hatcurLBiiieccenxxxxxD
\else
\ifnum#1=35 %
\hatcurLBiiieccenxxxxxE
\else
??????\fi
\fi
\fi
\fi
\fi
}
\newcommand{\hatcurLBiiIeccen}[1]{\ifnum#1=31 %
\hatcurLBiiIeccenxxxxxA
\else
\ifnum#1=32 %
\hatcurLBiiIeccenxxxxxB
\else
\ifnum#1=33 %
\hatcurLBiiIeccenxxxxxC
\else
\ifnum#1=34 %
\hatcurLBiiIeccenxxxxxD
\else
\ifnum#1=35 %
\hatcurLBiiIeccenxxxxxE
\else
??????\fi
\fi
\fi
\fi
\fi
}
\newcommand{\hatcurLBiikepeccen}[1]{\ifnum#1=31 %
\hatcurLBiikepeccenxxxxxA
\else
\ifnum#1=33 %
\hatcurLBiikepeccenxxxxxC
\else
\ifnum#1=34 %
\hatcurLBiikepeccenxxxxxD
\else
\ifnum#1=35 %
\hatcurLBiikepeccenxxxxxE
\else
??????\fi
\fi
\fi
\fi
}
\newcommand{\hatcurLBiireccen}[1]{\ifnum#1=31 %
\hatcurLBiireccenxxxxxA
\else
\ifnum#1=32 %
\hatcurLBiireccenxxxxxB
\else
\ifnum#1=33 %
\hatcurLBiireccenxxxxxC
\else
\ifnum#1=34 %
\hatcurLBiireccenxxxxxD
\else
\ifnum#1=35 %
\hatcurLBiireccenxxxxxE
\else
??????\fi
\fi
\fi
\fi
\fi
}
\newcommand{\hatcurLBiiReccen}[1]{\ifnum#1=31 %
\hatcurLBiiReccenxxxxxA
\else
\ifnum#1=32 %
\hatcurLBiiReccenxxxxxB
\else
\ifnum#1=33 %
\hatcurLBiiReccenxxxxxC
\else
\ifnum#1=34 %
\hatcurLBiiReccenxxxxxD
\else
\ifnum#1=35 %
\hatcurLBiiReccenxxxxxE
\else
??????\fi
\fi
\fi
\fi
\fi
}
\newcommand{\hatcurLBiizeccen}[1]{\ifnum#1=31 %
\hatcurLBiizeccenxxxxxA
\else
\ifnum#1=32 %
\hatcurLBiizeccenxxxxxB
\else
\ifnum#1=33 %
\hatcurLBiizeccenxxxxxC
\else
\ifnum#1=34 %
\hatcurLBiizeccenxxxxxD
\else
\ifnum#1=35 %
\hatcurLBiizeccenxxxxxE
\else
??????\fi
\fi
\fi
\fi
\fi
}
\newcommand{\hatcurLBikepeccen}[1]{\ifnum#1=31 %
\hatcurLBikepeccenxxxxxA
\else
\ifnum#1=33 %
\hatcurLBikepeccenxxxxxC
\else
\ifnum#1=34 %
\hatcurLBikepeccenxxxxxD
\else
\ifnum#1=35 %
\hatcurLBikepeccenxxxxxE
\else
??????\fi
\fi
\fi
\fi
}
\newcommand{\hatcurLBireccen}[1]{\ifnum#1=31 %
\hatcurLBireccenxxxxxA
\else
\ifnum#1=32 %
\hatcurLBireccenxxxxxB
\else
\ifnum#1=33 %
\hatcurLBireccenxxxxxC
\else
\ifnum#1=34 %
\hatcurLBireccenxxxxxD
\else
\ifnum#1=35 %
\hatcurLBireccenxxxxxE
\else
??????\fi
\fi
\fi
\fi
\fi
}
\newcommand{\hatcurLBiReccen}[1]{\ifnum#1=31 %
\hatcurLBiReccenxxxxxA
\else
\ifnum#1=32 %
\hatcurLBiReccenxxxxxB
\else
\ifnum#1=33 %
\hatcurLBiReccenxxxxxC
\else
\ifnum#1=34 %
\hatcurLBiReccenxxxxxD
\else
\ifnum#1=35 %
\hatcurLBiReccenxxxxxE
\else
??????\fi
\fi
\fi
\fi
\fi
}
\newcommand{\hatcurLBizeccen}[1]{\ifnum#1=31 %
\hatcurLBizeccenxxxxxA
\else
\ifnum#1=32 %
\hatcurLBizeccenxxxxxB
\else
\ifnum#1=33 %
\hatcurLBizeccenxxxxxC
\else
\ifnum#1=34 %
\hatcurLBizeccenxxxxxD
\else
\ifnum#1=35 %
\hatcurLBizeccenxxxxxE
\else
??????\fi
\fi
\fi
\fi
\fi
}
\newcommand{\hatcurLCbsqeccen}[1]{\ifnum#1=31 %
\hatcurLCbsqeccenxxxxxA
\else
\ifnum#1=32 %
\hatcurLCbsqeccenxxxxxB
\else
\ifnum#1=33 %
\hatcurLCbsqeccenxxxxxC
\else
\ifnum#1=34 %
\hatcurLCbsqeccenxxxxxD
\else
\ifnum#1=35 %
\hatcurLCbsqeccenxxxxxE
\else
??????\fi
\fi
\fi
\fi
\fi
}
\newcommand{\hatcurLCdipeccen}[1]{\ifnum#1=31 %
\hatcurLCdipeccenxxxxxA
\else
\ifnum#1=32 %
\hatcurLCdipeccenxxxxxB
\else
\ifnum#1=33 %
\hatcurLCdipeccenxxxxxC
\else
\ifnum#1=34 %
\hatcurLCdipeccenxxxxxD
\else
\ifnum#1=35 %
\hatcurLCdipeccenxxxxxE
\else
??????\fi
\fi
\fi
\fi
\fi
}
\newcommand{\hatcurLCdureccen}[1]{\ifnum#1=31 %
\hatcurLCdureccenxxxxxA
\else
\ifnum#1=32 %
\hatcurLCdureccenxxxxxB
\else
\ifnum#1=33 %
\hatcurLCdureccenxxxxxC
\else
\ifnum#1=34 %
\hatcurLCdureccenxxxxxD
\else
\ifnum#1=35 %
\hatcurLCdureccenxxxxxE
\else
??????\fi
\fi
\fi
\fi
\fi
}
\newcommand{\hatcurLCdurhreccen}[1]{\ifnum#1=31 %
\hatcurLCdurhreccenxxxxxA
\else
\ifnum#1=32 %
\hatcurLCdurhreccenxxxxxB
\else
\ifnum#1=33 %
\hatcurLCdurhreccenxxxxxC
\else
\ifnum#1=34 %
\hatcurLCdurhreccenxxxxxD
\else
\ifnum#1=35 %
\hatcurLCdurhreccenxxxxxE
\else
??????\fi
\fi
\fi
\fi
\fi
}
\newcommand{\hatcurLCdurhrshorteccen}[1]{\ifnum#1=31 %
\hatcurLCdurhrshorteccenxxxxxA
\else
\ifnum#1=32 %
\hatcurLCdurhrshorteccenxxxxxB
\else
\ifnum#1=33 %
\hatcurLCdurhrshorteccenxxxxxC
\else
\ifnum#1=34 %
\hatcurLCdurhrshorteccenxxxxxD
\else
\ifnum#1=35 %
\hatcurLCdurhrshorteccenxxxxxE
\else
??????\fi
\fi
\fi
\fi
\fi
}
\newcommand{\hatcurLCdurshorteccen}[1]{\ifnum#1=31 %
\hatcurLCdurshorteccenxxxxxA
\else
\ifnum#1=32 %
\hatcurLCdurshorteccenxxxxxB
\else
\ifnum#1=33 %
\hatcurLCdurshorteccenxxxxxC
\else
\ifnum#1=34 %
\hatcurLCdurshorteccenxxxxxD
\else
\ifnum#1=35 %
\hatcurLCdurshorteccenxxxxxE
\else
??????\fi
\fi
\fi
\fi
\fi
}
\newcommand{\hatcurLChatnetmeccen}[1]{\ifnum#1=31 %
\hatcurLChatnetmeccenxxxxxA
\else
\ifnum#1=32 %
\hatcurLChatnetmeccenxxxxxB
\else
\ifnum#1=33 %
\hatcurLChatnetmeccenxxxxxC
\else
\ifnum#1=34 %
\hatcurLChatnetmeccenxxxxxD
\else
\ifnum#1=35 %
\hatcurLChatnetmeccenxxxxxE
\else
??????\fi
\fi
\fi
\fi
\fi
}
\newcommand{\hatcurLCiblendeccen}[1]{\ifnum#1=31 %
\hatcurLCiblendeccenxxxxxA
\else
\ifnum#1=32 %
\hatcurLCiblendeccenxxxxxB
\else
\ifnum#1=33 %
\hatcurLCiblendeccenxxxxxC
\else
\ifnum#1=34 %
\hatcurLCiblendeccenxxxxxD
\else
\ifnum#1=35 %
\hatcurLCiblendeccenxxxxxE
\else
??????\fi
\fi
\fi
\fi
\fi
}
\newcommand{\hatcurLCimpeccen}[1]{\ifnum#1=31 %
\hatcurLCimpeccenxxxxxA
\else
\ifnum#1=32 %
\hatcurLCimpeccenxxxxxB
\else
\ifnum#1=33 %
\hatcurLCimpeccenxxxxxC
\else
\ifnum#1=34 %
\hatcurLCimpeccenxxxxxD
\else
\ifnum#1=35 %
\hatcurLCimpeccenxxxxxE
\else
??????\fi
\fi
\fi
\fi
\fi
}
\newcommand{\hatcurLCingdureccen}[1]{\ifnum#1=31 %
\hatcurLCingdureccenxxxxxA
\else
\ifnum#1=32 %
\hatcurLCingdureccenxxxxxB
\else
\ifnum#1=33 %
\hatcurLCingdureccenxxxxxC
\else
\ifnum#1=34 %
\hatcurLCingdureccenxxxxxD
\else
\ifnum#1=35 %
\hatcurLCingdureccenxxxxxE
\else
??????\fi
\fi
\fi
\fi
\fi
}
\newcommand{\hatcurLCPeccen}[1]{\ifnum#1=31 %
\hatcurLCPeccenxxxxxA
\else
\ifnum#1=32 %
\hatcurLCPeccenxxxxxB
\else
\ifnum#1=33 %
\hatcurLCPeccenxxxxxC
\else
\ifnum#1=34 %
\hatcurLCPeccenxxxxxD
\else
\ifnum#1=35 %
\hatcurLCPeccenxxxxxE
\else
??????\fi
\fi
\fi
\fi
\fi
}
\newcommand{\hatcurLCPprececcen}[1]{\ifnum#1=31 %
\hatcurLCPprececcenxxxxxA
\else
\ifnum#1=32 %
\hatcurLCPprececcenxxxxxB
\else
\ifnum#1=33 %
\hatcurLCPprececcenxxxxxC
\else
\ifnum#1=34 %
\hatcurLCPprececcenxxxxxD
\else
\ifnum#1=35 %
\hatcurLCPprececcenxxxxxE
\else
??????\fi
\fi
\fi
\fi
\fi
}
\newcommand{\hatcurLCPshorteccen}[1]{\ifnum#1=31 %
\hatcurLCPshorteccenxxxxxA
\else
\ifnum#1=32 %
\hatcurLCPshorteccenxxxxxB
\else
\ifnum#1=33 %
\hatcurLCPshorteccenxxxxxC
\else
\ifnum#1=34 %
\hatcurLCPshorteccenxxxxxD
\else
\ifnum#1=35 %
\hatcurLCPshorteccenxxxxxE
\else
??????\fi
\fi
\fi
\fi
\fi
}
\newcommand{\hatcurLCqeccen}[1]{\ifnum#1=31 %
\hatcurLCqeccenxxxxxA
\else
\ifnum#1=32 %
\hatcurLCqeccenxxxxxB
\else
\ifnum#1=33 %
\hatcurLCqeccenxxxxxC
\else
\ifnum#1=34 %
\hatcurLCqeccenxxxxxD
\else
\ifnum#1=35 %
\hatcurLCqeccenxxxxxE
\else
??????\fi
\fi
\fi
\fi
\fi
}
\newcommand{\hatcurLCqshorteccen}[1]{\ifnum#1=31 %
\hatcurLCqshorteccenxxxxxA
\else
\ifnum#1=32 %
\hatcurLCqshorteccenxxxxxB
\else
\ifnum#1=33 %
\hatcurLCqshorteccenxxxxxC
\else
\ifnum#1=34 %
\hatcurLCqshorteccenxxxxxD
\else
\ifnum#1=35 %
\hatcurLCqshorteccenxxxxxE
\else
??????\fi
\fi
\fi
\fi
\fi
}
\newcommand{\hatcurLCrhoeccen}[1]{\ifnum#1=31 %
\hatcurLCrhoeccenxxxxxA
\else
\ifnum#1=32 %
\hatcurLCrhoeccenxxxxxB
\else
\ifnum#1=33 %
\hatcurLCrhoeccenxxxxxC
\else
\ifnum#1=34 %
\hatcurLCrhoeccenxxxxxD
\else
\ifnum#1=35 %
\hatcurLCrhoeccenxxxxxE
\else
??????\fi
\fi
\fi
\fi
\fi
}
\newcommand{\hatcurLCrprstareccen}[1]{\ifnum#1=31 %
\hatcurLCrprstareccenxxxxxA
\else
\ifnum#1=32 %
\hatcurLCrprstareccenxxxxxB
\else
\ifnum#1=33 %
\hatcurLCrprstareccenxxxxxC
\else
\ifnum#1=34 %
\hatcurLCrprstareccenxxxxxD
\else
\ifnum#1=35 %
\hatcurLCrprstareccenxxxxxE
\else
??????\fi
\fi
\fi
\fi
\fi
}
\newcommand{\hatcurLCTAeccen}[1]{\ifnum#1=31 %
\hatcurLCTAeccenxxxxxA
\else
\ifnum#1=32 %
\hatcurLCTAeccenxxxxxB
\else
\ifnum#1=33 %
\hatcurLCTAeccenxxxxxC
\else
\ifnum#1=34 %
\hatcurLCTAeccenxxxxxD
\else
\ifnum#1=35 %
\hatcurLCTAeccenxxxxxE
\else
??????\fi
\fi
\fi
\fi
\fi
}
\newcommand{\hatcurLCTBeccen}[1]{\ifnum#1=31 %
\hatcurLCTBeccenxxxxxA
\else
\ifnum#1=32 %
\hatcurLCTBeccenxxxxxB
\else
\ifnum#1=33 %
\hatcurLCTBeccenxxxxxC
\else
\ifnum#1=34 %
\hatcurLCTBeccenxxxxxD
\else
\ifnum#1=35 %
\hatcurLCTBeccenxxxxxE
\else
??????\fi
\fi
\fi
\fi
\fi
}
\newcommand{\hatcurLCTeccen}[1]{\ifnum#1=31 %
\hatcurLCTeccenxxxxxA
\else
\ifnum#1=32 %
\hatcurLCTeccenxxxxxB
\else
\ifnum#1=33 %
\hatcurLCTeccenxxxxxC
\else
\ifnum#1=34 %
\hatcurLCTeccenxxxxxD
\else
\ifnum#1=35 %
\hatcurLCTeccenxxxxxE
\else
??????\fi
\fi
\fi
\fi
\fi
}
\newcommand{\hatcurLCzetaeccen}[1]{\ifnum#1=31 %
\hatcurLCzetaeccenxxxxxA
\else
\ifnum#1=32 %
\hatcurLCzetaeccenxxxxxB
\else
\ifnum#1=33 %
\hatcurLCzetaeccenxxxxxC
\else
\ifnum#1=34 %
\hatcurLCzetaeccenxxxxxD
\else
\ifnum#1=35 %
\hatcurLCzetaeccenxxxxxE
\else
??????\fi
\fi
\fi
\fi
\fi
}
\newcommand{\hatcurPPaequiveccen}[1]{\ifnum#1=31 %
\hatcurPPaequiveccenxxxxxA
\else
\ifnum#1=32 %
\hatcurPPaequiveccenxxxxxB
\else
\ifnum#1=33 %
\hatcurPPaequiveccenxxxxxC
\else
\ifnum#1=34 %
\hatcurPPaequiveccenxxxxxD
\else
\ifnum#1=35 %
\hatcurPPaequiveccenxxxxxE
\else
??????\fi
\fi
\fi
\fi
\fi
}
\newcommand{\hatcurPPareccen}[1]{\ifnum#1=31 %
\hatcurPPareccenxxxxxA
\else
\ifnum#1=32 %
\hatcurPPareccenxxxxxB
\else
\ifnum#1=33 %
\hatcurPPareccenxxxxxC
\else
\ifnum#1=34 %
\hatcurPPareccenxxxxxD
\else
\ifnum#1=35 %
\hatcurPPareccenxxxxxE
\else
??????\fi
\fi
\fi
\fi
\fi
}
\newcommand{\hatcurPPareleccen}[1]{\ifnum#1=31 %
\hatcurPPareleccenxxxxxA
\else
\ifnum#1=32 %
\hatcurPPareleccenxxxxxB
\else
\ifnum#1=33 %
\hatcurPPareleccenxxxxxC
\else
\ifnum#1=34 %
\hatcurPPareleccenxxxxxD
\else
\ifnum#1=35 %
\hatcurPPareleccenxxxxxE
\else
??????\fi
\fi
\fi
\fi
\fi
}
\newcommand{\hatcurPPfluxapdimeccen}[1]{\ifnum#1=31 %
\hatcurPPfluxapdimeccenxxxxxA
\else
\ifnum#1=32 %
\hatcurPPfluxapdimeccenxxxxxB
\else
\ifnum#1=33 %
\hatcurPPfluxapdimeccenxxxxxC
\else
\ifnum#1=34 %
\hatcurPPfluxapdimeccenxxxxxD
\else
\ifnum#1=35 %
\hatcurPPfluxapdimeccenxxxxxE
\else
??????\fi
\fi
\fi
\fi
\fi
}
\newcommand{\hatcurPPfluxapeccen}[1]{\ifnum#1=31 %
\hatcurPPfluxapeccenxxxxxA
\else
\ifnum#1=32 %
\hatcurPPfluxapeccenxxxxxB
\else
\ifnum#1=33 %
\hatcurPPfluxapeccenxxxxxC
\else
\ifnum#1=34 %
\hatcurPPfluxapeccenxxxxxD
\else
\ifnum#1=35 %
\hatcurPPfluxapeccenxxxxxE
\else
??????\fi
\fi
\fi
\fi
\fi
}
\newcommand{\hatcurPPfluxavgdimeccen}[1]{\ifnum#1=31 %
\hatcurPPfluxavgdimeccenxxxxxA
\else
\ifnum#1=32 %
\hatcurPPfluxavgdimeccenxxxxxB
\else
\ifnum#1=33 %
\hatcurPPfluxavgdimeccenxxxxxC
\else
\ifnum#1=34 %
\hatcurPPfluxavgdimeccenxxxxxD
\else
\ifnum#1=35 %
\hatcurPPfluxavgdimeccenxxxxxE
\else
??????\fi
\fi
\fi
\fi
\fi
}
\newcommand{\hatcurPPfluxavgeccen}[1]{\ifnum#1=31 %
\hatcurPPfluxavgeccenxxxxxA
\else
\ifnum#1=32 %
\hatcurPPfluxavgeccenxxxxxB
\else
\ifnum#1=33 %
\hatcurPPfluxavgeccenxxxxxC
\else
\ifnum#1=34 %
\hatcurPPfluxavgeccenxxxxxD
\else
\ifnum#1=35 %
\hatcurPPfluxavgeccenxxxxxE
\else
??????\fi
\fi
\fi
\fi
\fi
}
\newcommand{\hatcurPPfluxavglogeccen}[1]{\ifnum#1=31 %
\hatcurPPfluxavglogeccenxxxxxA
\else
\ifnum#1=32 %
\hatcurPPfluxavglogeccenxxxxxB
\else
\ifnum#1=33 %
\hatcurPPfluxavglogeccenxxxxxC
\else
\ifnum#1=34 %
\hatcurPPfluxavglogeccenxxxxxD
\else
\ifnum#1=35 %
\hatcurPPfluxavglogeccenxxxxxE
\else
??????\fi
\fi
\fi
\fi
\fi
}
\newcommand{\hatcurPPfluxperidimeccen}[1]{\ifnum#1=31 %
\hatcurPPfluxperidimeccenxxxxxA
\else
\ifnum#1=32 %
\hatcurPPfluxperidimeccenxxxxxB
\else
\ifnum#1=33 %
\hatcurPPfluxperidimeccenxxxxxC
\else
\ifnum#1=34 %
\hatcurPPfluxperidimeccenxxxxxD
\else
\ifnum#1=35 %
\hatcurPPfluxperidimeccenxxxxxE
\else
??????\fi
\fi
\fi
\fi
\fi
}
\newcommand{\hatcurPPfluxperieccen}[1]{\ifnum#1=31 %
\hatcurPPfluxperieccenxxxxxA
\else
\ifnum#1=32 %
\hatcurPPfluxperieccenxxxxxB
\else
\ifnum#1=33 %
\hatcurPPfluxperieccenxxxxxC
\else
\ifnum#1=34 %
\hatcurPPfluxperieccenxxxxxD
\else
\ifnum#1=35 %
\hatcurPPfluxperieccenxxxxxE
\else
??????\fi
\fi
\fi
\fi
\fi
}
\newcommand{\hatcurPPgeccen}[1]{\ifnum#1=31 %
\hatcurPPgeccenxxxxxA
\else
\ifnum#1=32 %
\hatcurPPgeccenxxxxxB
\else
\ifnum#1=33 %
\hatcurPPgeccenxxxxxC
\else
\ifnum#1=34 %
\hatcurPPgeccenxxxxxD
\else
\ifnum#1=35 %
\hatcurPPgeccenxxxxxE
\else
??????\fi
\fi
\fi
\fi
\fi
}
\newcommand{\hatcurPPieccen}[1]{\ifnum#1=31 %
\hatcurPPieccenxxxxxA
\else
\ifnum#1=32 %
\hatcurPPieccenxxxxxB
\else
\ifnum#1=33 %
\hatcurPPieccenxxxxxC
\else
\ifnum#1=34 %
\hatcurPPieccenxxxxxD
\else
\ifnum#1=35 %
\hatcurPPieccenxxxxxE
\else
??????\fi
\fi
\fi
\fi
\fi
}
\newcommand{\hatcurPPloggeccen}[1]{\ifnum#1=31 %
\hatcurPPloggeccenxxxxxA
\else
\ifnum#1=32 %
\hatcurPPloggeccenxxxxxB
\else
\ifnum#1=33 %
\hatcurPPloggeccenxxxxxC
\else
\ifnum#1=34 %
\hatcurPPloggeccenxxxxxD
\else
\ifnum#1=35 %
\hatcurPPloggeccenxxxxxE
\else
??????\fi
\fi
\fi
\fi
\fi
}
\newcommand{\hatcurPPmeccen}[1]{\ifnum#1=31 %
\hatcurPPmeccenxxxxxA
\else
\ifnum#1=32 %
\hatcurPPmeccenxxxxxB
\else
\ifnum#1=33 %
\hatcurPPmeccenxxxxxC
\else
\ifnum#1=34 %
\hatcurPPmeccenxxxxxD
\else
\ifnum#1=35 %
\hatcurPPmeccenxxxxxE
\else
??????\fi
\fi
\fi
\fi
\fi
}
\newcommand{\hatcurPPmeeccen}[1]{\ifnum#1=31 %
\hatcurPPmeeccenxxxxxA
\else
\ifnum#1=32 %
\hatcurPPmeeccenxxxxxB
\else
\ifnum#1=33 %
\hatcurPPmeeccenxxxxxC
\else
\ifnum#1=34 %
\hatcurPPmeeccenxxxxxD
\else
\ifnum#1=35 %
\hatcurPPmeeccenxxxxxE
\else
??????\fi
\fi
\fi
\fi
\fi
}
\newcommand{\hatcurPPmelongeccen}[1]{\ifnum#1=31 %
\hatcurPPmelongeccenxxxxxA
\else
\ifnum#1=32 %
\hatcurPPmelongeccenxxxxxB
\else
\ifnum#1=33 %
\hatcurPPmelongeccenxxxxxC
\else
\ifnum#1=34 %
\hatcurPPmelongeccenxxxxxD
\else
\ifnum#1=35 %
\hatcurPPmelongeccenxxxxxE
\else
??????\fi
\fi
\fi
\fi
\fi
}
\newcommand{\hatcurPPmeshorteccen}[1]{\ifnum#1=31 %
\hatcurPPmeshorteccenxxxxxA
\else
\ifnum#1=32 %
\hatcurPPmeshorteccenxxxxxB
\else
\ifnum#1=33 %
\hatcurPPmeshorteccenxxxxxC
\else
\ifnum#1=34 %
\hatcurPPmeshorteccenxxxxxD
\else
\ifnum#1=35 %
\hatcurPPmeshorteccenxxxxxE
\else
??????\fi
\fi
\fi
\fi
\fi
}
\newcommand{\hatcurPPmlongeccen}[1]{\ifnum#1=31 %
\hatcurPPmlongeccenxxxxxA
\else
\ifnum#1=32 %
\hatcurPPmlongeccenxxxxxB
\else
\ifnum#1=33 %
\hatcurPPmlongeccenxxxxxC
\else
\ifnum#1=34 %
\hatcurPPmlongeccenxxxxxD
\else
\ifnum#1=35 %
\hatcurPPmlongeccenxxxxxE
\else
??????\fi
\fi
\fi
\fi
\fi
}
\newcommand{\hatcurPPmrcorreccen}[1]{\ifnum#1=31 %
\hatcurPPmrcorreccenxxxxxA
\else
\ifnum#1=32 %
\hatcurPPmrcorreccenxxxxxB
\else
\ifnum#1=33 %
\hatcurPPmrcorreccenxxxxxC
\else
\ifnum#1=34 %
\hatcurPPmrcorreccenxxxxxD
\else
\ifnum#1=35 %
\hatcurPPmrcorreccenxxxxxE
\else
??????\fi
\fi
\fi
\fi
\fi
}
\newcommand{\hatcurPPmshorteccen}[1]{\ifnum#1=31 %
\hatcurPPmshorteccenxxxxxA
\else
\ifnum#1=32 %
\hatcurPPmshorteccenxxxxxB
\else
\ifnum#1=33 %
\hatcurPPmshorteccenxxxxxC
\else
\ifnum#1=34 %
\hatcurPPmshorteccenxxxxxD
\else
\ifnum#1=35 %
\hatcurPPmshorteccenxxxxxE
\else
??????\fi
\fi
\fi
\fi
\fi
}
\newcommand{\hatcurPPperieccen}[1]{\ifnum#1=31 %
\hatcurPPperieccenxxxxxA
\else
\ifnum#1=32 %
\hatcurPPperieccenxxxxxB
\else
\ifnum#1=33 %
\hatcurPPperieccenxxxxxC
\else
\ifnum#1=34 %
\hatcurPPperieccenxxxxxD
\else
\ifnum#1=35 %
\hatcurPPperieccenxxxxxE
\else
??????\fi
\fi
\fi
\fi
\fi
}
\newcommand{\hatcurPPphiconjeccen}[1]{\ifnum#1=31 %
\hatcurPPphiconjeccenxxxxxA
\else
\ifnum#1=32 %
\hatcurPPphiconjeccenxxxxxB
\else
\ifnum#1=33 %
\hatcurPPphiconjeccenxxxxxC
\else
\ifnum#1=34 %
\hatcurPPphiconjeccenxxxxxD
\else
\ifnum#1=35 %
\hatcurPPphiconjeccenxxxxxE
\else
??????\fi
\fi
\fi
\fi
\fi
}
\newcommand{\hatcurPPreccen}[1]{\ifnum#1=31 %
\hatcurPPreccenxxxxxA
\else
\ifnum#1=32 %
\hatcurPPreccenxxxxxB
\else
\ifnum#1=33 %
\hatcurPPreccenxxxxxC
\else
\ifnum#1=34 %
\hatcurPPreccenxxxxxD
\else
\ifnum#1=35 %
\hatcurPPreccenxxxxxE
\else
??????\fi
\fi
\fi
\fi
\fi
}
\newcommand{\hatcurPPreeccen}[1]{\ifnum#1=31 %
\hatcurPPreeccenxxxxxA
\else
\ifnum#1=32 %
\hatcurPPreeccenxxxxxB
\else
\ifnum#1=33 %
\hatcurPPreeccenxxxxxC
\else
\ifnum#1=34 %
\hatcurPPreeccenxxxxxD
\else
\ifnum#1=35 %
\hatcurPPreeccenxxxxxE
\else
??????\fi
\fi
\fi
\fi
\fi
}
\newcommand{\hatcurPPrelongeccen}[1]{\ifnum#1=31 %
\hatcurPPrelongeccenxxxxxA
\else
\ifnum#1=32 %
\hatcurPPrelongeccenxxxxxB
\else
\ifnum#1=33 %
\hatcurPPrelongeccenxxxxxC
\else
\ifnum#1=34 %
\hatcurPPrelongeccenxxxxxD
\else
\ifnum#1=35 %
\hatcurPPrelongeccenxxxxxE
\else
??????\fi
\fi
\fi
\fi
\fi
}
\newcommand{\hatcurPPreshorteccen}[1]{\ifnum#1=31 %
\hatcurPPreshorteccenxxxxxA
\else
\ifnum#1=32 %
\hatcurPPreshorteccenxxxxxB
\else
\ifnum#1=33 %
\hatcurPPreshorteccenxxxxxC
\else
\ifnum#1=34 %
\hatcurPPreshorteccenxxxxxD
\else
\ifnum#1=35 %
\hatcurPPreshorteccenxxxxxE
\else
??????\fi
\fi
\fi
\fi
\fi
}
\newcommand{\hatcurPPrhoeccen}[1]{\ifnum#1=31 %
\hatcurPPrhoeccenxxxxxA
\else
\ifnum#1=32 %
\hatcurPPrhoeccenxxxxxB
\else
\ifnum#1=33 %
\hatcurPPrhoeccenxxxxxC
\else
\ifnum#1=34 %
\hatcurPPrhoeccenxxxxxD
\else
\ifnum#1=35 %
\hatcurPPrhoeccenxxxxxE
\else
??????\fi
\fi
\fi
\fi
\fi
}
\newcommand{\hatcurPPrlongeccen}[1]{\ifnum#1=31 %
\hatcurPPrlongeccenxxxxxA
\else
\ifnum#1=32 %
\hatcurPPrlongeccenxxxxxB
\else
\ifnum#1=33 %
\hatcurPPrlongeccenxxxxxC
\else
\ifnum#1=34 %
\hatcurPPrlongeccenxxxxxD
\else
\ifnum#1=35 %
\hatcurPPrlongeccenxxxxxE
\else
??????\fi
\fi
\fi
\fi
\fi
}
\newcommand{\hatcurPPrshorteccen}[1]{\ifnum#1=31 %
\hatcurPPrshorteccenxxxxxA
\else
\ifnum#1=32 %
\hatcurPPrshorteccenxxxxxB
\else
\ifnum#1=33 %
\hatcurPPrshorteccenxxxxxC
\else
\ifnum#1=34 %
\hatcurPPrshorteccenxxxxxD
\else
\ifnum#1=35 %
\hatcurPPrshorteccenxxxxxE
\else
??????\fi
\fi
\fi
\fi
\fi
}
\newcommand{\hatcurPPtcirceccen}[1]{\ifnum#1=31 %
\hatcurPPtcirceccenxxxxxA
\else
\ifnum#1=32 %
\hatcurPPtcirceccenxxxxxB
\else
\ifnum#1=33 %
\hatcurPPtcirceccenxxxxxC
\else
\ifnum#1=34 %
\hatcurPPtcirceccenxxxxxD
\else
\ifnum#1=35 %
\hatcurPPtcirceccenxxxxxE
\else
??????\fi
\fi
\fi
\fi
\fi
}
\newcommand{\hatcurPPteffeccen}[1]{\ifnum#1=31 %
\hatcurPPteffeccenxxxxxA
\else
\ifnum#1=32 %
\hatcurPPteffeccenxxxxxB
\else
\ifnum#1=33 %
\hatcurPPteffeccenxxxxxC
\else
\ifnum#1=34 %
\hatcurPPteffeccenxxxxxD
\else
\ifnum#1=35 %
\hatcurPPteffeccenxxxxxE
\else
??????\fi
\fi
\fi
\fi
\fi
}
\newcommand{\hatcurPPthetaeccen}[1]{\ifnum#1=31 %
\hatcurPPthetaeccenxxxxxA
\else
\ifnum#1=32 %
\hatcurPPthetaeccenxxxxxB
\else
\ifnum#1=33 %
\hatcurPPthetaeccenxxxxxC
\else
\ifnum#1=34 %
\hatcurPPthetaeccenxxxxxD
\else
\ifnum#1=35 %
\hatcurPPthetaeccenxxxxxE
\else
??????\fi
\fi
\fi
\fi
\fi
}
\newcommand{\hatcurPPtinfalleccen}[1]{\ifnum#1=31 %
\hatcurPPtinfalleccenxxxxxA
\else
\ifnum#1=32 %
\hatcurPPtinfalleccenxxxxxB
\else
\ifnum#1=33 %
\hatcurPPtinfalleccenxxxxxC
\else
\ifnum#1=34 %
\hatcurPPtinfalleccenxxxxxD
\else
\ifnum#1=35 %
\hatcurPPtinfalleccenxxxxxE
\else
??????\fi
\fi
\fi
\fi
\fi
}
\newcommand{\hatcurRVecceneccen}[1]{\ifnum#1=31 %
\hatcurRVecceneccenxxxxxA
\else
\ifnum#1=32 %
\hatcurRVecceneccenxxxxxB
\else
\ifnum#1=33 %
\hatcurRVecceneccenxxxxxC
\else
\ifnum#1=34 %
\hatcurRVecceneccenxxxxxD
\else
\ifnum#1=35 %
\hatcurRVecceneccenxxxxxE
\else
??????\fi
\fi
\fi
\fi
\fi
}
\newcommand{\hatcurRVeccentwosiglimeccen}[1]{\ifnum#1=31 %
\hatcurRVeccentwosiglimeccenxxxxxA
\else
\ifnum#1=32 %
\hatcurRVeccentwosiglimeccenxxxxxB
\else
\ifnum#1=33 %
\hatcurRVeccentwosiglimeccenxxxxxC
\else
\ifnum#1=34 %
\hatcurRVeccentwosiglimeccenxxxxxD
\else
\ifnum#1=35 %
\hatcurRVeccentwosiglimeccenxxxxxE
\else
??????\fi
\fi
\fi
\fi
\fi
}
\newcommand{\hatcurRVfitrmsAeccen}[1]{\ifnum#1=33 %
\hatcurRVfitrmsAeccenxxxxxC
\else
\ifnum#1=35 %
\hatcurRVfitrmsAeccenxxxxxE
\else
??????\fi
\fi
}
\newcommand{\hatcurRVfitrmsBeccen}[1]{\ifnum#1=33 %
\hatcurRVfitrmsBeccenxxxxxC
\else
\ifnum#1=35 %
\hatcurRVfitrmsBeccenxxxxxE
\else
??????\fi
\fi
}
\newcommand{\hatcurRVfitrmsCeccen}[1]{\ifnum#1=33 %
\hatcurRVfitrmsCeccenxxxxxC
\else
\ifnum#1=35 %
\hatcurRVfitrmsCeccenxxxxxE
\else
??????\fi
\fi
}
\newcommand{\hatcurRVfitrmsDeccen}[1]{\ifnum#1=33 %
\hatcurRVfitrmsDeccenxxxxxC
\else
??????\fi
}
\newcommand{\hatcurRVfitrmseccen}[1]{\ifnum#1=31 %
\hatcurRVfitrmseccenxxxxxA
\else
\ifnum#1=32 %
\hatcurRVfitrmseccenxxxxxB
\else
\ifnum#1=34 %
\hatcurRVfitrmseccenxxxxxD
\else
??????\fi
\fi
\fi
}
\newcommand{\hatcurRVgammaAeccen}[1]{\ifnum#1=33 %
\hatcurRVgammaAeccenxxxxxC
\else
\ifnum#1=35 %
\hatcurRVgammaAeccenxxxxxE
\else
??????\fi
\fi
}
\newcommand{\hatcurRVgammaBeccen}[1]{\ifnum#1=33 %
\hatcurRVgammaBeccenxxxxxC
\else
\ifnum#1=35 %
\hatcurRVgammaBeccenxxxxxE
\else
??????\fi
\fi
}
\newcommand{\hatcurRVgammaCeccen}[1]{\ifnum#1=33 %
\hatcurRVgammaCeccenxxxxxC
\else
\ifnum#1=35 %
\hatcurRVgammaCeccenxxxxxE
\else
??????\fi
\fi
}
\newcommand{\hatcurRVgammaDeccen}[1]{\ifnum#1=33 %
\hatcurRVgammaDeccenxxxxxC
\else
??????\fi
}
\newcommand{\hatcurRVgammaeccen}[1]{\ifnum#1=31 %
\hatcurRVgammaeccenxxxxxA
\else
\ifnum#1=32 %
\hatcurRVgammaeccenxxxxxB
\else
\ifnum#1=34 %
\hatcurRVgammaeccenxxxxxD
\else
??????\fi
\fi
\fi
}
\newcommand{\hatcurRVheccen}[1]{\ifnum#1=31 %
\hatcurRVheccenxxxxxA
\else
\ifnum#1=32 %
\hatcurRVheccenxxxxxB
\else
\ifnum#1=33 %
\hatcurRVheccenxxxxxC
\else
\ifnum#1=34 %
\hatcurRVheccenxxxxxD
\else
\ifnum#1=35 %
\hatcurRVheccenxxxxxE
\else
??????\fi
\fi
\fi
\fi
\fi
}
\newcommand{\hatcurRVjitterAeccen}[1]{\ifnum#1=33 %
\hatcurRVjitterAeccenxxxxxC
\else
\ifnum#1=35 %
\hatcurRVjitterAeccenxxxxxE
\else
??????\fi
\fi
}
\newcommand{\hatcurRVjitterBeccen}[1]{\ifnum#1=33 %
\hatcurRVjitterBeccenxxxxxC
\else
\ifnum#1=35 %
\hatcurRVjitterBeccenxxxxxE
\else
??????\fi
\fi
}
\newcommand{\hatcurRVjitterCeccen}[1]{\ifnum#1=33 %
\hatcurRVjitterCeccenxxxxxC
\else
\ifnum#1=35 %
\hatcurRVjitterCeccenxxxxxE
\else
??????\fi
\fi
}
\newcommand{\hatcurRVjitterDeccen}[1]{\ifnum#1=33 %
\hatcurRVjitterDeccenxxxxxC
\else
??????\fi
}
\newcommand{\hatcurRVjittereccen}[1]{\ifnum#1=31 %
\hatcurRVjittereccenxxxxxA
\else
\ifnum#1=32 %
\hatcurRVjittereccenxxxxxB
\else
\ifnum#1=34 %
\hatcurRVjittereccenxxxxxD
\else
??????\fi
\fi
\fi
}
\newcommand{\hatcurRVjittertwosiglimAeccen}[1]{\ifnum#1=33 %
\hatcurRVjittertwosiglimAeccenxxxxxC
\else
\ifnum#1=35 %
\hatcurRVjittertwosiglimAeccenxxxxxE
\else
??????\fi
\fi
}
\newcommand{\hatcurRVjittertwosiglimBeccen}[1]{\ifnum#1=33 %
\hatcurRVjittertwosiglimBeccenxxxxxC
\else
\ifnum#1=35 %
\hatcurRVjittertwosiglimBeccenxxxxxE
\else
??????\fi
\fi
}
\newcommand{\hatcurRVjittertwosiglimCeccen}[1]{\ifnum#1=33 %
\hatcurRVjittertwosiglimCeccenxxxxxC
\else
\ifnum#1=35 %
\hatcurRVjittertwosiglimCeccenxxxxxE
\else
??????\fi
\fi
}
\newcommand{\hatcurRVjittertwosiglimDeccen}[1]{\ifnum#1=33 %
\hatcurRVjittertwosiglimDeccenxxxxxC
\else
??????\fi
}
\newcommand{\hatcurRVjittertwosiglimeccen}[1]{\ifnum#1=31 %
\hatcurRVjittertwosiglimeccenxxxxxA
\else
\ifnum#1=32 %
\hatcurRVjittertwosiglimeccenxxxxxB
\else
\ifnum#1=34 %
\hatcurRVjittertwosiglimeccenxxxxxD
\else
??????\fi
\fi
\fi
}
\newcommand{\hatcurRVkeccen}[1]{\ifnum#1=31 %
\hatcurRVkeccenxxxxxA
\else
\ifnum#1=32 %
\hatcurRVkeccenxxxxxB
\else
\ifnum#1=33 %
\hatcurRVkeccenxxxxxC
\else
\ifnum#1=34 %
\hatcurRVkeccenxxxxxD
\else
\ifnum#1=35 %
\hatcurRVkeccenxxxxxE
\else
??????\fi
\fi
\fi
\fi
\fi
}
\newcommand{\hatcurRVKeccen}[1]{\ifnum#1=31 %
\hatcurRVKeccenxxxxxA
\else
\ifnum#1=32 %
\hatcurRVKeccenxxxxxB
\else
\ifnum#1=33 %
\hatcurRVKeccenxxxxxC
\else
\ifnum#1=34 %
\hatcurRVKeccenxxxxxD
\else
\ifnum#1=35 %
\hatcurRVKeccenxxxxxE
\else
??????\fi
\fi
\fi
\fi
\fi
}
\newcommand{\hatcurRVomegaeccen}[1]{\ifnum#1=31 %
\hatcurRVomegaeccenxxxxxA
\else
\ifnum#1=32 %
\hatcurRVomegaeccenxxxxxB
\else
\ifnum#1=33 %
\hatcurRVomegaeccenxxxxxC
\else
\ifnum#1=34 %
\hatcurRVomegaeccenxxxxxD
\else
\ifnum#1=35 %
\hatcurRVomegaeccenxxxxxE
\else
??????\fi
\fi
\fi
\fi
\fi
}
\newcommand{\hatcurRVrheccen}[1]{\ifnum#1=31 %
\hatcurRVrheccenxxxxxA
\else
\ifnum#1=32 %
\hatcurRVrheccenxxxxxB
\else
\ifnum#1=33 %
\hatcurRVrheccenxxxxxC
\else
\ifnum#1=34 %
\hatcurRVrheccenxxxxxD
\else
\ifnum#1=35 %
\hatcurRVrheccenxxxxxE
\else
??????\fi
\fi
\fi
\fi
\fi
}
\newcommand{\hatcurRVrkeccen}[1]{\ifnum#1=31 %
\hatcurRVrkeccenxxxxxA
\else
\ifnum#1=32 %
\hatcurRVrkeccenxxxxxB
\else
\ifnum#1=33 %
\hatcurRVrkeccenxxxxxC
\else
\ifnum#1=34 %
\hatcurRVrkeccenxxxxxD
\else
\ifnum#1=35 %
\hatcurRVrkeccenxxxxxE
\else
??????\fi
\fi
\fi
\fi
\fi
}
\newcommand{\hatcurRVtroneeccen}[1]{\ifnum#1=31 %
\hatcurRVtroneeccenxxxxxA
\else
\ifnum#1=32 %
\hatcurRVtroneeccenxxxxxB
\else
\ifnum#1=33 %
\hatcurRVtroneeccenxxxxxC
\else
\ifnum#1=34 %
\hatcurRVtroneeccenxxxxxD
\else
\ifnum#1=35 %
\hatcurRVtroneeccenxxxxxE
\else
??????\fi
\fi
\fi
\fi
\fi
}
\newcommand{\hatcurRVtrtwoeccen}[1]{\ifnum#1=31 %
\hatcurRVtrtwoeccenxxxxxA
\else
\ifnum#1=32 %
\hatcurRVtrtwoeccenxxxxxB
\else
\ifnum#1=33 %
\hatcurRVtrtwoeccenxxxxxC
\else
\ifnum#1=34 %
\hatcurRVtrtwoeccenxxxxxD
\else
\ifnum#1=35 %
\hatcurRVtrtwoeccenxxxxxE
\else
??????\fi
\fi
\fi
\fi
\fi
}
\newcommand{\hatcurSMEiiloggeccen}[1]{\ifnum#1=31 %
\hatcurSMEiiloggeccenxxxxxA
\else
\ifnum#1=32 %
\hatcurSMEiiloggeccenxxxxxB
\else
\ifnum#1=33 %
\hatcurSMEiiloggeccenxxxxxC
\else
\ifnum#1=34 %
\hatcurSMEiiloggeccenxxxxxD
\else
\ifnum#1=35 %
\hatcurSMEiiloggeccenxxxxxE
\else
??????\fi
\fi
\fi
\fi
\fi
}
\newcommand{\hatcurSMEiiteffeccen}[1]{\ifnum#1=31 %
\hatcurSMEiiteffeccenxxxxxA
\else
\ifnum#1=32 %
\hatcurSMEiiteffeccenxxxxxB
\else
\ifnum#1=33 %
\hatcurSMEiiteffeccenxxxxxC
\else
\ifnum#1=34 %
\hatcurSMEiiteffeccenxxxxxD
\else
\ifnum#1=35 %
\hatcurSMEiiteffeccenxxxxxE
\else
??????\fi
\fi
\fi
\fi
\fi
}
\newcommand{\hatcurSMEiivsineccen}[1]{\ifnum#1=31 %
\hatcurSMEiivsineccenxxxxxA
\else
\ifnum#1=32 %
\hatcurSMEiivsineccenxxxxxB
\else
\ifnum#1=33 %
\hatcurSMEiivsineccenxxxxxC
\else
\ifnum#1=34 %
\hatcurSMEiivsineccenxxxxxD
\else
\ifnum#1=35 %
\hatcurSMEiivsineccenxxxxxE
\else
??????\fi
\fi
\fi
\fi
\fi
}
\newcommand{\hatcurSMEiizfeheccen}[1]{\ifnum#1=31 %
\hatcurSMEiizfeheccenxxxxxA
\else
\ifnum#1=32 %
\hatcurSMEiizfeheccenxxxxxB
\else
\ifnum#1=33 %
\hatcurSMEiizfeheccenxxxxxC
\else
\ifnum#1=34 %
\hatcurSMEiizfeheccenxxxxxD
\else
\ifnum#1=35 %
\hatcurSMEiizfeheccenxxxxxE
\else
??????\fi
\fi
\fi
\fi
\fi
}
\newcommand{\hatcurSMEiizfehshorteccen}[1]{\ifnum#1=31 %
\hatcurSMEiizfehshorteccenxxxxxA
\else
\ifnum#1=32 %
\hatcurSMEiizfehshorteccenxxxxxB
\else
\ifnum#1=33 %
\hatcurSMEiizfehshorteccenxxxxxC
\else
\ifnum#1=34 %
\hatcurSMEiizfehshorteccenxxxxxD
\else
\ifnum#1=35 %
\hatcurSMEiizfehshorteccenxxxxxE
\else
??????\fi
\fi
\fi
\fi
\fi
}
\newcommand{\hatcurSMEiloggeccen}[1]{\ifnum#1=31 %
\hatcurSMEiloggeccenxxxxxA
\else
\ifnum#1=32 %
\hatcurSMEiloggeccenxxxxxB
\else
\ifnum#1=33 %
\hatcurSMEiloggeccenxxxxxC
\else
\ifnum#1=34 %
\hatcurSMEiloggeccenxxxxxD
\else
\ifnum#1=35 %
\hatcurSMEiloggeccenxxxxxE
\else
??????\fi
\fi
\fi
\fi
\fi
}
\newcommand{\hatcurSMEiteffeccen}[1]{\ifnum#1=31 %
\hatcurSMEiteffeccenxxxxxA
\else
\ifnum#1=32 %
\hatcurSMEiteffeccenxxxxxB
\else
\ifnum#1=33 %
\hatcurSMEiteffeccenxxxxxC
\else
\ifnum#1=34 %
\hatcurSMEiteffeccenxxxxxD
\else
\ifnum#1=35 %
\hatcurSMEiteffeccenxxxxxE
\else
??????\fi
\fi
\fi
\fi
\fi
}
\newcommand{\hatcurSMEivmaceccen}[1]{\ifnum#1=31 %
\hatcurSMEivmaceccenxxxxxA
\else
\ifnum#1=32 %
\hatcurSMEivmaceccenxxxxxB
\else
\ifnum#1=33 %
\hatcurSMEivmaceccenxxxxxC
\else
\ifnum#1=34 %
\hatcurSMEivmaceccenxxxxxD
\else
\ifnum#1=35 %
\hatcurSMEivmaceccenxxxxxE
\else
??????\fi
\fi
\fi
\fi
\fi
}
\newcommand{\hatcurSMEivmiceccen}[1]{\ifnum#1=31 %
\hatcurSMEivmiceccenxxxxxA
\else
\ifnum#1=32 %
\hatcurSMEivmiceccenxxxxxB
\else
\ifnum#1=33 %
\hatcurSMEivmiceccenxxxxxC
\else
\ifnum#1=34 %
\hatcurSMEivmiceccenxxxxxD
\else
\ifnum#1=35 %
\hatcurSMEivmiceccenxxxxxE
\else
??????\fi
\fi
\fi
\fi
\fi
}
\newcommand{\hatcurSMEivsineccen}[1]{\ifnum#1=31 %
\hatcurSMEivsineccenxxxxxA
\else
\ifnum#1=32 %
\hatcurSMEivsineccenxxxxxB
\else
\ifnum#1=33 %
\hatcurSMEivsineccenxxxxxC
\else
\ifnum#1=34 %
\hatcurSMEivsineccenxxxxxD
\else
\ifnum#1=35 %
\hatcurSMEivsineccenxxxxxE
\else
??????\fi
\fi
\fi
\fi
\fi
}
\newcommand{\hatcurSMEizfeheccen}[1]{\ifnum#1=31 %
\hatcurSMEizfeheccenxxxxxA
\else
\ifnum#1=32 %
\hatcurSMEizfeheccenxxxxxB
\else
\ifnum#1=33 %
\hatcurSMEizfeheccenxxxxxC
\else
\ifnum#1=34 %
\hatcurSMEizfeheccenxxxxxD
\else
\ifnum#1=35 %
\hatcurSMEizfeheccenxxxxxE
\else
??????\fi
\fi
\fi
\fi
\fi
}
\newcommand{\hatcurSMEizfehshorteccen}[1]{\ifnum#1=31 %
\hatcurSMEizfehshorteccenxxxxxA
\else
\ifnum#1=32 %
\hatcurSMEizfehshorteccenxxxxxB
\else
\ifnum#1=33 %
\hatcurSMEizfehshorteccenxxxxxC
\else
\ifnum#1=34 %
\hatcurSMEizfehshorteccenxxxxxD
\else
\ifnum#1=35 %
\hatcurSMEizfehshorteccenxxxxxE
\else
??????\fi
\fi
\fi
\fi
\fi
}
\newcommand{\hatcurXAveccen}[1]{\ifnum#1=31 %
\hatcurXAveccenxxxxxA
\else
\ifnum#1=32 %
\hatcurXAveccenxxxxxB
\else
\ifnum#1=33 %
\hatcurXAveccenxxxxxC
\else
\ifnum#1=34 %
\hatcurXAveccenxxxxxD
\else
\ifnum#1=35 %
\hatcurXAveccenxxxxxE
\else
??????\fi
\fi
\fi
\fi
\fi
}
\newcommand{\hatcurXdisteccen}[1]{\ifnum#1=31 %
\hatcurXdisteccenxxxxxA
\else
\ifnum#1=32 %
\hatcurXdisteccenxxxxxB
\else
\ifnum#1=33 %
\hatcurXdisteccenxxxxxC
\else
\ifnum#1=34 %
\hatcurXdisteccenxxxxxD
\else
\ifnum#1=35 %
\hatcurXdisteccenxxxxxE
\else
??????\fi
\fi
\fi
\fi
\fi
}
\newcommand{\hatcurXdistredeccen}[1]{\ifnum#1=31 %
\hatcurXdistredeccenxxxxxA
\else
\ifnum#1=32 %
\hatcurXdistredeccenxxxxxB
\else
\ifnum#1=33 %
\hatcurXdistredeccenxxxxxC
\else
\ifnum#1=34 %
\hatcurXdistredeccenxxxxxD
\else
\ifnum#1=35 %
\hatcurXdistredeccenxxxxxE
\else
??????\fi
\fi
\fi
\fi
\fi
}
\newcommand{\hatcurXEBVeccen}[1]{\ifnum#1=31 %
\hatcurXEBVeccenxxxxxA
\else
\ifnum#1=32 %
\hatcurXEBVeccenxxxxxB
\else
\ifnum#1=33 %
\hatcurXEBVeccenxxxxxC
\else
\ifnum#1=34 %
\hatcurXEBVeccenxxxxxD
\else
\ifnum#1=35 %
\hatcurXEBVeccenxxxxxE
\else
??????\fi
\fi
\fi
\fi
\fi
}
\newcommand{\hatcurXjhisoredeccen}[1]{\ifnum#1=31 %
\hatcurXjhisoredeccenxxxxxA
\else
\ifnum#1=32 %
\hatcurXjhisoredeccenxxxxxB
\else
\ifnum#1=33 %
\hatcurXjhisoredeccenxxxxxC
\else
\ifnum#1=34 %
\hatcurXjhisoredeccenxxxxxD
\else
\ifnum#1=35 %
\hatcurXjhisoredeccenxxxxxE
\else
??????\fi
\fi
\fi
\fi
\fi
}
\newcommand{\hatcurXjkisoredeccen}[1]{\ifnum#1=31 %
\hatcurXjkisoredeccenxxxxxA
\else
\ifnum#1=32 %
\hatcurXjkisoredeccenxxxxxB
\else
\ifnum#1=33 %
\hatcurXjkisoredeccenxxxxxC
\else
\ifnum#1=34 %
\hatcurXjkisoredeccenxxxxxD
\else
\ifnum#1=35 %
\hatcurXjkisoredeccenxxxxxE
\else
??????\fi
\fi
\fi
\fi
\fi
}
\newcommand{\hatcurXmhisoredeccen}[1]{\ifnum#1=31 %
\hatcurXmhisoredeccenxxxxxA
\else
\ifnum#1=32 %
\hatcurXmhisoredeccenxxxxxB
\else
\ifnum#1=33 %
\hatcurXmhisoredeccenxxxxxC
\else
\ifnum#1=34 %
\hatcurXmhisoredeccenxxxxxD
\else
\ifnum#1=35 %
\hatcurXmhisoredeccenxxxxxE
\else
??????\fi
\fi
\fi
\fi
\fi
}
\newcommand{\hatcurXmiisoredeccen}[1]{\ifnum#1=31 %
\hatcurXmiisoredeccenxxxxxA
\else
\ifnum#1=32 %
\hatcurXmiisoredeccenxxxxxB
\else
\ifnum#1=33 %
\hatcurXmiisoredeccenxxxxxC
\else
\ifnum#1=34 %
\hatcurXmiisoredeccenxxxxxD
\else
\ifnum#1=35 %
\hatcurXmiisoredeccenxxxxxE
\else
??????\fi
\fi
\fi
\fi
\fi
}
\newcommand{\hatcurXmjisoredeccen}[1]{\ifnum#1=31 %
\hatcurXmjisoredeccenxxxxxA
\else
\ifnum#1=32 %
\hatcurXmjisoredeccenxxxxxB
\else
\ifnum#1=33 %
\hatcurXmjisoredeccenxxxxxC
\else
\ifnum#1=34 %
\hatcurXmjisoredeccenxxxxxD
\else
\ifnum#1=35 %
\hatcurXmjisoredeccenxxxxxE
\else
??????\fi
\fi
\fi
\fi
\fi
}
\newcommand{\hatcurXmkisoredeccen}[1]{\ifnum#1=31 %
\hatcurXmkisoredeccenxxxxxA
\else
\ifnum#1=32 %
\hatcurXmkisoredeccenxxxxxB
\else
\ifnum#1=33 %
\hatcurXmkisoredeccenxxxxxC
\else
\ifnum#1=34 %
\hatcurXmkisoredeccenxxxxxD
\else
\ifnum#1=35 %
\hatcurXmkisoredeccenxxxxxE
\else
??????\fi
\fi
\fi
\fi
\fi
}
\newcommand{\hatcurXmvisoredeccen}[1]{\ifnum#1=31 %
\hatcurXmvisoredeccenxxxxxA
\else
\ifnum#1=32 %
\hatcurXmvisoredeccenxxxxxB
\else
\ifnum#1=33 %
\hatcurXmvisoredeccenxxxxxC
\else
\ifnum#1=34 %
\hatcurXmvisoredeccenxxxxxD
\else
\ifnum#1=35 %
\hatcurXmvisoredeccenxxxxxE
\else
??????\fi
\fi
\fi
\fi
\fi
}
\newcommand{\hatcurXsecdureccen}[1]{\ifnum#1=31 %
\hatcurXsecdureccenxxxxxA
\else
\ifnum#1=32 %
\hatcurXsecdureccenxxxxxB
\else
\ifnum#1=33 %
\hatcurXsecdureccenxxxxxC
\else
\ifnum#1=34 %
\hatcurXsecdureccenxxxxxD
\else
\ifnum#1=35 %
\hatcurXsecdureccenxxxxxE
\else
??????\fi
\fi
\fi
\fi
\fi
}
\newcommand{\hatcurXsecingdureccen}[1]{\ifnum#1=31 %
\hatcurXsecingdureccenxxxxxA
\else
\ifnum#1=32 %
\hatcurXsecingdureccenxxxxxB
\else
\ifnum#1=33 %
\hatcurXsecingdureccenxxxxxC
\else
\ifnum#1=34 %
\hatcurXsecingdureccenxxxxxD
\else
\ifnum#1=35 %
\hatcurXsecingdureccenxxxxxE
\else
??????\fi
\fi
\fi
\fi
\fi
}
\newcommand{\hatcurXsecondaryeccen}[1]{\ifnum#1=31 %
\hatcurXsecondaryeccenxxxxxA
\else
\ifnum#1=32 %
\hatcurXsecondaryeccenxxxxxB
\else
\ifnum#1=33 %
\hatcurXsecondaryeccenxxxxxC
\else
\ifnum#1=34 %
\hatcurXsecondaryeccenxxxxxD
\else
\ifnum#1=35 %
\hatcurXsecondaryeccenxxxxxE
\else
??????\fi
\fi
\fi
\fi
\fi
}
\newcommand{\hatcurXsecphaseeccen}[1]{\ifnum#1=31 %
\hatcurXsecphaseeccenxxxxxA
\else
\ifnum#1=32 %
\hatcurXsecphaseeccenxxxxxB
\else
\ifnum#1=33 %
\hatcurXsecphaseeccenxxxxxC
\else
\ifnum#1=34 %
\hatcurXsecphaseeccenxxxxxD
\else
\ifnum#1=35 %
\hatcurXsecphaseeccenxxxxxE
\else
??????\fi
\fi
\fi
\fi
\fi
}
\newcommand{\hatcurXviisoredeccen}[1]{\ifnum#1=31 %
\hatcurXviisoredeccenxxxxxA
\else
\ifnum#1=32 %
\hatcurXviisoredeccenxxxxxB
\else
\ifnum#1=33 %
\hatcurXviisoredeccenxxxxxC
\else
\ifnum#1=34 %
\hatcurXviisoredeccenxxxxxD
\else
\ifnum#1=35 %
\hatcurXviisoredeccenxxxxxE
\else
??????\fi
\fi
\fi
\fi
\fi
}
\newcommand{\hatcurXvkisoredeccen}[1]{\ifnum#1=31 %
\hatcurXvkisoredeccenxxxxxA
\else
\ifnum#1=32 %
\hatcurXvkisoredeccenxxxxxB
\else
\ifnum#1=33 %
\hatcurXvkisoredeccenxxxxxC
\else
\ifnum#1=34 %
\hatcurXvkisoredeccenxxxxxD
\else
\ifnum#1=35 %
\hatcurXvkisoredeccenxxxxxE
\else
??????\fi
\fi
\fi
\fi
\fi
}

%% file: HATS566-004.eccen.tex
\newcommand{\hatcurhtreccenxxxxxA}{HATS566-004}                        
\newcommand{\hatcurfieldeccenxxxxxA}{\ensuremath{string}}              
\newcommand{\hatcurCCraeccenxxxxxA}{\ensuremath{12^{\mathrm h}46^{\mathrm m}48.72{\mathrm s}}}                       
\newcommand{\hatcurCCdececcenxxxxxA}{\ensuremath{-24{\arcdeg}25{\arcmin}38.5{\arcsec}}}                      
\newcommand{\hatcurCCmageccenxxxxxA}{13.105}                           
\newcommand{\hatcurCCtwomasseccenxxxxxA}{2MASS~12464866-2425385}       
\newcommand{\hatcurCCgsceccenxxxxxA}{GSC~6688-00298}                   
\newcommand{\hatcurCCtassmveccenxxxxxA}{\ensuremath{13.105\pm0.030}}   
\newcommand{\hatcurCCtassmvshorteccenxxxxxA}{\ensuremath{13.1}}        
\newcommand{\hatcurCCtassmBeccenxxxxxA}{\ensuremath{13.687\pm0.030}}   
\newcommand{\hatcurCCtassmBshorteccenxxxxxA}{\ensuremath{13.7}}        
\newcommand{\hatcurCCtassmIeccenxxxxxA}{\ensuremath{100\pm1000}}       
\newcommand{\hatcurCCtassmIshorteccenxxxxxA}{\ensuremath{100.0}}       
\newcommand{\hatcurCCtassmgeccenxxxxxA}{\ensuremath{13.356\pm0.030}}   
\newcommand{\hatcurCCtassmgshorteccenxxxxxA}{\ensuremath{13.4}}        
\newcommand{\hatcurCCtassmreccenxxxxxA}{\ensuremath{12.950\pm0.030}}   
\newcommand{\hatcurCCtassmrshorteccenxxxxxA}{\ensuremath{12.9}}        
\newcommand{\hatcurCCtassmieccenxxxxxA}{\ensuremath{12.775\pm0.080}}   
\newcommand{\hatcurCCtassmishorteccenxxxxxA}{\ensuremath{12.8}}        
\newcommand{\hatcurCCtwomassJmageccenxxxxxA}{\ensuremath{11.912\pm0.021}} 
\newcommand{\hatcurCCtwomassHmageccenxxxxxA}{\ensuremath{11.618\pm0.025}} 
\newcommand{\hatcurCCtwomassKmageccenxxxxxA}{\ensuremath{11.572\pm0.023}} 
\newcommand{\hatcurCCcitJmageccenxxxxxA}{\ensuremath{11.931\pm0.021}}  
\newcommand{\hatcurCCcitHmageccenxxxxxA}{\ensuremath{11.613\pm0.025}}  
\newcommand{\hatcurCCcitKmageccenxxxxxA}{\ensuremath{11.596\pm0.023}}  
\newcommand{\hatcurCCbbJmageccenxxxxxA}{\ensuremath{11.977\pm0.023}}   
\newcommand{\hatcurCCbbHmageccenxxxxxA}{\ensuremath{11.634\pm0.026}}   
\newcommand{\hatcurCCbbKmageccenxxxxxA}{\ensuremath{11.616\pm0.023}}   
\newcommand{\hatcurCCesoJmageccenxxxxxA}{\ensuremath{11.979\pm0.024}}  
\newcommand{\hatcurCCesoHmageccenxxxxxA}{\ensuremath{11.628\pm0.029}}  
\newcommand{\hatcurCCesoKmageccenxxxxxA}{\ensuremath{11.615\pm0.024}}  
\newcommand{\hatcurCCesoJHmageccenxxxxxA}{\ensuremath{0.351\pm0.035}}  
\newcommand{\hatcurCCesoJKmageccenxxxxxA}{\ensuremath{0.365\pm0.034}}  
\newcommand{\hatcurCCesoHKmageccenxxxxxA}{\ensuremath{0.014\pm0.038}}  
\newcommand{\hatcurLCdipeccenxxxxxA}{\ensuremath{6.1}}                 
\newcommand{\hatcurLCrprstareccenxxxxxA}{\ensuremath{0.0915\pm0.0050}} 
\newcommand{\hatcurLCbsqeccenxxxxxA}{\ensuremath{0.24_{-0.13}^{+0.13}}} 
\newcommand{\hatcurLCimpeccenxxxxxA}{\ensuremath{0.49_{-0.16}^{+0.12}}} 
\newcommand{\hatcurLCzetaeccenxxxxxA}{\ensuremath{11.68\pm0.16}}       
\newcommand{\hatcurLCdureccenxxxxxA}{\ensuremath{0.1915\pm0.0045}}     
\newcommand{\hatcurLCdurshorteccenxxxxxA}{\ensuremath{0.1915}}         
\newcommand{\hatcurLCdurhreccenxxxxxA}{\ensuremath{4.59\pm0.11}}       
\newcommand{\hatcurLCdurhrshorteccenxxxxxA}{\ensuremath{4.595}}        
\newcommand{\hatcurLCqeccenxxxxxA}{\ensuremath{0.0567\pm0.0013}}       
\newcommand{\hatcurLCqshorteccenxxxxxA}{\ensuremath{0.057}}            
\newcommand{\hatcurLCingdureccenxxxxxA}{\ensuremath{0.0206\pm0.0042}}  
\newcommand{\hatcurLCPeccenxxxxxA}{\ensuremath{3.377960\pm0.000012}}   
\newcommand{\hatcurLCPprececcenxxxxxA}{\ensuremath{3.3779599}}         
\newcommand{\hatcurLCPshorteccenxxxxxA}{\ensuremath{3.3780}}           
\newcommand{\hatcurLCTeccenxxxxxA}{\ensuremath{2456922.9901\pm0.0013}} 
\newcommand{\hatcurLCTAeccenxxxxxA}{\ensuremath{2456281.1776\pm0.0028}} 
\newcommand{\hatcurLCTBeccenxxxxxA}{\ensuremath{2457115.5338\pm0.0014}} 
\newcommand{\hatcurLChatnetmeccenxxxxxA}{\ensuremath{12.943640\pm0.000064}} 
\newcommand{\hatcurLCiblendeccenxxxxxA}{\ensuremath{0.651\pm0.072}}    
\newcommand{\hatcurLCrhoeccenxxxxxA}{\ensuremath{0.27\pm0.11}}         
\newcommand{\hatcurSMEiteffeccenxxxxxA}{\ensuremath{6060\pm120}}       
\newcommand{\hatcurSMEizfeheccenxxxxxA}{\ensuremath{0.000\pm0.070}}    
\newcommand{\hatcurSMEizfehshorteccenxxxxxA}{\ensuremath{0.00}}        
\newcommand{\hatcurSMEiloggeccenxxxxxA}{\ensuremath{4.05\pm0.26}}      
\newcommand{\hatcurSMEivsineccenxxxxxA}{\ensuremath{7.00\pm0.50}}      
\newcommand{\hatcurSMEivmaceccenxxxxxA}{\ensuremath{0.0}}              
\newcommand{\hatcurSMEivmiceccenxxxxxA}{\ensuremath{0.0}}              
\newcommand{\hatcurSMEiiteffeccenxxxxxA}{\ensuremath{6050\pm120}}      
\newcommand{\hatcurSMEiizfeheccenxxxxxA}{\ensuremath{0.000\pm0.070}}   
\newcommand{\hatcurSMEiizfehshorteccenxxxxxA}{\ensuremath{0.0}}        
\newcommand{\hatcurSMEiiloggeccenxxxxxA}{\ensuremath{4.012\pm0.060}}   
\newcommand{\hatcurSMEiivsineccenxxxxxA}{\ensuremath{7.01\pm0.50}}     
\newcommand{\hatcurLBizeccenxxxxxA}{\ensuremath{0.1658}}               
\newcommand{\hatcurLBiizeccenxxxxxA}{\ensuremath{0.3498}}              
\newcommand{\hatcurLBiieccenxxxxxA}{\ensuremath{0.2195}}               
\newcommand{\hatcurLBiiieccenxxxxxA}{\ensuremath{0.3558}}              
\newcommand{\hatcurLBiIeccenxxxxxA}{\ensuremath{0.2003}}               
\newcommand{\hatcurLBiiIeccenxxxxxA}{\ensuremath{0.3553}}              
\newcommand{\hatcurLBigeccenxxxxxA}{\ensuremath{0.4810}}               
\newcommand{\hatcurLBiigeccenxxxxxA}{\ensuremath{0.2874}}              
\newcommand{\hatcurLBireccenxxxxxA}{\ensuremath{0.3000}}               
\newcommand{\hatcurLBiireccenxxxxxA}{\ensuremath{0.3587}}              
\newcommand{\hatcurLBiReccenxxxxxA}{\ensuremath{0.2774}}               
\newcommand{\hatcurLBiiReccenxxxxxA}{\ensuremath{0.3592}}              
\newcommand{\hatcurLBikepeccenxxxxxA}{\ensuremath{0.1000}}             
\newcommand{\hatcurLBiikepeccenxxxxxA}{\ensuremath{0.1000}}            
\newcommand{\hatcurISOmeccenxxxxxA}{\ensuremath{1.28\pm0.11}}          
\newcommand{\hatcurISOmshorteccenxxxxxA}{\ensuremath{1.28}}            
\newcommand{\hatcurISOmlongeccenxxxxxA}{\ensuremath{1.28\pm0.11}}      
\newcommand{\hatcurISOreccenxxxxxA}{\ensuremath{1.87_{-0.22}^{+0.31}}} 
\newcommand{\hatcurISOrshorteccenxxxxxA}{\ensuremath{1.87}}            
\newcommand{\hatcurISOrlongeccenxxxxxA}{\ensuremath{1.87_{-0.22}^{+0.31}}} 
\newcommand{\hatcurISOrhoeccenxxxxxA}{\ensuremath{0.28\pm0.11}}        
\newcommand{\hatcurISOrholongeccenxxxxxA}{\ensuremath{0.28\pm0.11}}    
\newcommand{\hatcurISOloggeccenxxxxxA}{\ensuremath{4.00\pm0.10}}       
\newcommand{\hatcurISOlumeccenxxxxxA}{\ensuremath{4.2_{-1.0}^{+1.6}}}  
\newcommand{\hatcurISOlumshorteccenxxxxxA}{\ensuremath{4.20}}          
\newcommand{\hatcurISOmveccenxxxxxA}{\ensuremath{3.23\pm0.34}}         
\newcommand{\hatcurISOvieccenxxxxxA}{\ensuremath{0.615\pm0.033}}       
\newcommand{\hatcurISOageeccenxxxxxA}{\ensuremath{3.98_{-0.87}^{+1.39}}} 
\newcommand{\hatcurISOsigmaeccenxxxxxA}{\ensuremath{0.00030\pm0.00013}} 
\newcommand{\hatcurISOMJeccenxxxxxA}{\ensuremath{2.22\pm0.32}}         
\newcommand{\hatcurISOMHeccenxxxxxA}{\ensuremath{1.92\pm0.32}}         
\newcommand{\hatcurISOMKeccenxxxxxA}{\ensuremath{1.87\pm0.32}}         
\newcommand{\hatcurISOJKeccenxxxxxA}{\ensuremath{0.350\pm0.020}}       
\newcommand{\hatcurISOspececcenxxxxxA}{F}                              
\newcommand{\hatcurRVKeccenxxxxxA}{\ensuremath{103\pm14}}              
\newcommand{\hatcurRVrkeccenxxxxxA}{\ensuremath{-0.09_{-0.14}^{+0.22}}} 
\newcommand{\hatcurRVrheccenxxxxxA}{\ensuremath{0.01\pm0.24}}          
\newcommand{\hatcurRVkeccenxxxxxA}{\ensuremath{-0.020\pm0.063}}        
\newcommand{\hatcurRVheccenxxxxxA}{\ensuremath{0.001\pm0.097}}         
\newcommand{\hatcurRVtroneeccenxxxxxA}{\ensuremath{0\pm0}}             
\newcommand{\hatcurRVtrtwoeccenxxxxxA}{\ensuremath{0\pm0}}             
\newcommand{\hatcurRVgammaeccenxxxxxA}{\ensuremath{-8704\pm10}}        
\newcommand{\hatcurRVjittereccenxxxxxA}{\ensuremath{0.0\pm5.0}}        
\newcommand{\hatcurRVjittertwosiglimeccenxxxxxA}{\ensuremath{<11.1}}   
\newcommand{\hatcurRVfitrmseccenxxxxxA}{\ensuremath{.1fym}}            %
\newcommand{\hatcurRVecceneccenxxxxxA}{\ensuremath{0.072\pm0.075}}     
\newcommand{\hatcurRVeccentwosiglimeccenxxxxxA}{\ensuremath{<0.233}}   
\newcommand{\hatcurRVomegaeccenxxxxxA}{\ensuremath{177\pm89}}          
\newcommand{\hatcurPPieccenxxxxxA}{\ensuremath{85.0\pm2.2}}            
\newcommand{\hatcurPPgeccenxxxxxA}{\ensuremath{8.0\pm3.3}}             
\newcommand{\hatcurPPloggeccenxxxxxA}{\ensuremath{2.90\pm0.16}}        
\newcommand{\hatcurPPareccenxxxxxA}{\ensuremath{5.50\pm0.69}}          
\newcommand{\hatcurPPareleccenxxxxxA}{\ensuremath{0.0479\pm0.0013}}    
\newcommand{\hatcurPPrhoeccenxxxxxA}{\ensuremath{0.24_{-0.11}^{+0.15}}} 
\newcommand{\hatcurPPmeccenxxxxxA}{\ensuremath{0.91\pm0.12}}           
\newcommand{\hatcurPPmshorteccenxxxxxA}{\ensuremath{0.91}}             
\newcommand{\hatcurPPmlongeccenxxxxxA}{\ensuremath{0.91\pm0.12}}       
\newcommand{\hatcurPPmeeccenxxxxxA}{\ensuremath{288\pm39}}             
\newcommand{\hatcurPPmeshorteccenxxxxxA}{\ensuremath{288.0}}           
\newcommand{\hatcurPPmelongeccenxxxxxA}{\ensuremath{288\pm39}}         
\newcommand{\hatcurPPreccenxxxxxA}{\ensuremath{1.67_{-0.25}^{+0.35}}}  
\newcommand{\hatcurPPrshorteccenxxxxxA}{\ensuremath{1.67}}             
\newcommand{\hatcurPPrlongeccenxxxxxA}{\ensuremath{1.67_{-0.25}^{+0.35}}} 
\newcommand{\hatcurPPreeccenxxxxxA}{\ensuremath{18.7_{-2.8}^{+3.9}}}   
\newcommand{\hatcurPPreshorteccenxxxxxA}{\ensuremath{18.7}}            
\newcommand{\hatcurPPrelongeccenxxxxxA}{\ensuremath{18.7_{-2.8}^{+3.9}}} 
\newcommand{\hatcurPPmrcorreccenxxxxxA}{\ensuremath{0.11}}             
\newcommand{\hatcurPPteffeccenxxxxxA}{\ensuremath{1830\pm120}}         
\newcommand{\hatcurPPthetaeccenxxxxxA}{\ensuremath{0.040\pm0.010}}     
\newcommand{\hatcurPPfluxperieccenxxxxxA}{\ensuremath{2.92_{-0.56}^{+1.07}}} 
\newcommand{\hatcurPPfluxperidimeccenxxxxxA}{\ensuremath{9}}           
\newcommand{\hatcurPPfluxapeccenxxxxxA}{\ensuremath{2.22\pm0.60}}      
\newcommand{\hatcurPPfluxapdimeccenxxxxxA}{\ensuremath{9}}             
\newcommand{\hatcurPPfluxavgeccenxxxxxA}{\ensuremath{2.51_{-0.52}^{+0.81}}} 
\newcommand{\hatcurPPfluxavgdimeccenxxxxxA}{\ensuremath{9}}            
\newcommand{\hatcurPPfluxavglogeccenxxxxxA}{\ensuremath{9.40\pm0.11}}  
\newcommand{\hatcurXsecphaseeccenxxxxxA}{\ensuremath{0.487\pm0.040}}   
\newcommand{\hatcurXsecondaryeccenxxxxxA}{\ensuremath{2456924.64\pm0.14}} 
\newcommand{\hatcurXsecdureccenxxxxxA}{\ensuremath{0.192\pm0.026}}     
\newcommand{\hatcurXsecingdureccenxxxxxA}{\ensuremath{0.020\pm0.013}}  
\newcommand{\hatcurPPphiconjeccenxxxxxA}{\ensuremath{-0.10\pm0.26}}    
\newcommand{\hatcurPPperieccenxxxxxA}{\ensuremath{2456923.32\pm0.87}}  
\newcommand{\hatcurPPaequiveccenxxxxxA}{\ensuremath{0.0233\pm0.0031}}  
\newcommand{\hatcurPPtcirceccenxxxxxA}{\ensuremath{47_{-29}^{+48}}}    
\newcommand{\hatcurPPtinfalleccenxxxxxA}{\ensuremath{230_{-110}^{+160}}} 
\newcommand{\hatcurXdisteccenxxxxxA}{\ensuremath{890_{-110}^{+150}}}   
\newcommand{\hatcurXAveccenxxxxxA}{\ensuremath{0.155\pm0.092}}         
\newcommand{\hatcurXdistredeccenxxxxxA}{\ensuremath{870_{-100}^{+150}}} 
\newcommand{\hatcurXEBVeccenxxxxxA}{\ensuremath{0.050\pm0.030}}        
\newcommand{\hatcurXmvisoredeccenxxxxxA}{\ensuremath{13.105\pm0.029}}  
\newcommand{\hatcurXmiisoredeccenxxxxxA}{\ensuremath{12.407\pm0.019}}  
\newcommand{\hatcurXmjisoredeccenxxxxxA}{\ensuremath{11.972\pm0.015}}  
\newcommand{\hatcurXmhisoredeccenxxxxxA}{\ensuremath{11.662\pm0.016}}  
\newcommand{\hatcurXmkisoredeccenxxxxxA}{\ensuremath{11.598\pm0.016}}  
\newcommand{\hatcurXviisoredeccenxxxxxA}{\ensuremath{0.698\pm0.024}}   
\newcommand{\hatcurXvkisoredeccenxxxxxA}{\ensuremath{1.506\pm0.036}}   
\newcommand{\hatcurXjhisoredeccenxxxxxA}{\ensuremath{0.310\pm0.012}}   
\newcommand{\hatcurXjkisoredeccenxxxxxA}{\ensuremath{0.374\pm0.010}}   
\newcommand{\hatcurCCpmraeccenxxxxxA}{\ensuremath{1.2\pm1.0}}          
\newcommand{\hatcurCCpmdececcenxxxxxA}{\ensuremath{-0.4\pm1.1}}        
\newcommand{\hatcurCCpmeccenxxxxxA}{\ensuremath{1.3\pm1.5}}            

%% file: HATS586-004.eccen.tex
\newcommand{\hatcurhtreccenxxxxxB}{HATS586-004}                        
\newcommand{\hatcurfieldeccenxxxxxB}{\ensuremath{string}}              
\newcommand{\hatcurCCraeccenxxxxxB}{\ensuremath{23^{\mathrm h}04^{\mathrm m}18.12{\mathrm s}}}                       
\newcommand{\hatcurCCdececcenxxxxxB}{\ensuremath{-21{\arcdeg}16{\arcmin}19.0{\arcsec}}}                      
\newcommand{\hatcurCCmageccenxxxxxB}{14.384}                           
\newcommand{\hatcurCCtwomasseccenxxxxxB}{2MASS~23041801-2116189}       
\newcommand{\hatcurCCgsceccenxxxxxB}{GSC~6400-00924}                   
\newcommand{\hatcurCCtassmveccenxxxxxB}{\ensuremath{14.384\pm0.010}}   
\newcommand{\hatcurCCtassmvshorteccenxxxxxB}{\ensuremath{14.4}}        
\newcommand{\hatcurCCtassmBeccenxxxxxB}{\ensuremath{15.106\pm0.030}}   
\newcommand{\hatcurCCtassmBshorteccenxxxxxB}{\ensuremath{15.1}}        
\newcommand{\hatcurCCtassmIeccenxxxxxB}{\ensuremath{100\pm1000}}       
\newcommand{\hatcurCCtassmIshorteccenxxxxxB}{\ensuremath{100.0}}       
\newcommand{\hatcurCCtassmgeccenxxxxxB}{\ensuremath{14.694\pm0.010}}   
\newcommand{\hatcurCCtassmgshorteccenxxxxxB}{\ensuremath{14.7}}        
\newcommand{\hatcurCCtassmreccenxxxxxB}{\ensuremath{14.130\pm0.010}}   
\newcommand{\hatcurCCtassmrshorteccenxxxxxB}{\ensuremath{14.1}}        
\newcommand{\hatcurCCtassmieccenxxxxxB}{\ensuremath{14.035\pm0.010}}   
\newcommand{\hatcurCCtassmishorteccenxxxxxB}{\ensuremath{14.0}}        
\newcommand{\hatcurCCtwomassJmageccenxxxxxB}{\ensuremath{13.090\pm0.022}} 
\newcommand{\hatcurCCtwomassHmageccenxxxxxB}{\ensuremath{12.768\pm0.027}} 
\newcommand{\hatcurCCtwomassKmageccenxxxxxB}{\ensuremath{12.699\pm0.033}} 
\newcommand{\hatcurCCcitJmageccenxxxxxB}{\ensuremath{13.106\pm0.022}}  
\newcommand{\hatcurCCcitHmageccenxxxxxB}{\ensuremath{12.763\pm0.027}}  
\newcommand{\hatcurCCcitKmageccenxxxxxB}{\ensuremath{12.723\pm0.033}}  
\newcommand{\hatcurCCbbJmageccenxxxxxB}{\ensuremath{13.157\pm0.024}}   
\newcommand{\hatcurCCbbHmageccenxxxxxB}{\ensuremath{12.784\pm0.028}}   
\newcommand{\hatcurCCbbKmageccenxxxxxB}{\ensuremath{12.743\pm0.033}}   
\newcommand{\hatcurCCesoJmageccenxxxxxB}{\ensuremath{13.159\pm0.026}}  
\newcommand{\hatcurCCesoHmageccenxxxxxB}{\ensuremath{12.779\pm0.031}}  
\newcommand{\hatcurCCesoKmageccenxxxxxB}{\ensuremath{12.742\pm0.034}}  
\newcommand{\hatcurCCesoJHmageccenxxxxxB}{\ensuremath{0.380\pm0.038}}  
\newcommand{\hatcurCCesoJKmageccenxxxxxB}{\ensuremath{0.418\pm0.042}}  
\newcommand{\hatcurCCesoHKmageccenxxxxxB}{\ensuremath{0.038\pm0.047}}  
\newcommand{\hatcurLCdipeccenxxxxxB}{\ensuremath{15.3}}                
\newcommand{\hatcurLCrprstareccenxxxxxB}{\ensuremath{0.1173\pm0.0039}} 
\newcommand{\hatcurLCbsqeccenxxxxxB}{\ensuremath{0.17_{-0.11}^{+0.12}}} 
\newcommand{\hatcurLCimpeccenxxxxxB}{\ensuremath{0.41_{-0.17}^{+0.13}}} 
\newcommand{\hatcurLCzetaeccenxxxxxB}{\ensuremath{19.28\pm0.56}}       
\newcommand{\hatcurLCdureccenxxxxxB}{\ensuremath{0.1186\pm0.0041}}     
\newcommand{\hatcurLCdurshorteccenxxxxxB}{\ensuremath{0.1186}}         
\newcommand{\hatcurLCdurhreccenxxxxxB}{\ensuremath{2.846\pm0.098}}     
\newcommand{\hatcurLCdurhrshorteccenxxxxxB}{\ensuremath{2.846}}        
\newcommand{\hatcurLCqeccenxxxxxB}{\ensuremath{0.0422\pm0.0015}}       
\newcommand{\hatcurLCqshorteccenxxxxxB}{\ensuremath{0.042}}            
\newcommand{\hatcurLCingdureccenxxxxxB}{\ensuremath{0.0147\pm0.0026}}  
\newcommand{\hatcurLCPeccenxxxxxB}{\ensuremath{2.8126551\pm0.0000056}} 
\newcommand{\hatcurLCPprececcenxxxxxB}{\ensuremath{2.8126551}}         
\newcommand{\hatcurLCPshorteccenxxxxxB}{\ensuremath{2.8127}}           
\newcommand{\hatcurLCTeccenxxxxxB}{\ensuremath{2456437.7492\pm0.0011}} 
\newcommand{\hatcurLCTAeccenxxxxxB}{\ensuremath{2455416.7554\pm0.0022}} 
\newcommand{\hatcurLCTBeccenxxxxxB}{\ensuremath{2457171.8522\pm0.0019}} 
\newcommand{\hatcurLChatnetmeccenxxxxxB}{\ensuremath{14.17329\pm0.00014}} 
\newcommand{\hatcurLCiblendeccenxxxxxB}{\ensuremath{0.822\pm0.069}}    
\newcommand{\hatcurLCrhoeccenxxxxxB}{\ensuremath{1.1\pm4.9}}           
\newcommand{\hatcurSMEiteffeccenxxxxxB}{\ensuremath{5750\pm110}}       
\newcommand{\hatcurSMEizfeheccenxxxxxB}{\ensuremath{0.380\pm0.070}}    
\newcommand{\hatcurSMEizfehshorteccenxxxxxB}{\ensuremath{0.38}}        
\newcommand{\hatcurSMEiloggeccenxxxxxB}{\ensuremath{4.49\pm0.21}}      
\newcommand{\hatcurSMEivsineccenxxxxxB}{\ensuremath{3.3\pm1.1}}        
\newcommand{\hatcurSMEivmaceccenxxxxxB}{\ensuremath{0.0}}              
\newcommand{\hatcurSMEivmiceccenxxxxxB}{\ensuremath{0.0}}              
\newcommand{\hatcurSMEiiteffeccenxxxxxB}{\ensuremath{5700\pm110}}      
\newcommand{\hatcurSMEiizfeheccenxxxxxB}{\ensuremath{0.390\pm0.050}}   
\newcommand{\hatcurSMEiizfehshorteccenxxxxxB}{\ensuremath{0.39}}       
\newcommand{\hatcurSMEiiloggeccenxxxxxB}{\ensuremath{4.395\pm0.055}}   
\newcommand{\hatcurSMEiivsineccenxxxxxB}{\ensuremath{3.56\pm0.69}}     
\newcommand{\hatcurLBizeccenxxxxxB}{\ensuremath{0.2272}}               
\newcommand{\hatcurLBiizeccenxxxxxB}{\ensuremath{0.3299}}              
\newcommand{\hatcurLBiieccenxxxxxB}{\ensuremath{0.2989}}               
\newcommand{\hatcurLBiiieccenxxxxxB}{\ensuremath{0.3243}}              
\newcommand{\hatcurLBiIeccenxxxxxB}{\ensuremath{0.2749}}               
\newcommand{\hatcurLBiiIeccenxxxxxB}{\ensuremath{0.3270}}              
\newcommand{\hatcurLBigeccenxxxxxB}{\ensuremath{0.6151}}               
\newcommand{\hatcurLBiigeccenxxxxxB}{\ensuremath{0.1921}}              
\newcommand{\hatcurLBireccenxxxxxB}{\ensuremath{0.4005}}               
\newcommand{\hatcurLBiireccenxxxxxB}{\ensuremath{0.3061}}              
\newcommand{\hatcurLBiReccenxxxxxB}{\ensuremath{0.3723}}               
\newcommand{\hatcurLBiiReccenxxxxxB}{\ensuremath{0.3122}}              
\newcommand{\hatcurISOmeccenxxxxxB}{\ensuremath{1.122_{-0.055}^{+0.088}}} 
\newcommand{\hatcurISOmshorteccenxxxxxB}{\ensuremath{1.12}}            
\newcommand{\hatcurISOmlongeccenxxxxxB}{\ensuremath{1.122_{-0.055}^{+0.088}}} 
\newcommand{\hatcurISOreccenxxxxxB}{\ensuremath{1.19_{-0.13}^{+0.32}}} 
\newcommand{\hatcurISOrshorteccenxxxxxB}{\ensuremath{1.19}}            
\newcommand{\hatcurISOrlongeccenxxxxxB}{\ensuremath{1.19_{-0.13}^{+0.32}}} 
\newcommand{\hatcurISOrhoeccenxxxxxB}{\ensuremath{0.93\pm0.36}}        
\newcommand{\hatcurISOrholongeccenxxxxxB}{\ensuremath{0.93\pm0.36}}    
\newcommand{\hatcurISOloggeccenxxxxxB}{\ensuremath{4.33\pm0.14}}       
\newcommand{\hatcurISOlumeccenxxxxxB}{\ensuremath{1.35_{-0.32}^{+0.82}}} 
\newcommand{\hatcurISOlumshorteccenxxxxxB}{\ensuremath{1.35}}          
\newcommand{\hatcurISOmveccenxxxxxB}{\ensuremath{4.50\pm0.44}}         
\newcommand{\hatcurISOvieccenxxxxxB}{\ensuremath{0.733\pm0.036}}       
\newcommand{\hatcurISOageeccenxxxxxB}{\ensuremath{4.2\pm1.8}}          
\newcommand{\hatcurISOsigmaeccenxxxxxB}{\ensuremath{nff\pmnff}}        
\newcommand{\hatcurISOMJeccenxxxxxB}{\ensuremath{3.32\pm0.42}}         
\newcommand{\hatcurISOMHeccenxxxxxB}{\ensuremath{2.98\pm0.42}}         
\newcommand{\hatcurISOMKeccenxxxxxB}{\ensuremath{2.92\pm0.42}}         
\newcommand{\hatcurISOJKeccenxxxxxB}{\ensuremath{0.400\pm0.030}}       
\newcommand{\hatcurISOspececcenxxxxxB}{G}                              
\newcommand{\hatcurRVKeccenxxxxxB}{\ensuremath{134\pm48}}              
\newcommand{\hatcurRVrkeccenxxxxxB}{\ensuremath{-0.04\pm0.22}}         
\newcommand{\hatcurRVrheccenxxxxxB}{\ensuremath{0.16\pm0.28}}          
\newcommand{\hatcurRVkeccenxxxxxB}{\ensuremath{-0.007_{-0.111}^{+0.070}}} 
\newcommand{\hatcurRVheccenxxxxxB}{\ensuremath{0.041_{-0.073}^{+0.228}}} 
\newcommand{\hatcurRVtroneeccenxxxxxB}{\ensuremath{0\pm0}}             
\newcommand{\hatcurRVtrtwoeccenxxxxxB}{\ensuremath{0\pm0}}             
\newcommand{\hatcurRVgammaeccenxxxxxB}{\ensuremath{12419\pm19}}        
\newcommand{\hatcurRVjittereccenxxxxxB}{\ensuremath{31\pm15}}          
\newcommand{\hatcurRVjittertwosiglimeccenxxxxxB}{\ensuremath{<61.6}}   
\newcommand{\hatcurRVfitrmseccenxxxxxB}{\ensuremath{.1fym}}            %
\newcommand{\hatcurRVecceneccenxxxxxB}{\ensuremath{0.11\pm0.15}}       
\newcommand{\hatcurRVeccentwosiglimeccenxxxxxB}{\ensuremath{<0.471}}   
\newcommand{\hatcurRVomegaeccenxxxxxB}{\ensuremath{126\pm95}}          
\newcommand{\hatcurPPieccenxxxxxB}{\ensuremath{86.4_{-2.0}^{+1.5}}}    
\newcommand{\hatcurPPgeccenxxxxxB}{\ensuremath{12.9\pm3.5}}            
\newcommand{\hatcurPPloggeccenxxxxxB}{\ensuremath{3.112_{-0.120}^{+0.063}}} 
\newcommand{\hatcurPPareccenxxxxxB}{\ensuremath{7.29_{-1.39}^{+0.85}}} 
\newcommand{\hatcurPPareleccenxxxxxB}{\ensuremath{0.04053_{-0.00067}^{+0.00104}}} 
\newcommand{\hatcurPPrhoeccenxxxxxB}{\ensuremath{0.47\pm0.19}}         
\newcommand{\hatcurPPmeccenxxxxxB}{\ensuremath{1.00_{-0.16}^{+0.34}}}  
\newcommand{\hatcurPPmshorteccenxxxxxB}{\ensuremath{1.00}}             
\newcommand{\hatcurPPmlongeccenxxxxxB}{\ensuremath{1.00_{-0.16}^{+0.34}}} 
\newcommand{\hatcurPPmeeccenxxxxxB}{\ensuremath{317_{-51}^{+108}}}     
\newcommand{\hatcurPPmeshorteccenxxxxxB}{\ensuremath{317.0}}           
\newcommand{\hatcurPPmelongeccenxxxxxB}{\ensuremath{317_{-51}^{+108}}} 
\newcommand{\hatcurPPreccenxxxxxB}{\ensuremath{1.37_{-0.17}^{+0.36}}}  
\newcommand{\hatcurPPrshorteccenxxxxxB}{\ensuremath{1.37}}             
\newcommand{\hatcurPPrlongeccenxxxxxB}{\ensuremath{1.37_{-0.17}^{+0.36}}} 
\newcommand{\hatcurPPreeccenxxxxxB}{\ensuremath{15.3_{-1.9}^{+4.0}}}   
\newcommand{\hatcurPPreshorteccenxxxxxB}{\ensuremath{15.3}}            
\newcommand{\hatcurPPrelongeccenxxxxxB}{\ensuremath{15.3_{-1.9}^{+4.0}}} 
\newcommand{\hatcurPPmrcorreccenxxxxxB}{\ensuremath{0.27}}             
\newcommand{\hatcurPPteffeccenxxxxxB}{\ensuremath{1498_{-91}^{+186}}}  
\newcommand{\hatcurPPthetaeccenxxxxxB}{\ensuremath{0.052\pm0.010}}     
\newcommand{\hatcurPPfluxperieccenxxxxxB}{\ensuremath{1.37_{-0.36}^{+2.21}}} 
\newcommand{\hatcurPPfluxperidimeccenxxxxxB}{\ensuremath{9}}           
\newcommand{\hatcurPPfluxapeccenxxxxxB}{\ensuremath{9.2\pm3.0}}        
\newcommand{\hatcurPPfluxapdimeccenxxxxxB}{\ensuremath{8}}             
\newcommand{\hatcurPPfluxavgeccenxxxxxB}{\ensuremath{1.14_{-0.25}^{+0.68}}} 
\newcommand{\hatcurPPfluxavgdimeccenxxxxxB}{\ensuremath{9}}            
\newcommand{\hatcurPPfluxavglogeccenxxxxxB}{\ensuremath{9.06_{-0.11}^{+0.20}}} 
\newcommand{\hatcurXsecphaseeccenxxxxxB}{\ensuremath{0.495\pm0.072}}   
\newcommand{\hatcurXsecondaryeccenxxxxxB}{\ensuremath{2456439.14\pm0.20}} 
\newcommand{\hatcurXsecdureccenxxxxxB}{\ensuremath{0.127\pm0.089}}     
\newcommand{\hatcurXsecingdureccenxxxxxB}{\ensuremath{0.017\pm0.048}}  
\newcommand{\hatcurPPphiconjeccenxxxxxB}{\ensuremath{-0.02\pm0.22}}    
\newcommand{\hatcurPPperieccenxxxxxB}{\ensuremath{2456437.80\pm0.61}}  
\newcommand{\hatcurPPaequiveccenxxxxxB}{\ensuremath{0.0348_{-0.0066}^{+0.0046}}} 
\newcommand{\hatcurPPtcirceccenxxxxxB}{\ensuremath{52\pm42}}           
\newcommand{\hatcurPPtinfalleccenxxxxxB}{\ensuremath{630\pm470}}       
\newcommand{\hatcurXdisteccenxxxxxB}{\ensuremath{920_{-110}^{+240}}}   
\newcommand{\hatcurXAveccenxxxxxB}{\ensuremath{0.068\pm0.072}}         
\newcommand{\hatcurXdistredeccenxxxxxB}{\ensuremath{910_{-110}^{+240}}} 
\newcommand{\hatcurXEBVeccenxxxxxB}{\ensuremath{0.022\pm0.023}}        
\newcommand{\hatcurXmvisoredeccenxxxxxB}{\ensuremath{14.386\pm0.012}}  
\newcommand{\hatcurXmiisoredeccenxxxxxB}{\ensuremath{13.613\pm0.013}}  
\newcommand{\hatcurXmjisoredeccenxxxxxB}{\ensuremath{13.141\pm0.020}}  
\newcommand{\hatcurXmhisoredeccenxxxxxB}{\ensuremath{12.795_{-0.028}^{+0.020}}} 
\newcommand{\hatcurXmkisoredeccenxxxxxB}{\ensuremath{12.732_{-0.029}^{+0.021}}} 
\newcommand{\hatcurXviisoredeccenxxxxxB}{\ensuremath{0.773\pm0.015}}   
\newcommand{\hatcurXvkisoredeccenxxxxxB}{\ensuremath{1.654_{-0.025}^{+0.034}}} 
\newcommand{\hatcurXjhisoredeccenxxxxxB}{\ensuremath{0.347_{-0.011}^{+0.017}}} 
\newcommand{\hatcurXjkisoredeccenxxxxxB}{\ensuremath{0.4090_{-0.0100}^{+0.0160}}} 
\newcommand{\hatcurCCpmraeccenxxxxxB}{\ensuremath{2.9\pm1.9}}          
\newcommand{\hatcurCCpmdececcenxxxxxB}{\ensuremath{-20.3\pm1.9}}       
\newcommand{\hatcurCCpmeccenxxxxxB}{\ensuremath{20.5\pm2.7}}           

%% file: HATS747-032.eccen.tex
\newcommand{\hatcurhtreccenxxxxxC}{HATS747-032}                      
\newcommand{\hatcurfieldeccenxxxxxC}{\ensuremath{string}}            
\newcommand{\hatcurCCraeccenxxxxxC}{\ensuremath{19^{\mathrm h}38^{\mathrm m}31.92{\mathrm s}}}                     
\newcommand{\hatcurCCdececcenxxxxxC}{\ensuremath{-55{\arcdeg}19{\arcmin}48.4{\arcsec}}}                    
\newcommand{\hatcurCCmageccenxxxxxC}{11.911}                         
\newcommand{\hatcurCCtwomasseccenxxxxxC}{2MASS~19383207-5519483}     
\newcommand{\hatcurCCgsceccenxxxxxC}{GSC~8778-01635}                 
\newcommand{\hatcurCCtassmveccenxxxxxC}{\ensuremath{11.911\pm0.070}} 
\newcommand{\hatcurCCtassmvshorteccenxxxxxC}{\ensuremath{11.9}}      
\newcommand{\hatcurCCtassmBeccenxxxxxC}{\ensuremath{12.633\pm0.070}} 
\newcommand{\hatcurCCtassmBshorteccenxxxxxC}{\ensuremath{12.6}}      
\newcommand{\hatcurCCtassmIeccenxxxxxC}{\ensuremath{100\pm1000}}     
\newcommand{\hatcurCCtassmIshorteccenxxxxxC}{\ensuremath{100.0}}     
\newcommand{\hatcurCCtassmgeccenxxxxxC}{\ensuremath{12.174\pm0.040}} 
\newcommand{\hatcurCCtassmgshorteccenxxxxxC}{\ensuremath{12.2}}      
\newcommand{\hatcurCCtassmreccenxxxxxC}{\ensuremath{11.704\pm0.040}} 
\newcommand{\hatcurCCtassmrshorteccenxxxxxC}{\ensuremath{11.7}}      
\newcommand{\hatcurCCtassmieccenxxxxxC}{\ensuremath{11.52\pm0.11}}   
\newcommand{\hatcurCCtassmishorteccenxxxxxC}{\ensuremath{11.5}}      
\newcommand{\hatcurCCtwomassJmageccenxxxxxC}{\ensuremath{10.659\pm0.024}} 
\newcommand{\hatcurCCtwomassHmageccenxxxxxC}{\ensuremath{10.337\pm0.026}} 
\newcommand{\hatcurCCtwomassKmageccenxxxxxC}{\ensuremath{10.287\pm0.027}} 
\newcommand{\hatcurCCcitJmageccenxxxxxC}{\ensuremath{10.676\pm0.024}} 
\newcommand{\hatcurCCcitHmageccenxxxxxC}{\ensuremath{10.332\pm0.026}} 
\newcommand{\hatcurCCcitKmageccenxxxxxC}{\ensuremath{10.311\pm0.027}} 
\newcommand{\hatcurCCbbJmageccenxxxxxC}{\ensuremath{10.725\pm0.026}} 
\newcommand{\hatcurCCbbHmageccenxxxxxC}{\ensuremath{10.353\pm0.027}} 
\newcommand{\hatcurCCbbKmageccenxxxxxC}{\ensuremath{10.331\pm0.027}} 
\newcommand{\hatcurCCesoJmageccenxxxxxC}{\ensuremath{10.727\pm0.027}} 
\newcommand{\hatcurCCesoHmageccenxxxxxC}{\ensuremath{10.347\pm0.030}} 
\newcommand{\hatcurCCesoKmageccenxxxxxC}{\ensuremath{10.330\pm0.028}} 
\newcommand{\hatcurCCesoJHmageccenxxxxxC}{\ensuremath{0.380\pm0.038}} 
\newcommand{\hatcurCCesoJKmageccenxxxxxC}{\ensuremath{0.398\pm0.039}} 
\newcommand{\hatcurCCesoHKmageccenxxxxxC}{\ensuremath{0.018\pm0.041}} 
\newcommand{\hatcurLCdipeccenxxxxxC}{\ensuremath{14.0}}              
\newcommand{\hatcurLCrprstareccenxxxxxC}{\ensuremath{0.1286\pm0.0086}} 
\newcommand{\hatcurLCbsqeccenxxxxxC}{\ensuremath{0.112_{-0.070}^{+0.104}}} 
\newcommand{\hatcurLCimpeccenxxxxxC}{\ensuremath{0.34_{-0.13}^{+0.13}}} 
\newcommand{\hatcurLCzetaeccenxxxxxC}{\ensuremath{20.45\pm0.25}}     
\newcommand{\hatcurLCdureccenxxxxxC}{\ensuremath{0.1122\pm0.0019}}   
\newcommand{\hatcurLCdurshorteccenxxxxxC}{\ensuremath{0.1122}}       
\newcommand{\hatcurLCdurhreccenxxxxxC}{\ensuremath{2.692\pm0.046}}   
\newcommand{\hatcurLCdurhrshorteccenxxxxxC}{\ensuremath{2.692}}      
\newcommand{\hatcurLCqeccenxxxxxC}{\ensuremath{0.04400\pm0.00075}}   
\newcommand{\hatcurLCqshorteccenxxxxxC}{\ensuremath{0.044}}          
\newcommand{\hatcurLCingdureccenxxxxxC}{\ensuremath{0.0143\pm0.0018}} 
\newcommand{\hatcurLCPeccenxxxxxC}{\ensuremath{2.5495539\pm0.0000058}} 
\newcommand{\hatcurLCPprececcenxxxxxC}{\ensuremath{2.5495539}}       
\newcommand{\hatcurLCPshorteccenxxxxxC}{\ensuremath{2.5496}}         
\newcommand{\hatcurLCTeccenxxxxxC}{\ensuremath{2456484.48415\pm0.00050}} 
\newcommand{\hatcurLCTAeccenxxxxxC}{\ensuremath{2456367.20464\pm0.00063}} 
\newcommand{\hatcurLCTBeccenxxxxxC}{\ensuremath{2457162.6655\pm0.0015}} 
\newcommand{\hatcurLChatnetmeccenxxxxxC}{\ensuremath{11.783440\pm0.000046}} 
\newcommand{\hatcurLCiblendeccenxxxxxC}{\ensuremath{0.72\pm0.10}}    
\newcommand{\hatcurLCrhoeccenxxxxxC}{\ensuremath{1.49\pm0.24}}       
\newcommand{\hatcurSMEiteffeccenxxxxxC}{\ensuremath{5700\pm100}}     
\newcommand{\hatcurSMEizfeheccenxxxxxC}{\ensuremath{0.280\pm0.050}}  
\newcommand{\hatcurSMEizfehshorteccenxxxxxC}{\ensuremath{0.28}}      
\newcommand{\hatcurSMEiloggeccenxxxxxC}{\ensuremath{4.54\pm0.15}}    
\newcommand{\hatcurSMEivsineccenxxxxxC}{\ensuremath{3.42\pm0.66}}    
\newcommand{\hatcurSMEivmaceccenxxxxxC}{\ensuremath{0.0}}            
\newcommand{\hatcurSMEivmiceccenxxxxxC}{\ensuremath{0.0}}            
\newcommand{\hatcurSMEiiteffeccenxxxxxC}{\ensuremath{5659\pm85}}     
\newcommand{\hatcurSMEiizfeheccenxxxxxC}{\ensuremath{0.290\pm0.050}} 
\newcommand{\hatcurSMEiizfehshorteccenxxxxxC}{\ensuremath{0.29}}     
\newcommand{\hatcurSMEiiloggeccenxxxxxC}{\ensuremath{4.443\pm0.041}} 
\newcommand{\hatcurSMEiivsineccenxxxxxC}{\ensuremath{3.87\pm0.42}}   
\newcommand{\hatcurLBizeccenxxxxxC}{\ensuremath{0.2308}}             
\newcommand{\hatcurLBiizeccenxxxxxC}{\ensuremath{0.3242}}            
\newcommand{\hatcurLBiieccenxxxxxC}{\ensuremath{0.3013}}             
\newcommand{\hatcurLBiiieccenxxxxxC}{\ensuremath{0.3193}}            
\newcommand{\hatcurLBiIeccenxxxxxC}{\ensuremath{0.2778}}             
\newcommand{\hatcurLBiiIeccenxxxxxC}{\ensuremath{0.3216}}            
\newcommand{\hatcurLBigeccenxxxxxC}{\ensuremath{0.6140}}             
\newcommand{\hatcurLBiigeccenxxxxxC}{\ensuremath{0.1919}}            
\newcommand{\hatcurLBireccenxxxxxC}{\ensuremath{0.4016}}             
\newcommand{\hatcurLBiireccenxxxxxC}{\ensuremath{0.3030}}            
\newcommand{\hatcurLBiReccenxxxxxC}{\ensuremath{0.3737}}             
\newcommand{\hatcurLBiiReccenxxxxxC}{\ensuremath{0.3085}}            
\newcommand{\hatcurLBikepeccenxxxxxC}{\ensuremath{0.1000}}           
\newcommand{\hatcurLBiikepeccenxxxxxC}{\ensuremath{0.1000}}          
\newcommand{\hatcurISOmeccenxxxxxC}{\ensuremath{1.061\pm0.032}}      
\newcommand{\hatcurISOmshorteccenxxxxxC}{\ensuremath{1.06}}          
\newcommand{\hatcurISOmlongeccenxxxxxC}{\ensuremath{1.061\pm0.032}}  
\newcommand{\hatcurISOreccenxxxxxC}{\ensuremath{1.018_{-0.037}^{+0.051}}} 
\newcommand{\hatcurISOrshorteccenxxxxxC}{\ensuremath{1.02}}          
\newcommand{\hatcurISOrlongeccenxxxxxC}{\ensuremath{1.018_{-0.037}^{+0.051}}} 
\newcommand{\hatcurISOrhoeccenxxxxxC}{\ensuremath{1.42\pm0.17}}      
\newcommand{\hatcurISOrholongeccenxxxxxC}{\ensuremath{1.42\pm0.17}}  
\newcommand{\hatcurISOloggeccenxxxxxC}{\ensuremath{4.449\pm0.037}}   
\newcommand{\hatcurISOlumeccenxxxxxC}{\ensuremath{0.95\pm0.12}}      
\newcommand{\hatcurISOlumshorteccenxxxxxC}{\ensuremath{0.95}}        
\newcommand{\hatcurISOmveccenxxxxxC}{\ensuremath{4.90\pm0.15}}       
\newcommand{\hatcurISOvieccenxxxxxC}{\ensuremath{0.745\pm0.026}}     
\newcommand{\hatcurISOageeccenxxxxxC}{\ensuremath{2.9\pm1.7}}        
\newcommand{\hatcurISOsigmaeccenxxxxxC}{\ensuremath{0.00150\pm0.00020}} 
\newcommand{\hatcurISOMJeccenxxxxxC}{\ensuremath{3.69\pm0.12}}       
\newcommand{\hatcurISOMHeccenxxxxxC}{\ensuremath{3.34\pm0.10}}       
\newcommand{\hatcurISOMKeccenxxxxxC}{\ensuremath{3.28\pm0.10}}       
\newcommand{\hatcurISOJKeccenxxxxxC}{\ensuremath{0.420\pm0.020}}     
\newcommand{\hatcurISOspececcenxxxxxC}{G}                            
\newcommand{\hatcurRVKeccenxxxxxC}{\ensuremath{166.7\pm8.7}}         
\newcommand{\hatcurRVrkeccenxxxxxC}{\ensuremath{-0.105_{-0.062}^{+0.091}}} 
\newcommand{\hatcurRVrheccenxxxxxC}{\ensuremath{-0.07\pm0.13}}       
\newcommand{\hatcurRVkeccenxxxxxC}{\ensuremath{-0.019\pm0.016}}      
\newcommand{\hatcurRVheccenxxxxxC}{\ensuremath{-0.009_{-0.033}^{+0.022}}} 
\newcommand{\hatcurRVtroneeccenxxxxxC}{\ensuremath{0\pm0}}           
\newcommand{\hatcurRVtrtwoeccenxxxxxC}{\ensuremath{0\pm0}}           
\newcommand{\hatcurRVgammaAeccenxxxxxC}{\ensuremath{11064\pm22}}     
\newcommand{\hatcurRVjitterAeccenxxxxxC}{\ensuremath{37\pm23}}       
\newcommand{\hatcurRVjittertwosiglimAeccenxxxxxC}{\ensuremath{<89.3}} 
\newcommand{\hatcurRVfitrmsAeccenxxxxxC}{\ensuremath{0.0}}           
\newcommand{\hatcurRVgammaBeccenxxxxxC}{\ensuremath{11063\pm25}}     
\newcommand{\hatcurRVjitterBeccenxxxxxC}{\ensuremath{49\pm28}}       
\newcommand{\hatcurRVjittertwosiglimBeccenxxxxxC}{\ensuremath{<106.6}} 
\newcommand{\hatcurRVfitrmsBeccenxxxxxC}{\ensuremath{0.0}}           
\newcommand{\hatcurRVgammaCeccenxxxxxC}{\ensuremath{11079\pm12}}     
\newcommand{\hatcurRVjitterCeccenxxxxxC}{\ensuremath{2\pm20}}        
\newcommand{\hatcurRVjittertwosiglimCeccenxxxxxC}{\ensuremath{<45.6}} 
\newcommand{\hatcurRVfitrmsCeccenxxxxxC}{\ensuremath{0.0}}           
\newcommand{\hatcurRVgammaDeccenxxxxxC}{\ensuremath{11064.9\pm6.7}}  
\newcommand{\hatcurRVjitterDeccenxxxxxC}{\ensuremath{11\pm12}}       
\newcommand{\hatcurRVjittertwosiglimDeccenxxxxxC}{\ensuremath{<33.1}} 
\newcommand{\hatcurRVfitrmsDeccenxxxxxC}{\ensuremath{.1fym}}         %
\newcommand{\hatcurRVecceneccenxxxxxC}{\ensuremath{0.031\pm0.025}}   
\newcommand{\hatcurRVeccentwosiglimeccenxxxxxC}{\ensuremath{<0.080}} 
\newcommand{\hatcurRVomegaeccenxxxxxC}{\ensuremath{213\pm63}}        
\newcommand{\hatcurPPieccenxxxxxC}{\ensuremath{87.61\pm0.92}}        
\newcommand{\hatcurPPgeccenxxxxxC}{\ensuremath{17.5_{-2.3}^{+3.8}}}  
\newcommand{\hatcurPPloggeccenxxxxxC}{\ensuremath{3.243_{-0.062}^{+0.085}}} 
\newcommand{\hatcurPPareccenxxxxxC}{\ensuremath{7.87_{-0.37}^{+0.28}}} 
\newcommand{\hatcurPPareleccenxxxxxC}{\ensuremath{0.03725\pm0.00037}} 
\newcommand{\hatcurPPrhoeccenxxxxxC}{\ensuremath{0.68_{-0.12}^{+0.23}}} 
\newcommand{\hatcurPPmeccenxxxxxC}{\ensuremath{1.165\pm0.065}}       
\newcommand{\hatcurPPmshorteccenxxxxxC}{\ensuremath{1.16}}           
\newcommand{\hatcurPPmlongeccenxxxxxC}{\ensuremath{1.165\pm0.065}}   
\newcommand{\hatcurPPmeeccenxxxxxC}{\ensuremath{370\pm21}}           
\newcommand{\hatcurPPmeshorteccenxxxxxC}{\ensuremath{370.1}}         
\newcommand{\hatcurPPmelongeccenxxxxxC}{\ensuremath{370\pm21}}       
\newcommand{\hatcurPPreccenxxxxxC}{\ensuremath{1.282_{-0.116}^{+0.083}}} 
\newcommand{\hatcurPPrshorteccenxxxxxC}{\ensuremath{1.28}}           
\newcommand{\hatcurPPrlongeccenxxxxxC}{\ensuremath{1.282_{-0.116}^{+0.083}}} 
\newcommand{\hatcurPPreeccenxxxxxC}{\ensuremath{14.37_{-1.30}^{+0.93}}} 
\newcommand{\hatcurPPreshorteccenxxxxxC}{\ensuremath{14.4}}          
\newcommand{\hatcurPPrelongeccenxxxxxC}{\ensuremath{14.37_{-1.30}^{+0.93}}} 
\newcommand{\hatcurPPmrcorreccenxxxxxC}{\ensuremath{0.08}}           
\newcommand{\hatcurPPteffeccenxxxxxC}{\ensuremath{1426\pm40}}        
\newcommand{\hatcurPPthetaeccenxxxxxC}{\ensuremath{0.0637\pm0.0064}} 
\newcommand{\hatcurPPfluxperieccenxxxxxC}{\ensuremath{9.93_{-1.00}^{+1.47}}} 
\newcommand{\hatcurPPfluxperidimeccenxxxxxC}{\ensuremath{8}}         
\newcommand{\hatcurPPfluxapeccenxxxxxC}{\ensuremath{8.8\pm1.0}}      
\newcommand{\hatcurPPfluxapdimeccenxxxxxC}{\ensuremath{8}}           
\newcommand{\hatcurPPfluxavgeccenxxxxxC}{\ensuremath{9.3\pm1.1}}     
\newcommand{\hatcurPPfluxavgdimeccenxxxxxC}{\ensuremath{8}}          
\newcommand{\hatcurPPfluxavglogeccenxxxxxC}{\ensuremath{8.970\pm0.048}} 
\newcommand{\hatcurXsecphaseeccenxxxxxC}{\ensuremath{0.488\pm0.010}} 
\newcommand{\hatcurXsecondaryeccenxxxxxC}{\ensuremath{2456485.728\pm0.026}} 
\newcommand{\hatcurXsecdureccenxxxxxC}{\ensuremath{0.1104\pm0.0063}} 
\newcommand{\hatcurXsecingdureccenxxxxxC}{\ensuremath{0.0140\pm0.0017}} 
\newcommand{\hatcurPPphiconjeccenxxxxxC}{\ensuremath{-0.29_{-0.15}^{+0.23}}} 
\newcommand{\hatcurPPperieccenxxxxxC}{\ensuremath{2456485.22\pm0.64}} 
\newcommand{\hatcurPPaequiveccenxxxxxC}{\ensuremath{0.0383\pm0.0021}} 
\newcommand{\hatcurPPtcirceccenxxxxxC}{\ensuremath{59_{-15}^{+37}}}  
\newcommand{\hatcurPPtinfalleccenxxxxxC}{\ensuremath{670\pm140}}     
\newcommand{\hatcurXdisteccenxxxxxC}{\ensuremath{257\pm13}}          
\newcommand{\hatcurXAveccenxxxxxC}{\ensuremath{0.000\pm0.056}}       
\newcommand{\hatcurXdistredeccenxxxxxC}{\ensuremath{254\pm13}}       
\newcommand{\hatcurXEBVeccenxxxxxC}{\ensuremath{0.000\pm0.018}}      
\newcommand{\hatcurXmvisoredeccenxxxxxC}{\ensuremath{11.955\pm0.050}} 
\newcommand{\hatcurXmiisoredeccenxxxxxC}{\ensuremath{11.197\pm0.029}} 
\newcommand{\hatcurXmjisoredeccenxxxxxC}{\ensuremath{10.726\pm0.018}} 
\newcommand{\hatcurXmhisoredeccenxxxxxC}{\ensuremath{10.367\pm0.017}} 
\newcommand{\hatcurXmkisoredeccenxxxxxC}{\ensuremath{10.306\pm0.019}} 
\newcommand{\hatcurXviisoredeccenxxxxxC}{\ensuremath{0.758\pm0.026}} 
\newcommand{\hatcurXvkisoredeccenxxxxxC}{\ensuremath{1.649\pm0.057}} 
\newcommand{\hatcurXjhisoredeccenxxxxxC}{\ensuremath{0.359\pm0.015}} 
\newcommand{\hatcurXjkisoredeccenxxxxxC}{\ensuremath{0.419\pm0.017}} 
\newcommand{\hatcurCCpmraeccenxxxxxC}{\ensuremath{7.1\pm1.1}}        
\newcommand{\hatcurCCpmdececcenxxxxxC}{\ensuremath{-40.6\pm1.3}}     
\newcommand{\hatcurCCpmeccenxxxxxC}{\ensuremath{41.2\pm1.7}}         

%% file: HATS754-010.eccen.tex
\newcommand{\hatcurhtreccenxxxxxD}{HATS754-010}                        
\newcommand{\hatcurfieldeccenxxxxxD}{\ensuremath{string}}              
\newcommand{\hatcurCCraeccenxxxxxD}{\ensuremath{00^{\mathrm h}03^{\mathrm m}05.88{\mathrm s}}}                       
\newcommand{\hatcurCCdececcenxxxxxD}{\ensuremath{-62{\arcdeg}28{\arcmin}09.6{\arcsec}}}                      
\newcommand{\hatcurCCmageccenxxxxxD}{13.849}                           
\newcommand{\hatcurCCtwomasseccenxxxxxD}{2MASS~00030587-6228096}       
\newcommand{\hatcurCCgsceccenxxxxxD}{GSC~8840-01777}                   
\newcommand{\hatcurCCtassmveccenxxxxxD}{\ensuremath{13.849\pm0.010}}   
\newcommand{\hatcurCCtassmvshorteccenxxxxxD}{\ensuremath{13.8}}        
\newcommand{\hatcurCCtassmBeccenxxxxxD}{\ensuremath{14.595\pm0.010}}   
\newcommand{\hatcurCCtassmBshorteccenxxxxxD}{\ensuremath{14.6}}        
\newcommand{\hatcurCCtassmIeccenxxxxxD}{\ensuremath{nff\pmnff}}        
\newcommand{\hatcurCCtassmIshorteccenxxxxxD}{\ensuremath{0.0}}         
\newcommand{\hatcurCCtassmgeccenxxxxxD}{\ensuremath{14.182\pm0.010}}   
\newcommand{\hatcurCCtassmgshorteccenxxxxxD}{\ensuremath{14.2}}        
\newcommand{\hatcurCCtassmreccenxxxxxD}{\ensuremath{13.650\pm0.010}}   
\newcommand{\hatcurCCtassmrshorteccenxxxxxD}{\ensuremath{13.7}}        
\newcommand{\hatcurCCtassmieccenxxxxxD}{\ensuremath{13.459\pm0.050}}   
\newcommand{\hatcurCCtassmishorteccenxxxxxD}{\ensuremath{13.5}}        
\newcommand{\hatcurCCtwomassJmageccenxxxxxD}{\ensuremath{12.513\pm0.024}} 
\newcommand{\hatcurCCtwomassHmageccenxxxxxD}{\ensuremath{12.139\pm0.022}} 
\newcommand{\hatcurCCtwomassKmageccenxxxxxD}{\ensuremath{12.095\pm0.019}} 
\newcommand{\hatcurCCcitJmageccenxxxxxD}{\ensuremath{12.527\pm0.024}}  
\newcommand{\hatcurCCcitHmageccenxxxxxD}{\ensuremath{12.134\pm0.023}}  
\newcommand{\hatcurCCcitKmageccenxxxxxD}{\ensuremath{12.119\pm0.019}}  
\newcommand{\hatcurCCbbJmageccenxxxxxD}{\ensuremath{12.580\pm0.026}}   
\newcommand{\hatcurCCbbHmageccenxxxxxD}{\ensuremath{12.155\pm0.023}}   
\newcommand{\hatcurCCbbKmageccenxxxxxD}{\ensuremath{12.139\pm0.019}}   
\newcommand{\hatcurCCesoJmageccenxxxxxD}{\ensuremath{12.583\pm0.028}}  
\newcommand{\hatcurCCesoHmageccenxxxxxD}{\ensuremath{12.149\pm0.026}}  
\newcommand{\hatcurCCesoKmageccenxxxxxD}{\ensuremath{12.138\pm0.020}}  
\newcommand{\hatcurCCesoJHmageccenxxxxxD}{\ensuremath{0.434\pm0.035}}  
\newcommand{\hatcurCCesoJKmageccenxxxxxD}{\ensuremath{0.446\pm0.033}}  
\newcommand{\hatcurCCesoHKmageccenxxxxxD}{\ensuremath{0.011\pm0.032}}  
\newcommand{\hatcurLCdipeccenxxxxxD}{\ensuremath{13.4}}                
\newcommand{\hatcurLCrprstareccenxxxxxD}{\ensuremath{0.1532\pm0.0086}} 
\newcommand{\hatcurLCbsqeccenxxxxxD}{\ensuremath{0.885_{-0.012}^{+0.012}}} 
\newcommand{\hatcurLCimpeccenxxxxxD}{\ensuremath{0.9407_{-0.0066}^{+0.0065}}} 
\newcommand{\hatcurLCzetaeccenxxxxxD}{\ensuremath{60.8\pm4.6}}         
\newcommand{\hatcurLCdureccenxxxxxD}{\ensuremath{0.0650\pm0.0030}}     
\newcommand{\hatcurLCdurshorteccenxxxxxD}{\ensuremath{0.0650}}         
\newcommand{\hatcurLCdurhreccenxxxxxD}{\ensuremath{1.559\pm0.073}}     
\newcommand{\hatcurLCdurhrshorteccenxxxxxD}{\ensuremath{1.559}}        
\newcommand{\hatcurLCqeccenxxxxxD}{\ensuremath{0.0308\pm0.0014}}       
\newcommand{\hatcurLCqshorteccenxxxxxD}{\ensuremath{0.031}}            
\newcommand{\hatcurLCingdureccenxxxxxD}{\ensuremath{0.0811\pm0.0037}}  
\newcommand{\hatcurLCPeccenxxxxxD}{\ensuremath{2.1061607\pm0.0000044}} 
\newcommand{\hatcurLCPprececcenxxxxxD}{\ensuremath{2.1061607}}         
\newcommand{\hatcurLCPshorteccenxxxxxD}{\ensuremath{2.1062}}           
\newcommand{\hatcurLCTeccenxxxxxD}{\ensuremath{2456689.61745\pm0.00083}} 
\newcommand{\hatcurLCTAeccenxxxxxD}{\ensuremath{2456179.9265\pm0.0014}} 
\newcommand{\hatcurLCTBeccenxxxxxD}{\ensuremath{2457289.8733\pm0.0015}} 
\newcommand{\hatcurLChatnetmeccenxxxxxD}{\ensuremath{13.68766\pm0.00010}} 
\newcommand{\hatcurLCiblendeccenxxxxxD}{\ensuremath{0.820\pm0.088}}    
\newcommand{\hatcurLCrhoeccenxxxxxD}{\ensuremath{1.45_{-0.23}^{+0.36}}} 
\newcommand{\hatcurSMEiteffeccenxxxxxD}{\ensuremath{5650\pm140}}       
\newcommand{\hatcurSMEizfeheccenxxxxxD}{\ensuremath{0.380\pm0.070}}    
\newcommand{\hatcurSMEizfehshorteccenxxxxxD}{\ensuremath{0.38}}        
\newcommand{\hatcurSMEiloggeccenxxxxxD}{\ensuremath{4.69\pm0.17}}      
\newcommand{\hatcurSMEivsineccenxxxxxD}{\ensuremath{3.78\pm0.76}}      
\newcommand{\hatcurSMEivmaceccenxxxxxD}{\ensuremath{0.0}}              
\newcommand{\hatcurSMEivmiceccenxxxxxD}{\ensuremath{0.0}}              
\newcommand{\hatcurSMEiiteffeccenxxxxxD}{\ensuremath{5380\pm73}}       
\newcommand{\hatcurSMEiizfeheccenxxxxxD}{\ensuremath{0.250\pm0.070}}   
\newcommand{\hatcurSMEiizfehshorteccenxxxxxD}{\ensuremath{0.25}}       
\newcommand{\hatcurSMEiiloggeccenxxxxxD}{\ensuremath{4.425\pm0.046}}   
\newcommand{\hatcurSMEiivsineccenxxxxxD}{\ensuremath{4.07\pm0.58}}     
\newcommand{\hatcurLBizeccenxxxxxD}{\ensuremath{0.2673}}               
\newcommand{\hatcurLBiizeccenxxxxxD}{\ensuremath{0.3043}}              
\newcommand{\hatcurLBiieccenxxxxxD}{\ensuremath{0.3462}}               
\newcommand{\hatcurLBiiieccenxxxxxD}{\ensuremath{0.2910}}              
\newcommand{\hatcurLBiIeccenxxxxxD}{\ensuremath{0.3203}}               
\newcommand{\hatcurLBiiIeccenxxxxxD}{\ensuremath{0.2955}}              
\newcommand{\hatcurLBigeccenxxxxxD}{\ensuremath{0.6869}}               
\newcommand{\hatcurLBiigeccenxxxxxD}{\ensuremath{0.1335}}              
\newcommand{\hatcurLBireccenxxxxxD}{\ensuremath{0.4588}}               
\newcommand{\hatcurLBiireccenxxxxxD}{\ensuremath{0.2646}}              
\newcommand{\hatcurLBiReccenxxxxxD}{\ensuremath{0.4277}}               
\newcommand{\hatcurLBiiReccenxxxxxD}{\ensuremath{0.2728}}              
\newcommand{\hatcurLBikepeccenxxxxxD}{\ensuremath{0.1000}}             
\newcommand{\hatcurLBiikepeccenxxxxxD}{\ensuremath{0.1000}}            
\newcommand{\hatcurISOmeccenxxxxxD}{\ensuremath{0.956\pm0.031}}        
\newcommand{\hatcurISOmshorteccenxxxxxD}{\ensuremath{0.96}}            
\newcommand{\hatcurISOmlongeccenxxxxxD}{\ensuremath{0.956\pm0.031}}    
\newcommand{\hatcurISOreccenxxxxxD}{\ensuremath{0.977\pm0.060}}        
\newcommand{\hatcurISOrshorteccenxxxxxD}{\ensuremath{0.98}}            
\newcommand{\hatcurISOrlongeccenxxxxxD}{\ensuremath{0.977\pm0.060}}    
\newcommand{\hatcurISOrhoeccenxxxxxD}{\ensuremath{1.45_{-0.22}^{+0.30}}} 
\newcommand{\hatcurISOrholongeccenxxxxxD}{\ensuremath{1.45_{-0.22}^{+0.30}}} 
\newcommand{\hatcurISOloggeccenxxxxxD}{\ensuremath{4.438\pm0.053}}     
\newcommand{\hatcurISOlumeccenxxxxxD}{\ensuremath{0.72\pm0.11}}        
\newcommand{\hatcurISOlumshorteccenxxxxxD}{\ensuremath{0.72}}          
\newcommand{\hatcurISOmveccenxxxxxD}{\ensuremath{5.27\pm0.17}}         
\newcommand{\hatcurISOvieccenxxxxxD}{\ensuremath{0.821\pm0.019}}       
\newcommand{\hatcurISOageeccenxxxxxD}{\ensuremath{7.4\pm3.0}}          
\newcommand{\hatcurISOsigmaeccenxxxxxD}{\ensuremath{0.00130\pm0.00028}} 
\newcommand{\hatcurISOMJeccenxxxxxD}{\ensuremath{3.90\pm0.14}}         
\newcommand{\hatcurISOMHeccenxxxxxD}{\ensuremath{3.49\pm0.14}}         
\newcommand{\hatcurISOMKeccenxxxxxD}{\ensuremath{3.42\pm0.14}}         
\newcommand{\hatcurISOJKeccenxxxxxD}{\ensuremath{0.24\pm0.24}}         
\newcommand{\hatcurISOspececcenxxxxxD}{G}                              
\newcommand{\hatcurRVKeccenxxxxxD}{\ensuremath{151\pm12}}              
\newcommand{\hatcurRVrkeccenxxxxxD}{\ensuremath{0.09\pm0.11}}          
\newcommand{\hatcurRVrheccenxxxxxD}{\ensuremath{-0.06\pm0.14}}         
\newcommand{\hatcurRVkeccenxxxxxD}{\ensuremath{0.015_{-0.018}^{+0.031}}} 
\newcommand{\hatcurRVheccenxxxxxD}{\ensuremath{-0.008_{-0.037}^{+0.025}}} 
\newcommand{\hatcurRVtroneeccenxxxxxD}{\ensuremath{0\pm0}}             
\newcommand{\hatcurRVtrtwoeccenxxxxxD}{\ensuremath{0\pm0}}             
\newcommand{\hatcurRVgammaeccenxxxxxD}{\ensuremath{17733.3\pm7.5}}     
\newcommand{\hatcurRVjittereccenxxxxxD}{\ensuremath{0\pm18}}           
\newcommand{\hatcurRVjittertwosiglimeccenxxxxxD}{\ensuremath{<33.8}}   
\newcommand{\hatcurRVfitrmseccenxxxxxD}{\ensuremath{.1fym}}            %
\newcommand{\hatcurRVecceneccenxxxxxD}{\ensuremath{0.036\pm0.041}}     
\newcommand{\hatcurRVeccentwosiglimeccenxxxxxD}{\ensuremath{<0.108}}   
\newcommand{\hatcurRVomegaeccenxxxxxD}{\ensuremath{270\pm140}}         
\newcommand{\hatcurPPieccenxxxxxD}{\ensuremath{82.32\pm0.87}}          
\newcommand{\hatcurPPgeccenxxxxxD}{\ensuremath{10.7_{-1.7}^{+2.4}}}    
\newcommand{\hatcurPPloggeccenxxxxxD}{\ensuremath{3.028\pm0.087}}      
\newcommand{\hatcurPPareccenxxxxxD}{\ensuremath{6.98\pm0.41}}          
\newcommand{\hatcurPPareleccenxxxxxD}{\ensuremath{0.03168\pm0.00034}}  
\newcommand{\hatcurPPrhoeccenxxxxxD}{\ensuremath{0.363_{-0.081}^{+0.128}}} 
\newcommand{\hatcurPPmeccenxxxxxD}{\ensuremath{0.930\pm0.075}}         
\newcommand{\hatcurPPmshorteccenxxxxxD}{\ensuremath{0.93}}             
\newcommand{\hatcurPPmlongeccenxxxxxD}{\ensuremath{0.930\pm0.075}}     
\newcommand{\hatcurPPmeeccenxxxxxD}{\ensuremath{295\pm24}}             
\newcommand{\hatcurPPmeshorteccenxxxxxD}{\ensuremath{295.5}}           
\newcommand{\hatcurPPmelongeccenxxxxxD}{\ensuremath{295\pm24}}         
\newcommand{\hatcurPPreccenxxxxxD}{\ensuremath{1.47\pm0.14}}           
\newcommand{\hatcurPPrshorteccenxxxxxD}{\ensuremath{1.47}}             
\newcommand{\hatcurPPrlongeccenxxxxxD}{\ensuremath{1.47\pm0.14}}       
\newcommand{\hatcurPPreeccenxxxxxD}{\ensuremath{16.4\pm1.5}}           
\newcommand{\hatcurPPreshorteccenxxxxxD}{\ensuremath{16.4}}            
\newcommand{\hatcurPPrelongeccenxxxxxD}{\ensuremath{16.4\pm1.5}}       
\newcommand{\hatcurPPmrcorreccenxxxxxD}{\ensuremath{0.03}}             
\newcommand{\hatcurPPteffeccenxxxxxD}{\ensuremath{1442\pm51}}          
\newcommand{\hatcurPPthetaeccenxxxxxD}{\ensuremath{0.0414\pm0.0050}}   
\newcommand{\hatcurPPfluxperieccenxxxxxD}{\ensuremath{1.05\pm0.33}}    
\newcommand{\hatcurPPfluxperidimeccenxxxxxD}{\ensuremath{9}}           
\newcommand{\hatcurPPfluxapeccenxxxxxD}{\ensuremath{9.1\pm1.3}}        
\newcommand{\hatcurPPfluxapdimeccenxxxxxD}{\ensuremath{8}}             
\newcommand{\hatcurPPfluxavgeccenxxxxxD}{\ensuremath{9.8\pm1.5}}       
\newcommand{\hatcurPPfluxavgdimeccenxxxxxD}{\ensuremath{8}}            
\newcommand{\hatcurPPfluxavglogeccenxxxxxD}{\ensuremath{8.990\pm0.060}} 
\newcommand{\hatcurXsecphaseeccenxxxxxD}{\ensuremath{0.509\pm0.023}}   
\newcommand{\hatcurXsecondaryeccenxxxxxD}{\ensuremath{2456690.690\pm0.048}} 
\newcommand{\hatcurXsecdureccenxxxxxD}{\ensuremath{0.0751\pm0.0074}}   
\newcommand{\hatcurXsecingdureccenxxxxxD}{\ensuremath{0.0375\pm0.0060}} 
\newcommand{\hatcurPPphiconjeccenxxxxxD}{\ensuremath{0.23_{-0.45}^{+0.18}}} 
\newcommand{\hatcurPPperieccenxxxxxD}{\ensuremath{2456689.14\pm0.63}}  
\newcommand{\hatcurPPaequiveccenxxxxxD}{\ensuremath{0.0374\pm0.0025}}  
\newcommand{\hatcurPPtcirceccenxxxxxD}{\ensuremath{10.0_{-3.5}^{+6.3}}} 
\newcommand{\hatcurPPtinfalleccenxxxxxD}{\ensuremath{338_{-83}^{+132}}} 
\newcommand{\hatcurXdisteccenxxxxxD}{\ensuremath{554\pm36}}            
\newcommand{\hatcurXAveccenxxxxxD}{\ensuremath{0.0000\pm0.0061}}       
\newcommand{\hatcurXdistredeccenxxxxxD}{\ensuremath{529\pm39}}         
\newcommand{\hatcurXEBVeccenxxxxxD}{\ensuremath{0.0000\pm0.0020}}      
\newcommand{\hatcurXmvisoredeccenxxxxxD}{\ensuremath{13.887\pm0.020}}  
\newcommand{\hatcurXmiisoredeccenxxxxxD}{\ensuremath{13.0670\pm0.0087}} 
\newcommand{\hatcurXmjisoredeccenxxxxxD}{\ensuremath{12.525\pm0.024}}  
\newcommand{\hatcurXmhisoredeccenxxxxxD}{\ensuremath{12.108\pm0.038}}  
\newcommand{\hatcurXmkisoredeccenxxxxxD}{\ensuremath{12.038\pm0.040}}  
\newcommand{\hatcurXviisoredeccenxxxxxD}{\ensuremath{0.821\pm0.018}}   
\newcommand{\hatcurXvkisoredeccenxxxxxD}{\ensuremath{1.849\pm0.058}}   
\newcommand{\hatcurXjhisoredeccenxxxxxD}{\ensuremath{0.417\pm0.016}}   
\newcommand{\hatcurXjkisoredeccenxxxxxD}{\ensuremath{0.486\pm0.018}}   
\newcommand{\hatcurCCpmraeccenxxxxxD}{\ensuremath{2.1\pm1.4}}          
\newcommand{\hatcurCCpmdececcenxxxxxD}{\ensuremath{4.7\pm1.5}}         
\newcommand{\hatcurCCpmeccenxxxxxD}{\ensuremath{5.1\pm2.1}}            

%% file: HATS779-003.eccen.tex
\newcommand{\hatcurhtreccenxxxxxE}{HATS779-003}                      
\newcommand{\hatcurfieldeccenxxxxxE}{\ensuremath{string}}            
\newcommand{\hatcurCCraeccenxxxxxE}{\ensuremath{19^{\mathrm h}46^{\mathrm m}45.12{\mathrm s}}}                     
\newcommand{\hatcurCCdececcenxxxxxE}{\ensuremath{-63{\arcdeg}33{\arcmin}56.2{\arcsec}}}                    
\newcommand{\hatcurCCmageccenxxxxxE}{12.56}                          
\newcommand{\hatcurCCtwomasseccenxxxxxE}{2MASS~19464518-6333561}     
\newcommand{\hatcurCCgsceccenxxxxxE}{GSC~9089-00775}                 
\newcommand{\hatcurCCtassmveccenxxxxxE}{\ensuremath{12.56\pm0.10}}   
\newcommand{\hatcurCCtassmvshorteccenxxxxxE}{\ensuremath{12.6}}      
\newcommand{\hatcurCCtassmBeccenxxxxxE}{\ensuremath{13.26\pm0.10}}   
\newcommand{\hatcurCCtassmBshorteccenxxxxxE}{\ensuremath{13.3}}      
\newcommand{\hatcurCCtassmIeccenxxxxxE}{\ensuremath{nff\pmnff}}      
\newcommand{\hatcurCCtassmIshorteccenxxxxxE}{\ensuremath{0.0}}       
\newcommand{\hatcurCCtassmgeccenxxxxxE}{\ensuremath{nff\pmnff}}      
\newcommand{\hatcurCCtassmgshorteccenxxxxxE}{\ensuremath{0.0}}       
\newcommand{\hatcurCCtassmreccenxxxxxE}{\ensuremath{nff\pmnff}}      
\newcommand{\hatcurCCtassmrshorteccenxxxxxE}{\ensuremath{0.0}}       
\newcommand{\hatcurCCtassmieccenxxxxxE}{\ensuremath{nff\pmnff}}      
\newcommand{\hatcurCCtassmishorteccenxxxxxE}{\ensuremath{0.0}}       
\newcommand{\hatcurCCtwomassJmageccenxxxxxE}{\ensuremath{11.439\pm0.025}} 
\newcommand{\hatcurCCtwomassHmageccenxxxxxE}{\ensuremath{11.192\pm0.023}} 
\newcommand{\hatcurCCtwomassKmageccenxxxxxE}{\ensuremath{11.118\pm0.025}} 
\newcommand{\hatcurCCcitJmageccenxxxxxE}{\ensuremath{11.458\pm0.026}} 
\newcommand{\hatcurCCcitHmageccenxxxxxE}{\ensuremath{11.187\pm0.023}} 
\newcommand{\hatcurCCcitKmageccenxxxxxE}{\ensuremath{11.142\pm0.025}} 
\newcommand{\hatcurCCbbJmageccenxxxxxE}{\ensuremath{11.504\pm0.026}} 
\newcommand{\hatcurCCbbHmageccenxxxxxE}{\ensuremath{11.208\pm0.024}} 
\newcommand{\hatcurCCbbKmageccenxxxxxE}{\ensuremath{11.162\pm0.025}} 
\newcommand{\hatcurCCesoJmageccenxxxxxE}{\ensuremath{11.505\pm0.027}} 
\newcommand{\hatcurCCesoHmageccenxxxxxE}{\ensuremath{11.203\pm0.027}} 
\newcommand{\hatcurCCesoKmageccenxxxxxE}{\ensuremath{11.161\pm0.026}} 
\newcommand{\hatcurCCesoJHmageccenxxxxxE}{\ensuremath{0.302\pm0.037}} 
\newcommand{\hatcurCCesoJKmageccenxxxxxE}{\ensuremath{0.3440\pm0.0090}} 
\newcommand{\hatcurCCesoHKmageccenxxxxxE}{\ensuremath{0.042\pm0.038}} 
\newcommand{\hatcurLCdipeccenxxxxxE}{\ensuremath{13.1}}              
\newcommand{\hatcurLCrprstareccenxxxxxE}{\ensuremath{0.1047\pm0.0012}} 
\newcommand{\hatcurLCbsqeccenxxxxxE}{\ensuremath{0.074_{-0.053}^{+0.041}}} 
\newcommand{\hatcurLCimpeccenxxxxxE}{\ensuremath{0.272_{-0.128}^{+0.068}}} 
\newcommand{\hatcurLCzetaeccenxxxxxE}{\ensuremath{17.19\pm0.16}}     
\newcommand{\hatcurLCdureccenxxxxxE}{\ensuremath{0.1295\pm0.0014}}   
\newcommand{\hatcurLCdurshorteccenxxxxxE}{\ensuremath{0.1295}}       
\newcommand{\hatcurLCdurhreccenxxxxxE}{\ensuremath{3.107\pm0.033}}   
\newcommand{\hatcurLCdurhrshorteccenxxxxxE}{\ensuremath{3.107}}      
\newcommand{\hatcurLCqeccenxxxxxE}{\ensuremath{0.07110\pm0.00076}}   
\newcommand{\hatcurLCqshorteccenxxxxxE}{\ensuremath{0.071}}          
\newcommand{\hatcurLCingdureccenxxxxxE}{\ensuremath{0.01313\pm0.00081}} 
\newcommand{\hatcurLCPeccenxxxxxE}{\ensuremath{1.8209993\pm0.0000015}} 
\newcommand{\hatcurLCPprececcenxxxxxE}{\ensuremath{1.8209993}}       
\newcommand{\hatcurLCPshorteccenxxxxxE}{\ensuremath{1.8210}}         
\newcommand{\hatcurLCTeccenxxxxxE}{\ensuremath{2456981.80210\pm0.00044}} 
\newcommand{\hatcurLCTAeccenxxxxxE}{\ensuremath{2455679.7876\pm0.0012}} 
\newcommand{\hatcurLCTBeccenxxxxxE}{\ensuremath{2457227.63697\pm0.00049}} 
\newcommand{\hatcurLCrhoeccenxxxxxE}{\ensuremath{0.50\pm0.20}}       
\newcommand{\hatcurSMEiteffeccenxxxxxE}{\ensuremath{6250\pm160}}     
\newcommand{\hatcurSMEizfeheccenxxxxxE}{\ensuremath{0.180\pm0.090}}  
\newcommand{\hatcurSMEizfehshorteccenxxxxxE}{\ensuremath{0.18}}      
\newcommand{\hatcurSMEiloggeccenxxxxxE}{\ensuremath{4.20\pm0.23}}    
\newcommand{\hatcurSMEivsineccenxxxxxE}{\ensuremath{8.66\pm0.32}}    
\newcommand{\hatcurSMEivmaceccenxxxxxE}{\ensuremath{0.0}}            
\newcommand{\hatcurSMEivmiceccenxxxxxE}{\ensuremath{0.0}}            
\newcommand{\hatcurSMEiiteffeccenxxxxxE}{\ensuremath{6300\pm100}}    
\newcommand{\hatcurSMEiizfeheccenxxxxxE}{\ensuremath{0.210\pm0.060}} 
\newcommand{\hatcurSMEiizfehshorteccenxxxxxE}{\ensuremath{0.21}}     
\newcommand{\hatcurSMEiiloggeccenxxxxxE}{\ensuremath{4.241\pm0.020}} 
\newcommand{\hatcurSMEiivsineccenxxxxxE}{\ensuremath{8.66\pm0.34}}   
\newcommand{\hatcurLBizeccenxxxxxE}{\ensuremath{0.1430}}             
\newcommand{\hatcurLBiizeccenxxxxxE}{\ensuremath{0.3653}}            
\newcommand{\hatcurLBiieccenxxxxxE}{\ensuremath{0.1962}}             
\newcommand{\hatcurLBiiieccenxxxxxE}{\ensuremath{0.3739}}            
\newcommand{\hatcurLBiIeccenxxxxxE}{\ensuremath{0.1770}}             
\newcommand{\hatcurLBiiIeccenxxxxxE}{\ensuremath{0.3724}}            
\newcommand{\hatcurLBigeccenxxxxxE}{\ensuremath{0.4494}}             
\newcommand{\hatcurLBiigeccenxxxxxE}{\ensuremath{0.3131}}            
\newcommand{\hatcurLBireccenxxxxxE}{\ensuremath{0.2746}}             
\newcommand{\hatcurLBiireccenxxxxxE}{\ensuremath{0.3780}}            
\newcommand{\hatcurLBiReccenxxxxxE}{\ensuremath{0.2524}}             
\newcommand{\hatcurLBiiReccenxxxxxE}{\ensuremath{0.3784}}            
\newcommand{\hatcurLBikepeccenxxxxxE}{\ensuremath{0.1000}}           
\newcommand{\hatcurLBiikepeccenxxxxxE}{\ensuremath{0.1000}}          
\newcommand{\hatcurISOmeccenxxxxxE}{\ensuremath{1.358_{-0.062}^{+0.082}}} 
\newcommand{\hatcurISOmshorteccenxxxxxE}{\ensuremath{1.36}}          
\newcommand{\hatcurISOmlongeccenxxxxxE}{\ensuremath{1.358_{-0.062}^{+0.082}}} 
\newcommand{\hatcurISOreccenxxxxxE}{\ensuremath{1.56_{-0.15}^{+0.23}}} 
\newcommand{\hatcurISOrshorteccenxxxxxE}{\ensuremath{1.56}}          
\newcommand{\hatcurISOrlongeccenxxxxxE}{\ensuremath{1.56_{-0.15}^{+0.23}}} 
\newcommand{\hatcurISOrhoeccenxxxxxE}{\ensuremath{0.50\pm0.16}}      
\newcommand{\hatcurISOrholongeccenxxxxxE}{\ensuremath{0.50\pm0.16}}  
\newcommand{\hatcurISOloggeccenxxxxxE}{\ensuremath{4.181\pm0.089}}   
\newcommand{\hatcurISOlumeccenxxxxxE}{\ensuremath{3.45_{-0.72}^{+1.16}}} 
\newcommand{\hatcurISOlumshorteccenxxxxxE}{\ensuremath{3.45}}        
\newcommand{\hatcurISOmveccenxxxxxE}{\ensuremath{3.40\pm0.29}}       
\newcommand{\hatcurISOvieccenxxxxxE}{\ensuremath{0.550\pm0.027}}     
\newcommand{\hatcurISOageeccenxxxxxE}{\ensuremath{2.30\pm0.56}}      
\newcommand{\hatcurISOsigmaeccenxxxxxE}{\ensuremath{0.00070\pm0.00020}} 
\newcommand{\hatcurISOMJeccenxxxxxE}{\ensuremath{2.50\pm0.28}}       
\newcommand{\hatcurISOMHeccenxxxxxE}{\ensuremath{2.26\pm0.27}}       
\newcommand{\hatcurISOMKeccenxxxxxE}{\ensuremath{2.21\pm0.27}}       
\newcommand{\hatcurISOJKeccenxxxxxE}{\ensuremath{0.290\pm0.020}}     
\newcommand{\hatcurISOspececcenxxxxxE}{F}                            
\newcommand{\hatcurRVKeccenxxxxxE}{\ensuremath{185.2\pm4.8}}         
\newcommand{\hatcurRVrkeccenxxxxxE}{\ensuremath{0.274_{-0.116}^{+0.066}}} 
\newcommand{\hatcurRVrheccenxxxxxE}{\ensuremath{0.20_{-0.29}^{+0.22}}} 
\newcommand{\hatcurRVkeccenxxxxxE}{\ensuremath{0.099\pm0.050}}       
\newcommand{\hatcurRVheccenxxxxxE}{\ensuremath{0.066_{-0.089}^{+0.124}}} 
\newcommand{\hatcurRVtroneeccenxxxxxE}{\ensuremath{0\pm0}}           
\newcommand{\hatcurRVtrtwoeccenxxxxxE}{\ensuremath{0\pm0}}           
\newcommand{\hatcurRVgammaAeccenxxxxxE}{\ensuremath{-14182\pm10}}    
\newcommand{\hatcurRVjitterAeccenxxxxxE}{\ensuremath{0.3\pm8.0}}     
\newcommand{\hatcurRVjittertwosiglimAeccenxxxxxE}{\ensuremath{<20.8}} 
\newcommand{\hatcurRVfitrmsAeccenxxxxxE}{\ensuremath{0.0}}           
\newcommand{\hatcurRVgammaBeccenxxxxxE}{\ensuremath{-14227\pm56}}    
\newcommand{\hatcurRVjitterBeccenxxxxxE}{\ensuremath{3\pm96}}        
\newcommand{\hatcurRVjittertwosiglimBeccenxxxxxE}{\ensuremath{<200.3}} 
\newcommand{\hatcurRVfitrmsBeccenxxxxxE}{\ensuremath{0.0}}           
\newcommand{\hatcurRVgammaCeccenxxxxxE}{\ensuremath{-14241\pm53}}    
\newcommand{\hatcurRVjitterCeccenxxxxxE}{\ensuremath{15\pm100}}      
\newcommand{\hatcurRVjittertwosiglimCeccenxxxxxE}{\ensuremath{<245.8}} 
\newcommand{\hatcurRVfitrmsCeccenxxxxxE}{\ensuremath{0.0}}           
\newcommand{\hatcurRVecceneccenxxxxxE}{\ensuremath{0.140\pm0.081}}   
\newcommand{\hatcurRVeccentwosiglimeccenxxxxxE}{\ensuremath{<0.306}} 
\newcommand{\hatcurRVomegaeccenxxxxxE}{\ensuremath{50\pm130}}        
\newcommand{\hatcurPPieccenxxxxxE}{\ensuremath{86.3\pm1.7}}          
\newcommand{\hatcurPPgeccenxxxxxE}{\ensuremath{13.1\pm3.0}}          
\newcommand{\hatcurPPloggeccenxxxxxE}{\ensuremath{3.118_{-0.110}^{+0.078}}} 
\newcommand{\hatcurPPareccenxxxxxE}{\ensuremath{4.43\pm0.48}}        
\newcommand{\hatcurPPareleccenxxxxxE}{\ensuremath{0.03232\pm0.00058}} 
\newcommand{\hatcurPPrhoeccenxxxxxE}{\ensuremath{0.41\pm0.15}}       
\newcommand{\hatcurPPmeccenxxxxxE}{\ensuremath{1.350\pm0.053}}       
\newcommand{\hatcurPPmshorteccenxxxxxE}{\ensuremath{1.35}}           
\newcommand{\hatcurPPmlongeccenxxxxxE}{\ensuremath{1.350\pm0.053}}   
\newcommand{\hatcurPPmeeccenxxxxxE}{\ensuremath{429\pm17}}           
\newcommand{\hatcurPPmeshorteccenxxxxxE}{\ensuremath{429.1}}         
\newcommand{\hatcurPPmelongeccenxxxxxE}{\ensuremath{429\pm17}}       
\newcommand{\hatcurPPreccenxxxxxE}{\ensuremath{1.60_{-0.16}^{+0.24}}} 
\newcommand{\hatcurPPrshorteccenxxxxxE}{\ensuremath{1.60}}           
\newcommand{\hatcurPPrlongeccenxxxxxE}{\ensuremath{1.60_{-0.16}^{+0.24}}} 
\newcommand{\hatcurPPreeccenxxxxxE}{\ensuremath{17.9_{-1.7}^{+2.7}}} 
\newcommand{\hatcurPPreshorteccenxxxxxE}{\ensuremath{17.9}}          
\newcommand{\hatcurPPrelongeccenxxxxxE}{\ensuremath{17.9_{-1.7}^{+2.7}}} 
\newcommand{\hatcurPPmrcorreccenxxxxxE}{\ensuremath{0.80}}           
\newcommand{\hatcurPPteffeccenxxxxxE}{\ensuremath{2120_{-110}^{+150}}} 
\newcommand{\hatcurPPthetaeccenxxxxxE}{\ensuremath{0.0400\pm0.0052}} 
\newcommand{\hatcurPPfluxperieccenxxxxxE}{\ensuremath{5.9_{-1.4}^{+3.7}}} 
\newcommand{\hatcurPPfluxperidimeccenxxxxxE}{\ensuremath{9}}         
\newcommand{\hatcurPPfluxapeccenxxxxxE}{\ensuremath{3.55\pm0.54}}    
\newcommand{\hatcurPPfluxapdimeccenxxxxxE}{\ensuremath{9}}           
\newcommand{\hatcurPPfluxavgeccenxxxxxE}{\ensuremath{4.56_{-0.85}^{+1.39}}} 
\newcommand{\hatcurPPfluxavgdimeccenxxxxxE}{\ensuremath{9}}          
\newcommand{\hatcurPPfluxavglogeccenxxxxxE}{\ensuremath{9.66\pm0.11}} 
\newcommand{\hatcurXsecphaseeccenxxxxxE}{\ensuremath{0.563\pm0.033}} 
\newcommand{\hatcurXsecondaryeccenxxxxxE}{\ensuremath{2456982.828\pm0.059}} 
\newcommand{\hatcurXsecdureccenxxxxxE}{\ensuremath{0.146\pm0.033}}   
\newcommand{\hatcurXsecingdureccenxxxxxE}{\ensuremath{0.0154\pm0.0046}} 
\newcommand{\hatcurPPphiconjeccenxxxxxE}{\ensuremath{0.125_{-0.091}^{+0.136}}} 
\newcommand{\hatcurPPperieccenxxxxxE}{\ensuremath{2456981.58\pm0.21}} 
\newcommand{\hatcurPPaequiveccenxxxxxE}{\ensuremath{0.0174\pm0.0020}} 
\newcommand{\hatcurPPtcirceccenxxxxxE}{\ensuremath{6.0\pm3.9}}       
\newcommand{\hatcurPPtinfalleccenxxxxxE}{\ensuremath{30\pm17}}       
\newcommand{\hatcurXdisteccenxxxxxE}{\ensuremath{616_{-62}^{+91}}}   
\newcommand{\hatcurXAveccenxxxxxE}{\ensuremath{0.26\pm0.14}}         
\newcommand{\hatcurXdistredeccenxxxxxE}{\ensuremath{607_{-60}^{+90}}} 
\newcommand{\hatcurXEBVeccenxxxxxE}{\ensuremath{0.082\pm0.045}}      
\newcommand{\hatcurXmvisoredeccenxxxxxE}{\ensuremath{12.572\pm0.085}} 
\newcommand{\hatcurXmiisoredeccenxxxxxE}{\ensuremath{11.888\pm0.035}} 
\newcommand{\hatcurXmjisoredeccenxxxxxE}{\ensuremath{11.493\pm0.018}} 
\newcommand{\hatcurXmhisoredeccenxxxxxE}{\ensuremath{11.220\pm0.016}} 
\newcommand{\hatcurXmkisoredeccenxxxxxE}{\ensuremath{11.155\pm0.018}} 
\newcommand{\hatcurXviisoredeccenxxxxxE}{\ensuremath{0.684\pm0.059}} 
\newcommand{\hatcurXvkisoredeccenxxxxxE}{\ensuremath{1.416\pm0.094}} 
\newcommand{\hatcurXjhisoredeccenxxxxxE}{\ensuremath{0.273\pm0.015}} 
\newcommand{\hatcurXjkisoredeccenxxxxxE}{\ensuremath{0.337\pm0.019}} 
\newcommand{\hatcurCCpmraeccenxxxxxE}{\ensuremath{16.7\pm1.3}}       
\newcommand{\hatcurCCpmdececcenxxxxxE}{\ensuremath{-12.1\pm1.3}}     
\newcommand{\hatcurCCpmeccenxxxxxE}{\ensuremath{20.6\pm1.8}}         

%% file: newcommand_spec_HATS566-004.tex
\newcommand{\hatcurxxxxxA}{HATS-31}
\newcommand{\hatcurbxxxxxA}{HATS-31b}
\newcommand{\hatcurcxxxxxA}{HATS-31c}

\newcommand{\hatcurplanetnumxxxxxA}{31}

\newcommand{\hatcurRVgammaabsxxxxxA}{\hatcurRVgamma{\hatcurplanetnumxxxxxA}}                           

\newcommand{\hatcurRVgammaabsinstxxxxxA}{HARPS}                           

\newcommand{\hatcurRVgammarelxxxxxA}{\hatcurRVgamma{\hatcurplanetnumxxxxxA}}                           

\newcommand{\hatcurCCtassvixxxxxA}{NULL}                  

\newcommand{\hatcurSMEversionxxxxxA}{ii}                                       

\newcommand{\hatcurisoshortxxxxxA}{YY}
\newcommand{\hatcurisofullxxxxxA}{Yonsei-Yale (YY)}
\newcommand{\hatcurisocitexxxxxA}{yi:2001}

\newcommand{\hatcurlumindxxxxxA}{\arstar}

\newcommand{\hatcurjhkfilsetxxxxxA}{ESO}

\newcommand{\hatcurSMEteffxxxxxA}{\ifthenelse{\equal{\hatcurSMEversionxxxxxA}{i}}{\hatcurSMEiteff{\hatcurplanetnumxxxxxA}}{\hatcurSMEiiteff{\hatcurplanetnumxxxxxA}}}
\newcommand{\hatcurSMEzfehxxxxxA}{\ifthenelse{\equal{\hatcurSMEversionxxxxxA}{i}}{\hatcurSMEizfeh{\hatcurplanetnumxxxxxA}}{\hatcurSMEiizfeh{\hatcurplanetnumxxxxxA}}}
\newcommand{\hatcurSMEzfehshortxxxxxA}{\ifthenelse{\equal{\hatcurSMEversionxxxxxA}{i}}{\hatcurSMEizfehshort{\hatcurplanetnumxxxxxA}}{\hatcurSMEiizfehshort{\hatcurplanetnumxxxxxA}}}
\newcommand{\hatcurSMEloggxxxxxA}{\ifthenelse{\equal{\hatcurSMEversionxxxxxA}{i}}{\hatcurSMEilogg{\hatcurplanetnumxxxxxA}}{\hatcurSMEiilogg{\hatcurplanetnumxxxxxA}}}
\newcommand{\hatcurSMEvsinxxxxxA}{\ifthenelse{\equal{\hatcurSMEversionxxxxxA}{i}}{\hatcurSMEivsin{\hatcurplanetnumxxxxxA}}{\hatcurSMEiivsin{\hatcurplanetnumxxxxxA}}}
\newcommand{\hatcurSMEvmacxxxxxA}{\ifthenelse{\equal{\hatcurSMEversionxxxxxA}{i}}{\hatcurSMEivmac{\hatcurplanetnumxxxxxA}}{\hatcurSMEiivmac{\hatcurplanetnumxxxxxA}}}
\newcommand{\hatcurSMEvmicxxxxxA}{\ifthenelse{\equal{\hatcurSMEversionxxxxxA}{i}}{\hatcurSMEivmic{\hatcurplanetnumxxxxxA}}{\hatcurSMEiivmic{\hatcurplanetnumxxxxxA}}}

%% file: newcommand_spec_HATS586-004.tex
\newcommand{\hatcurxxxxxB}{HATS-32}
\newcommand{\hatcurbxxxxxB}{HATS-32b}
\newcommand{\hatcurcxxxxxB}{HATS-32c}

\newcommand{\hatcurplanetnumxxxxxB}{32}

\newcommand{\hatcurRVgammaabsxxxxxB}{\hatcurRVgamma{\hatcurplanetnumxxxxxB}}                           

\newcommand{\hatcurRVgammaabsinstxxxxxB}{FEROS}                           

\newcommand{\hatcurRVgammarelxxxxxB}{\hatcurRVgamma{\hatcurplanetnumxxxxxB}}                           

\newcommand{\hatcurCCtassvixxxxxB}{NULL}                  

\newcommand{\hatcurSMEversionxxxxxB}{ii}                                       

\newcommand{\hatcurisoshortxxxxxB}{YY}
\newcommand{\hatcurisofullxxxxxB}{Yonsei-Yale (YY)}
\newcommand{\hatcurisocitexxxxxB}{yi:2001}

\newcommand{\hatcurlumindxxxxxB}{\arstar}

\newcommand{\hatcurjhkfilsetxxxxxB}{ESO}

\newcommand{\hatcurSMEteffxxxxxB}{\ifthenelse{\equal{\hatcurSMEversionxxxxxB}{i}}{\hatcurSMEiteff{\hatcurplanetnumxxxxxB}}{\hatcurSMEiiteff{\hatcurplanetnumxxxxxB}}}
\newcommand{\hatcurSMEzfehxxxxxB}{\ifthenelse{\equal{\hatcurSMEversionxxxxxB}{i}}{\hatcurSMEizfeh{\hatcurplanetnumxxxxxB}}{\hatcurSMEiizfeh{\hatcurplanetnumxxxxxB}}}
\newcommand{\hatcurSMEzfehshortxxxxxB}{\ifthenelse{\equal{\hatcurSMEversionxxxxxB}{i}}{\hatcurSMEizfehshort{\hatcurplanetnumxxxxxB}}{\hatcurSMEiizfehshort{\hatcurplanetnumxxxxxB}}}
\newcommand{\hatcurSMEloggxxxxxB}{\ifthenelse{\equal{\hatcurSMEversionxxxxxB}{i}}{\hatcurSMEilogg{\hatcurplanetnumxxxxxB}}{\hatcurSMEiilogg{\hatcurplanetnumxxxxxB}}}
\newcommand{\hatcurSMEvsinxxxxxB}{\ifthenelse{\equal{\hatcurSMEversionxxxxxB}{i}}{\hatcurSMEivsin{\hatcurplanetnumxxxxxB}}{\hatcurSMEiivsin{\hatcurplanetnumxxxxxB}}}
\newcommand{\hatcurSMEvmacxxxxxB}{\ifthenelse{\equal{\hatcurSMEversionxxxxxB}{i}}{\hatcurSMEivmac{\hatcurplanetnumxxxxxB}}{\hatcurSMEiivmac{\hatcurplanetnumxxxxxB}}}
\newcommand{\hatcurSMEvmicxxxxxB}{\ifthenelse{\equal{\hatcurSMEversionxxxxxB}{i}}{\hatcurSMEivmic{\hatcurplanetnumxxxxxB}}{\hatcurSMEiivmic{\hatcurplanetnumxxxxxB}}}

%% file: newcommand_spec_HATS747-032.tex
\newcommand{\hatcurxxxxxC}{HATS-33}
\newcommand{\hatcurbxxxxxC}{HATS-33b}
\newcommand{\hatcurcxxxxxC}{HATS-33c}

\newcommand{\hatcurplanetnumxxxxxC}{33}

\newcommand{\hatcurRVgammaabsxxxxxC}{\hatcurRVgammaC{\hatcurplanetnumxxxxxC}}                           

\newcommand{\hatcurRVgammaabsinstxxxxxC}{HARPS}                           

\newcommand{\hatcurRVgammarelxxxxxC}{\hatcurRVgamma{\hatcurplanetnumxxxxxC}}                           

\newcommand{\hatcurCCtassvixxxxxC}{NULL}                  

\newcommand{\hatcurSMEversionxxxxxC}{ii}                                       

\newcommand{\hatcurisoshortxxxxxC}{YY}
\newcommand{\hatcurisofullxxxxxC}{Yonsei-Yale (YY)}
\newcommand{\hatcurisocitexxxxxC}{yi:2001}

\newcommand{\hatcurlumindxxxxxC}{\arstar}

\newcommand{\hatcurjhkfilsetxxxxxC}{ESO}

\newcommand{\hatcurSMEteffxxxxxC}{\ifthenelse{\equal{\hatcurSMEversionxxxxxC}{i}}{\hatcurSMEiteff{\hatcurplanetnumxxxxxC}}{\hatcurSMEiiteff{\hatcurplanetnumxxxxxC}}}
\newcommand{\hatcurSMEzfehxxxxxC}{\ifthenelse{\equal{\hatcurSMEversionxxxxxC}{i}}{\hatcurSMEizfeh{\hatcurplanetnumxxxxxC}}{\hatcurSMEiizfeh{\hatcurplanetnumxxxxxC}}}
\newcommand{\hatcurSMEzfehshortxxxxxC}{\ifthenelse{\equal{\hatcurSMEversionxxxxxC}{i}}{\hatcurSMEizfehshort{\hatcurplanetnumxxxxxC}}{\hatcurSMEiizfehshort{\hatcurplanetnumxxxxxC}}}
\newcommand{\hatcurSMEloggxxxxxC}{\ifthenelse{\equal{\hatcurSMEversionxxxxxC}{i}}{\hatcurSMEilogg{\hatcurplanetnumxxxxxC}}{\hatcurSMEiilogg{\hatcurplanetnumxxxxxC}}}
\newcommand{\hatcurSMEvsinxxxxxC}{\ifthenelse{\equal{\hatcurSMEversionxxxxxC}{i}}{\hatcurSMEivsin{\hatcurplanetnumxxxxxC}}{\hatcurSMEiivsin{\hatcurplanetnumxxxxxC}}}
\newcommand{\hatcurSMEvmacxxxxxC}{\ifthenelse{\equal{\hatcurSMEversionxxxxxC}{i}}{\hatcurSMEivmac{\hatcurplanetnumxxxxxC}}{\hatcurSMEiivmac{\hatcurplanetnumxxxxxC}}}
\newcommand{\hatcurSMEvmicxxxxxC}{\ifthenelse{\equal{\hatcurSMEversionxxxxxC}{i}}{\hatcurSMEivmic{\hatcurplanetnumxxxxxC}}{\hatcurSMEiivmic{\hatcurplanetnumxxxxxC}}}

%% file: newcommand_spec_HATS754-010.tex
\newcommand{\hatcurxxxxxD}{HATS-34}
\newcommand{\hatcurbxxxxxD}{HATS-34b}
\newcommand{\hatcurcxxxxxD}{HATS-34c}

\newcommand{\hatcurplanetnumxxxxxD}{34}

\newcommand{\hatcurRVgammaabsxxxxxD}{\hatcurRVgamma{\hatcurplanetnumxxxxxD}}                           

\newcommand{\hatcurRVgammaabsinstxxxxxD}{FEROS}                           

\newcommand{\hatcurRVgammarelxxxxxD}{\hatcurRVgamma{\hatcurplanetnumxxxxxD}}                           

\newcommand{\hatcurCCtassvixxxxxD}{NULL}                  

\newcommand{\hatcurSMEversionxxxxxD}{ii}                                       

\newcommand{\hatcurisoshortxxxxxD}{YY}
\newcommand{\hatcurisofullxxxxxD}{Yonsei-Yale (YY)}
\newcommand{\hatcurisocitexxxxxD}{yi:2001}

\newcommand{\hatcurlumindxxxxxD}{\arstar}

\newcommand{\hatcurjhkfilsetxxxxxD}{ESO}

\newcommand{\hatcurSMEteffxxxxxD}{\ifthenelse{\equal{\hatcurSMEversionxxxxxD}{i}}{\hatcurSMEiteff{\hatcurplanetnumxxxxxD}}{\hatcurSMEiiteff{\hatcurplanetnumxxxxxD}}}
\newcommand{\hatcurSMEzfehxxxxxD}{\ifthenelse{\equal{\hatcurSMEversionxxxxxD}{i}}{\hatcurSMEizfeh{\hatcurplanetnumxxxxxD}}{\hatcurSMEiizfeh{\hatcurplanetnumxxxxxD}}}
\newcommand{\hatcurSMEzfehshortxxxxxD}{\ifthenelse{\equal{\hatcurSMEversionxxxxxD}{i}}{\hatcurSMEizfehshort{\hatcurplanetnumxxxxxD}}{\hatcurSMEiizfehshort{\hatcurplanetnumxxxxxD}}}
\newcommand{\hatcurSMEloggxxxxxD}{\ifthenelse{\equal{\hatcurSMEversionxxxxxD}{i}}{\hatcurSMEilogg{\hatcurplanetnumxxxxxD}}{\hatcurSMEiilogg{\hatcurplanetnumxxxxxD}}}
\newcommand{\hatcurSMEvsinxxxxxD}{\ifthenelse{\equal{\hatcurSMEversionxxxxxD}{i}}{\hatcurSMEivsin{\hatcurplanetnumxxxxxD}}{\hatcurSMEiivsin{\hatcurplanetnumxxxxxD}}}
\newcommand{\hatcurSMEvmacxxxxxD}{\ifthenelse{\equal{\hatcurSMEversionxxxxxD}{i}}{\hatcurSMEivmac{\hatcurplanetnumxxxxxD}}{\hatcurSMEiivmac{\hatcurplanetnumxxxxxD}}}
\newcommand{\hatcurSMEvmicxxxxxD}{\ifthenelse{\equal{\hatcurSMEversionxxxxxD}{i}}{\hatcurSMEivmic{\hatcurplanetnumxxxxxD}}{\hatcurSMEiivmic{\hatcurplanetnumxxxxxD}}}

%% file: newcommand_spec_HATS779-003.tex
\newcommand{\hatcurxxxxxE}{HATS-35}
\newcommand{\hatcurbxxxxxE}{HATS-35b}
\newcommand{\hatcurcxxxxxE}{HATS-35c}

\newcommand{\hatcurplanetnumxxxxxE}{35}

\newcommand{\hatcurRVgammaabsxxxxxE}{\hatcurRVgammaA{\hatcurplanetnumxxxxxE}}                           

\newcommand{\hatcurRVgammaabsinstxxxxxE}{FEROS}                           

\newcommand{\hatcurRVgammarelxxxxxE}{\hatcurRVgammaA{\hatcurplanetnumxxxxxE}}                           

\newcommand{\hatcurCCtassvixxxxxE}{NULL}                  

\newcommand{\hatcurSMEversionxxxxxE}{ii}                                       

\newcommand{\hatcurisoshortxxxxxE}{YY}
\newcommand{\hatcurisofullxxxxxE}{Yonsei-Yale (YY)}
\newcommand{\hatcurisocitexxxxxE}{yi:2001}

\newcommand{\hatcurlumindxxxxxE}{\arstar}

\newcommand{\hatcurjhkfilsetxxxxxE}{ESO}

\newcommand{\hatcurSMEteffxxxxxE}{\ifthenelse{\equal{\hatcurSMEversionxxxxxE}{i}}{\hatcurSMEiteff{\hatcurplanetnumxxxxxE}}{\hatcurSMEiiteff{\hatcurplanetnumxxxxxE}}}
\newcommand{\hatcurSMEzfehxxxxxE}{\ifthenelse{\equal{\hatcurSMEversionxxxxxE}{i}}{\hatcurSMEizfeh{\hatcurplanetnumxxxxxE}}{\hatcurSMEiizfeh{\hatcurplanetnumxxxxxE}}}
\newcommand{\hatcurSMEzfehshortxxxxxE}{\ifthenelse{\equal{\hatcurSMEversionxxxxxE}{i}}{\hatcurSMEizfehshort{\hatcurplanetnumxxxxxE}}{\hatcurSMEiizfehshort{\hatcurplanetnumxxxxxE}}}
\newcommand{\hatcurSMEloggxxxxxE}{\ifthenelse{\equal{\hatcurSMEversionxxxxxE}{i}}{\hatcurSMEilogg{\hatcurplanetnumxxxxxE}}{\hatcurSMEiilogg{\hatcurplanetnumxxxxxE}}}
\newcommand{\hatcurSMEvsinxxxxxE}{\ifthenelse{\equal{\hatcurSMEversionxxxxxE}{i}}{\hatcurSMEivsin{\hatcurplanetnumxxxxxE}}{\hatcurSMEiivsin{\hatcurplanetnumxxxxxE}}}
\newcommand{\hatcurSMEvmacxxxxxE}{\ifthenelse{\equal{\hatcurSMEversionxxxxxE}{i}}{\hatcurSMEivmac{\hatcurplanetnumxxxxxE}}{\hatcurSMEiivmac{\hatcurplanetnumxxxxxE}}}
\newcommand{\hatcurSMEvmicxxxxxE}{\ifthenelse{\equal{\hatcurSMEversionxxxxxE}{i}}{\hatcurSMEivmic{\hatcurplanetnumxxxxxE}}{\hatcurSMEiivmic{\hatcurplanetnumxxxxxE}}}

%% file: data/phfu_tab_combined_short.tex
HATS-31 & $ 56441.63105 $ & $   0.00775 $ & $   0.00473 $ & $ \cdots $ & $ r$ &         HS\\
   HATS-31 & $ 56330.15845 $ & $  -0.00143 $ & $   0.00435 $ & $ \cdots $ & $ r$ &         HS\\
   HATS-31 & $ 56424.74146 $ & $   0.00333 $ & $   0.00446 $ & $ \cdots $ & $ r$ &         HS\\
   HATS-31 & $ 56448.38748 $ & $  -0.00568 $ & $   0.00425 $ & $ \cdots $ & $ r$ &         HS\\
   HATS-31 & $ 56357.18277 $ & $   0.01570 $ & $   0.00450 $ & $ \cdots $ & $ r$ &         HS\\
   HATS-31 & $ 56417.98607 $ & $  -0.00175 $ & $   0.00432 $ & $ \cdots $ & $ r$ &         HS\\
   HATS-31 & $ 56404.47434 $ & $  -0.00078 $ & $   0.00556 $ & $ \cdots $ & $ r$ &         HS\\
   HATS-31 & $ 56401.09654 $ & $  -0.00630 $ & $   0.00440 $ & $ \cdots $ & $ r$ &         HS\\
   HATS-31 & $ 56414.60840 $ & $   0.00526 $ & $   0.00417 $ & $ \cdots $ & $ r$ &         HS\\
   HATS-31 & $ 56390.96315 $ & $  -0.00847 $ & $   0.00457 $ & $ \cdots $ & $ r$ &         HS\\

%% file: hats31-35.bbl
\begin{thebibliography}{}
\expandafter\ifx\csname natexlab\endcsname\relax\def\natexlab#1{#1}\fi

\bibitem[{{Addison} {et~al.}(2013){Addison}, {Tinney}, {Wright}, {Bayliss},
  {Zhou}, {Hartman}, {Bakos}, \& {Schmidt}}]{2013ApJ...774L...9A}
{Addison}, B.~C., {Tinney}, C.~G., {Wright}, D.~J., {et~al.} 2013, \apjl, 774,
  L9

\bibitem[{{Bakos} {et~al.}(2004){Bakos}, {Noyes}, {Kov{\'a}cs}, {Stanek},
  {Sasselov}, \& {Domsa}}]{2004PASP..116..266B}
{Bakos}, G., {Noyes}, R.~W., {Kov{\'a}cs}, G., {et~al.} 2004, \pasp, 116, 266

\bibitem[{{Bakos} {et~al.}(2010){Bakos}, {Torres}, {P{\'a}l}, {Hartman},
  {Kov{\'a}cs}, {Noyes}, {Latham}, {Sasselov}, {Sip{\H o}cz}, {Esquerdo},
  {Fischer}, {Johnson}, {Marcy}, {Butler}, {Isaacson}, {Howard}, {Vogt},
  {Kov{\'a}cs}, {Fernandez}, {Mo{\'o}r}, {Stefanik}, {L{\'a}z{\'a}r}, {Papp},
  \& {S{\'a}ri}}]{bakos:2010:hat11}
{Bakos}, G.~{\'A}., {Torres}, G., {P{\'a}l}, A., {et~al.} 2010, \apj, 710, 1724

\bibitem[{{Bakos} {et~al.}(2013){Bakos}, {Csubry}, {Penev}, {Bayliss},
  {Jord{\'a}n}, {Afonso}, {Hartman}, {Henning}, {Kov{\'a}cs}, {Noyes},
  {B{\'e}ky}, {Suc}, {Cs{\'a}k}, {Rabus}, {L{\'a}z{\'a}r}, {Papp}, {S{\'a}ri},
  {Conroy}, {Zhou}, {Sackett}, {Schmidt}, {Mancini}, {Sasselov}, \&
  {Ueltzhoeffer}}]{bakos:2013:hatsouth}
{Bakos}, G.~{\'A}., {Csubry}, Z., {Penev}, K., {et~al.} 2013, \pasp, 125, 154

\bibitem[{{Barge} {et~al.}(2008){Barge}, {Baglin}, {Auvergne}, {Rauer},
  {L{\'e}ger}, {Schneider}, {Pont}, {Aigrain}, {Almenara}, {Alonso},
  {Barbieri}, {Bord{\'e}}, {Bouchy}, {Deeg}, {La Reza}, {Deleuil}, {Dvorak},
  {Erikson}, {Fridlund}, {Gillon}, {Gondoin}, {Guillot}, {Hatzes}, {Hebrard},
  {Jorda}, {Kabath}, {Lammer}, {Llebaria}, {Loeillet}, {Magain}, {Mazeh},
  {Moutou}, {Ollivier}, {P{\"a}tzold}, {Queloz}, {Rouan}, {Shporer}, \&
  {Wuchterl}}]{2008A&A...482L..17B}
{Barge}, P., {Baglin}, A., {Auvergne}, M., {et~al.} 2008, \aap, 482, L17

\bibitem[{{Bayliss} {et~al.}(2013){Bayliss}, {Zhou}, {Penev}, {Bakos},
  {Hartman}, {Jord{\'a}n}, {Mancini}, {Mohler-Fischer}, {Suc}, {Rabus},
  {B{\'e}ky}, {Csubry}, {Buchhave}, {Henning}, {Nikolov}, {Cs{\'a}k}, {Brahm},
  {Espinoza}, {Noyes}, {Schmidt}, {Conroy}, {Wright}, {Tinney}, {Addison},
  {Sackett}, {Sasselov}, {L{\'a}z{\'a}r}, {Papp}, \&
  {S{\'a}ri}}]{bayliss:2013:hats3}
{Bayliss}, D., {Zhou}, G., {Penev}, K., {et~al.} 2013, \aj, 146, 113

\bibitem[{{Borucki} {et~al.}(2010){Borucki}, {Koch}, {Basri}, {Batalha},
  {Brown}, {Caldwell}, {Caldwell}, {Christensen-Dalsgaard}, {Cochran},
  {DeVore}, {Dunham}, {Dupree}, {Gautier}, {Geary}, {Gilliland}, {Gould},
  {Howell}, {Jenkins}, {Kondo}, {Latham}, {Marcy}, {Meibom}, {Kjeldsen},
  {Lissauer}, {Monet}, {Morrison}, {Sasselov}, {Tarter}, {Boss}, {Brownlee},
  {Owen}, {Buzasi}, {Charbonneau}, {Doyle}, {Fortney}, {Ford}, {Holman},
  {Seager}, {Steffen}, {Welsh}, {Rowe}, {Anderson}, {Buchhave}, {Ciardi},
  {Walkowicz}, {Sherry}, {Horch}, {Isaacson}, {Everett}, {Fischer}, {Torres},
  {Johnson}, {Endl}, {MacQueen}, {Bryson}, {Dotson}, {Haas}, {Kolodziejczak},
  {Van Cleve}, {Chandrasekaran}, {Twicken}, {Quintana}, {Clarke}, {Allen},
  {Li}, {Wu}, {Tenenbaum}, {Verner}, {Bruhweiler}, {Barnes}, \&
  {Prsa}}]{2010Sci...327..977B}
{Borucki}, W.~J., {Koch}, D., {Basri}, G., {et~al.} 2010, Science, 327, 977

\bibitem[{{Brahm} {et~al.}(2015){Brahm}, {Jord{\'a}n}, {Hartman}, {Bakos},
  {Bayliss}, {Penev}, {Zhou}, {Ciceri}, {Rabus}, {Espinoza}, {Mancini}, {de
  Val-Borro}, {Bhatti}, {Sato}, {Tan}, {Csubry}, {Buchhave}, {Henning},
  {Schmidt}, {Suc}, {Noyes}, {Papp}, {L{\'a}z{\'a}r}, \&
  {S{\'a}ri}}]{2015AJ....150...33B}
{Brahm}, R., {Jord{\'a}n}, A., {Hartman}, J.~D., {et~al.} 2015, \aj, 150, 33

\bibitem[{{Brown} {et~al.}(2013){Brown}, {Baliber}, {Bianco}, {Bowman},
  {Burleson}, {Conway}, {Crellin}, {Depagne}, {De Vera}, {Dilday}, {Dragomir},
  {Dubberley}, {Eastman}, {Elphick}, {Falarski}, {Foale}, {Ford}, {Fulton},
  {Garza}, {Gomez}, {Graham}, {Greene}, {Haldeman}, {Hawkins}, {Haworth},
  {Haynes}, {Hidas}, {Hjelstrom}, {Howell}, {Hygelund}, {Lister}, {Lobdill},
  {Martinez}, {Mullins}, {Norbury}, {Parrent}, {Paulson}, {Petry}, {Pickles},
  {Posner}, {Rosing}, {Ross}, {Sand}, {Saunders}, {Shobbrook}, {Shporer},
  {Street}, {Thomas}, {Tsapras}, {Tufts}, {Valenti}, {Vander Horst}, {Walker},
  {White}, \& {Willis}}]{brown:2013:lcogt}
{Brown}, T.~M., {Baliber}, N., {Bianco}, F.~B., {et~al.} 2013, \pasp, 125, 1031

\bibitem[{{Cardelli} {et~al.}(1989){Cardelli}, {Clayton}, \&
  {Mathis}}]{cardelli:1989}
{Cardelli}, J.~A., {Clayton}, G.~C., \& {Mathis}, J.~S. 1989, \apj, 345, 245

\bibitem[{{Claret}(2004)}]{claret:2004}
{Claret}, A. 2004, \aap, 428, 1001

\bibitem[{{Espinoza} {et~al.}(2016){Espinoza}, {Bayliss}, {Hartman}, {Bakos},
  {Jord{\'a}n}, {Zhou}, {Mancini}, {Brahm}, {Ciceri}, {Bhatti}, {Csubry},
  {Rabus}, {Penev}, {Bento}, {de Val-Borro}, {Henning}, {Schmidt}, {Suc},
  {Wright}, {Tinney}, {Tan}, \& {Noyes}}]{2016arXiv160600023E}
{Espinoza}, N., {Bayliss}, D., {Hartman}, J.~D., {et~al.} 2016, ArXiv e-prints,
  arXiv:1606.00023

\bibitem[{{Girardi} {et~al.}(2000){Girardi}, {Bressan}, {Bertelli}, \&
  {Chiosi}}]{girardi:2000}
{Girardi}, L., {Bressan}, A., {Bertelli}, G., \& {Chiosi}, C. 2000, \aaps, 141,
  371

\bibitem[{{Hansen} \& {Barman}(2007)}]{hansen:2007}
{Hansen}, B.~M.~S., \& {Barman}, T. 2007, \apj, 671, 861

\bibitem[{{Hartman} {et~al.}(2012){Hartman}, {Bakos}, {B{\'e}ky}, {Torres},
  {Latham}, {Csubry}, {Penev}, {Shporer}, {Fulton}, {Buchhave}, {Johnson},
  {Howard}, {Marcy}, {Fischer}, {Kov{\'a}cs}, {Noyes}, {Esquerdo}, {Everett},
  {Szklen{\'a}r}, {Quinn}, {Bieryla}, {Knox}, {Hinz}, {Sasselov}, {F{\H
  u}r{\'e}sz}, {Stefanik}, {L{\'a}z{\'a}r}, {Papp}, \&
  {S{\'a}ri}}]{hartman:2012:hat39hat41}
{Hartman}, J.~D., {Bakos}, G.~{\'A}., {B{\'e}ky}, B., {et~al.} 2012, \aj, 144,
  139

\bibitem[{{Hartman} {et~al.}(2015){Hartman}, {Bayliss}, {Brahm}, {Bakos},
  {Mancini}, {Jord{\'a}n}, {Penev}, {Rabus}, {Zhou}, {Butler}, {Espinoza}, {de
  Val-Borro}, {Bhatti}, {Csubry}, {Ciceri}, {Henning}, {Schmidt}, {Arriagada},
  {Shectman}, {Crane}, {Thompson}, {Suc}, {Cs{\'a}k}, {Tan}, {Noyes},
  {L{\'a}z{\'a}r}, {Papp}, \& {S{\'a}ri}}]{2015AJ....149..166H}
{Hartman}, J.~D., {Bayliss}, D., {Brahm}, R., {et~al.} 2015, \aj, 149, 166

\bibitem[{{Henden} {et~al.}(2009){Henden}, {Welch}, {Terrell}, \&
  {Levine}}]{henden:2009}
{Henden}, A.~A., {Welch}, D.~L., {Terrell}, D., \& {Levine}, S.~E. 2009, in
  American Astronomical Society Meeting Abstracts, Vol. 214, American
  Astronomical Society Meeting Abstracts \#214, \#407.02

\bibitem[{{Hippler} {et~al.}(2009){Hippler}, {Bergfors}, {Brandner Wolfgang},
  {Daemgen}, {Henning}, {Hormuth}, {Huber}, {Janson}, {Rochau}, {Rohloff}, \&
  {Wagner}}]{2009Msngr.137...14H}
{Hippler}, S., {Bergfors}, C., {Brandner Wolfgang}, {et~al.} 2009, The
  Messenger, 137, 14

\bibitem[{{Horton} {et~al.}(2012){Horton}, {Tinney}, {Case}, {Farrell}, {Gers},
  {Jones}, {Lawrence}, {Miziarski}, {Staszak}, {Orr}, {Vuong}, {Waller}, \&
  {Zhelem}}]{2012SPIE.8446E..3AH}
{Horton}, A., {Tinney}, C.~G., {Case}, S., {et~al.} 2012, in \procspie, Vol.
  8446, Ground-based and Airborne Instrumentation for Astronomy IV, 84463A

\bibitem[{{Jord{\'a}n} {et~al.}(2014){Jord{\'a}n}, {Brahm}, {Bakos}, {Bayliss},
  {Penev}, {Hartman}, {Zhou}, {Mancini}, {Mohler-Fischer}, {Ciceri}, {Sato},
  {Csubry}, {Rabus}, {Suc}, {Espinoza}, {Bhatti}, {Borro}, {Buchhave},
  {Cs{\'a}k}, {Henning}, {Schmidt}, {Tan}, {Noyes}, {B{\'e}ky}, {Butler},
  {Shectman}, {Crane}, {Thompson}, {Williams}, {Martin}, {Contreras},
  {L{\'a}z{\'a}r}, {Papp}, \& {S{\'a}ri}}]{jordan:2014:hats4}
{Jord{\'a}n}, A., {Brahm}, R., {Bakos}, G.~{\'A}., {et~al.} 2014, \aj, 148, 29

\bibitem[{{Kaufer} \& {Pasquini}(1998)}]{1998SPIE.3355..844K}
{Kaufer}, A., \& {Pasquini}, L. 1998, in \procspie, Vol. 3355, Optical
  Astronomical Instrumentation, ed. S.~{D'Odorico}, 844--854

\bibitem[{{Kov{\'a}cs} {et~al.}(2005){Kov{\'a}cs}, {Bakos}, \&
  {Noyes}}]{kovacs:2005:TFA}
{Kov{\'a}cs}, G., {Bakos}, G., \& {Noyes}, R.~W. 2005, \mnras, 356, 557

\bibitem[{{Kov{\'a}cs} {et~al.}(2002){Kov{\'a}cs}, {Zucker}, \&
  {Mazeh}}]{kovacs:2002:BLS}
{Kov{\'a}cs}, G., {Zucker}, S., \& {Mazeh}, T. 2002, \aap, 391, 369

\bibitem[{{Mancini} {et~al.}(2015){Mancini}, {Hartman}, {Penev}, {Bakos},
  {Brahm}, {Ciceri}, {Henning}, {Csubry}, {Bayliss}, {Zhou}, {Rabus}, {de
  Val-Borro}, {Espinoza}, {Jord{\'a}n}, {Suc}, {Bhatti}, {Schmidt}, {Sato},
  {Tan}, {Wright}, {Tinney}, {Addison}, {Noyes}, {L{\'a}z{\'a}r}, {Papp}, \&
  {S{\'a}ri}}]{2015A&A...580A..63M}
{Mancini}, L., {Hartman}, J.~D., {Penev}, K., {et~al.} 2015, \aap, 580, A63

\bibitem[{{Mandel} \& {Agol}(2002)}]{mandel:2002}
{Mandel}, K., \& {Agol}, E. 2002, \apjl, 580, L171

\bibitem[{{Mayor} {et~al.}(2003){Mayor}, {Pepe}, {Queloz}, {Bouchy},
  {Rupprecht}, {Lo Curto}, {Avila}, {Benz}, {Bertaux}, {Bonfils}, {Dall},
  {Dekker}, {Delabre}, {Eckert}, {Fleury}, {Gilliotte}, {Gojak}, {Guzman},
  {Kohler}, {Lizon}, {Longinotti}, {Lovis}, {Megevand}, {Pasquini}, {Reyes},
  {Sivan}, {Sosnowska}, {Soto}, {Udry}, {van Kesteren}, {Weber}, \&
  {Weilenmann}}]{2003Msngr.114...20M}
{Mayor}, M., {Pepe}, F., {Queloz}, D., {et~al.} 2003, The Messenger, 114, 20

\bibitem[{{P{\'a}l} {et~al.}(2008){P{\'a}l}, {Bakos}, {Torres}, {Noyes},
  {Latham}, {Kov{\'a}cs}, {Marcy}, {Fischer}, {Butler}, {Sasselov}, {Sip{\H
  o}cz}, {Esquerdo}, {Kov{\'a}cs}, {Stefanik}, {L{\'a}z{\'a}r}, {Papp}, \&
  {S{\'a}ri}}]{pal:2008:hat7}
{P{\'a}l}, A., {Bakos}, G.~{\'A}., {Torres}, G., {et~al.} 2008, \apj, 680, 1450

\bibitem[{{Penev} {et~al.}(2013){Penev}, {Bakos}, {Bayliss}, {Jord{\'a}n},
  {Mohler}, {Zhou}, {Suc}, {Rabus}, {Hartman}, {Mancini}, {B{\'e}ky}, {Csubry},
  {Buchhave}, {Henning}, {Nikolov}, {Cs{\'a}k}, {Brahm}, {Espinoza}, {Conroy},
  {Noyes}, {Sasselov}, {Schmidt}, {Wright}, {Tinney}, {Addison},
  {L{\'a}z{\'a}r}, {Papp}, \& {S{\'a}ri}}]{penev:2013:hats1}
{Penev}, K., {Bakos}, G.~{\'A}., {Bayliss}, D., {et~al.} 2013, \aj, 145, 5

\bibitem[{{Penev} {et~al.}(2016){Penev}, {Hartman}, {Bakos}, {Ciceri}, {Brahm},
  {Bayliss}, {Bento}, {Jord'an}, {Csubry}, {Bhatti}, {de Val-Borro},
  {Espinoza}, {Zhou}, {Mancini}, {Rabus}, {Suc}, {Henning}, {Schmidt}, {Noyes},
  {L'az'ar}, {Papp}, \& {S'ari}}]{2016arXiv160600848P}
{Penev}, K.~M., {Hartman}, J.~D., {Bakos}, G.~A., {et~al.} 2016, ArXiv
  e-prints, arXiv:1606.00848

\bibitem[{{Pollacco} {et~al.}(2006){Pollacco}, {Skillen}, {Collier Cameron},
  {Christian}, {Hellier}, {Irwin}, {Lister}, {Street}, {West}, {Anderson},
  {Clarkson}, {Deeg}, {Enoch}, {Evans}, {Fitzsimmons}, {Haswell}, {Hodgkin},
  {Horne}, {Kane}, {Keenan}, {Maxted}, {Norton}, {Osborne}, {Parley}, {Ryans},
  {Smalley}, {Wheatley}, \& {Wilson}}]{2006PASP..118.1407P}
{Pollacco}, D.~L., {Skillen}, I., {Collier Cameron}, A., {et~al.} 2006, \pasp,
  118, 1407

\bibitem[{{Queloz} {et~al.}(2001){Queloz}, {Mayor}, {Udry}, {Burnet},
  {Carrier}, {Eggenberger}, {Naef}, {Santos}, {Pepe}, {Rupprecht}, {Avila},
  {Baeza}, {Benz}, {Bertaux}, {Bouchy}, {Cavadore}, {Delabre}, {Eckert},
  {Fischer}, {Fleury}, {Gilliotte}, {Goyak}, {Guzman}, {Kohler}, {Lacroix},
  {Lizon}, {Megevand}, {Sivan}, {Sosnowska}, \&
  {Weilenmann}}]{2001Msngr.105....1Q}
{Queloz}, D., {Mayor}, M., {Udry}, S., {et~al.} 2001, The Messenger, 105, 1

\bibitem[{{Siverd} {et~al.}(2012){Siverd}, {Beatty}, {Pepper}, {Eastman},
  {Collins}, {Bieryla}, {Latham}, {Buchhave}, {Jensen}, {Crepp}, {Street},
  {Stassun}, {Gaudi}, {Berlind}, {Calkins}, {DePoy}, {Esquerdo}, {Fulton},
  {F{\H u}r{\'e}sz}, {Geary}, {Gould}, {Hebb}, {Kielkopf}, {Marshall}, {Pogge},
  {Stanek}, {Stefanik}, {Szentgyorgyi}, {Trueblood}, {Trueblood}, {Stutz}, \&
  {van Saders}}]{2012ApJ...761..123S}
{Siverd}, R.~J., {Beatty}, T.~G., {Pepper}, J., {et~al.} 2012, \apj, 761, 123

\bibitem[{{Sozzetti} {et~al.}(2007){Sozzetti}, {Torres}, {Charbonneau},
  {Latham}, {Holman}, {Winn}, {Laird}, \& {O'Donovan}}]{sozzetti:2007}
{Sozzetti}, A., {Torres}, G., {Charbonneau}, D., {et~al.} 2007, \apj, 664, 1190

\bibitem[{{Yi} {et~al.}(2001){Yi}, {Demarque}, {Kim}, {Lee}, {Ree}, {Lejeune},
  \& {Barnes}}]{yi:2001}
{Yi}, S., {Demarque}, P., {Kim}, Y.-C., {et~al.} 2001, \apjs, 136, 417

\bibitem[{{Zacharias} {et~al.}(2012){Zacharias}, {Finch}, {Girard}, {Henden},
  {Bartlett}, {Monet}, \& {Zacharias}}]{zacharias:2012:ucac4}
{Zacharias}, N., {Finch}, C.~T., {Girard}, T.~M., {et~al.} 2012, VizieR Online
  Data Catalog, 1322, 0

\bibitem[{{Zhou} {et~al.}(2015){Zhou}, {Bayliss}, {Kedziora-Chudczer},
  {Tinney}, {Bailey}, {Salter}, \& {Rodriguez}}]{2015MNRAS.454.3002Z}
{Zhou}, G., {Bayliss}, D.~D.~R., {Kedziora-Chudczer}, L., {et~al.} 2015,
  \mnras, 454, 3002

\end{thebibliography}
